\documentclass[10pt,epsf,preprint]{aastex}
\usepackage{epsf}
\input psfig 

\begin{document}
\title{Early-type galaxies in the SDSS}

\author{
Mariangela Bernardi\altaffilmark{\ref{Chicago}},
Ravi K. Sheth\altaffilmark{\ref{Fermilab}},
James Annis\altaffilmark{\ref{Fermilab}},
Scott Burles\altaffilmark{\ref{Fermilab}},
Daniel J. Eisenstein\altaffilmark{\ref{Arizona}},
Douglas P. Finkbeiner\altaffilmark{\ref{Berkeley},\ref{Princeton},\ref{HF}},
David W. Hogg\altaffilmark{\ref{NYU}},
Robert H. Lupton\altaffilmark{\ref{Princeton}},
David J. Schlegel\altaffilmark{\ref{Princeton}}, 
Mark Subbarao\altaffilmark{\ref{Chicago}},
Neta A. Bahcall\altaffilmark{\ref{Princeton}},
John P. Blakeslee\altaffilmark{\ref{JHU}},
J. Brinkmann\altaffilmark{\ref{APO}},
Francisco J. Castander\altaffilmark{\ref{yale},\ref{chile}},
Andrew J. Connolly\altaffilmark{\ref{Pitt}}, 
Istvan Csabai\altaffilmark{\ref{Eotvos},\ref{JHU}},
Mamoru Doi\altaffilmark{\ref{Tokyo}},
Masataka Fukugita\altaffilmark{\ref{ICRR},\ref{IAS}},
Joshua Frieman\altaffilmark{\ref{Chicago},\ref{Fermilab}},
Timothy Heckman\altaffilmark{\ref{JHU}},
Gregory S. Hennessy\altaffilmark{\ref{USNO}},
\v{Z}eljko Ivezi\'{c}\altaffilmark{\ref{Princeton}},
G. R. Knapp\altaffilmark{\ref{Princeton}},
Don Q. Lamb\altaffilmark{\ref{Chicago}},
Timothy McKay\altaffilmark{\ref{UMich}},
Jeffrey A. Munn\altaffilmark{\ref{USNO}},
Robert Nichol\altaffilmark{\ref{CMU}},
Sadanori Okamura\altaffilmark{\ref{Tokyo}}, 
Donald P. Schneider\altaffilmark{\ref{PSU}},
Aniruddha R. Thakar\altaffilmark{\ref{JHU}},
and Donald G.\ York\altaffilmark{\ref{Chicago}}
}

\newcounter{address}
\setcounter{address}{1}
\altaffiltext{\theaddress}{\stepcounter{address}
University of Chicago, Astronomy \& Astrophysics
Center, 5640 S. Ellis Ave., Chicago, IL 60637\label{Chicago}}
\altaffiltext{\theaddress}{\stepcounter{address}
Fermi National Accelerator Laboratory, P.O. Box 500,
Batavia, IL 60510\label{Fermilab}}
\altaffiltext{\theaddress}{\stepcounter{address}
Stewart Observatory, University of Arizona, 933 N. Clarry Ave., Tucson, AZ 85121\label{Arizona}}
\altaffiltext{\theaddress}{\stepcounter{address}
Department of Astronomy, University of California at Berkeley, 601 Campbell Hall, Berkeley, CA 94720\label{Berkeley}}
\altaffiltext{\theaddress}{\stepcounter{address}
Princeton University Observatory, Princeton, NJ 08544\label{Princeton}}
\altaffiltext{\theaddress}{\stepcounter{address}Hubble Fellow\label{HF}}
\altaffiltext{\theaddress}{\stepcounter{address}
Department of Physics, New York University, 4 Washington Place, New York, NY 10003\label{NYU}}
\altaffiltext{\theaddress}{\stepcounter{address}
Department of Physics \& Astronomy, The Johns Hopkins University, 3400 North Charles Street, Baltimore, MD 21218-2686\label{JHU}}
\altaffiltext{\theaddress}{\stepcounter{address}
Apache Point Observatory, 2001 Apache Point Road, P.O. Box 59, Sunspot, NM
88349-0059\label{APO}}
\altaffiltext{\theaddress}{\stepcounter{address} Yale University, P. O. Box
208101, New Haven, CT 06520\label{yale}}
\altaffiltext{\theaddress}{\stepcounter{address} Universidad de Chile, Casilla
36-D, Santiago, Chile\label{chile}}
\altaffiltext{\theaddress}{\stepcounter{address}
Department of Physics and Astronomy, University of Pittsburgh, Pittsburgh, PA 15620\label{Pitt}}
\altaffiltext{\theaddress}{\stepcounter{address}
Department of Physics of Complex Systems, E\"otv\"os University, Budapest, H-1117 Hungary\label{Eotvos}}
\altaffiltext{\theaddress}{\stepcounter{address}
Institute of Astronomy and Research Center for the Early Universe, School of Science, University of Tokyo, Tokyo 113-0033, Japan\label{Tokyo}}
\altaffiltext{\theaddress}{\stepcounter{address}
Institute for Cosmic Ray Research, University of Tokyo, Midori, Tanashi, Tokyo 188-8502, Japan\label{ICRR}}
\altaffiltext{\theaddress}{\stepcounter{address}
Institute for Advanced Study, Olden Lane, Princeton, NJ 08540\label{IAS}}
\altaffiltext{\theaddress}{\stepcounter{address}
U.S. Naval Observatory, 3450 Massachusetts Ave., NW, Washington, DC 20392-5420\label{USNO}}
\altaffiltext{\theaddress}{\stepcounter{address}
Department of Physics, University of Michigan, 500 East University, Ann Arbor, MI 48109\label{UMich}}
\altaffiltext{\theaddress}{\stepcounter{address}
Department of Physics, Carnegie Mellon University, Pittsburgh, PA 15213\label{CMU}}
\altaffiltext{\theaddress}{\stepcounter{address}
Department of Astronomy and Astrophysics, The Pennsylvania State University, University Park, PA 16802\label{PSU}}



\begin{abstract}
A sample of nearly 9000 early-type galaxies, in the redshift range 
$0.01 \le z \le 0.3$, was selected from the Sloan Digital Sky Survey  
using morphological and spectral criteria.  The sample was used to 
study how early-type galaxy observables, including luminosity $L$, 
effective radius $R$, surface brightness $I$, color, and velocity 
dispersion $\sigma$, are correlated with one another.  Measurement biases 
are understood with mock catalogs which reproduce all of the observed 
scaling relations and their dependences on fitting technique.  At any 
given redshift, the intrinsic distribution of luminosities, sizes and 
velocity dispersions in our sample are all approximately Gaussian.  
In the $r^*$ band $L\propto\sigma^{3.91\pm 0.20}$, 
$L\propto R^{1.58\pm 0.06}$, $R\propto I^{-0.75\pm 0.02}$, 
and the Fundamental Plane relation is 
 $R \propto \sigma^{1.49\pm 0.05}\,I^{-0.75\pm 0.01}$.  
These relations are approximately the same in the $g^*$, $i^*$ and 
$z^*$ bands.  The observed mass-to-light ratio scales as 
$M/L \propto L^{0.14\pm 0.02}$ at fixed luminosity or 
$M/L\propto M^{0.22\pm 0.05}$ at fixed mass.  

Relative to the population at the median redshift in the sample, 
galaxies at lower and higher redshifts have evolved little.  
The Fundamental Plane is used to quantify this evolution.
An apparent magnitude limit can masquerade as evolution; once this 
selection effect has been accounted for, the evolution is consistent 
with that of a passively evolving population which formed the bulk of 
its stars about 9~Gyrs ago.  

Chemical evolution and star formation histories of early-type galaxies 
are investigated using co-added spectra of similar objects in our sample.  
Chemical abundances correlate primarily with velocity dispersion:  
H$_\beta\propto\sigma^{-0.25\pm 0.02}$, 
Mg$_2\propto\sigma^{0.18\pm 0.02}$, Mg$b\propto\sigma^{0.28\pm 0.02}$, 
and $\langle{\rm Fe}\rangle\propto\sigma^{0.10\pm 0.03}$.  
At fixed $\sigma$, the population at $z\sim 0.2$ had weaker Mg$_2$ 
and stronger H$_\beta$ absorption compared to the population nearby.  
Comparison of these line-strengths and their evolution with single-burst 
stellar population models also suggests a formation time of 9~Gyrs ago.  

Redder galaxies have larger velocity dispersions:  
$g^*-r^*\propto\sigma^{0.26\pm 0.02}$.  Color also correlates with 
magnitude, $g^*-r^*\propto (-0.025\pm 0.003)\,M_{r_*}$, and size, 
but these correlations are entirely due to the $L-\sigma$ and $R_o-\sigma$ 
relations:  the primary correlation is color$-\sigma$.  
Correlations between color and chemical abundances are also presented.  
At fixed $\sigma$, the higher redshift population is bluer by an amount 
which is consistent with the Fundamental Plane and chemical abundance 
estimates of the ages of these galaxies.  

The red light in early-type galaxies is, on average, slightly more 
centrally concentrated than the blue.  Because of these color gradients, 
the strength of the color--magnitude relation depends on whether or 
not the colors are defined using a fixed metric aperture; 
the color$-\sigma$ relation is less sensitive to this choice.  

One of the principal advangtages of the SDSS sample over previous samples 
is that the galaxies in it lie in environments ranging from isolation in 
the field to the dense cores of clusters.  The Fundamental Plane shows 
that galaxies in dense regions are slightly different from galaxies in 
less dense regions, but the co-added spectra and color--magnitude 
relations show no statistically significant dependence on environment.  
\end{abstract}  
\keywords{galaxies: elliptical --- galaxies: evolution --- 
          galaxies: fundamental parameters --- galaxies: photometry --- 
          galaxies: stellar content}

\section{Introduction}
Galaxies have a wide range of luminosities, colors, masses, sizes,
surface brightnesses, morphologies, star formation histories and
environments.  This heterogeneity is not surprising, given the variety
of physical processes which likely influence their formation and
evolution, including gravitational collapse, hydrodynamics,
turbulence, magnetic fields, black-hole formation and accretion,
nuclear activity, tidal and merger interactions, and evolving and
inhomogeneous cosmic radiation fields.

What \emph{is} surprising is that populations of galaxies show several
very precise relationships among their measured properties.  The properties 
we use to describe galaxies span a large ``configuration space'', but 
galaxies do not fill it.  Galaxy spectral energy distributions, when 
scaled to a fixed broad-band luminosity, appear to occupy a thin, 
one-dimensional locus in color space or spectrum space 
(e.g., Connolly \& Szalay 1999).  Spiral galaxies 
show a good correlation between rotation velocity and luminosity 
(e.g., Tully \& Fisher 1977; Giovanelli et al. 1997).  
Galaxy morphology is strongly correlated with broad-band colors, 
strengths of spectral features, and inferred star-formation histories 
(e.g., Roberts \& Haynes 1994).  

Among all galaxy families, early-type (elliptical and S0) galaxies 
show the most precise regularities (Djorgovski \& Davis 1987; 
Burstein, Bender, Faber \& Nolthenius 1997).  Early-type galaxy 
surface-brightness distributions follow a very simple, universal 
``de~Vaucouleurs'' profile (de~Vaucouleurs 1948).  Their spectral energy 
distributions appear to be virtually universal, showing very little 
variation with mass, environment, or cosmic time (e.g., van~Dokkum \&
Franx 1996; Pahre 1998).  What variations they do show are measurable 
and precise.  Early-type galaxy colors, luminosities, half-light radii, 
velocity dispersions, and surface brightnesses are all correlated 
(Baum 1959; Fish 1964; Faber \& Jackson 1976; Kormendy 1977; 
Bingelli, Sandage \& Tarenghi 1984); they can be combined into a 
two-dimensional ``Fundamental Plane'' with very little scatter 
(e.g., Dressler et al. 1987; Djorgovski \& Davis 1987; Faber et al. 1987).  

The homogeneity of the early-type galaxy population is difficult to understand 
if early-type galaxies are assembled at late times by stochastic mergers 
of less-massive galaxies of, presumably, different ages, star formation 
histories, and gas contents, as many models postulate 
(e.g., Larson 1975; White \& Rees 1978; van Albada 1982; Kauffmann 1996; 
Kauffmann \& Charlot 1998).  It is possible that the homogeneity of 
early-type galaxies points to early formation (e.g., Worthey 1994; 
Bressan et al. 1994; Vazdekis et al. 1996; Tantalo, Chiosi \& Bressan 1998); 
certainly their stellar populations appear old (e.g., Bernardi et al. 1998; 
Colless et al. 1999; Trager et al. 2000a,b; Kuntschner et al. 2001).  
Alternatively, the observable properties of the stellar content 
of early-type galaxies are fixed entirely by the properties of the 
collisionless, self-gravitating, dark-matter haloes in which we
believe such galaxies lie (e.g., Hernquist 1990).  
These halos, almost by definition, are not subject to the vagaries of 
gas dynamics, star formation, and magnetic fields; they are influenced 
only by gravity.  

It is essentially a stated goal of the Sloan Digital Sky Survey (SDSS; 
York et al. 2000; Stoughton et al. 2002) to revolutionize the study of
galaxies.  The SDSS is imaging $\pi$ steradians of the sky (Northern 
Galactic Cap) in five bands and taking spectra of $\sim 10^6$ galaxies 
and $\sim 10^5$ QSOs.  Among the $10^6$ SDSS spectra there will be roughly 
$2\times 10^5$ spectra taken of early-type galaxies; in fact $10^5$ of the 
spectroscopic fibers are being used to assemble a sample of luminous 
early-type galaxies with a larger mean redshift than the main SDSS sample 
(Eisenstein et al. 2001).  The high quality of the SDSS 5-band CCD imaging 
(Gunn et al. 1998; Lupton et al. 2001) allows secure identification of 
early-type galaxies and precise measurements of their photometric 
properties; most spectroscopic targets in the SDSS are detected in the 
imaging at signal-to-noise ratios $(S/N)>100$.

Early-type galaxy studies in the past, for technical reasons, have
concentrated on galaxies in clusters at low (e.g., 
J{\o}rgensen, Franx \& Kj{\ae}rgaard 1996; Ellis et al. 1997; 
Pahre, Djorgovski \& de Carvalho 1998a,b; Scodeggio et al. 1998; 
Colless et al. 2001; Saglia et al. 2001; Kuntschner et al. 2001; 
Bernardi et al. 2001a,b) and intermediate redshifts 
(e.g., van Dokkum et al. 1998, 2001; Kelson et al. 2000; 
Ziegler et al. 2001).  Only the large area `Seven Samurai' 
(e.g., Faber et al. 1989) and ENEAR surveys (e.g., da Costa et al. 2000) 
of nearby early-types, recent work with galaxies in the SBF survey 
(Blakeslee et al. 2001), and some studies at intermediate redshifts 
by Schade et al. (1999), Treu et al. (1999, 2001a,b) and 
van Dokkum et al. (2001), are not restricted to cluster environments.  
In constrast, the SDSS is surveying a huge volume of the local Universe, 
so the sample includes early-type galaxies in every environment from 
voids to groups to rich clusters.  

The SDSS spectra are of high enough quality to measure early-type galaxy 
velocity dispersions with reasonable precision.  
As of writing, when only a small fraction of the planned SDSS imaging 
and spectroscopy has been taken, the number ($\sim 9000$) of early-type 
galaxies with well-measured velocity dispersions and surface-brightness 
profiles in the SDSS greatly exceeds the total number in the entire 
astronomical literature to date.  In this paper we use the SDSS sample 
to measure the Fundamental Plane and other early-type galaxy correlations 
in multiple bands.  We also investigate the dependences of colors and 
fundamental-plane residuals on early-type galaxy properties, redshift 
and environment.

The paper is organized as follows:  
Section~\ref{sample} describes the measurements and how the sample 
was selected.  Section~\ref{lf} shows the luminosity function.  
Section~\ref{fp} presents the Fundamental Plane, and studies its 
dependence on waveband, redshift and environment. 
The spectra of these galaxies are studied in Section~\ref{lindices}; 
these indicate that the chemical composition of the early-type galaxy 
population depends on redshift.  The chemical abundances and evolution 
are then combined with stellar population models to estimate the ages 
and metallicities of the galaxies in our sample.  
The color--magnitude and color--$\sigma$ relations provide evidence 
that the higher redshift population in our sample is bluer.  
This is shown in Section~\ref{cms}, which also shows that the 
color--magnitude and color--size relations are a result of the 
color$-\sigma$ correlation.  The section also discusses the effects 
of color gradients on measurements of the strength of the correlation 
between color and magnitude.  Our findings are summarized in 
Section~\ref{discuss}.  

Many details of this study are relegated to Appendices.  
Appendix~\ref{kcorrs} contains a discussion of the various K-corrections 
we have tried.  The way we estimate velocity dispersions is presented in 
Appendix~\ref{vmethods}, and aperture corrections are discussed in 
Appendix~\ref{vr}.  
Appendix~\ref{ML3d} presents the maximum-likelihood algorithm we 
use to estimate the correlations between early-type galaxy luminosities, 
sizes and velocity dispersions.  
Appendix~\ref{lx} studies the distributions of the velocity dispersions, 
sizes, surface-brightnesses, effective masses and effective densities at 
fixed luminosity; all of these are shown to be well described by 
Gaussian forms.  
Various projections of the Fundamental Plane are presented in 
Appendix~\ref{prjct}; these include the Faber--Jackson (1976) and 
Kormendy (1977) relations, and the $\kappa$--space projection of 
Bender, Burstein \& Faber (1992).  
A method for generating accurate mock complete and magnitude-limited 
galaxy catalogs is presented in Appendix~\ref{simul}, and the procedure 
used to estimate errors on our results is discussed in 
Appendix~\ref{compcat}.  

Except where stated otherwise, we write the Hubble constant as 
$H_0=100\,h~\mathrm{km\,s^{-1}\,Mpc^{-1}}$, and we perform our analysis 
in a cosmological world model with 
$(\Omega_{\rm M},\Omega_{\Lambda},h)=(0.3,0.7,0.7)$, where 
$\Omega_{\rm M}$ and $\Omega_{\Lambda}$ are the present-day scaled 
densities of matter and cosmological constant.  
In such a model, the age of the Universe at the present time is 
$t_0=9.43h^{-1}$ Gyr.  For comparison, an Einstein-de Sitter model has 
$(\Omega_{\rm M},\Omega_{\Lambda})=(1,0)$ and $t_0=6.52h^{-1}$~Gyr.  

\section{The sample}\label{sample}
The SDSS project is described in Stoughton et al. (2002).  
As of writing (summer 2001), the SDSS has imaged $\sim 1,500$ square 
degrees; $\sim 65,000$ galaxies and $\sim 8000$ QSOs have both photometric 
and spectroscopic information. For this work, a subsample of $\sim 9000$ 
early-type galaxies were selected, following objective morphological 
and spectroscopic criteria (see Section~\ref{selection}),
to investigate properties of this class of galaxies.  

\begin{figure}
\centering
\epsfxsize=\hsize\epsffile{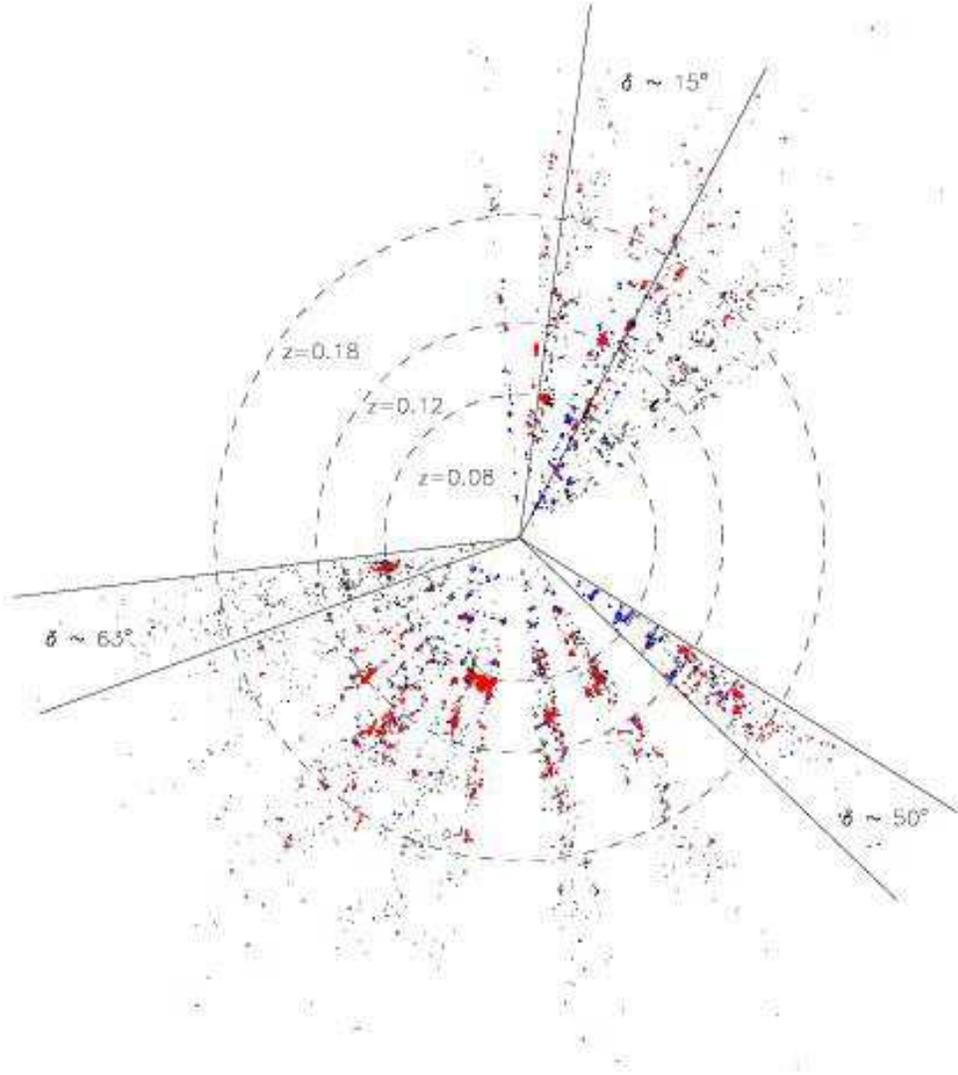}
\caption{Pie-diagram distribution of our sample.  Most of the sample 
is at low declination ($|\delta|\le 2^\circ$), but three wedges are at 
higher declinations (as indicated).  Right ascension increases 
clockwise, with the zero at twelve o'clock.  Galaxies with many ($\ge 15$) 
and a few ($\le 2$) near neighbours are shown with red and blue dots, 
whereas those in the intermediate regime, or for which the local number 
of neighbours was not determined, are shown with black dots.  }
\label{fig:pie}
\end{figure}

Figure~\ref{fig:pie} shows a redshift-space pie-diagram distribution 
of our sample.  Most of the sample is at low declination 
($|\delta|\le 2^\circ$); 
in addition, there are three wedges from three different disconnected 
regions on the sky.  Red and blue symbols denote galaxies which were 
classified as being in dense and underdense regions (as described in 
Section~\ref{environ}), whereas black symbols show galaxies in groups 
of intermediate richness, or for which the local density was not 
determined.  

\subsection{Imaging data}
The photometric and spectroscopic data were taken with the 2.5-m 
SDSS Telescope at the Apache Point Observatory (New Mexico) between
1999 March and 2000 October. Details of the photometric and 
spectroscopic observations and data reduction procedure will be 
presented elsewhere. Here we briefly summarize.

Images are obtained by drift scanning with a mosaic CCD camera 
(Gunn et al. 1998) which gives a field of view of 
$3\times 3~\mathrm{deg^2}$, with a spatial scale of 
$0.4~\mathrm{arcsec\,pix^{-1}}$ in five bandpasses 
($u$, $g$, $r$, $i$, $z$) with central wavelengths
(3560, 4680, 6180, 7500, 8870)\,\AA (Fukugita et al. 1996).
The errors in $u$ band measurements are larger than the others, 
so we will only present results in the other four bands.  
In addition, the photometric solutions we use 
in this paper are preliminary (for details, see discussion of the 
Early Data Release in Stoughton et al. 2002); 
we use $r^*$ rather than $r$, and similarly for the other 
bands, to denote this.

The effective integration time is 54~sec.  The raw CCD images are
bias-subtracted, flat-fielded and background-subtracted.  Pixels
contaminated by the light of cosmic rays and bad columns are masked.
Astronomical sources are detected and overlapping sources are
de-blended.  Surface photometry measurements are obtained by fitting a
set of two-dimensional models to the images.  All of this image
processing is performed with software specially designed for reducing
SDSS data (Lupton et al. 2001).  The uncertainty in the sky background 
subtraction is less than about 1\%.  The median effective seeing 
(the median FWHM of the stellar profiles) for the observations used 
here is $1.5~\mathrm{arcsec}$.

When obtaining the photometric parameters below, the light profiles
are corrected for the effects of seeing, atmospheric extinction,
Galactic extinction (this last uses the results of 
Schlegel, Finkbeiner, \& Davis 1998), and are flux-calibrated by 
comparison with a set of overlapping standard-star fields calibrated 
with a 0.5-m ``Photometric Telescope'' (Smith et al. 2002; 
Uomoto et al. 2002).  The uncertainty in the zero-point 
calibration in the $r^*$-band is $<0.01$~mag.  The Photometric Telescope 
is also used for measuring the atmospheric extinction coefficients 
in the five bands.

The SDSS image processing software provides several global photometric
parameters, for each object, which are obtained independently in each 
of the five bands.  Because we are interested in early-type galaxies, 
we use primarily the following: 
1) The ratio $b/a$ of the lengths of the minor and major axes 
of the observed surface brightness profile.  
2) The effective radius (or half-light radius) $r_{\rm dev}$ along the 
major axis and 
3) the total magnitude $m_{\rm dev}$; these are computed by fitting a 
two-dimensional version of de~Vaucouleurs (1948) $r^{1/4}$ model 
to the observed surface brightness profile.  
(The fitting procedure accounts for the effects of seeing.)  
4) A likelihood parameter that indicates which of the fitting models, 
the de Vaucouleurs or the exponential, provides a better fit to the 
observed light profile. 
5) The {\it model} magnitude $m_m$; this is the total magnitude calculated 
by using the (de Vaucouleurs or exponential) model which fits the galaxy 
profile best in the $r^*$-band.  The {\it model} magnitudes in the other 
four bands are computed using that $r^*$ fit as filter; in effect, this 
measures the colors of a galaxy through the same aperture. 
6) The Petrosian magnitude $m_p$ is also computed; this is the flux within 
$2r_p$, where $r_p$ is defined as the angular radius at which the ratio 
of the local surface brightness at $r$ to the mean surface brightness 
within a radius $r$ is 0.2 (Petrosian 1976).  
7,8) The Petrosian radii $r_{50}$ and $r_{90}$; these are the angular 
radii containing 50\% and 90\% of the Petrosian light, respectively. 

Although all the early-type galaxy analyses presented here have been 
performed with both the de~Vaucouleurs fit parameters and the Petrosian 
quantities, in most of the following only the results of the 
de~Vaucouleurs fits are presented.  This is because the
de~Vaucouleurs model appears to be a very good fit to the early-type
galaxy surface-brightness profiles in the SDSS sample and because it
is conventional, in the literature on early-type galaxies, to use these
quantities.  
In this paper, unless stated otherwise, galaxy colors are always 
computed using {\it model} magnitudes.

To convert the apparent magnitude $m$ to an absolute magnitude $M$ we 
must assume a particular cosmology and account for the fact that at 
different redshifts an observed bandpass corresponds to different 
restframe bands (the K-correction).  We write the Hubble constant today 
as $100h$ kms$^{-1}$Mpc$^{-1}$ and use 
$(\Omega_M,\Omega_{\Lambda},h)=(0.3,0.7,0.7)$.  
Most of our sample is at $cz\ge 3\times 10^5\times 0.03$ kms$^{-1}$;  
since line-of-sight peculiar velocities are not expected to exceed more 
than a few thousand kms$^{-1}$, we feel that it is reasonable to assume 
that all of a galaxy's redshift is due to the Hubble recession velocity.  
This means that we can compute the absolute magnitude in a given band by 
 $M = m -5\log_{10}[D_L(z;\Omega_M;\Omega_{\Lambda})] - 25 - K(z)$, 
where $m$ is the apparent magnitude, $D_{\rm L}$ is the luminosity 
distance in Mpc (from, e.g., Weinberg 1972; Hogg 1999), and 
$K(z)$ is the K-correction for the band.  

Because we have five colors and a spectrum for each galaxy, we could, 
in principle, compute an empirical K-correction for each galaxy.  
This requires a good understanding of the accuracy of the SDSS photometry 
and spectroscopy, and should be possible when the survey is closer to 
completion.  Rather than follow the procedure adopted by the 2dFGRS 
(Madgwick et al. 2001), or a procedure based on finding the closest 
template spectrum to each galaxy and using it to compute the K-correction 
(e.g., Lin et al. 1999 for the CNOC2 survey), we use a single redshift 
dependent template spectrum to estimate the K-correction.  In effect, 
although this allows galaxies at different redshift to be different, it 
ignores that fact that not all galaxies at the same redshift are alike.  
As a result, the absolute luminosities we compute are not as accurate as 
they could be, and this can introduce scatter in the various correlations 
we study below.  Of course, using a realistic K-correction is important, 
because inaccuracies in $K(z)$ can masquerade as evolutionary trends.  

For the reasons discussed more fully in Appendix~\ref{kcorrs}, our 
K-corrections are based on a combination of Bruzual \& Charlot (2002) 
and Coleman, Wu \& Weedman (1980) prescriptions.  Specifically, the 
K-corrections we apply were obtained by taking a Bruzual \& Charlot  
model for a $10^{11}M_\odot$ object which formed its stars with an 
IMF given by Kroupa (2000) in a single solar metallicity and abundance 
ratio burst 9~Gyr ago, computing the difference between the K-correction 
when evolution is allowed and ignored, and adding this difference to the 
K-corrections associated with Coleman, Wu \& Weedman early type galaxy 
template.  The results which follow are qualitatively similar for a 
number of other K-correction schemes (see Appendix~\ref{kcorrs} for details).  

\begin{figure}
\centering
\epsfxsize=\hsize\epsffile{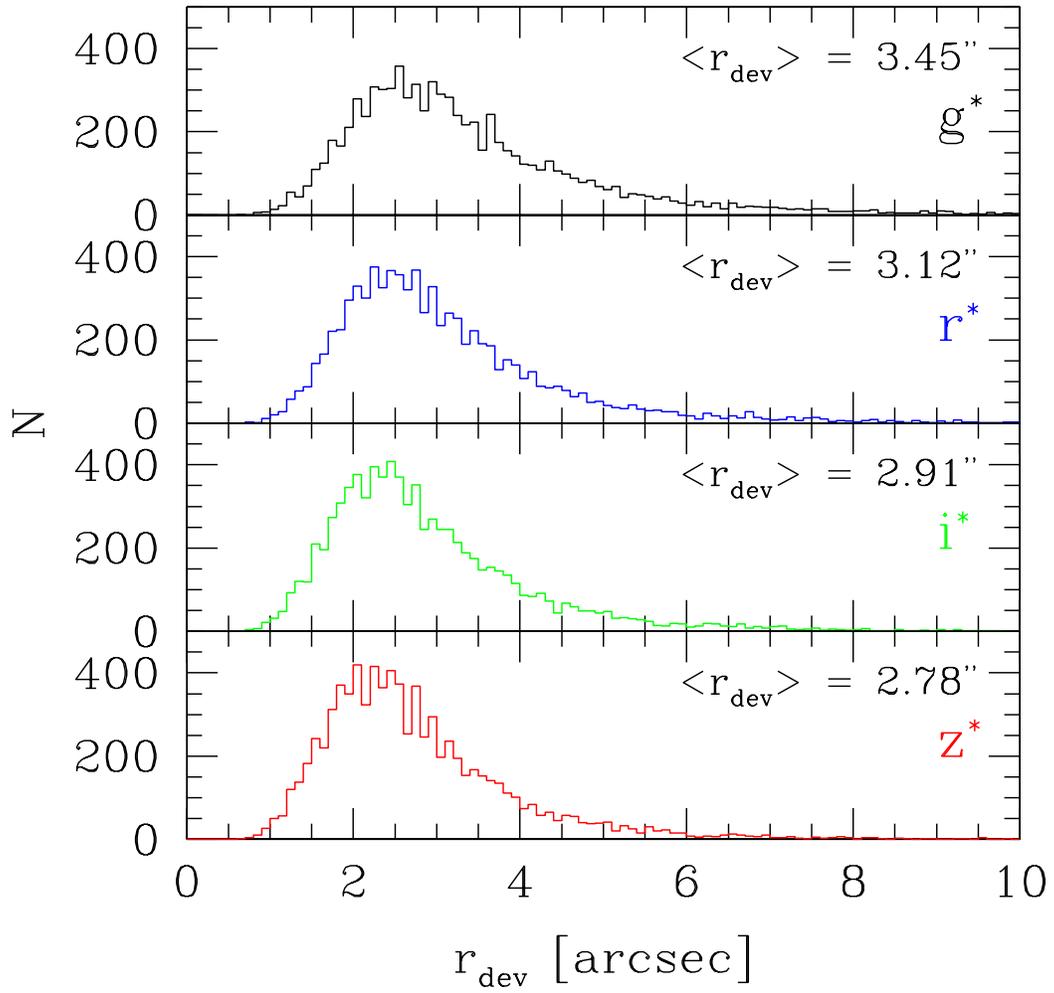}
\caption{Distribution of (seeing-corrected) effective angular sizes 
of galaxies in our sample.  Typical seeing is about 1.5 arcsec.  
The distribution of effective radii in all the bands are very 
similar, although the radii are slightly larger in the bluer bands.}
\label{angre}
\end{figure}

In addition to correcting the observed apparent magnitudes to 
absolute magnitudes, we must also apply two corrections to convert the 
(seeing corrected) effective angular radii, $r_{\rm dev}$, output by the 
SDSS pipeline to physical radii.  
First, we define the equivalent circular effective radius 
 $r_o\equiv\sqrt{b/a}\,r_{\rm dev}$.  
(Although the convention is to use $r_e$ to denote the effective radius, 
we feel that the notation $r_o$ is better, since it emphasizes that the 
radius is an effective circular, rather than elliptical aperture.)  
Figure~\ref{angre} shows the distribution of effective angular sizes 
$r_{\rm dev}$ of the galaxies in our sample before correcting them to $r_o$.  
Notice that $r_{\rm dev}$ for most of the objects is larger than the 
typical seeing scale of 1.5 arcsec, suggesting that seeing does not 
compromise our estimates of $r_o$.  

The reason we must make a second correction is also shown in 
Figure~\ref{angre}.  The different panels show that the effective radius 
of a galaxy depends on wavelength; galaxies appear slightly larger in the 
bluer bands.  Because our sample covers a reasonably large range in 
redshift, this trend means we must correct the effective sizes to a 
fixed restframe wavelength.  Therefore, when converting from effective 
angular size $r_o$ to effective physical size $R_o$ we correct $r_o$ 
(and the Petrosian radii $r_{50}$ and $r_{90}$) in each band by linearly 
interpolating from the observed bandpasses to the central rest wavelength 
of each filter.  The typical correction is of the order of 4\% (although 
it is sometimes as large as 10\%).  In this respect, this correction is 
analogous to the K-correction we would ideally have applied to the 
magnitude and surface brightness of each galaxy.  

\begin{table}[t]
\centering
\caption[]{Photometric parameters and median errors of the objects in 
our sample.\\}
\begin{tabular}{cccccccccc}
\hline &&&\\
Band & $m_{\rm min}$ & $m_{\rm max}$ & $\delta m_{\rm dev}$ 
     & $\delta\log r_o$ & $\delta\log {\rm I_e}$ 
     & $\delta m_{\rm m}$ & $\delta m_{\rm pet}$ 
     & $\delta\log r_{50}$ & $\delta\log r_{90}$\\ 
     & mag & mag & mag & dex & dex & mag & mag & dex & dex \\
\hline &&&\\
$g^*$ & 15.50 & 18.10 & 0.03 & 0.02 & 0.04 & 0.03 & 0.04 &  0.02 & 0.02 \\
$r^*$ & 14.50 & 17.45 & 0.02 & 0.01 & 0.03 & 0.02 & 0.02 &  0.01 & 0.01 \\  
$i^*$ & 14.50 & 17.00 & 0.02 & 0.01 & 0.03 & 0.02 & 0.02 &  0.01 & 0.01 \\ 
$z^*$ & 14.50 & 16.70 & 0.05 & 0.03 & 0.09 & 0.05 & 0.05 &  0.03 & 0.03 \\  
\hline &&&\\
\end{tabular}
\label{tab:photerr} 
\end{table}

Our study also requires the effective surface brightness 
$\mu_o\equiv -2.5\log_{10}I_o$, where $I_o$ is the mean surface 
brightness within the effective radius $R_o$ (as opposed to the surface 
brightness at $R_o$).  In particular, we set 
$\mu_o = m_{\rm dev} + 2.5\log_{10} (2\pi r^2_o) - K(z) - 10\log_{10}(1+z)$.  
Note that this quantity is K-corrected, and also corrected for the 
cosmological $(1+z)^4$ dimming. Our earlier remarks about the K-correction 
are also relevant here.

Table~\ref{tab:photerr} summarizes the uncertainties on the photometric 
parameters we use in this paper.  

\subsection{Spectroscopic data}\label{spectro}
As described in Stoughton et al. (2002), the SDSS takes spectra only 
for a target subsample of objects.  Spectra are obtained using a 
multi-object spectrograph which observes 640 objects at once. 
Each spectroscopic plug plate, 1.5 degrees in radius, has 640 fibers, 
each $3~\mathrm{arcsec}$ in diameter. Two fibers cannot be closer than 
$55~\mathrm{arcsec}$ due to the physical size of the fiber plug.  
Typically $\sim 500$ fibers per plate are used for galaxies, 
$\sim 90$ for QSOs, and the remaining for sky spectra and 
spectrophotometric standard stars.  

Each plate typically has three to five spectroscopic exposures of 
fifteen minutes, depending on the observing conditions (weather, moon); 
a minimum of three exposures is taken to ensure adequate cosmic ray 
rejection. For galaxies at $z \le 0.3$ the median spectrum $S/N$ 
per pixel is 16 (see Figure~\ref{fig:vmeth} in Appendix~\ref{vmethods}).
The wavelength range of each spectrum is $3900-9000$~\AA. 
The instrumental dispersion is $\log_{10}\lambda=10^{-4}$dex/pixel 
which corresponds to 69~kms$^{-1}$ per pixel.  (There is actually some 
variation in this instrumental dispersion with wavelength, which we 
account for; see Figure~\ref{fig:resol} and associated discussion 
in Appendix~\ref{vmethods}.)  
The instrumental resolution of galaxy spectra, measured from the 
autocorrelation of stellar template spectra, ranges from 85 to 
105 kms$^{-1}$, with a median value of 92 kms$^{-1}$.  

A highly automated software package has been designed for reducing 
SDSS spectral data. 
The raw data are bias-subtracted, flat-fielded, wavelength calibrated, 
sky-lines removed, co-added, cleaned from residual glitches 
(cosmic rays, bad pixels), and flux calibrated.  The spectro-software
classifies objects by spectral type and determines emission and absorption
redshifts. (Redshifts are corrected to the heliocentric reference frame.) 
The redshift success rate for objects targeted as galaxies is $>99$\% 
and errors in the measured redshift are less than about $10^{-4}$.  
Once the redshift as been determined the following quantites are 
computed:  Absorption-line strengths (Brodie \& Hance 1986; 
Diaz, Terlevich \& Terlevich 1989; Trager et al. 1998),
equivalent widths of the emission lines, and 
eigen-coefficients and classification numbers of a PCA analysis 
(Connolly \& Szalay 1999).  
Some information about the reliability of the redshift and the quality of 
the spectrum is also provided.

The SDSS pipeline does not provide an estimate of the line-of-sight 
velocity dispersion, $\sigma$, within a galaxy, so we must compute 
it separately.  The observed velocity dispersion $\sigma$ is the result 
of the superposition of many individual stellar spectra, each of which 
has been Doppler shifted because of the star's motion within the 
galaxy.  Therefore, it can be determined by analyzing the integrated 
spectrum of the whole galaxy.  A number of objective and accurate methods 
for making velocity dispersion measurements have been developed 
(Sargent et al. 1977; Tonry \& Davis 1979; 
Franx, Illingworth \& Heckman 1989; Bender 1990; Rix \& White 1992).  
Each of these methods has its own strengths, weaknesses, and biases.  
Appendix~\ref{vmethods} describes how we combined these different 
techniques to estimate $\sigma$ for the galaxies in our sample.  

The velocity dispersion estimates we use in what follows are obtained 
by fitting the wavelength range $4000- 7000$~\AA, and then averaging 
the estimates provided by the {\it Fourier-fitting} and 
{\it direct-fitting} methods to define what we call $\sigma_{\rm est}$.  
(We do not use the cross-correlation estimate because of its behavior 
at low $S/N$ as discussed in Appendix~\ref{vmethods}.)  The error on 
$\sigma_{\rm est}$ is determined by adding in quadrature the errors on 
the two estimates (i.e., the Fourier-fitting and direct-fitting) which 
we averaged.  The resulting error is between $\delta\log\sigma\sim 0.02$~dex
and 0.06~dex, depending on the signal-to-noise of the spectra, with 
a median value of $\sim 0.03$~dex. 
A few galaxies in our sample have been observed more than once.  
The scatter between different measurements is $\sim 0.04$~dex, 
consistent with the amplitude of the errors on the measurements 
(see Figure~\ref{fig:vrptd}).  Based on the typical $S/N$ of the SDSS 
spectra and the instrumental resolution, we chose 70 kms$^{-1}$ as a 
lower limit on the velocity dispersions we use in this paper.  

Following J{\o}rgensen et al. (1995) and Wegner et al. (1999), 
we correct $\sigma_{\rm est}$ to a standard relative circular 
aperture defined to be one-eighth of the effective radius:  
\begin{equation}
{\sigma_{\rm cor}\over\sigma_{\rm est}} = 
 \left(r_{\rm fiber}\over r_o/8\right)^{0.04},
 \label{appcorr}
\end{equation}
where $r_{\rm fiber}=1.5$ arcsec and $r_o$ is the effective 
radius of the galaxy measured in arcseconds.  In principle, we should 
also account for the effects of seeing on $\sigma_{\rm est}$, just 
as we do for $r_o$.  However, because the aperture correction depends 
so weakly on $r_o$ (as the 0.04 power), this is not likely to be a 
significant effect.  In any case, most galaxies in our sample have 
$r_o\ge 1.5$ arcsecs (see Figure~\ref{angre}).  

Note that this correction assumes that the velocity dispersion 
profiles of early-type galaxies having different $r_o$ are similar.  
At the present time, we do not have measurements of the profiles of 
any of the galaxies in our sample, so we cannot test this assumption.  
Later in this paper we will argue that the galaxies in our sample 
evolve very little; this means that if we select galaxies of the 
same luminosity and effective radius, then a plot of velocity dispersion 
versus redshift of these objects should allow us to determine if the 
aperture correction above is accurate.  The results of this exercise 
are presented in Appendix~\ref{vr}.  

\subsection{Sample selection}\label{selection}
The main goal of this paper is to investigate the properties of
early-type galaxies using the large SDSS database.  Therefore, one of
the crucial steps in our study is the separation of galaxies into early
and late types.  Furthermore, we want to select objects whose spectra 
are good enough to compute the central velocity dispersion.  In addition, 
because we wish to study the colors of the galaxies in our sample, 
we must not use color information to select the sample.  
To reach our goal, we have selected galaxies which satisfy the following 
criteria:
\begin{itemize}
 \item concentration index $r_{90}/r_{50} > 2.5$;
 \item the likelihood of the de Vaucouleurs model is at least 1.03 times 
       the likelihood of the exponential model;
 \item spectra without emission lines;
 \item spectra without masked regions (the SDSS spectroscopic pipeline 
  outputs a warning flag for spectra of low quality; 
  we only chose spectra for which this flag was set to zero); 
 \item redshift $< 0.3$ ;
 \item velocity dispersion larger than 70 kms$^{-1}$ and $S/N > 10$.
\end{itemize}

The SDSS pipeline does not output disk-to-bulge ratios from fits to 
the light profiles.  The first two of the criteria above attempt to 
select profile shapes which are likely to be those of spheroidal systems.  
The spectra of late-type galaxies show emission lines, so examining 
the spectra is a simple way of removing such objects from the sample.  

Because the aperture of an SDSS spectroscopic fiber ($3~\mathrm{arcsec}$) 
samples only the inner parts of nearby galaxies, and because the spectrum 
of the bulge of a nearby late-type galaxy can resemble that of an 
early-type galaxy, it is possible that some nearby late-type galaxies 
could be mistakenly included in the sample 
(e.g., Kochanek, Pahre \& Falco 2001).  
Most of these will have been excluded by the first two cuts on the 
shape of the light profile.  To check this, we visually inspected all 
galaxies with $r_{\rm dev} > 8~\mathrm{arcsec}$.  About $\sim 50$ 
of 225 (i.e., about 20\%) looked like late-types, and so we removed them.  

Most early-type galaxies in the SDSS database at $z\ge 0.3$ were 
targeted using different selection criteria than were used for the 
main SDSS galaxy sample (see Strauss et al. 2002; they make-up the 
Luminous Red Galaxy sample described by Eisenstein et al. 2001).  
In the interest of keeping our sample as close to being magnitude 
limited as possible, we restricted our sample to $z\le 0.3$.
In addition, one might expect an increasing fraction of the early-type 
population at higher redshifts to have emission lines:  if so, then 
our removal of emission line objects amounts to a small but redshift 
dependent selection effect.  Since our sample is restricted to $z\le 0.3$, 
this bias should be small.

In Appendix~\ref{vmethods} we discuss why we consider velocity 
dispersion estimates smaller than about 70~kms$^{-1}$ to be unreliable.  
The results of this paper are not significantly different if 
we change the cut-off on velocity dispersions to 100~kms$^{-1}$.  

\begin{figure}
\centering
\epsfxsize=\hsize\epsffile{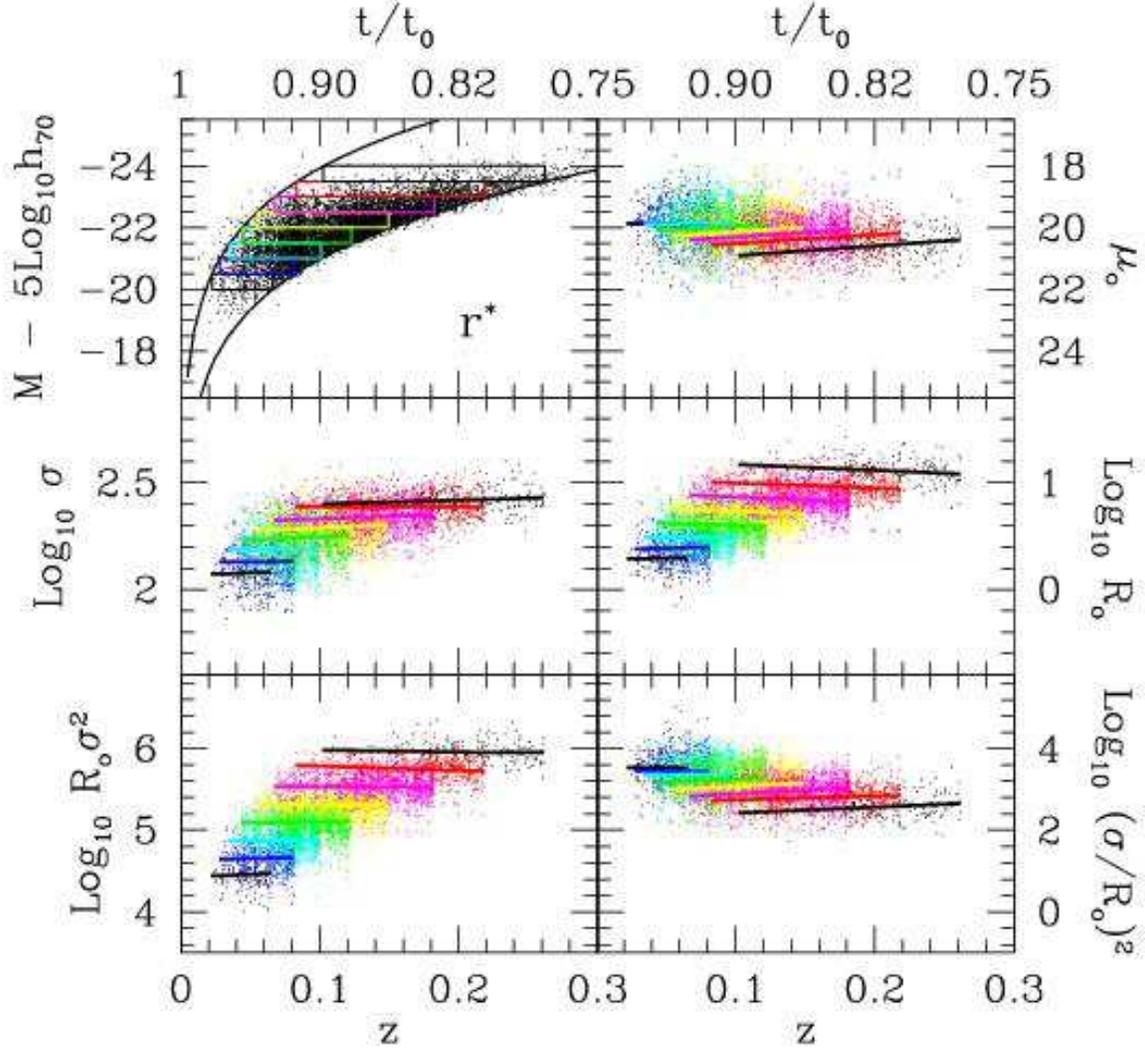}
\vspace{1cm}
\caption[]{Surface brightnesses, velocity dispersions, sizes, masses 
and densities of galaxies as a function of redshift, for a few bins 
in $r^*$ luminosity.  Top left panel shows volume limited catalogs 
which do not overlap in $r^*$ luminosity, dots in the other panels 
show the galaxies in the volume-limited subsamples defined by the 
top-left panel, and solid lines show the mean trend with redshift in 
each subsample.  Results in $g^*$, $i^*$ and $z^*$ are similar. }
\label{fig:XzR}
\end{figure}

Using the above criteria we have extracted a sample of $\sim 9,000$
early-type galaxies.  All the spectra in our sample have PCA 
classification numbers $a < -0.25$, typical of early-type galaxy 
spectra (Connolly \& Szalay 1999).  

Figure~\ref{fig:XzR} displays the properties of our $r^*$-sample 
as a function of redshift $z$.  The panels show the K-corrected 
absolute magnitude $M$, the K-corrected effective surface brightness 
$\mu_o$, also corrected for cosmological surface brightness dimming, 
the effective circular radius $R_o$ in $h^{-1}$kpc, corrected to a 
standard restframe wavelength, 
the aperture corrected velocity dispersion $\sigma$ in kms$^{-1}$, 
and two quantities which are related to an effective mass and density, 
all plotted as a function of redshift.  

The bold lines in the top left panel show the effect of the apparent 
magnitude cuts.  There is, in addition, a cut at small velocity 
dispersion ($\sim 70$~kms$^{-1}$) which, for our purposes here, is 
mostly irrelevant.  The apparent magnitude cuts imply complex 
$z$-dependent cuts on the other parameters we observe.  In what 
follows, we will attempt to account correctly for the selection effects 
that our magnitude cut implies.  

For small intervals in luminosity, our sample is complete over a 
reasonably large range in redshifts.  To illustrate, the thin boxes 
in the top-left panel show bins in absolute magnitude of width 0.5 mags 
over which the sample is complete.  The solid lines in the other panels 
show how the median surface-brightnesses, sizes, velocity dispersions, 
masses and densities of galaxies in each $r^*$ luminosity bin change as a 
function of redshift.  Although all these quantities depend on luminosity, 
the figure shows that, at fixed luminosity, there is some evidence for 
evolution:  at fixed luminosity the average surface brightness is 
brightening.  The size at fixed luminosity decreases at a rate which is 
about five times smaller than the rate of change of $\mu_o$.  This suggests 
that it is the luminosities which are changing and not the sizes.  
(To see why, suppose that the average size at fixed absolute magnitude is 
 $\langle \log_{10}[R_o/R_*(z)]\rangle=s[M-M_*(z)]$, where $R_*$ and $M_*$ 
are characteristic values, and $s$ is the slope of this mean relation.  
If the characteristic luminosity increases with $z$, but the 
characteristic size remains constant, $R_*(z)=R_*(0)$, and the 
slope of the relation also does not change, then the mean size at 
fixed $M$ decreases with $z$.  The surface brightness is 
$\mu_o\propto M + 5\log_{10}R_o$, so the mean $\mu_o$ at fixed $M$ 
changes five times faster than the mean $R_o$, at fixed luminosity.)  
We will argue later that this trend is qualitatively consistent with 
that expected of a passively evolving population.  

\begin{figure}
\centering
\epsfxsize=\hsize\epsffile{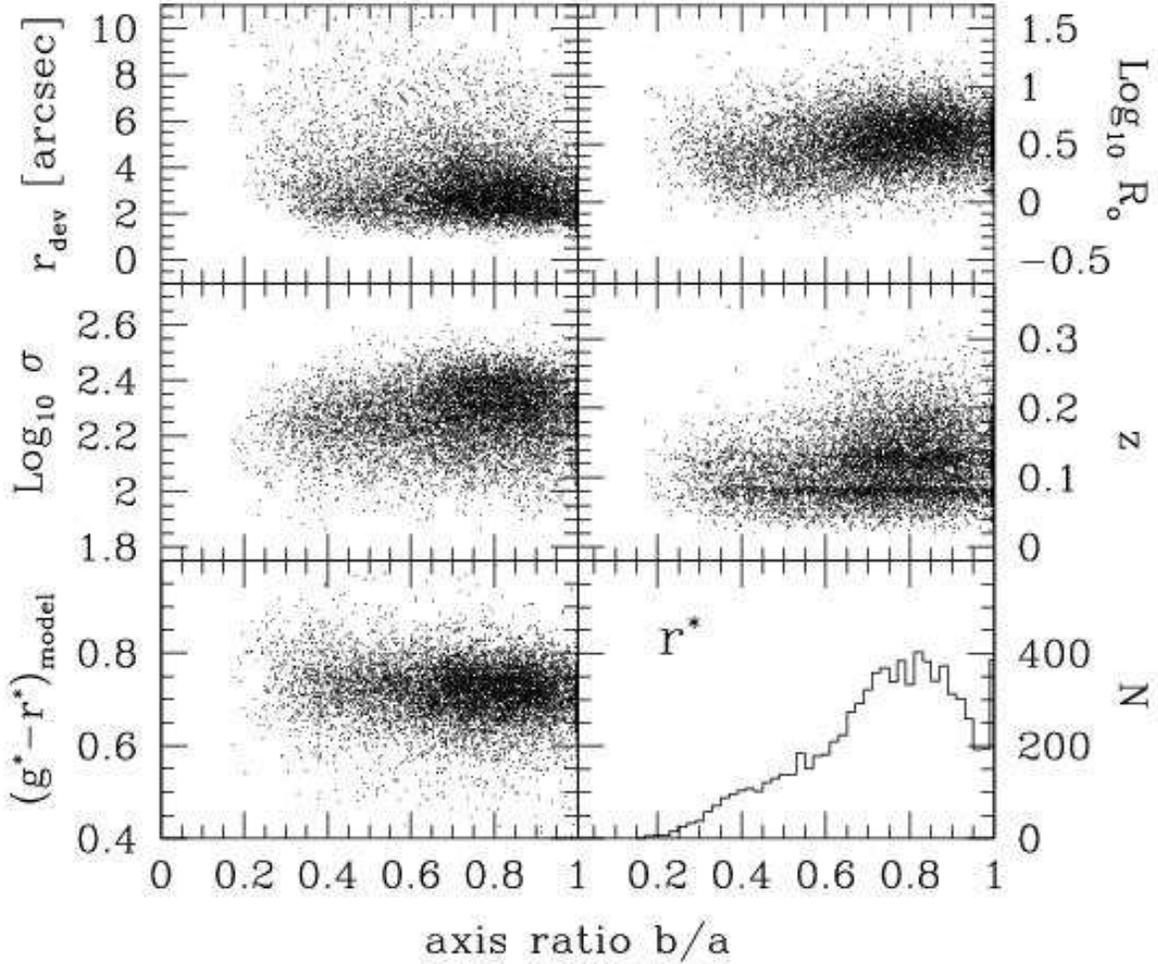}
\caption{Effective angular sizes $r_{\rm dev}$, effective 
circular physical sizes $R_o$, velocity dispersions $\sigma$, 
redshifts $z$, and $(g^*-r^*)$ colors as a function of axis ratio 
$b/a$ for the galaxies in our $r^*$ sample.  Bottom right panel 
shows that the typical axis ratio is $b/a\approx 0.8$.  
There is only a weak tendency for galaxies with small $r_{\rm dev}$ 
to be rounder, suggesting that the estimate of the shape is not 
compromised by seeing (typical seeing is about 1.5 arcsec).  
Results in the other bands are similar.}
\label{abscatter}
\end{figure}

Before moving on, it is worth pointing out that there is a 
morphologically based selection cut which we could have made but 
didn't.  Elliptical galaxies are expected to have axis ratios greater 
than about $0.6$ (e.g., Binney \& Tremaine 1987).  Since we have axis 
ratio measurements of all the objects in our sample, we could have 
included a cut on $b/a$.  The bottom right panel of 
Figure~\ref{abscatter} shows the distribution axis ratios $b/a$ in 
our $r^*$ sample:  about 20\% of the objects in it have $b/a\le 0.6$.  

Our combination of cuts on the shapes of the light profiles and spectral 
features mean that these objects are unlikely to be late-type galaxies.  
Indeed, a visual inspection of a random sample of the objects with axis 
ratios smaller than 0.6 shows that they look like S0s.  
The bottom left panel of Figure~\ref{abscatter} shows that $b/a$ does 
not correlate with color:  in particular, the colors of the most 
flattened objects are not bluer than in the rest of the sample.  
Also, recall that all objects with $r_{\rm dev}\ge 8$ arcsec were 
visually inspected and these, despite having $b/a\le 0.6$ (top left panel), 
did not appear peculiar.  
In addition, $b/a$ does not correlate with surface brightness or 
apparent magnitude.  However, there is a weak trend for the objects 
at higher $z$ to be rounder (middle right panel).  Galaxies with small 
angular sizes $r_{\rm dev}$ are assigned large values of $b/a$ only 
slightly more often than average (top left panel; the median 
$b/a$ is 0.79, 0.78, 0.76, and 0.7 for $r_{\rm dev}$ in the range  
1--2, 2--3, 3--4 and greater than 4 arcsec), the trend with $z$ 
may be related to the magnitude limit of our sample rather than 
reflecting problems associated with the fits to the observed light 
profiles.  For example, the smaller $R_o$ objects tend to be more 
flattened (top right panel), and to have slightly smaller velocity 
dispersions (middle left); because of the magnitude limit, these 
objects drop out of our sample at higher redshifts.  
Requiring that $b/a\ge 0.6$ would remove such objects from our sample 
completely.  

Later in this paper we will study the Fundamental Plane populated by 
the galaxies in this sample.  Excluding all objects with $b/a < 0.6$ 
has no effect on the shape of this Plane.  So, in the interests of 
keeping our sample as close to being magnitude limited as possible, 
we chose not to make an additional selection cut on $b/a$.  

\section{The luminosity function}\label{lf}
We measure the luminosity function of the galaxies in our sample using 
two techniques.  The first uses volume limited catalogs, and the 
second uses a maximum likelihood procedure 
(Sandage, Tammann \& Yahil 1979; Efstathiou, Ellis \& Peterson 1988).  

In the first method, we divide our parent catalog into many volume 
limited subsamples; this was possible because the parent catalog is so 
large.  When doing this, we must decide what size volumes to choose.  
We would like our volumes to be as large as possible so that each volume 
represents a fair sample of the Universe.  On the other hand, the volumes 
must not be so large that evolution effects are important.  In addition, 
because our catalog is cut at the bright as well as the faint end, 
large-volume subsamples span only a small range in luminosities.  
Therefore, we are forced to compromise:  we have chosen to make the 
volumes about $\Delta z=0.04$ thick, because 
$c\Delta z/H\approx 120h^{-1}$Mpc is larger than the largest 
structures seen in numerical simulations of the cold dark matter family 
of models (e.g., Colberg et al. 2000).  The catalogs are extracted from 
regions which cover a very wide angle on the sky, so the actual volume of 
any given volume limited catalog is considerably larger 
than ($120h^{-1}$Mpc)$^3$.  Therefore, this choice should provide volumes 
which are large enough in at least two of the three coordinate directions 
that they represent fair samples, but not so large in the redshift 
direction that the range in luminosities in any given catalog is small, 
or that evolution effects are washed out.  

The volume-limited subamples are constructed as follows.  
First, we specify the boundaries in redshift of the catalog:  
$z_{\rm min}$ and $z_{\rm max}=z_{\rm min} + 0.04$.  
In the context of a world model, these redshift limits, 
when combined with the angular size of the catalog, can be used to 
compute a volume.  This volume depends on $z_{\rm min}, z_{\rm max}$ 
and the world model:  as our fiducial model we set $\Omega_M = 0.3$ 
and $\Omega_\Lambda=1-\Omega_M$.  (Our results hardly change if we 
use an Einstein de-Sitter model instead.)  We then compute the 
K-corrected limiting luminosities $L_{\rm max}(z_{\rm min})$ and 
$L_{\rm min}(z_{\rm max})$ given the apparent magnitude limits, 
the redshift limits, and the assumed cosmology.  
A galaxy $i$ is included in the volume limited subsample if 
$z_{\rm min}\le z_i\le z_{\rm max}$ and 
$L_{\rm min}\le L_i\le L_{\rm max}$.  The luminosity function for the 
volume limited subsample is obtained by counting the number of galaxies 
in a luminosity bin and dividing by the volume of the subsample.  

\begin{figure}
\centering
\epsfxsize=\hsize\epsffile{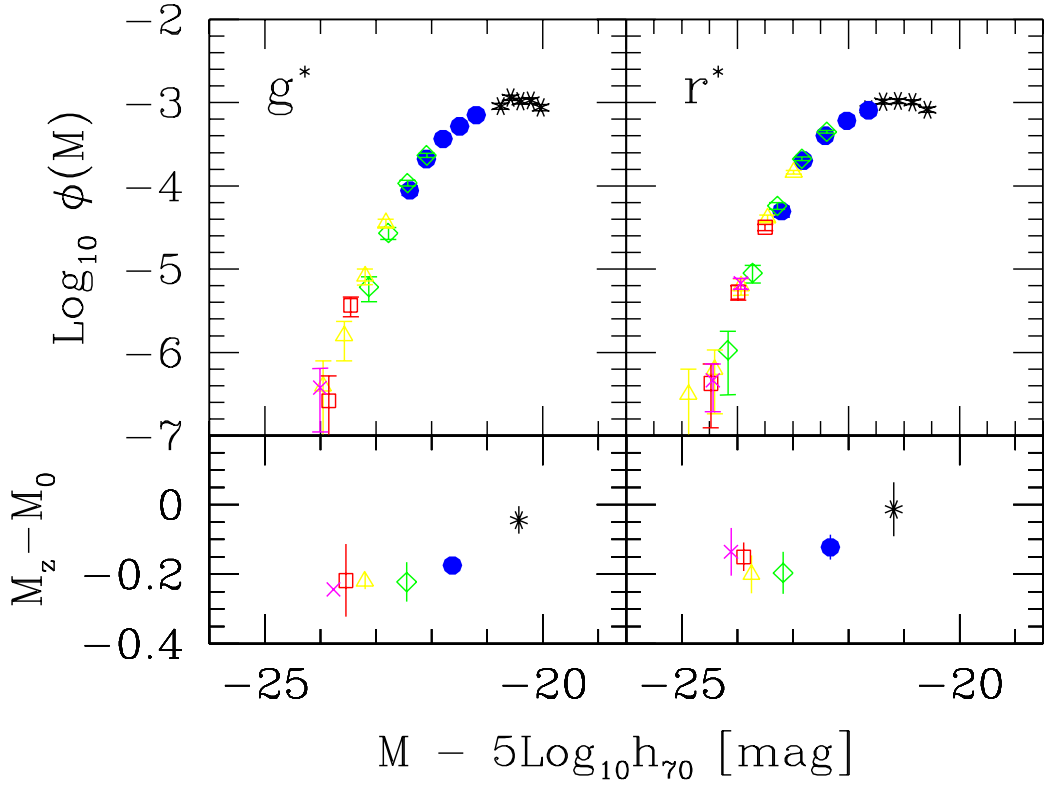}
\vspace{-3cm}
\caption{Luminosity functions in the $g^*$ and $r^*$ bands.  Stars, 
circles, diamonds, triangles, squares and crosses show measurements in 
volume limited catalogs which are adjacent in redshift of width 
$\Delta z=0.04$, starting from a minimum of $z_{\rm min}=0.04$.  
Top panels show that the higher redshift catalogs contribute at the 
bright end only.  At the same comoving density, the symbols which 
represent the higher redshift catalogs tend to be displaced slightly 
to the left of the those which represent the lower redshift catalogs.  
Bottom panels show this small mean shift towards increasing luminosity 
with increasing redshift.}  
\label{lfev}
\end{figure}

The top panels in Figure~\ref{lfev} show the result of doing this in the 
$g^*$ and $r^*$ bands.  Stars, circles, diamonds, triangles, squares and 
crosses show measurements in volume limited catalogs which have 
$z_{\rm min}=0.04$, 0.08, 0.12, 0.16, 0.20, and 0.24 and 
$z_{\rm max} = z_{\rm min}+0.04$.  Each subsample contains more than 
five hundred galaxies, except for the two most distant, which each 
contain about one hundred.  As one would expect, the nearby volumes 
provide the faint end of $\phi(M)$, and the more distant volumes show 
the bright end.  The extent to which the different volume limited 
catalogs all trace out the same curve is a measure of how little the 
luminosity function at low and high redshifts differs from that at the 
median redshift.  

The bottom panels in Figure~\ref{lfev} show evidence that, in fact, the 
galaxies in our data show evidence for a small amount of evolution:  at 
fixed comoving density, the higher redshift population is slightly brighter 
than that at lower redshifts.  Although volume-limited catalogs provide 
model-independent measures of this evolution, the test is most sensitive 
when a large range of luminosities can be probed at two different 
redshifts.  Because the SDSS catalogs are cut at both the faint and the 
bright ends, our test for evolution is severely limited.  Nevertheless, 
the small trends we see are both statistically significant, and 
qualitatively consistent with what one expects of a passively evolving 
population.  

Before we make more quantitative conclusions, notice that a bell-like 
Gaussian shape would provide a reasonable description of the luminosity 
function.  Although early-type galaxies are expected to have red colors, 
our sample was not selected using any color information.  It is reassuring, 
therefore, that the Gaussian shape we find here also provides a good fit to 
the luminosity function of the redder objects in the SDSS parent catalog 
(see the curves for the two reddest galaxy bins in Fig.14 of 
Blanton et al. 2001).  A Gaussian form also provides a reasonable 
description of the luminosity function of early-type galaxies in the 
CNOC2 survey (Lin et al. 1999, even though they actually fit a Schechter 
function to their measurements).  The 2DFGRS galaxies classified as being 
of Type 1 by Madgwick et al. (2001) should be similar to early-types.  
Their Type 1's extend to considerably fainter absolute magnitudes than 
our sample.  The population of early-type galaxies at faint absolute 
magnitudes is known to be very different from the brighter ones (e.g., 
Sandage \& Perelmuter 1990).  This is probably why the shape of the 
luminosity function they report is quite different from ours.  

Given that the Gaussian form provides a good description of our data, 
we use the maximum-likelihood method outlined by 
Sandage, Tammann \& Yahil (1979) to estimate the parameters of the 
best-fitting luminosity function.  For magnitude limited samples which 
are small and shallow, this is the method of choice.  For a sample such 
as ours, which spans a sufficiently wide range in redshifts that evolution 
effects might be important, the method requires a model for the evolution.  
We parametrize the luminosity evolution similarly to Lin et al. (1999).  
That is to say, if we were solving only for the luminosity function, 
then the likelihood function we maximize would be  
\begin{eqnarray}
{\cal L} &=& \prod_i{\phi(M_i,z_i|Q,M_*,\sigma_M)\over S(z_i|Q,M_*,\sigma_M)},
 \qquad {\rm where} \nonumber\\
\phi(M_i,z_i|Q,M_*,\sigma_M) &=& {\phi_*\over\sqrt{2\pi\sigma_M^2}}\,
           \exp\left(-{[M_i-M_*+Qz_i]^2\over 2\sigma_M^2}\right), \nonumber\\
S(z_i|Q,M_*,\sigma_M) &=& 
 \int_{M_{\rm min}(z_i)}^{M_{\rm max}(z_i)} dM\,\phi(M,z_i|Q,M_*,\sigma_M),
\label{mlphi}
\end{eqnarray}
$M_{\rm min}(z_i)$ and $M_{\rm max}(z_i)$ denote the minimum and maximum 
absolute magnitudes at $z_i$ which satisfy the apparent magnitude limits 
of the survey, and $i$ runs over all the galaxies in the catalog.  
(At small $z$, this parametrization of the evolution in absolute magnitude 
implies that the luminosity evolves as $L_*(z)/L_*(0) \approx (1+z)^q$, 
with $q = Q\,{\rm ln}(10)/2.5$.  Note that, in assuming that only $M_*$ 
evolves, this model assumes that there is no differential evolution in 
luminosities, i.e., that luminous and not so luminous galaxies evolve 
similarly.)  

\begin{figure}
\centering
\epsfxsize=\hsize\epsffile{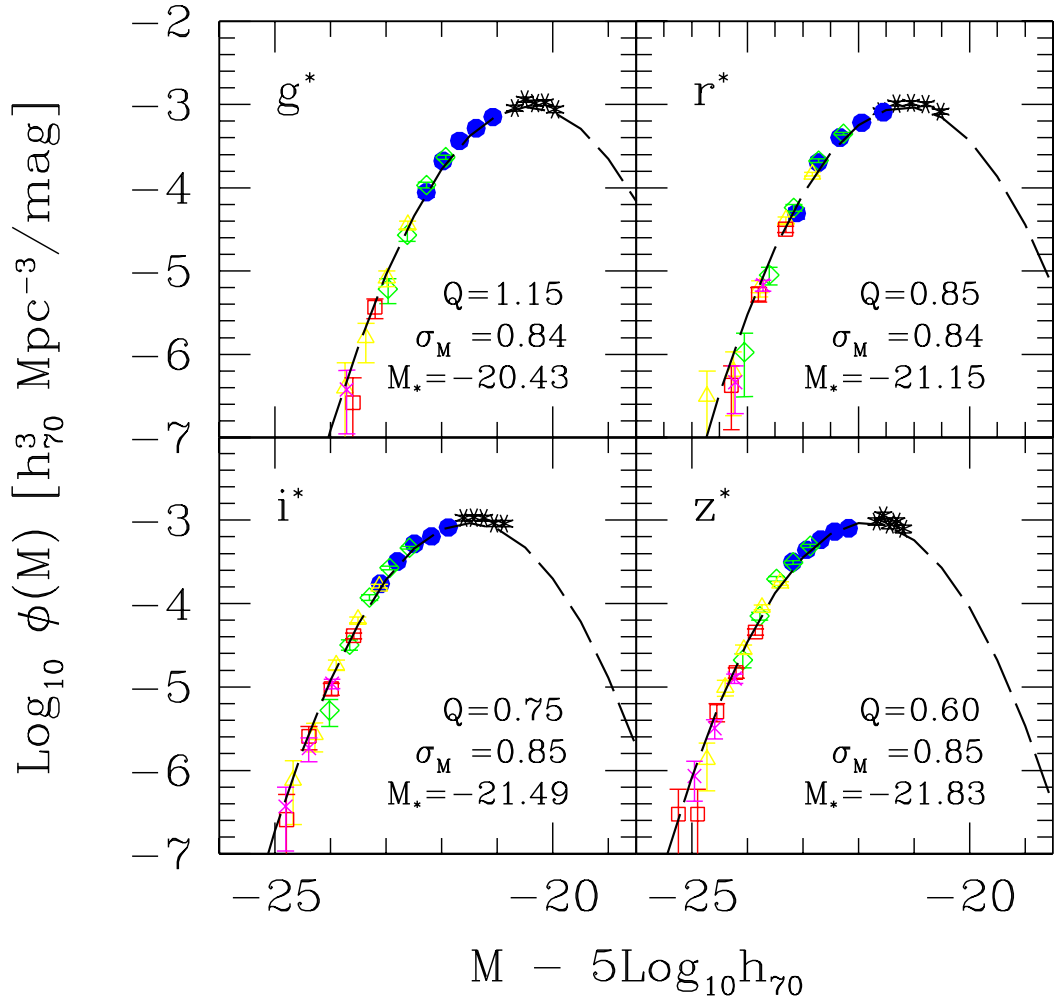}
\vspace{-1cm}
\caption{Luminosity functions in the $g^*$, $r^*$, $i^*$ and $z^*$ 
bands, corrected for pure luminosity evolution.  
Symbols with error bars show the estimates from our various 
volume limited catalogs; the higher redshift catalogs contribute at 
the bright end only.  Dashed curves show the shape of the Gaussian 
shaped luminosity function which maximizes the likelihood of seeing 
this data at redshift $z=0$.}  
\label{lf4}
\end{figure}

Figure~\ref{lf4} shows the result of estimating the luminosity function 
in this way in the $g^*$, $r^*$, $i^*$ and $z^*$ bands.  Later in this 
paper, we will solve simultaneously for the joint distribution of 
luminosity, size and velocity dispersion; it is the parameters which 
describe the luminosity function of this joint solution which are shown 
in Fig.~\ref{lf4}.  The dashed lines in each panel show the Gaussian shape 
of the luminosity function at redshift $z=0$.  For comparison, the symbols 
show the measurements in the same volume limited catalogs as before, 
except that now  we have subtracted the maximum likelihood estimate of 
the luminosity evolution from the absolute magnitudes $M$ before plotting 
them.  If the model for the evolution is accurate, then the different 
symbols should all trace out the same smooth dashed curve.  

\begin{figure}
\centering
\epsfxsize=\hsize\epsffile{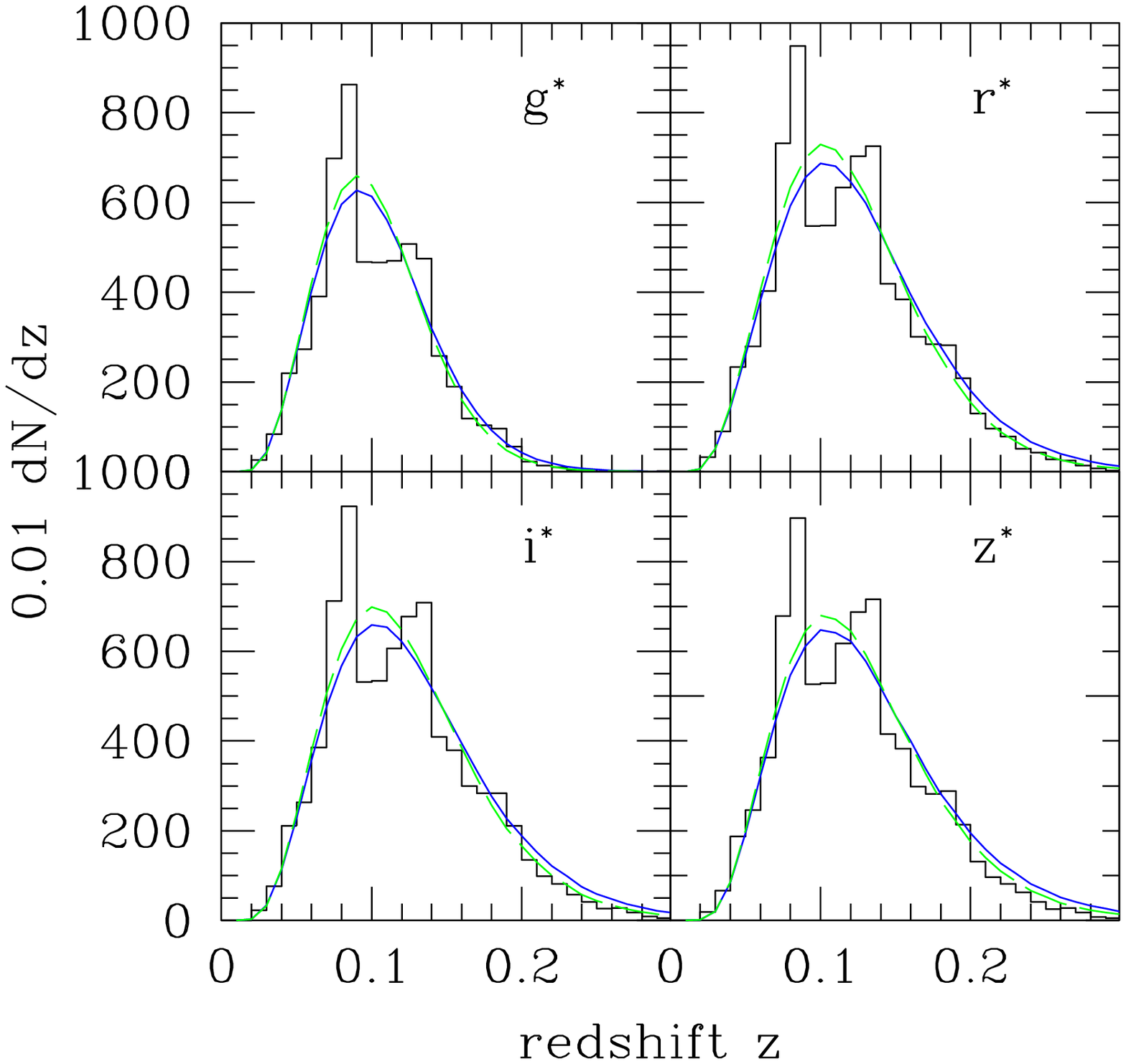}
\vspace{-0.5cm}
\caption{The number of galaxies as a function of redshift in our sample.  
Solid curves show the predicted counts if the comoving number densities 
are constant, but the luminosities brighten systematically with redshift: 
$M_*(z)= M_*(0)-Qz$ with $Q$ given by the previous figure.
Dashed curves show what one predicts if 
a) there is no evolution whatsoever, and the luminosity function is 
fixed to the value it has at the median redshift of our sample ($z=0.1$); or 
b) there is density as well as luminosity evolution, as reported by 
Lin et al. (1999); or 
c) the comoving number densities do not change, but the more luminous 
galaxies evolve less rapidly than less luminous galaxies.}
\label{nz4}
\end{figure}

The comoving number density of the galaxies in this sample is 
$\phi_* = 5.8\pm 0.3\times 10^{-3}h^3$Mpc$^{-3}{\rm mag}^{-1}$ in all 
four bands.  Because the different bands have different apparent magnitude 
limits, and they were fit independently of each other, it is reassuring 
that the same value of $\phi_*$ works for all the bands.  
For similar reasons, it is reassuring that the best-fit values of $M_*$ 
imply rest-frame colors at $z=0$ of $g^*-r^* = 0.72$, $r^*-i^* = 0.34$, 
and $r^*-z^* = 0.68$, which are close to those of the models which we used 
to compute our K-corrections, even though no a priori constraint was 
imposed on what these rest-frame colors should be.  

The histograms in each of the four panels of Figure~\ref{nz4} show the 
number of galaxies observed as a function of redshift in the four bands.  
The peak in the number counts at $z\sim 0.08$ is also present in the 
full SDSS sample, which includes late-types, and, perhaps more 
surprisingly, an overdensity at this same redshift is also present 
in the 2dF Galaxy Redshift Survey.  (The second bump at $z\sim 0.12$ is 
also present in the 2dFGRS counts.)    
The solid curves show what we expect to see for the evolving Gaussian 
function fits---the curves provide a reasonably good fit to the observed 
counts, although they slightly overestimate the numbers at high redshift 
in the redder wavebands.  
For comparison, the dashed curves show what is expected if the luminosities 
do not evolve and the no-evolution luminosity function is given by the one 
at the median redshift (i.e., a Gaussian with mean $M_*-0.1Q$).  
Although the fit to the high-redshift tail is slightly better, this 
no evolution model cannot explain the trends shown in the bottom panel 
of Figure~\ref{lfev}.  

The bottom panels in Figure~\ref{lfev} suggest two possible reasons why 
our model of pure luminosity evolution overestimates $dN/dz$ at higher 
$z$.  One possibility is that the comoving number densities are 
decreasing slightly with redshift.  A small amount of density evolution 
is not unexpected, because early-type galaxy morphologies may evolve 
(van Dokkum \& Franx 2001), and our sample is selected on the basis of 
a fixed morphology.  If we allow a small amount of density as well 
as luminosity evolution, and we use $\phi_*(z) = 10^{0.4Pz}\phi_*(0)$ 
with $P\approx -2$, as suggested by the results of Lin et al. (1999), 
then the resulting $dN/dz$ curves are also well fit by the dashed curves. 
A second possibility follows from the fact that we only observe the 
most luminous part of the higher redshift population.  If the most 
luminous galaxies at any given time are also the oldest, then one might 
expect the bright end of the luminosity function to evolve less rapidly 
than the fainter end.  Indeed, the Bruzual \& Charlot (2002) passive 
evolution model, described in Appendix~\ref{kcorrs}, predicts that the 
rest-frame luminosities at redshift $z=0.2$ should be brighter than 
those at $z=0$ by 0.3, 0.26, 0.24, and 0.21~mags in $g^*$, $r^*$, $i^*$ 
and $z^*$ respectively.  
The curvature seen in the bottom panel of Figure~\ref{lfev} suggests that 
although the evolution of the fainter objects in our sample (which we only 
see out to low redshifts) is consistent with this, the brighter objects 
are not.  Models of such differential evolution in the luminosities also 
predict $dN/dz$ distributions which are in better agreement with the 
observed counts at high redshift.  Since the evolution of the luminosity 
function is small, we prefer to wait until were are able to make more 
accurate K-corrections before accounting for either of these other 
possibilities more carefully.  Therefore, in what follows, we will 
continue to use the model with pure luminosity evolution.  

Repeating the exercise described above but for an Einstein--de-Sitter 
model yields qualitatively similar results, although the actual values 
of $M_*$ and $\phi_*$ are slightly different.  At face value, the fact 
that we see so little evolution in the luminosities argues for a 
relatively high formation redshift: the Bruzual \& Charlot (2002) 
models indicate that $t_{\rm form}\sim 9$Gyrs.  

\section{The Fundamental Plane}\label{fp}
In any given band, each galaxy in our sample is characterized by 
three numbers:  its luminosity, $L$, its size, $R_o$, and its velocity 
dispersion, $\sigma$.  
Correlations between these three observables are expected if 
early-type galaxies are in virial equilibrium, because 
\begin{equation}
\sigma_{vir}^2 \propto {GM_{vir}\over 2R_{vir}} 
                   \propto \left({M_{vir}\over L}\right)R_{vir}
\left({L/2\over R_{vir}^2}\right).
\end{equation}
If the size parameter $R_{vir}$ which enters the virial theorem is 
linearly proportional to the observed effective radius of the light, 
$R_o$, and if the observed line-of-sight velocity dispersion $\sigma$ 
is linearly proportional to $\sigma_{vir}$, then this relates the 
observed velocity dispersion to the product of the observed surface 
brightness and effective radius.  
Following Djorgovski \& Davis (1987), correlations involving all 
three variables are often called the Fundamental Plane (FP).  
In what follows, we will show how the surface brightness, $R_o$, and 
$\sigma$ are correlated.  Because both 
$\mu\propto -2.5\log_{10}[(L/2)/R_o^2]$ and $\sigma$ are distance 
independent quantities, it is in these variables that studies of 
early-type galaxies are usually presented.  

\subsection{Finding the best-fitting plane}\label{fits}
The Fundamental Plane is defined by:
\begin{equation}
\log_{10} R_o = a\,\log_{10}\sigma + b\,\log_{10}I_o + c
\end{equation}
where the coefficients $a$, $b$, and $c$ are determined by minimizing 
the residuals from the plane.  There are a number of ways in which 
this is usually done.  Let 
\begin{eqnarray}
\Delta_1 &\equiv& 
 \log_{10}R_o - a\,\log_{10}\sigma - b\,\log_{10}I_o - c
 \qquad {\rm and}\nonumber \\ \nonumber \\
\Delta_o &\equiv& {\Delta_1\over (1 + a^2 + b^2)^{1/2}}.  
\label{deltas}
\end{eqnarray}
Then summing $\Delta_1^2$ over all $N$ galaxies and finding that set of 
$a$, $b$ and $c$ for which the sum is minimized gives what is often 
called the direct fit, whereas minimizing the sum of $\Delta_o^2$ 
instead gives the orthogonal fit.  Although the orthogonal fit is, 
perhaps, the more physically meaningful, the direct fit is of more 
interest if the FP is to be used as a distance indicator.  

A little algebra shows that the direct fit coefficients are 
\begin{eqnarray}
a &=& {\sigma_{II}^2\sigma_{RV}^2-\sigma_{IR}^2\sigma_{IV}^2\over
\sigma_{II}^2\sigma_{VV}^2-\sigma_{IV}^4}, \qquad 
b = {\sigma_{VV}^2\sigma_{IR}^2-\sigma_{RV}^2\sigma_{IV}^2\over
\sigma_{II}^2\sigma_{VV}^2-\sigma_{IV}^4}, \nonumber\\ 
c &=& \overline{\log_{10}R_o} - a\,\overline{\log_{10}\sigma} - 
b\,\overline{\log_{10}I_o}, \qquad {\rm and} \nonumber \\
\langle \Delta_1^2\rangle &=& 
{\sigma_{II}^2\sigma_{RR}^2\sigma_{VV}^2 
- \sigma_{II}^2 \sigma_{RV}^4 - \sigma_{RR}^2 \sigma_{IV}^4 - 
\sigma_{VV}^2 \sigma_{IR}^4 + 2 \sigma_{IR}^2\sigma_{IV}^2\sigma_{RV}^2
\over \sigma_{II}^2\sigma_{VV}^2-\sigma_{IV}^4},
\label{dfit}
\end{eqnarray}
where $\overline{\log_{10}X}\equiv \sum_i \log_{10}X_i/N$ and 
$\sigma^2_{xy}\equiv \sum_i (\log_{10}X_i - \overline{\log_{10}X})
(\log_{10}Y_i - \overline{\log_{10}Y})/N$.  For what follows, it is 
also convenient to define 
 $r_{xy} = \sigma_{xy}^2/(\sigma_{xx}\sigma_{yy})$.  
The final expression above gives the scatter around the relation.  
If surface brightness and velocity dispersion are uncorrelated 
(we will show below that, indeed, $\sigma_{IV}\approx 0$), then $a$ 
equals the slope of the relation between velocity dispersion and the 
mean size at fixed velocity dispersion, $b$ is the slope of the relation 
between surface-brightness and the mean size at fixed surface-brightness, 
and the rms scatter is $\sigma_{RR}\sqrt{1 - r_{RV}^2 - r_{IR}^2}$.  
Errors in the observables affect the measured $\sigma_{xy}^2$, and thus 
will bias the determination of the best-fit coefficients and the intrinsic 
scatter around the fit.  If $\epsilon_{xy}$ is the rms error in the joint 
measurement of $\log_{10} X$ and $\log_{10}Y$, then subtracting the 
appropriate $\epsilon^2_{xy}$ from each $\sigma_{xy}^2$ before using them 
provides estimates of the error-corrected values of $a$, $b$ and $c$.  
Expressions for the orthogonal fit coefficients can be derived similarly, 
although, because they require solving a cubic equation, they are lengthy, 
so we have not included them here.  

Neither minimization procedure above accounts for the fact that the 
sample is magnitude-limited, and has a cut at small velocity dispersions.  
In addition, because our sample spans a wide range of redshifts, we must 
worry about effects which may be due to evolution.  The magnitude limit 
means that we cannot simply divide our sample up into small redshift 
ranges (over which evolution is negligible), because a small redshift 
range probes only a limited range of luminosities, sizes and velocity 
dispersions.  To account for all these effects, we use the 
maximum-likelihood approach described in Appendix~\ref{ML3d}.  
This method is the natural choice given that the joint distribution of 
$M\equiv -2.5\log_{10}L$, $R\equiv\log_{10}R_o$ and $V\equiv\log_{10}\sigma$ 
is quite well described by a multivariate Gaussian.  In Appendix~\ref{ML3d} 
we show how to compute the maximum-likelihood estimate of this distribution 
(the covariance matrix ${\cal C}$).  What remains is to write down how the 
covariance matrix changes when we change variables from $(M,R,V)$ to 
$(\mu,R,V)$.  Because $(\mu_o-\mu_*) \equiv (M-M_*) + 5(R-R_*)$, the 
covariance matrix is 
\begin{displaymath}
{\cal F} \equiv \left( \begin{array}{ccc} 
     \sigma^2_M + 10\sigma_M\sigma_R\rho_{RM} + 25\sigma_R^2 & 
     \sigma_R\sigma_M\,\rho_{RM} + 5\sigma^2_R & 
     \sigma_V\sigma_M\,\rho_{VM} + 5\sigma_R\sigma_V\,\rho_{RV}\\
     \sigma_R\sigma_M\,\rho_{RM} + 5\sigma^2_R & 
     \sigma^2_R & \sigma_R\sigma_V\,\rho_{RV}\\
     \sigma_V\sigma_M\,\rho_{VM} + 5\sigma_R\sigma_V\,\rho_{RV} & 
     \sigma_R\sigma_V\,\rho_{RV} & \sigma^2_V\\
           \end{array}\right) ,
\end{displaymath}
the coefficients of which are given in Table~\ref{MLcov}.  

\begin{table}[t]
\centering
\caption[]{Maximum-likelihood estimates, in the four SDSS bands, of the 
joint distribution of luminosities, sizes and velocity dispersions.  
Table shows the mean values of the variables at redshift $z$, 
$M_*-Qz$, $R_*$, $V_*$, and the elements of the covariance matrix 
${\cal C}$ defined by the various pairwise correlations between the 
variables (see Appendix~\ref{ML3d}).  These coefficients are also used 
in computing the matrix ${\cal F}$ in the main text.\\}
\begin{tabular}{cccccccccccc}
\tableline 
Band & $N_{\rm gals}$ & $M_*$ & $\sigma_M$ & $R_*$ & $\sigma_R$ &
$V_*$ & $\sigma_V$ & $\rho_{RM}$ & $\rho_{VM}$ & $\rho_{RV}$ & Q\\
\hline\\
$g^*$ & 5825 & $-20.43$ & 0.844 & 0.520 & 0.254 & 2.197 & 0.113 & $-0.886$ &
$-0.750$ & 0.536 & 1.15 \\
$r^*$ & 8228 & $-21.15$ & 0.841 & 0.490 & 0.241 & 2.200 & 0.111 & $-0.882$ &
$-0.774$ & 0.543 & 0.85 \\
$i^*$ & 8022 & $-21.49$ & 0.851 & 0.465 & 0.241 & 2.201 & 0.110 & $-0.886$ &
$-0.781$ & 0.542 & 0.75 \\
$z^*$ & 7914 & $-21.83$ & 0.845 & 0.450 & 0.241 & 2.200 & 0.110 & $-0.885$ &
$-0.782$ & 0.543 & 0.60 \\
\tableline
\end{tabular}
\label{MLcov}
\end{table}

This matrix is fundamentally useful because it describes the intrinsic 
correlations between the sizes, surface-brightnesses and velocity 
dispersions of early-type galaxies, i.e., the effects of how the sample 
was selected and observational errors have been accounted for.  For example, 
when the values from Table~\ref{MLcov} are inserted, the coefficients in 
the top right (and bottom left) of ${\cal F}$ are very close to zero, 
indicating that surface brightness and velocity dispersion are uncorrelated.  
In addition, the eigenvalues and vectors of ${\cal F}$ give information 
about the shape and thickness of the Fundamental Plane.  
For example, the smallest eigenvalue is considerably smaller than the other 
two; this says that, when viewed in the appropriate projection, the plane 
is quite thin.  The associated eigenvector gives the coefficients of the 
`orthogonal' fit, and the rms scatter around this orthogonal fit is given 
by the (square root of the) smallest eigenvalue.  

If we wish to use the FP as a distance indicator, then we are more 
interested in finding those coefficients which minimize the scatter in 
$R_o$.  This means that we would like to find that pair $(a,b)$ which 
minimize $\langle\Delta_1^2\rangle$, where $\Delta_1$ is given by 
equation~(\ref{deltas}).  A little algebra shows that the solution is 
given by inserting the maximum likelihood estimates of the scatter in 
surface-brightnesses, sizes and velocity dispersions into 
equation~(\ref{dfit}).  

\begin{table}
\centering
\caption[]{Coefficients of the FP in the four SDSS bands.  For each 
set of coefficients, the scatter orthogonal to the plane and in the 
direction of $R_o$ are also given.  
\\}
\begin{tabular}{cccccc}
\tableline 
Band & $a$ & $b$ & $c$ & rms$_{\rm orth}^{\rm int}$ & rms$_{R_o}^{\rm int}$ \\
\hline\\
\\[-8mm]
\multicolumn{6}{c}{\bf Orthogonal fits} \\
\multicolumn{6}{l}{\bf Maximum Likelihood} \\
\\[-6mm]
$g^*$ & 1.45$\pm 0.06$  & $-0.74\pm 0.01$  & $-8.779\pm 0.029$ & 0.056 & 0.100 \\
$r^*$ & 1.49$\pm 0.05$  & $-0.75\pm 0.01$  & $-8.778\pm 0.020$ & 0.052 & 0.094 \\
$i^*$ & 1.52$\pm 0.05$  & $-0.78\pm 0.01$  & $-8.895\pm 0.021$ & 0.049 & 0.091 \\
$z^*$ & 1.51$\pm 0.05$  & $-0.77\pm 0.01$  & $-8.707\pm 0.023$ & 0.049 & 0.089 \\
\multicolumn{6}{l}{\bf $\chi^2 - $ Evolution $-$ Selection effects}\\
\\[-5mm]
$g^*$ & 1.43$\pm 0.06$  & $-0.78\pm 0.01$  & $-9.057\pm 0.032$ & 0.058 & 0.101 \\
$r^*$ & 1.45$\pm 0.05$  & $-0.76\pm 0.01$  & $-8.719\pm 0.020$ & 0.052 & 0.094 \\
$i^*$ & 1.48$\pm 0.05$  & $-0.77\pm 0.01$  & $-8.699\pm 0.024$ & 0.050 & 0.090 \\
$z^*$ & 1.48$\pm 0.05$  & $-0.77\pm 0.01$  & $-8.577\pm 0.025$ & 0.049 & 0.089\\
\multicolumn{6}{l}{\bf $\chi^2 - $ Evolution}\\
\\[-5mm]
$g^*$ & 1.35$\pm 0.06$  & $-0.77\pm 0.01$  & $-8.820\pm 0.033$ & 0.058 & 0.100 \\
$r^*$ & 1.40$\pm 0.05$  & $-0.77\pm 0.01$  & $-8.678\pm 0.023$ & 0.053 & 0.092 \\
$i^*$ & 1.41$\pm 0.05$  & $-0.78\pm 0.01$  & $-8.688\pm 0.024$ & 0.050 & 0.090 \\
$z^*$ & 1.41$\pm 0.05$  & $-0.78\pm 0.01$  & $-8.566\pm 0.026$ & 0.048 & 0.089\\
\hline\\
\\[-8mm]
\multicolumn{6}{c}{\bf Direct fits} \\
\multicolumn{6}{l}{\bf Maximum Likelihood} \\
\\[-6mm]
$g^*$ & 1.08$\pm 0.05$ & $-0.74\pm 0.01$ & $-8.033\pm 0.024$ & 0.061 & 0.092\\
$r^*$ & 1.17$\pm 0.04$ & $-0.75\pm 0.01$ & $-8.022\pm 0.020$ & 0.056 & 0.088\\
$i^*$ & 1.21$\pm 0.04$ & $-0.77\pm 0.01$ & $-8.164\pm 0.018$ & 0.053 & 0.085\\
$z^*$ & 1.20$\pm 0.04$ & $-0.76\pm 0.01$ & $-7.995\pm 0.021$ & 0.053 & 0.084\\
\multicolumn{6}{l}{\bf $\chi^2 - $ Evolution $-$ Selection effects}\\
\\[-5mm]
$g^*$ & 1.05$\pm 0.05$ & $-0.79\pm 0.01$ & $-8.268\pm 0.026$ & 0.063 & 0.094\\
$r^*$ & 1.12$\pm 0.04$ & $-0.76\pm 0.01$ & $-7.932\pm 0.020$ & 0.057 & 0.088\\
$i^*$ & 1.14$\pm 0.04$ & $-0.76\pm 0.01$ & $-7.904\pm 0.019$ & 0.054 & 0.085\\
$z^*$ & 1.14$\pm 0.04$ & $-0.76\pm 0.01$ & $-7.784\pm 0.021$ & 0.053 & 0.084\\
\multicolumn{6}{l}{\bf $\chi^2 - $ Evolution}\\
\\[-5mm]
$g^*$ & 0.99$\pm 0.05$ & $-0.76\pm 0.01$ & $-7.921\pm 0.026$ & 0.065 & 0.093\\
$r^*$ & 1.06$\pm 0.04$ & $-0.75\pm 0.01$ & $-7.775\pm 0.020$ & 0.059 & 0.088\\
$i^*$ & 1.09$\pm 0.04$ & $-0.77\pm 0.01$ & $-7.823\pm 0.018$ & 0.056 & 0.085\\
$z^*$ & 1.09$\pm 0.04$ & $-0.78\pm 0.01$ & $-7.818\pm 0.020$ & 0.053 & 0.083\\
\tableline
\end{tabular}
\label{fpcoeffs} 
\end{table}

\begin{figure}
\centering
\epsfxsize=\hsize\epsffile{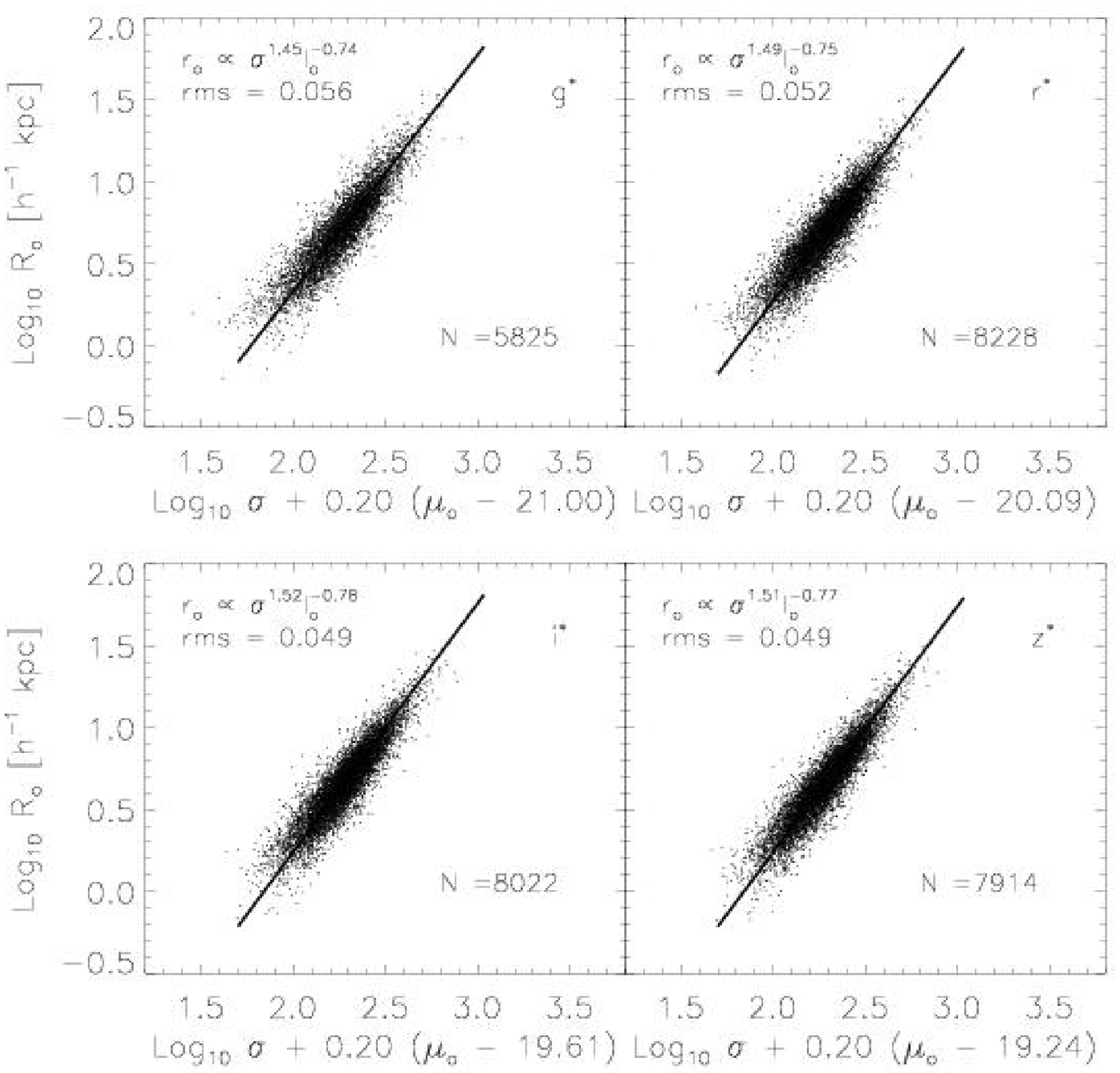}
\caption{The Fundamental Plane in the four SDSS bands.  Coefficients 
shown are those which minimize the scatter orthogonal to the plane, as 
determined by the maximum-likelihood method.  Surface-brightnesses have 
been corrected for evolution.}  
\label{fig:FPgriz}
\end{figure}

The maximum likelihood ${\cal F}$ can be used to provide estimates of 
the direct and orthogonal fit coefficients, as well as the intrinsic 
scatter around the mean relations (orthogonal to the plane as well 
as in the direction of $R_o$).  These are given in Table~\ref{fpcoeffs}.  
Although $b$ is approximately the same both for the `orthogonal' and 
the `direct' fits, $a$ from the direct fit is always about 25\% smaller 
than from the orthogonal fit.  
In either case, note how similar $a$ and $b$ are in all four bands.  
This similarity, and the fact that the thickness of the FP decreases 
slightly with increasing wavelength, can be used to constrain models of 
how different stellar populations (which may contribute more or less to 
the different bands) are distributed in early-type galaxies.  If the direct 
fit is used as a distance indicator, then the thickness of the FP translates 
into an uncertainty in derived distances of about 20\%.  

Table~\ref{fpcoeffs} also shows results from the more traditional 
$\chi^2-$fitting techniques, which were obtained as follows.  (These 
fits were not weighted by errors, and the intrinsic scatter with 
respect to the fits was estimated by subtracting the measurement 
errors in quadrature from the observed scatter.)  
Ignoring evolution and selection effects when minimizing 
$\langle\Delta_1^2\rangle$ and $\langle\Delta_o^2\rangle$, results in 
coefficients $a$ which are about 10\% larger than those we obtained from 
the maximum likelihood method.  We have not shown these in the Table 
for the following reason.  If the population at high redshift is more 
luminous than that nearby, as expected if the evolution is passive, then 
the higher redshift population would have systematically smaller values of 
$\mu_o$.  Since the higher redshift population makes up most of the 
large $R_o$ part of our sample, this could make the Plane appear 
steeper, i.e., it could cause the best-fit $a$ to be biased to a 
larger value.  
If we use the maximum-likelihood estimate of how the luminosities brighten 
with redshift, then we can subtract off the brightening from $\mu_o$ 
before minimizing $\langle\Delta_1^2\rangle$ and $\langle\Delta_o^2\rangle$.  
This reduces the best-fit value of $a$ so that it is closer to that of 
the maximum likelihood method.  The coefficients obtained in this way 
are labeled `$\chi^2$ $-$ Evolution' in Table~\ref{fpcoeffs}; they are 
statistically different from the maximum likelihood estimates, presumably 
because they do not account for selection effects or for the effects of 
observational errors.  
If we weight each galaxy by the inverse of $S(z_i|M_*,Q)$ (the selection 
function defined in equation~\ref{mlphi}), when minimizing, then this 
should at least partially account for selection effects.  The resulting 
estimates of $a$, $b$ and $c$ are labeled 
`$\chi^2$ $-$ Evolution $-$ Selection effects' in Table~\ref{fpcoeffs}.   
The small remaining difference between these and the maximum likelihood 
estimates is likely due to the fact that the likelihood analysis 
accounts more consistently for errors.  

Figure~\ref{fig:FPgriz} shows the FP in the four SDSS bands.  
We have chosen to present the plane using the coefficients, 
obtained using the maximum-likelihood method, which minimize the 
scatter orthogonal to the plane.  (In all cases, the evolution of 
the luminosities has been subtracted from the surface brightnesses.)   
The results to follow regarding the shape of the FP, and estimates of how 
the mean properties of early-types depend on redshift and environment, are 
independent of which fits we use.  In addition, recall that a fair number 
of the galaxies in our sample have velocity dispersion measurements with 
small S/N (e.g., Figure~\ref{fig:vmeth}).  The FP is relatively insensitive 
to these objects:  removing objects with S/N $<$ 15 had little effect on 
the best fit values of $a$, $b$.  Removing objects with small axis ratios 
also had little effect on the maximum likelihood coefficients.  

In principle, the likelihood analysis provides an estimate of the error on 
each of the derived coefficients.  However, this estimate assumes that the 
parametric fit was indeed good.  (Although we have evidence that the fit 
is good, we emphasize that, when the data set is larger a non-parametric 
fit should be performed.)  Therefore, we have estimated errors on the 
numbers quoted in Table~\ref{fpcoeffs} as follows.  
The large size of our sample allows us to construct many random 
subsamples, each of which is substantially larger than most of the 
samples available in the literature.  Estimating the elements of the 
covariance matrix presented in Table~\ref{MLcov}, and then transforming 
to get the FP coefficients in Table~\ref{fpcoeffs}, in each of these 
subsamples provides an estimate of how well we have determined 
$a$, $b$ and $c$.  (Note that the errors we find in this way are 
comparable to those sometimes quoted in the literature, even though 
each of the subsamples we generated is an order of magnitude larger 
than any sample available in the literature.)  
Because each subsample contains fewer galaxies than our full sample, 
this procedure is likely to provide an overestimate of the true formal 
error for our sample.  However, the formal error does not account for 
the uncertainties in our K-corrections and velocity dispersion aperture 
corrections, so an overestimate is probably more realistic.  

\begin{table}[t]
\centering
\caption[]{Coefficients of the FP in the complete and magnitude-limited 
simulated catalogs, obtained by minimizing a $\chi^2$ in which evolution 
in the surface brightnesses has been removed, and which weights objects 
by the inverse of the selection function.\\}
\begin{tabular}{cccccc}
\tableline 
Band & $a$ & $b$ & $c$ & rms$_{\rm orth}$ & rms$_{R_o}$ \\
\hline\\
\\[-8mm]
\multicolumn{6}{c}{\bf Orthogonal fits} \\
\multicolumn{6}{l}{\bf Complete} \\
\\[-5mm]
$g^*$ & 1.44$\pm 0.05$ & $-0.74\pm 0.01$ & $-8.763\pm 0.028$ & 0.056 & 0.100 \\
$r^*$ & 1.48$\pm 0.05$ & $-0.75\pm 0.01$ & $-8.722\pm 0.020$ & 0.052 & 0.094 \\
\\[-4mm]
\multicolumn{6}{l}{\bf Magnitude limited}\\
\\[-5mm]
$g^*$ & 1.39$\pm 0.06$ & $-0.74\pm 0.01$ & $-8.643\pm 0.028$ & 0.056 & 0.100 \\
$r^*$ & 1.43$\pm 0.05$ & $-0.76\pm 0.01$ & $-8.721\pm 0.021$ & 0.052 & 0.093 \\
\hline\\
\\[-8mm]
\multicolumn{6}{c}{\bf Direct fits} \\
\multicolumn{6}{l}{\bf Complete} \\
\\[-5mm]
$g^*$ & 1.09$\pm 0.04$ & $-0.74\pm 0.01$ & $-7.992\pm 0.023$ & 0.061 & 0.091\\
$r^*$ & 1.16$\pm 0.04$ & $-0.75\pm 0.01$ & $-8.005\pm 0.020$ & 0.056 & 0.088\\
\\[-4mm]
\multicolumn{6}{l}{\bf Magnitude limited}\\
\\[-5mm]
$g^*$ & 1.04$\pm 0.05$ & $-0.74\pm 0.01$ & $-7.817\pm 0.025$ & 0.061 & 0.090 \\
$r^*$ & 1.11$\pm 0.04$ & $-0.75\pm 0.01$ & $-7.895\pm 0.020$ & 0.056 & 0.087 \\
\tableline
\end{tabular}
\label{simcoeffs} 
\end{table}

As a check on the relative roles of evolution and selection effects, 
we simulated complete and magnitude-limited samples (with a velocity 
dispersion cut) following the procedures outlined in Appendix~\ref{simul}.  
We then estimated the coefficients of the FP in the simulated catalogs 
using the different methods.  The results are summarized in 
Table~\ref{simcoeffs}.  When applied to the complete simulations, the 
$\chi^2-$minimization method yields estimates of $a$ which are biased 
high; it yields the input Fundamental Plane coefficients only after 
evolution has been subtracted from the surface brightnesses.  
However, in the magnitude limited simulations, once evolution has been 
subtracted, it provides an estimate of $a$ which is biased low, unless 
selection effects are also accounted for.  Note that this is similar to 
what we found with the data.  The maximum-likelihood method successfully 
recovers the same intrinsic covariance matrix and evolution as the one 
used to generate the simulations, both for the complete and the 
magnitude-limited mock catalogs, and so it recovers the same correct 
coefficients for the FP in both cases.  (We have not shown these estimates 
in the Table.)

\begin{table}[t]
\centering
\caption[]{Selection of Fundamental Plane coefficients from the literature.\\}
\begin{tabular}{lllllll}
\tableline
Source & Band & N$_{\rm gal}$ & $a$ & $b$ & $\Delta_{R_o}$ & Fit method\\
\hline\\
\\[-4mm]
Dressler et al. (1987) & $B$ & 97 & 1.33$\pm 0.05$ & $-0.83\pm 0.03$ & 20\% & inverse\\
Lucey et al. (1991) & $B$ & 26 & 1.27$\pm0.07$ & $-0.78\pm 0.09$ & 13\% & inverse\\
Guzm\'an et al. (1993) & $V$ & 37 & 1.14$\pm$ -- & $-0.79\pm$ -- & 17\% &
direct \\
Kelson et al. (2000) & $V$ & 30 & 1.31$\pm 0.13$ & $-0.86\pm 0.10$ & 14\% &
orthogonal \\
Djorgovski \& Davis (1987) & $r_{G}$ & 106 & 1.39$\pm 0.14$ & $-0.90\pm 0.09$ & 20\% &
2-step inverse \\
J{\o}rgensen et al. (1996) & $r$ & 226 & 1.24$\pm 0.07$ & $-0.82\pm
0.02$ & 19\% & orthogonal \\
Hudson et al. (1997) & $R$ & 352 & 1.38$\pm 0.04$ & $-0.82\pm 0.03$ & 20\% & inverse\\
Gibbons et al. (2001) & $R$ & 428 & 1.37$\pm 0.04$ & $-0.825\pm 0.01$ & 20\% & inverse \\
Colless et al. (2001) & $R$ & 255 & 1.22$\pm 0.09$ & $-0.84\pm 0.03$ & 20\% &
ML \\ 
Scodeggio (1997) & $I$ & 294 & 1.55$\pm 0.05$ & $-0.80\pm 0.02$ & 22\% & orthogonal \\
Pahre et al. (1998a) & $K$ & 251 & 1.53$\pm 0.08$ & $-0.79\pm 0.03$ & 21\% &
orthogonal \\
\tableline
\end{tabular}
\label{fpcoeffslit}
\end{table}

A selection of results from the literature is presented in 
Table~\ref{fpcoeffslit}.  Many of these samples were constructed by 
combining new measurements with previously published photometric and 
velocity dispersion measurements, often made by other authors.  
(Exceptions are J{\o}rgensen et al. 1996, Scodeggio 1997, 
and Colless et al. 2001.)  With respect to previous samples, the SDSS 
sample presented here is both extremely large and homogeneous.  

Notice the relatively large spread in published values of $a$, and the 
fact that $a$ is larger at longer wavelengths.  In contrast, the 
Fundamental Plane we obtain in this paper is remarkably similar in all 
wavebands---although our value of $b$ is consistent with those in the 
literature, the value of $a$ we find in all wavebands is close to the 
largest published values.  In addition, the eigenvectors of our covariance 
matrix satisfy the same relations presented by Saglia et al. (2001).  
Namely, ${\bf\hat v}_1 = {\hat R}_o - a{\hat V} - b{\hat I}_o$, 
${\bf\hat v}_2\approx -{\hat R}_o/b - {\hat V}(1+b^2)/(ab) + {\hat I}_o$ 
and ${\bf\hat v}_3 \approx {\hat R}_o + {\hat I}_o/b$.  
And, when used as a distance indicator, the FP we find is as accurate 
as most of the samples containing more than $\sim 100$ galaxies 
in the literature.  
Unfortunately, at the present time, we have no galaxies in common
with those in any of the samples listed in Table~\ref{fpcoeffslit}, 
so it is difficult to say why our FP coefficients appear to show 
so little dependence on wavelength, or why $a$ is higher than it 
is in the literature.  

The fact that $a\ne 2$ means that the FP is tilted relative to the 
simplest virial theorem prediction $R_o\propto \sigma^2/I_o$.  
One of the assumptions of this simplest prediction is that the 
kinetic energy which enters the virial theorem is proportional 
to the square of the observed central velocity dispersion.  
Busarello et al. (1997) argue that, in fact, the kinetic energy is 
proportional to $\sigma^{1.6}$ rather than to $\sigma^2$.  Since this is 
close to the $\sigma^{1.5}$ scaling we see, it would be interesting to 
see if the kinetic energy scales with $\sigma$ for the galaxies in our 
sample similarly to how it does in Busarello et al.'s sample.  
This requires measurements of the velocity dispersion profiles 
of (a subsample of) the galaxies in our sample, and has yet to 
be done.

Correlations between pairs of observables, such as the 
Faber--Jackson (1976) relation between luminosity and velocity 
dispersion, and the Kormendy (1977) relation between the size and 
the surface brightness can be thought of as projections of the 
Fundamental Plane.  These, along with the $\kappa$--space projection 
of Bender, Burstein \& Faber (1992), are presented in 
Appendix~\ref{prjct}.  

\subsection{Residuals and the shape of the FP}\label{fpscat}
Once the FP has been obtained, there are at least two definitions 
of its thickness which are of interest.  
If the FP is to be used as a distance indicator, then the quantity of 
interest is the scatter around the relation in the $R_o$ direction 
only.  On the other hand, if the FP is to be used to constrain models of 
stellar evolution, then one is more interested in the scatter orthogonal 
to the plane.  We discuss both of these below.  

The thickness of the FP is some combination of residuals which are 
intrinsic and those coming from measurement errors.  We would like to 
verify that the thickness is not dominated by measurement errors.  
The residuals from the FP in the different bands are highly correlated; 
a galaxy which scatters above the FP in $g^*$ also scatters above the 
FP in, say, $z^*$.  Although the errors in the photometry in the different 
bands are not completely independent, this suggests that the scatter around 
the FP has a real, intrinsic component.  It is this intrinsic thickness 
which the maximum likelihood analysis is supposed to have estimated.  
The intrinsic scatter may be somewhat smaller than the maximum likelihood 
estimates because there is a contribution to the scatter which comes from 
our assumption that all early-type galaxies are identical when we apply 
the K-correction, for which we have not accounted.  


\begin{figure}
\centering
\epsfxsize=\hsize\epsffile{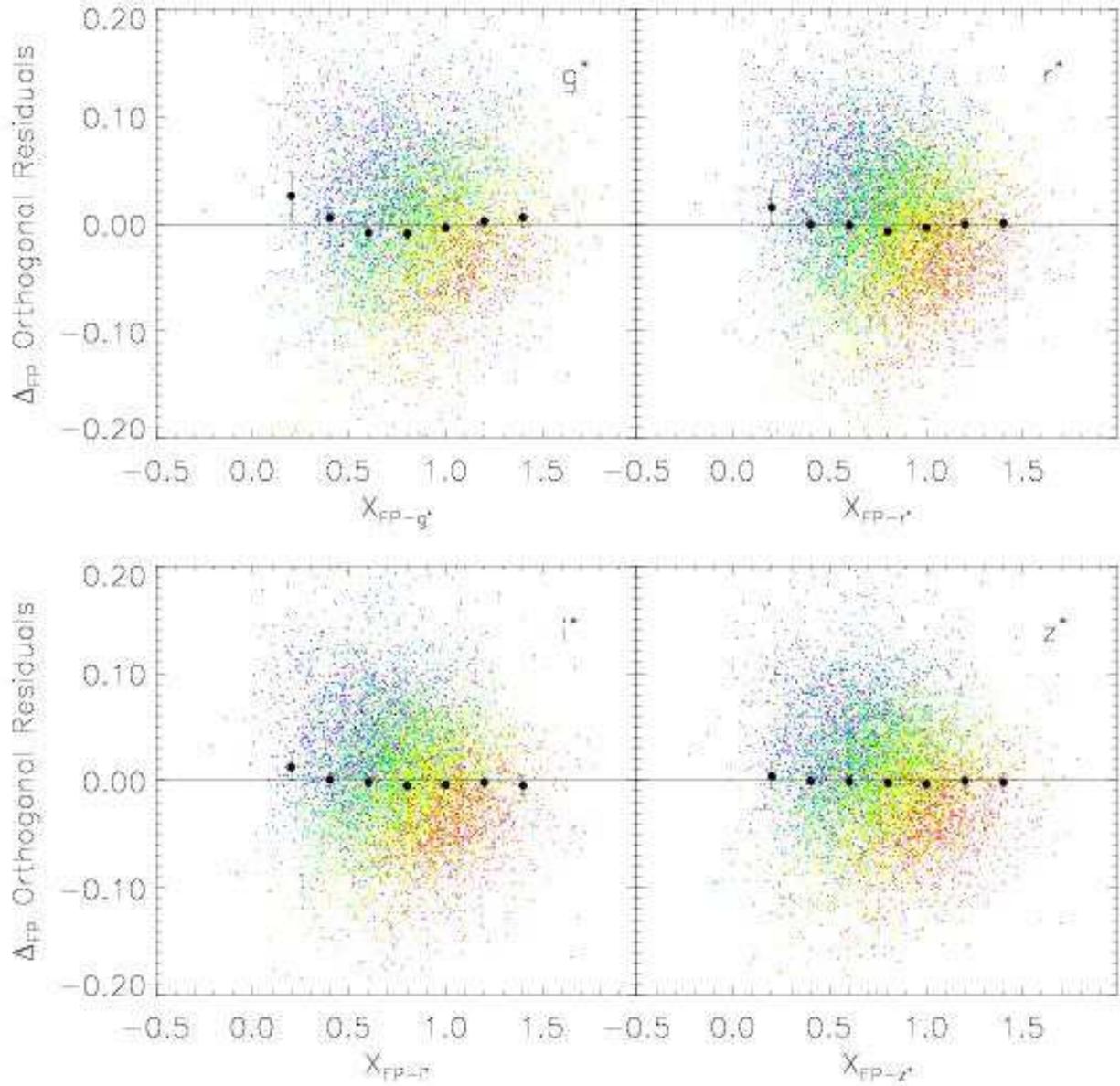}
\caption{Residuals orthogonal to the maximum-likelihood FP fit 
as a function of distance along the fit (the long axis of the plane).  
Error bars show the mean plus and minus three times the error in the 
mean in each bin.  Galaxies with low/high velocity dispersions populate 
the upper-left/lower-right of each panel, but the full sample shows 
little curvature. } 
\label{fig:warped}
\end{figure}

All our estimates of the scatter around the FP show that the FP appears 
to become thicker at shorter wavelengths.  Presumably, this is because 
the light in the redder bands, being less affected by recent 
star-formation and extinction by dust, is a more faithful tracer of 
the dynamical state of the galaxy.  
The orthogonal scatter in our sample, which spans a 
wide range of environments, is comparable to the values given in the 
literature obtained from cluster samples (e.g., Pahre et al. 1998a); 
this constrains models of how the stellar populations of early-type 
galaxies depend on environment.  
If the direct fit to the FP is used as a distance indicator, then 
the intrinsic scatter introduces an uncertainty in distance estimates 
of $\sim {\rm ln}(10)\times 0.09\sim$ 20\%.  

Our next step is to check that the FP really is a plane, and not,
for example, a saddle.  To do this, we should show the residuals 
from the orthogonal fit as a function of distance along the 
long axis of the plane.  Specifically, if 
$X\equiv \log_{10}\sigma + (b/a)\log_{10}I_o + (c/a)$, then 
\begin{equation}
X_{FP} \equiv X \sqrt{1+a^2} + (\log_{10}R_o - aX) {a\over\sqrt{1+a^2}}
       =  {X + a\log_{10}R_o \over \sqrt{1+a^2}},
\label{Xfp}
\end{equation}
and we would like to know if the residuals $\Delta_o$ defined earlier  
correlate with $X_{FP}$.  A scatter plot of these residuals versus 
$X_{FP}$ is shown in Figure~\ref{fig:warped} (we have first subtracted 
off the weak evolution in the surface brightnesses).  The symbols 
superimposed on the scatter plot show the mean value of the residuals 
and plus and minus three times the error in the mean, for a few small 
bins along $X_{FP}$.  The figure shows that the FP is reasonably 
flat; it is slightly more warped in the shorter wavelenghts.  

\begin{figure}
\centering
\epsfxsize=\hsize\epsffile{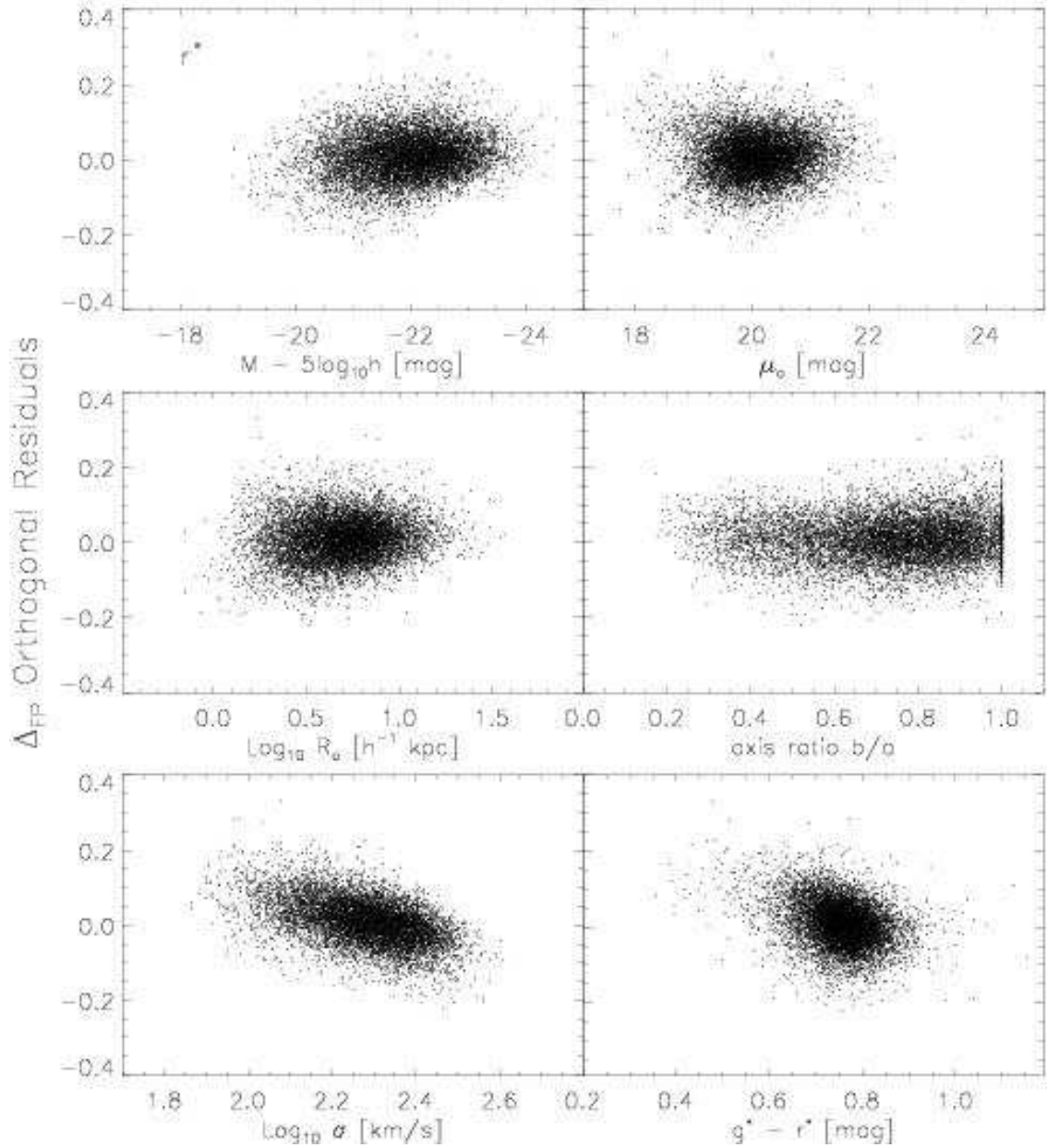}
\caption{Residuals orthogonal to the FP in $r^*$ versus absolute 
magnitude $M$, surface brightness $\mu_o$, 
effective radius $\log_{10} R_o$, axis ratio $b/a$, 
velocity dispersion $\log_{10} \sigma$, and ($g^*-r^*$) color.  
Note the absence of correlation with all parameters other than 
velocity dispersion and color.  }
\label{fig:FPresidR}
\end{figure}

\begin{figure}
\centering
\epsfxsize=\hsize\epsffile{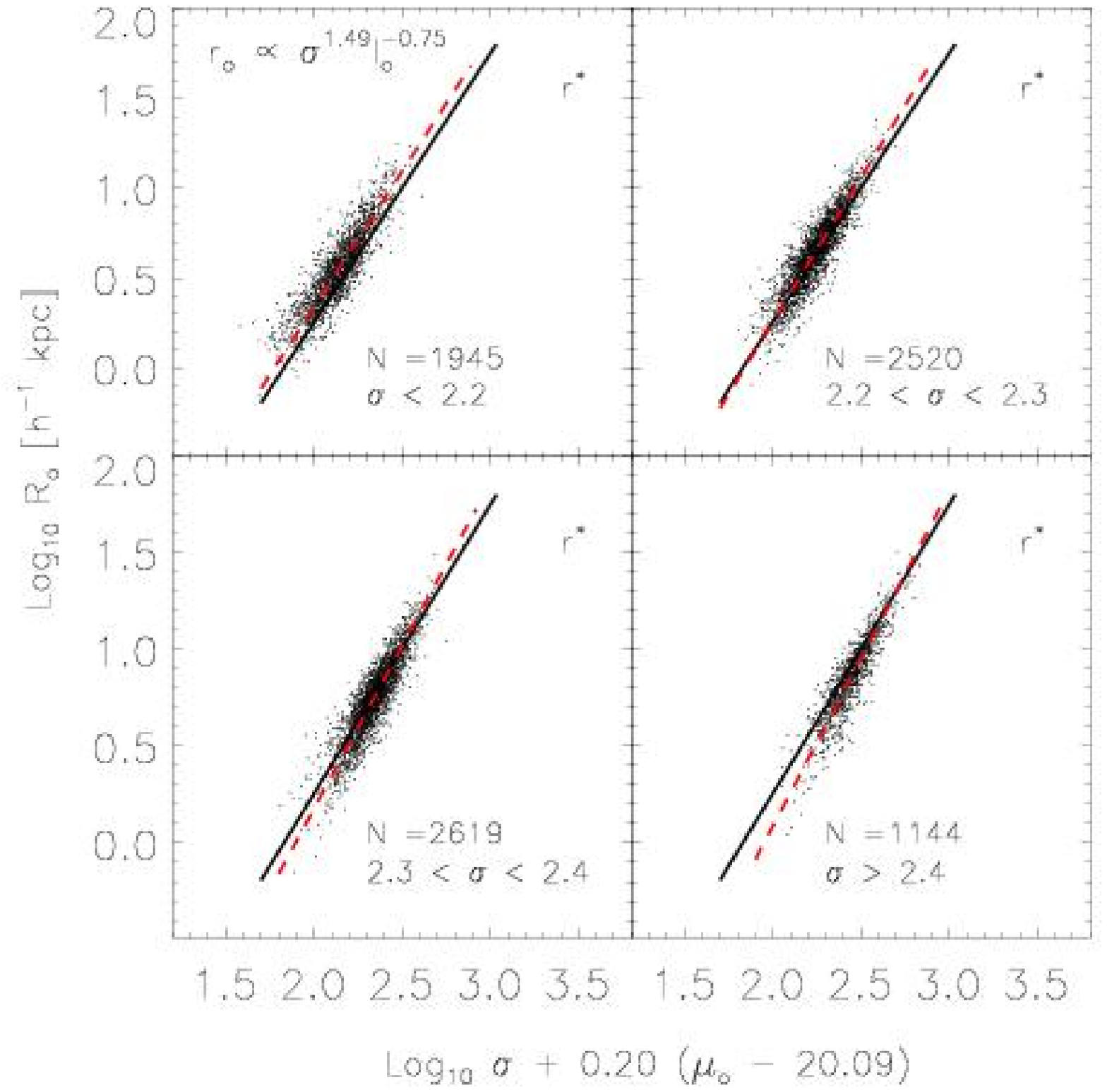}
\caption{The FP in four subsamples defined by velocity dispersion.  
Solid curve (same in all four panels) shows the maximum likelihood 
relation of the parent $r^*$ sample and dashed lines show the 
best-fit, obtained by minimization the residuals orthogonal to 
the plane, using only the galaxies in each subsample.  
The slope of the minimization fit increases with increasing 
velocity dispersion, whereas maximum-likelihood fits to the subsamples 
give the same slope as for the full sample.  } 
\label{fig:FPsigR}
\end{figure}

Given that the FP is not significantly warped, we would like to know 
if deviations from the Plane correlate with any of the three physical 
parameters used to define it.  When the plane is defined by minimizing 
with respect to $\log_{10} R_o$, there is little if any correlation of 
the residuals with absolute magnitude, surface brightness, 
effective radius, axis-ratio, velocity dispersion, or color so we 
have chosen to not present them here.  
Instead, Figure~\ref{fig:FPresidR} shows the result of plotting the 
residuals orthogonal to the plane when the plane is defined by the 
orthogonal fit.  The residuals show no correlation with 
$M$, $\mu_o$, $\log_{10} R_o$, or axis ratio (we have subtracted the 
weak evolution in $M$ and $\mu_o$ when making the scatter plots).  
The residuals are anti-correlated with $\log_{10}\sigma$ and 
slightly less anti-correlated with $(g^*-r^*)$ color.  The correlation 
with color is due to the fact that velocity dispersion and color are 
tightly correlated (see Section~\ref{cms} below).  
The correlation with velocity dispersion is not a selection effect, 
nor is it associated with evolution; we see a similar trend with 
velocity dispersion in both the complete and the magnitude-limited 
simulated catalogs.  

Figure~\ref{fig:FPsigR} shows why this happens.  The four panels show 
the FP in four subsamples of the full $r^*$ sample, divided according to 
velocity dispersion.  
Notice how the different scatter plots in Figure~\ref{fig:FPsigR} 
show sharp cut-offs approximately perpendicular to the x-axis:  
Lines of constant $\sigma$ are approximately perpendicular to the 
x-axis.  Whereas the direct fit is not affected by a cut-off which 
is perpendicular to the x-axis (for the same reason that the 
$\langle X|M\rangle$ relations presented in Appendix~\ref{lx} were 
not affected by the fact that the sample is magnitude limited), the 
orthogonal fit is.  Hence, the residuals with respect to the 
orthogonal fit show a correlation with velocity dispersion, 
whereas those from the direct fit do not.  (Indeed, by using the 
coefficients provided in Tables~\ref{MLcov} and~\ref{fpcoeffs} 
and the definition of the residuals $\Delta_1$ and $\Delta_2$,
one can compute the mean residual at fixed velocity dispersion, 
$\langle \Delta_1|\log_{10}\sigma \rangle$.  The result is proportional 
to $\log_{10}\sigma$, with a constant of proportionality which is close 
to zero when the parameters for the direct fit are inserted, but is 
significantly larger than zero when the parameters for the orthogonal 
fit are used.)  

To illustrate, the solid curves in Figure~\ref{fig:FPsigR} (the same in 
each panel) show the maximum likelihood FP for the full sample.  The 
dashed curves show the FP, determined by using the $\chi^2-$method to 
minimize the residuals orthogonal to the plane, in various subsamples 
defined by velocity dispersion.  
The panels for larger velocity dispersions show steeper relations.  
Evidence for a steepening of the relation with 
increasing velocity dispersion was seen by J{\o}rgensen et al. (1996).  
Their sample was considerably smaller than ours, and so they ruled the 
trend they saw as only marginal.  Our much larger sample shows this trend 
clearly.  We have already argued that this steepening is an artifact of the 
fact that lines of constant velocity dispersion are perpendicular to the 
$x$-axis.  The maximum-likelihood fit to the subsamples is virtually the 
same as that for the full sample, provided we include the correct velocity 
dispersion cuts in the normalization $S_i$ (equation~\ref{mlphi}) of the 
likelihood.  In other words, the maximum-likelihood fit is able to account 
for the bias introduced by making a cut in velocity dispersion as well as 
apparent magnitude.  

\subsection{The mass-to-light ratio}\label{m2l}
The Fundamental Plane is sometimes used to make inferences about how 
the mass-to-light ratio depends on different observed or physical 
parameters.  For example, the scaling required by the virial theorem, 
$M_o\propto R_o\sigma^2$, 
combined with the assumption that the mass-to-light ratio scales as 
$M_o/L\propto M_o^\gamma$ yields a Fundamental Plane like relation 
of the form:
\begin{equation}
R_o\propto \sigma^{2(1-\gamma)/(1+\gamma)}I_o^{-1/(1+\gamma)}.
\end{equation}
The observed Fundamental Plane is $R_o\propto \sigma^a\,I_o^b$.  
If the relation above is to satisfy the observations, then 
$\gamma$ must simultaneously satisfy two relations:  
$\gamma = (2-a)/(2+a)$, and $\gamma=-(1+b)/b$.  
The values of $b$ in the literature are all about $-0.8$; setting 
$\gamma$ equal to the value required by $b$ and then writing $a$ in 
terms of $b$ gives $a = -2(1+2b)$.  Most of the values of $a$ and $b$ 
in the shorter wavebands reported in the literature (see, e.g., 
Table~\ref{fpcoeffslit}) are consistent with this scaling, whereas the 
higher values of $a$ found at longer wavelengths are not.  Although the 
direct fits to our sample have small values of $a$, the orthogonal fits 
give high values in all four bands.  These fits do not support the 
assumption that $M_o/L$ can be parametrized as a function of $M_o$ alone.  

\begin{figure}
\centering
\epsfxsize=\hsize\epsffile{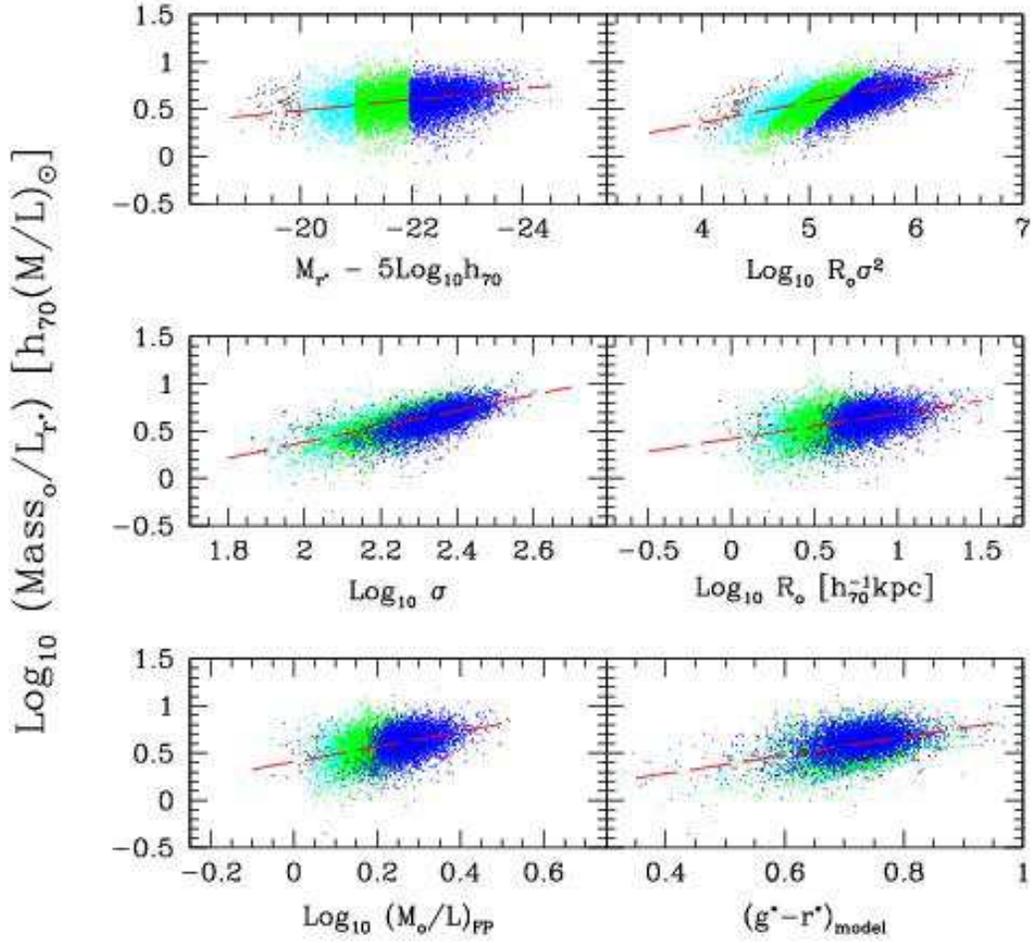}
\vspace{1cm}
\caption{Ratio of effective mass $R_o\sigma^2$ to effective luminosity 
$(L/2)$ as a function of luminosity (top left), mass (top right), 
velocity dispersion (middle left), surface brightness (middle right), 
the combination of velocity dispersion and size suggested by the 
Fundamental Plane (bottom left), and color (bottom right).  
Notice the substantial scatter around the best fit linear relation 
in the bottom left panel, the slope of which is shallower than unity.}
\label{fig:m2l}
\end{figure}

Another way to phrase this is to note that, when combined with the 
virial theorem requirement that $(M_o/L)\propto \sigma^2/(R_oI_o)$, 
the Fundamental Plane relation $R_o\propto \sigma^aI_o^b$ yields 
\begin{equation}
\left({M_o\over L}\right)_{\rm FP} \propto\sigma^{2 + a/b}\,R_o^{-(1+b)/b} 
\label{m2lfp}
\end{equation}
(e.g. J{\o}rgensen et al. 1996; Kelson et al. 2000).
The quantity on the right hand side is the mass-to-light ratio 
`predicted' by the Fundamental Plane, if $\sigma$ and $R_o$ 
are given, and the scatter in the Fundamental Plane is ignored.  
This is a function of $M_o$ alone only if $a = -2(1+2b)$.  
Our orthogonal fit coefficients $a$ and $b$ are not related in this 
way.  Rather, for our Fundamental Plane, the dependence on 
$\sigma$ in equation~(\ref{m2lfp}) cancels out almost exactly:  
to a very good approximation, we find 
 $(M_o/L)_{\rm FP}\propto R_o^{-(1+b)/b}\propto R_o^{0.33}$.  

Substituting the Fundamental Plane relation for $R_o$ rather 
than $I_o$ in the virial theorem yields 
 $(M_o/L)\propto (\sigma^2/I_o)^{1-a/2}/I_o^{-a/2-b}$.  
Inserting  $a = -2(1+2b)$ shows yet again that $(M_o/L)\propto M_o$.  
In constrast, our values for $a$ and $b$ show that the dependence 
on $I_o$ alone is weak, and the mass-to-light ratio is determined 
mainly by the combination $(\sigma^2/I_o)^{0.25}$, which is consistent 
with the $R_o^{0.33}$ scaling above.  Whether there is a simple physical 
reason for this is an open question. 

In contrast to the predicted ratio, $(M_o/L)_{\rm FP}$, the combination 
$R_o\sigma^2/L$ is the `observed' mass-to-light ratio.  The ratio of the 
observed value to the FP prediction of equation~(\ref{m2lfp}) is 
$(R_o/I_o^b\sigma^a)^{1/b}$.  The scatter in the logarithm of this ratio 
is $1/b$ times the scatter in Fundamental Plane in the direction of $R_o$ 
(i.e., it is the scatter in the quantity we called $\Delta_1$ in the 
previous subsections, divided by $b$).  
Inserting the values from Table~\ref{fpcoeffs} shows that if the values of 
$\sigma$ and the effective radius in $r^*$ are used to predict the values 
of the mass-to-light ratio in $r^*$, then the uncertainty in the predicted 
ratio is 26\%.  This is larger than the values quoted in the literature 
for early-type galaxies in clusters (e.g., J{\o}rgensen et al. 1986; 
Kelson et al. 2000).  

Unfortunately, this is somewhat confusing terminology, because 
the two mass-to-light ratios are not proportional to each other.  
This can be seen by using the maximum-likelihood results of 
Table~\ref{MLcov} to compute the mean of the observed mass to light 
ratio $R_o\sigma^2/L$ at fixed predicted $(M/L)_{\rm FP}$, 
or simply by plotting the two quantities against one another.  
Figure~\ref{fig:m2l} shows how $R_o\sigma^2/L$ correlates with 
luminosity, mass $R_o\sigma^2$, velocity dispersion, surface brightness, 
the ratio predicted by the Fundamental Plane, and color.  The different 
panels show obvious correlations; the maximum likelihood predictions 
for these correlations can be derived from the coefficients in 
Table~\ref{MLcov}:  
$(R_o\sigma^2/L)\propto L^{0.14\pm 0.02}$, 
$(R_o\sigma^2/L)\propto (R_o\sigma^2)^{0.22\pm 0.05}$, 
$(R_o\sigma^2/L)\propto \sigma^{0.84\pm 0.1}$, and 
$(R_o\sigma^2/L)\propto R_o^{0.27\pm 0.06}$.  
These are shown as dashed lines in the top four panels.  
A linear fit to the scatter plot in the bottom left panel gives 
$(R_o\sigma^2/L)\propto (M/L)_{\rm FP}^{0.80\pm 0.05}$, with an 
rms scatter around the fit of 0.14:  the ratio predicted by the 
Fundamental Plane is not proportional to the observed ratio.  
A scatter plot of $(M/L)_{\rm FP}$ against all these quantities is 
tighter, of course (recall the scatter around the FP has been removed), 
although some of the slopes are significantly different.  
For example, $(M/L)_{\rm FP}\propto L^{0.16\pm 0.04}$, 
$(M/L)_{\rm FP}\propto (R_o\sigma^2)^{0.13\pm 0.03}$, and
$(M/L)_{\rm FP}\propto \sigma^{0.21\pm 0.03}$:  
the `observed' and `predicted' slopes of the mass-to-light ratio 
versus $\sigma$ relations are very different.   
For this reason, one should be careful in interpretting what 
is meant by the `predicted' mass-to-light ratio.  

\subsection{The Fundamental Plane:  Evidence for evolution?}\label{evolve}
\begin{figure}
\centering
\epsfxsize=\hsize\epsffile{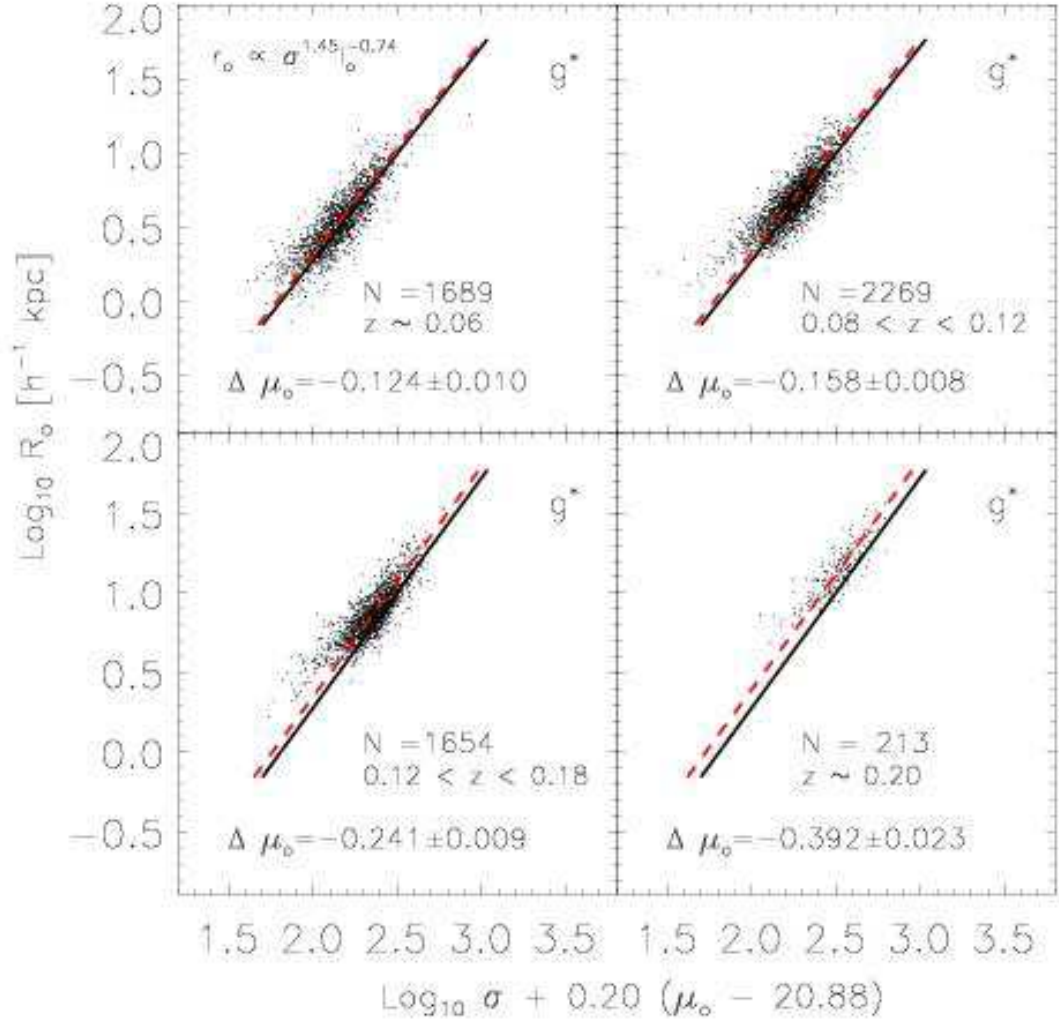}
\vspace{1cm}
\caption{The $g^*$ FP in four redshift bins. The slope of the FP is 
fixed to that at zero redshift; only the zero-point is allowed 
to vary.  The zero-point shifts systematically with redshift.  
The same plot for $r^*$ shows similar but smaller shifts.  }
\label{fig:FPzG}
\end{figure}

The Fundamental Plane is sometimes used to test for evolution.  
This is done by plotting $R_o$ versus the combination of $\mu_o$ and 
$\sigma$ which defines the Fundamental Plane at low redshift, and then 
seeing if the high-redshift population traces the same locus as the low 
redshift population.  Figure~\ref{fig:FPzG} shows this test for our 
$g^*$ band sample:  solid lines (same in each panel) show the relation 
which fits the zero-redshift sample; dashed lines show a line with the 
same slope which best-fits the higher redshift sample.  The population at 
higher redshift is displaced slightly to the left of the low redshift 
population; the text in the bottom of each panel shows this shift, 
expressed as a change in the surface brightness $\mu_o$.  
The plot appears to show that, on average, the higher redshift galaxies 
are brighter, with the brightening scaling approximately as 
$\Delta\mu_o \approx -2z$.  

\begin{figure}
\centering
\epsfxsize=\hsize\epsffile{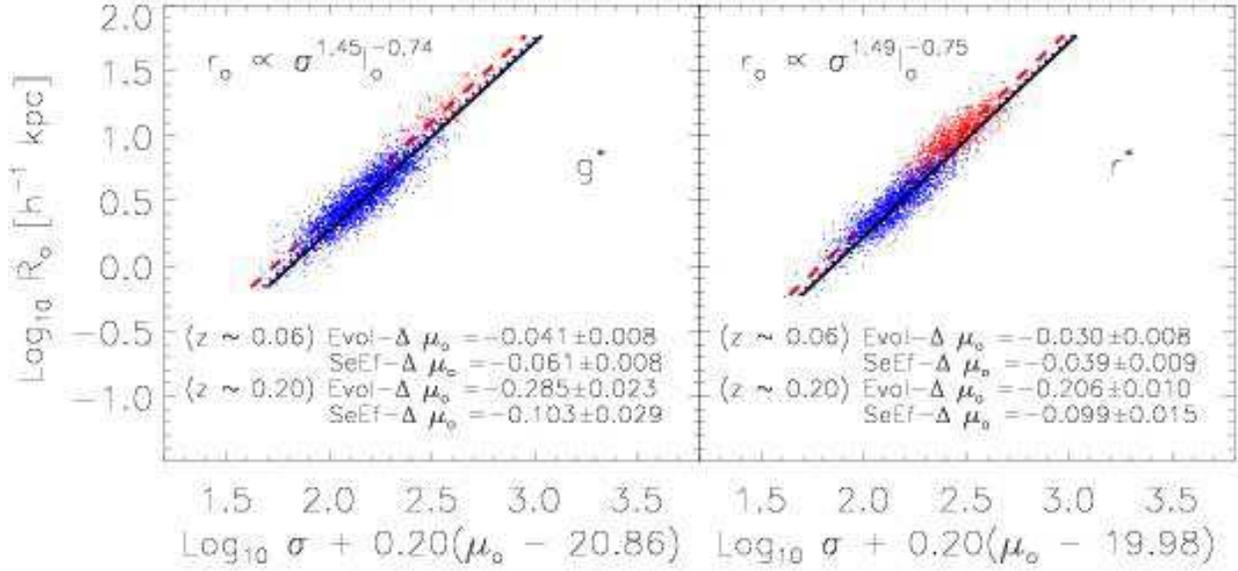}
\vspace{0cm}
\caption{The FP in the $g^*$ (left panel) and $r^*$ (right panel)
magnitude-limited mock catalogs.  Solid line shows the FP at $z=0$.  
Dotted and dashed lines show fits using a low and high redshift 
subsample only.  For these fits, the slope of the FP is required to be 
the same as the solid line; only the zero-point is allowed to vary.  
The shift seen in the complete simulations is labeled 
`Evol$-\Delta\mu_o$', whereas the shift seen in the magnitude limited 
simulations is the sum of this and the quantity labeled 
`SeEf$-\Delta\mu_o$'.  This sum is similar to the shift seen in 
the SDSS data, suggesting that selection effects are not negligble.}
\label{fig:FPsimz}
\end{figure}

How much of this apparent brightening is really due to evolution, and 
how much is an artifact of the fact that our sample is magnitude limited?  
To address this, we generated complete and magnitude limited mock galaxy 
catalogs as described in Appendix~\ref{simul}, and then performed the same 
test for evolution.  Comparing the shifts in the two simulations allows 
us to estimate how much of the shift is due to the selection effect.  
Figure~\ref{fig:FPsimz} shows the results in our simulated $g^*$ (left) 
and $r^*$ (right) catalogs.  The solid lines in each panel show the 
zero-redshift relation, and the dotted and dashed lines show lines of the 
same slope which best-fit the points at low and high redshift, respectively.  
The text in the bottom shows how much of the shift in $\mu_o$ is due 
to the magnitude limit, and how much to evolution.  The sum of the two 
contributions is the total shift seen in the magnitude limited simulations.
Notice that this sum is similar to that seen in the data 
(Figure~\ref{fig:FPzG}), both at low and high redshifts, suggesting 
that our simulations describe the varying roles played by evolution and 
selection effects accurately.  Since the parameters of the simulations 
were set by the maximum likelihood analysis, we conclude that the 
likelihood analysis of the evolution in luminosities is reasonably 
accurate ($\Delta\mu_o\approx -1.15z$) in $g^*$, but we note that this 
evolution is less than one would have infered if selection effects 
were ignored ($\Delta\mu_o\approx -2z$).  

The importance of selection effects in our sample has implications 
for another way in which studies of evolution are presented.  
If galaxies do not evolve, then the FP can be used to define a standard 
candle, so the test checks if residuals from the FP in the direction of 
the surface-brightness variable, when plotted versus redshift, follow 
Tolman's $(1+z)^4$ cosmological dimming law.  
If Friedmann-Robertson-Walker models are correct, then departures from 
this $(1+z)^4$ dimming trend can be used to test for evolution.  
This can be done if one assumes that the main effect of evolution is 
to change the luminosities of galaxies.  If so, then evolution will 
show up as a tendency for the residuals from the FP, in the $\mu$ 
direction, to drift away from the $(1+z)^4$ dimming 
(e.g., Sandage \& Perelmuter 1990; Pahre, Djorgovski \& de Carvalho 1996).  

\begin{figure}
\centering
\epsfxsize=\hsize\epsffile{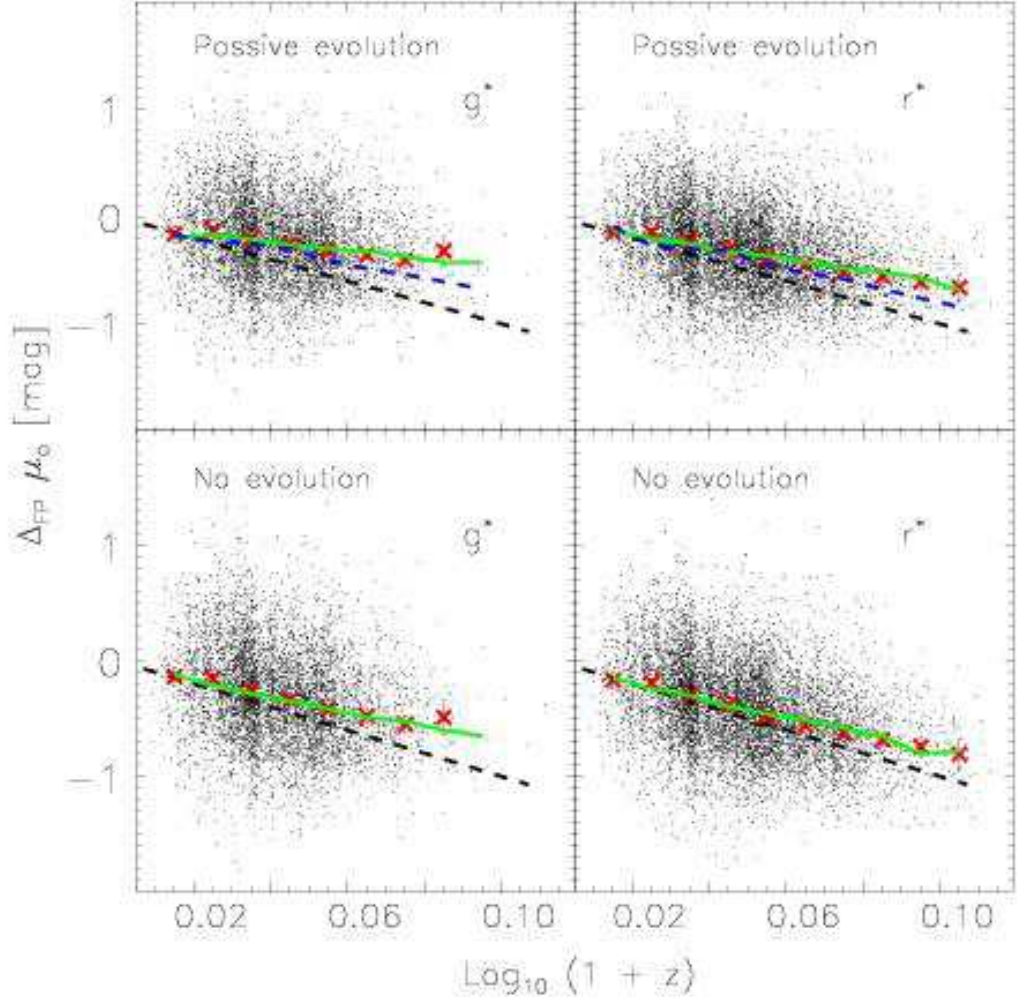}
\caption{Residuals of the zero-redshift FP with respect to the 
surface brightness, before correcting for cosmological dimming, 
versus redshift in the four bands.  Lowest dashed line in all panels 
shows the $(1+z)^4$ dimming expected if there is no evolution.  Solid 
curves in top panels show the same measurement in mock simulations of 
a magnitude limited sample of a passively evolving population.  Dashed 
lines in between show the actual evolution in surface brightness---the 
difference between these and the solid curves is an artifact of the 
magnitude limit.  Bottom panel shows the same test applied using the 
parameters of the Fundamental Plane which best describes the data if 
there is required to be no evolution whatsoever.  Solid lines show what 
one would observe in a magnitude limited sample of such a population.  
In this case, the entire trend away from the $(1+z)^4$ dimming is a 
selection effect.  Note how, once the magnitude limit has been applied, 
both the evolving (top) and non-evolving populations (bottom) appear 
very similar to our observed sample.  }
\label{fig:FPresidmuz}
\end{figure}

Figure~\ref{fig:FPresidmuz} shows this trend in our dataset.  
The lowest dashed lines in all panels show the expected $(1+z)^4$ 
dimming; panels on the left/right show results in $g^*$/$r^*$.  
Consider the top two panels first.  The points show residuals with respect 
to the zero-redshift Fundamental Plane in our sample.  The crosses show the 
median residual in a small redshift bin.  The galaxies do not quite follow 
the expected $(1+z)^4$ dimming.  The similarity to the $(1+z)^4$ dimming 
argues in favour of standard cosmological models, whereas the small 
difference from the expected trend is sometimes interpretted as evidence 
for evolution (e.g., J{\o}rgensen et al. 1999; van Dokkum et al. 1998, 2001; 
Treu et al. 1999, 2001a,b).  

Of course, to correctly quantify this evolution, we must 
account for selection effects.  The dashed lines which lie between the $(1+z)^4$ 
scaling and the data (i.e., the crosses) show how the surface brightness 
should scale if there were passive evolution of the form suggested 
by the maximum likelihood analysis, but there were no magnitude 
limit.  That is, if $M_*(z)=M_*(0)-Qz$, then the surface brightnesses 
should scale as $(1+z)^{4-0.92\,Q}$.
The solid curves show the result of making the measurement in simulated 
magnitude limited catalogs which include this passive evolution.  
Notice how different these solid curves are from the dashed curves (they 
imply $Q$ about twice the correct value), but note how similar they are 
to the data.  This shows that about half of the evolution one would 
naively have infered from such a plot is a consequence of the magnitude 
limit.  

To further emphasize the strength of this effect, we constructed simulations 
in which there was no evolution whatsoever.  We did this by first making 
maximum likelihood estimates of the joint luminosity, size and velocity 
dispersion distribution in which no evolution was allowed.  (For the 
reasons discussed earlier, the associated no-evolution Fundamental Plane 
coefficient $a$ is steeper by about 10\%.)  This was then used to generate 
mock catalogs in which there is no evolution.  The crosses in the bottom 
panels show the result of repeating the same procedure as in the top panels, 
but now using the parameters of the no-evolution Fundamental Plane, and the 
solid line shows the measurement in the no-evolution simulations in which, 
by construction, the population of galaxies at all redshifts is identical.  
Therefore, the shifts from the $(1+z)^4$ dimming we see in the magnitude 
limited no-evolution catalogs (solid curves in bottom panels) are entirely 
due to the magnitude limit.  Notice how similar the solid lines from our 
no-evolution simulations are to the actual data.  If we believed there 
really were no evolution, then the results shown in the bottom panel would 
lead us to conclude that much of the trend away from the $(1+z)^4$ dimming 
is a selection effect---it is not evidence for evolution.  

(The fact that we were able to find a non-evolving population which mimics 
the observations so well suggests that the population of early-type 
galaxies at the median redshift of our sample must be rather similar 
to the population at lower and at higher redshifts.  This, in turn, can 
constrain models of when the stars in these galaxies must have formed.)

We view our no-evolution simulations as a warning about the accuracy of 
this particular test of evolution.  If the evolution is weak, then it 
appears that the results of this test depend critically on how the catalog 
was selected, and on what one uses as the fiducial Fundamental Plane.  
To make this second point, we followed the procedure adopted by many other 
recent publications.  Namely, we assumed that the zero-redshift Fundamental 
Plane has the shape reported by J{\o}rgensen et al. (1996) for Coma, for 
which $a$ is about 15\% smaller than what we find in $g^*$.  If no account 
is taken of selection effects, then the inferred evolution in $\mu_o$ 
results in a value of $Q$ which is about a factor of four times larger 
than the one we report in Table~\ref{MLcov}!  

Our results indicate that inferences about evolution which are based on 
this test depend uncomfortably strongly on the strength of selection 
effects, and on what one assumes for the fiducial shape of the Fundamental 
Plane.  In this respect, our findings about the role of, and the need to 
account for selection effects are consistent with those reported by 
Simard et al. (1999).  While we believe we have strong evidence that 
the early-type population is evolving, we do not believe that the 
strongest evidence of this evolution comes from either of the tests 
presented in this subsection.  Later in this paper we will present 
evidence of evolution which we believe is more reliable.  Nevertheless,
it is reassuring that the evolution we see from these Fundamental Plane 
tests is consistent with that which we estimated using the likelihood 
analysis, and is also consistent with what we use to make our 
K-corrections.  Namely, a passively evolving population which formed 
the bulk of its stars about 9~Gyrs ago appears to provide a reasonable 
description of the evolution of the surface brightnesses in our 
sample.  

\subsection{The Fundamental Plane:  Dependence on environment}\label{environ}

\begin{figure}
\centering
\epsfxsize=1.2\hsize\epsffile{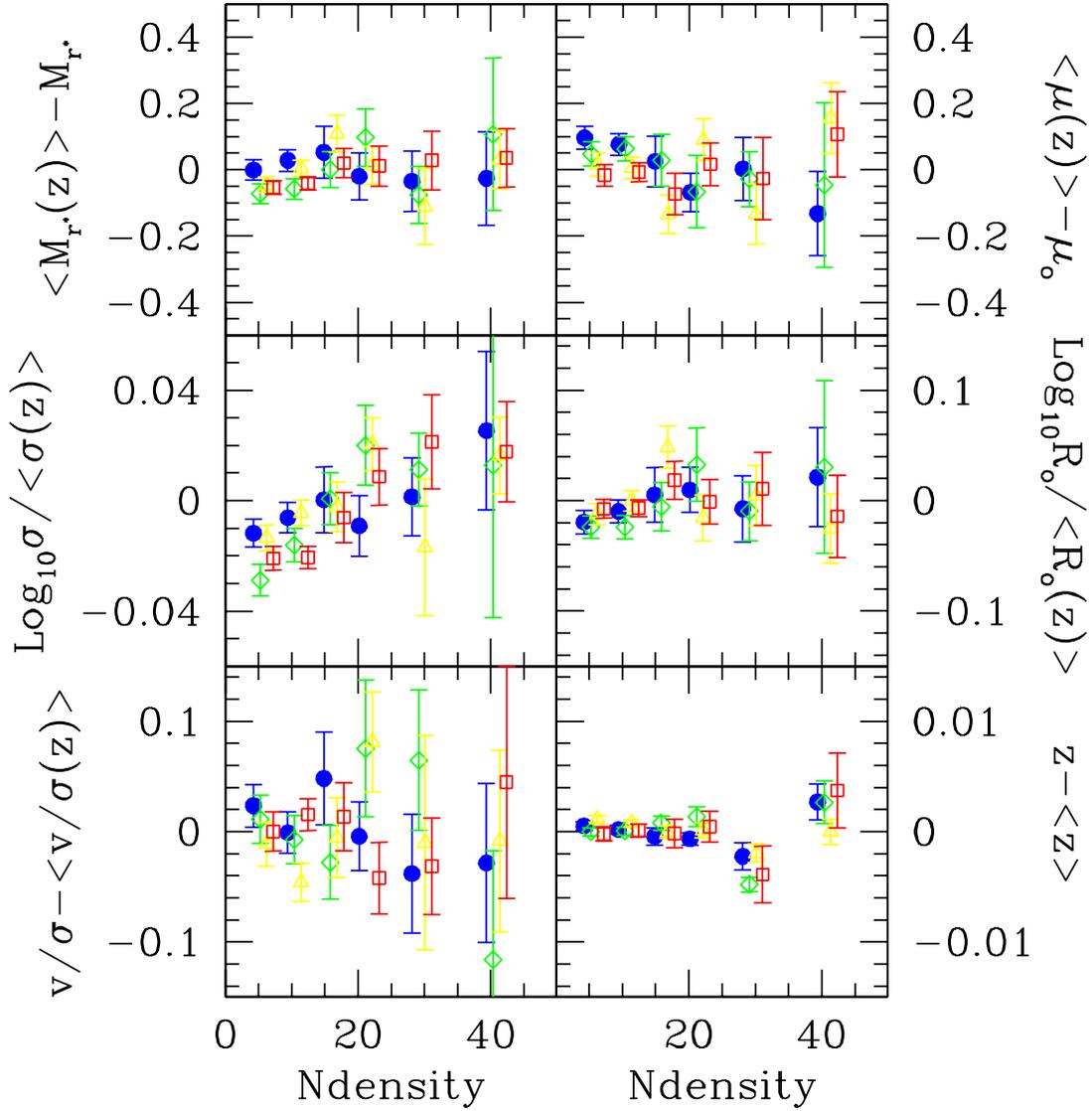}
\vspace{-1.cm}
\caption{Luminosities, surface brightnesses, velocity dispersions, 
sizes, axis ratios, and mean redshifts, as a function of nearby 
early-type neighbours.  The different symbols for each bin in density 
show averages over galaxies in different redshift bins:  circles, 
diamonds, triangles, and squares are for galaxies with redshifts in the 
range $0.075< z\le 0.1$, $0.1< z\le 0.12$, $0.12< z\le 0.14$, and 
$0.14< z\le 0.18$. Although the velocity dispersions appear to increase 
with increasing local density, the increase is small.}  
\label{lrvden}
\end{figure}

This section is devoted to a study of if and how the properties of 
early-type galaxies depend on environment.  To do so, we must come up 
with a working definition of environment.  The set of galaxies in the 
SDSS photometric database is much larger than those for which the survey 
actually measures redshifts.  Some of these galaxies may well lie close 
to galaxies in our sample, in which case they will contribute 
to the local density.  We would like to find some way of accounting for 
such objects when we estimate the local density.  

For a subset of the galaxies in our sample, the colors expected of a 
passively evolving early-type were used to select a region in 
$g^*$--$r^*$ versus $r^*$--$i^*$ color space at the redshift of the 
galaxy of interest.  
All galaxies within 0.1 magnitudes in color of this point were included 
if they were: 
a) within $1h^{-1}$ Mpc of the main galaxy, and 
b) brighter than $-20.25$ in $M_{i^*}$. 
(The box in color space is sufficiently large that the difference 
between this techinique, and using the observed colors themselves 
to define the selection box is not important.)  
These two cuts are made assuming every galaxy in the color-color 
range is at the redshift of the galaxy of interest.  
The end result of this is that each galaxy in the subsample is 
assigned a number of neighbors.  Note that, because of the selection 
on color, our estimate of the local density is actually an estimate 
of the number of neighbors which have the same colors as early-type 
galaxies.  

In what follows, we will often present results for different redshift bins.  
When we do, it is important to bear in mind that our procedure for 
assigning neighbours is least secure in the lowest redshift 
bin (typically $z\le 0.08$).  

Figure~\ref{lrvden} shows how the luminosities, surface brightnesses, 
sizes, velocity dispersions and (a combination of the) axis ratios depend 
on environment.  The different symbols for each bin in density show 
averages over galaxies in different redshift bins:  circles, diamonds, 
triangles, and squares are for galaxies with redshifts in the range 
$0.075< z\le 0.1$, $0.1< z\le 0.12$, $0.12< z\le 0.14$, and 
$0.14< z\le 0.18$.  Error bars show the error in the mean value for 
each bin.  Symbols for the higher redshift bins have been offset slightly 
to the right.  

For any given set of symbols, the bottom right panel shows that the 
mean redshift in each bin in density is not very different from the 
mean redshift averaged over all bins.  This suggests that our procedure 
for estimating the local densities is not biased.  The other panels 
show corresponding plots for the other observed parameters.  
When the number of near neighbours is small, the luminosities, sizes 
and velocity dispersions all increase slightly as the local density 
increases, whereas the surface brightnesses decrease slightly.  
All these trends are very weak.  The bottom right panel shows 
$v/\sigma \equiv \sqrt{(b/a)/[(b/a) - 1]}$, where $b/a$ is the axis 
ratio; $v/\sigma$ is a measure of the ratio of rotational to random 
motions within the galaxy (e.g., Binney 1982).  There are no obvious 
trends with environment.  

\begin{figure}
\centering
\epsfxsize=\hsize\epsffile{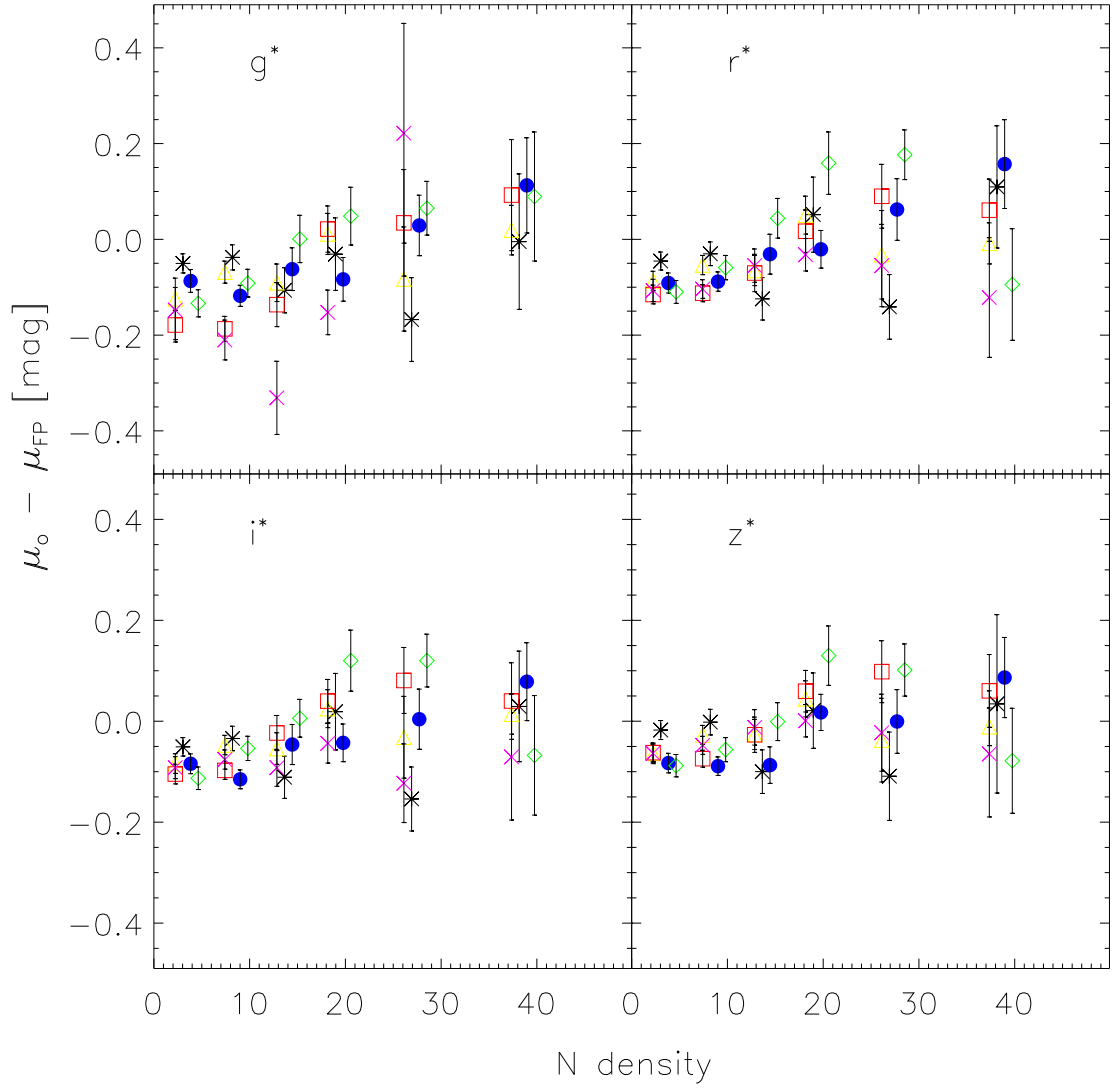}
\vspace{-1.cm}
\caption{Residuals from the FP as a function of number of nearby 
neighbours.  Stars, circles, diamonds, triangles, squares and crosses show 
averages over galaxies in the redshift ranges $z\le 0.075$, $0.075< z\le 0.1$, 
$0.1< z\le 0.12$, $0.12< z\le 0.14$, $0.14< z\le 0.18$ and $z>0.18$.}  
\label{FPenviron}
\end{figure}

It is difficult to say with certainty that the trends with environment 
in the top four panels are significant.  A more efficient way of seeing 
if the properties of galaxies depend on environment is to show the 
residuals from the Fundamental Plane.  As we argue below, this efficiency 
comes at a cost:  if the residuals correlate with environment, it is 
difficult to decide if the correlation is due to changes in luminosity, 
size or velocity dispersion.  

Figure~\ref{FPenviron} shows the differences between galaxy surface 
brightnesses and those predicted by the zero-redshift maximum likelihood 
FP given their sizes and velocity dispersions, as a function of local 
density.  Stars, circles, diamonds, triangles, squares and crosses show 
averages over galaxies in the redshift ranges $z\le 0.075$, $0.075< z\le 0.1$, 
$0.1< z\le 0.12$, $0.12< z\le 0.14$, $0.14< z\le 0.18$ and $z>0.18$.  
Error bars show the error in determining the mean.  (For clarity, the 
symbols have been offset slightly from each other.)  
The plot shows that the residuals depend on redshift---we have already 
argued that this is a combination of evolution and selection effects.  
Notice, that in all redshift bins, the residuals tend to 
increase as local density increases.  This suggests that the scatter 
from the Fundamental Plane depends on environment.  
If the offset in surface brightness is interpretted as evidence that 
galaxies in denser regions are slightly less luminous than their 
counterparts in less dense regions, then this might be evidence 
that they formed at higher redshift.  While this is a reasonable 
conclusion, we should be cautious:  because 
$\mu_o - \mu_{\rm FP}(R_o,\sigma) = -\Delta_1/b$, what we 
have really found is that the residuals in the direction of $R_o$ 
correlate with environment.  
Because $\sigma - \sigma_{\rm FP}(R_o,\mu_o) = -\Delta_1/a$, 
we might also have concluded that the velocity dispersions of galaxies 
in dense regions are systematically different from those of galaxies 
which have the same sizes and luminosities but are in the field.  
Thus, while the Fundamental Plane suggests that the properties of 
galaxies depend on environment, it does not say how.  
For this reason, it will be interesting to remake Figure~\ref{lrvden} 
with a larger sample when it becomes available.  

\section{Line-indices: Chemical evolution and environment}\label{lindices}
The previous section used the Fundamental Plane to estimate how the 
properties of the galaxies in our sample depend on redshift and 
environment.  This section shows that the chemical composition of the 
population at high redshift is different from that nearby.  It also 
presents evidence that galaxies in dense and underdense environments 
are not dramatically different.  

We have chosen to present results for Mg$_2$ (measured in magnitudes), 
and Mg$b$, $\langle{\rm Fe}\rangle$ and H$_\beta$ (measured in Angstroms), 
where $\langle{\rm Fe}\rangle$ represents an average over Fe5270 and Fe5335.  
Mg$_2$ and Mg$b$ are alpha elements, so, roughly speaking, they reflect 
the occurence of Type II supernovae, whereas Fe is produced in SN~Ia.  
All these line indices depend both on the age and the metallicity of 
the stellar population (e.g., Worthey 1994), although Mg and Fe are 
more closely related to the metallicity, whereas the equivalent width 
of H$_\beta$ is an indicator of recent star formation.  

These line indices correlate with velocity dispersion $\sigma$.  
Because $\sigma$ correlates with luminosity, the magnitude limit of our 
sample means that we have no objects with low velocity dispersions 
at high redshifts (Figure~\ref{fig:XzR}).  For this reason, we first 
present results at fixed velocity dispersion.  

\subsection{Correlations with velocity dispersion}
To measure spectral features reliably requires a spectrum with a higher 
signal-to-noise ratio than we have for any individual galaxy in our 
sample.  So we have adopted the following procedure.  
The previous subsection described our criteria for assigning a local 
density to each galaxy.  For a number of small bins in local density, 
we further divided the galaxies in our sample into five bins each of 
redshift, luminosity, velocity dispersion, and effective radius.
The small scatter around the Fundamental Plane implies that galaxies 
in the same bin are very similar to each other.  Therefore, we 
co-added the spectra of all the galaxies in each bin to increase the 
signal-to-noise ratio, and then estimated the Mg$_2$, Mg$b$, H$_\beta$ 
and $\langle{\rm Fe}\rangle$ line-indices in the higher signal-to-noise 
composite spectra following methods outlined by Trager et al. (1998).  
(Analysis of the properties of early-type galaxies using these higher 
signal-to-noise composite spectra is on-going.)  The estimated indices 
were aperture-corrected following J{\o}rgensen (1997):
Mg$_2$ = Mg$_2^{\rm est}$ + $0.04\log_{10}[1.5/(r_o/8)]$, 
${\rm H}_\beta = {\rm H}_\beta^{\rm est} [1.5/(r_o/8)]^{-0.005}$, and 
$\langle{\rm Fe}\rangle = 
    \langle{\rm Fe}\rangle^{\rm est}[1.5/(r_o/8)]^{0.05}$ and 
$\langle{\rm Mg}b\rangle = 
    \langle{\rm Mg}b\rangle^{\rm est}[1.5/(r_o/8)]^{0.05}$.  
(Because the indices were measured for co-added spectra, we use the mean 
values of $r_o$ in each bin to make the aperture correction.)  
In addition, the observed line indices of an individual galaxy are 
broadened by the velocity dispersion of the galaxy.  Simulations similar 
to those we used to estimate the velocity dispersion itself 
(see Appendix~\ref{vmethods}) were used to estimate and correct for 
the effect of the broadening.  For all the indices presented here, 
the required corrections increase with increasing $\sigma$.  
(We use the mean value of $\sigma$ in each bin to make the corrections.)
Whereas the corrections to Mg$_2$ and H$_\beta$ are small (on the order 
of a percent), the corrections to Mg$b$ and Fe are larger (on the order 
of ten percent).  

\begin{figure}[t]
\centering
\epsfxsize=\hsize\epsffile{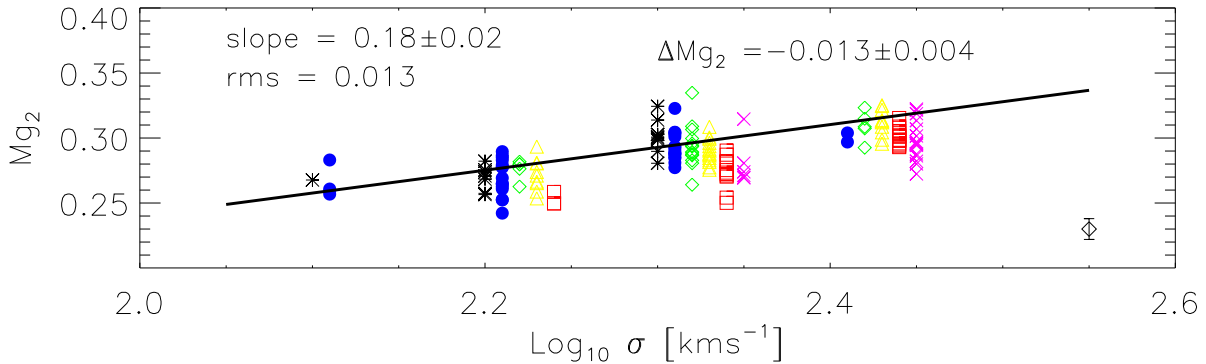}
\vspace{-10cm}
\caption{Mg$_2$ as a function of $\sigma$.  
Stars, filled circles, diamonds, triangles, squares and crosses show 
results from coadded spectra of similar galaxies in successively higher 
redshift bins ($z<0.075$, $0.075<z\le 0.1$, $0.1<z\le 0.12$, 
$0.12<z\le 0.14$, $0.14<z\le 0.18$, and $z>0.18$).  
Symbol with bar in bottom corner shows the typical uncertainty on the 
measurements.  At fixed redshift, Mg$_2$ increases with increasing 
$\sigma$.  At fixed $\sigma$, the spectra from higher redshift galaxies 
are weaker in Mg$_2$.  Text at top right shows the shift between the 
lowest and highest redshift bins averaged over the mean shifts at 
$\log_{10}\sigma=2.2$, 2.3 and 2.4.  
We also performed linear fits to the relations at each redshift, and then 
averaged the slopes, zero-points and rms scatter around the fit.  
Solid line shows the mean relation obtained in this way, and text at 
top shows the averaged slope and averaged scatter.  }
\label{fig:Mg2sigma}
\end{figure}

\begin{figure}
\centering
\epsfxsize=\hsize\epsffile{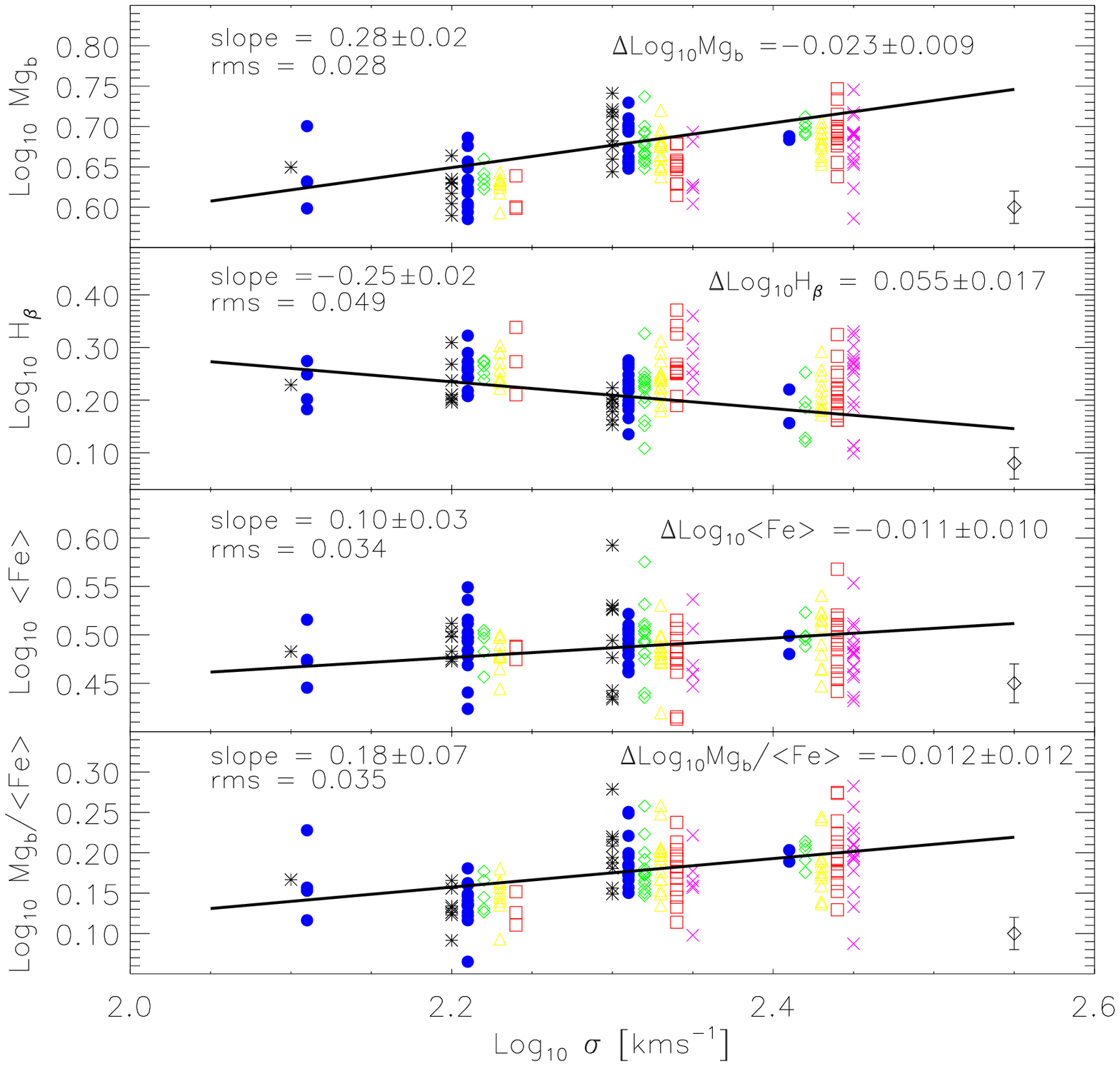}
\caption{Same as previous figure, but now showing the spectral line-indices 
Mg$b$, H$_\beta$, $\langle{\rm Fe}\rangle$, and the ratio [Mg$b$/Fe] 
(top to bottom) as functions of $\sigma$.  
At fixed redshift, Mg$b$ and $\langle{\rm Fe}\rangle$ increase, 
whereas H$_\beta$ decreases with increasing $\sigma$.  
At fixed $\sigma$, the spectra from higher redshift galaxies are weaker 
in both Mg$_2$ and $\langle{\rm Fe}\rangle$, but stronger in H$_\beta$.  
Text at top right shows the shift between the lowest and highest redshfit 
bins averaged over the values at $\log_{10}\sigma=2.2$, 2.3 and 2.4.  }
\label{fig:lindices}
\end{figure}

Figures~\ref{fig:Mg2sigma} and~\ref{fig:lindices} show the results.  
In all panels, stars, filled circles, diamonds, triangles, squares and 
crosses show the redshift bins $z<0.075$, $0.075<z\le 0.1$, $0.1<z\le 0.12$, 
$0.12<z\le 0.14$, $0.14<z\le 0.18$, and $z>0.18$.  The median redshifts 
in these bins are 0.062, 0.086, 0.110, 0.130, 0.156 and 0.200.  For 
clarity, at each bin in velocity dispersion, the symbols for successive 
redshift bins have been offset slightly to the right from each other.  
This should help to separate out the effects of evolution from those which 
are due to the correlation with $\sigma$.  The solid line and text in each 
panel shows the relation which is obtained by performing simple linear 
fits at each redshift, and then averaging the slopes, zero-points, 
and rms scatter around the fit at each redshift.  Text at top right of 
each panel shows the shift between the lowest and highest redshift 
bins, averaged over the values at $\log_{10}\sigma=2.2$, 
$\log_{10}\sigma=2.3$ and $\log_{10}\sigma=2.4$.  Roughly speaking, 
this means that the shifts occur over a range of about $0.2-0.06=0.14$ 
in redshift, which corresponds to a time interval of 1.63~Gyr.  

Figure~\ref{fig:Mg2sigma} shows the well-known correlation between 
Mg$_2$ and $\sigma$ (e.g., Bender 1992): at fixed redshift, 
${\rm Mg}_2\propto \sigma^{0.18\pm 0.02}$ with a scatter around the 
mean relation at each redshift of 0.013~mags.  The fit we find is similar 
to that found in previous work based on spectra of individual (as opposed 
to coadded) galaxies (e.g., J{\o}rgensen 1997; Bernardi et al. 1998; 
Pahre et al. 1998; Kuntschner 2000; Blakeslee et al. 2001), although the 
scatter we find is somewhat smaller.  The slope of our fit is shallower 
than that reported by Colless et al. (1999), but this is probably a 
consequence of our decision to perform linear regression, rather than 
maximum-likelihood, fits.  (Maximum-likelihood fits are difficult at the 
present time because our bins in luminosity are rather large.  We plan to 
make the maximum-likelihood estimate when the sample is larger, so that 
finer bins in luminosity can be made.)  

Although the magnitude limit of our sample makes it difficult to study 
the evolution of the Mg$_2 - \sigma$ relation, a few bins in $\sigma$ 
do have galaxies from a range of different redshifts.  Recall that, for 
the purposes of presentation, the points in each bin in $\sigma$ have been 
shifted to the right by an amount which depends on the redshift bin they 
represent.  When plotted in this way, the fact that the points associated 
with each bin in $\sigma$ slope down and to the right suggests that, at 
fixed $\sigma$, the higher redshift galaxies have smaller values of Mg$_2$.  
Large values of Mg$_2$ are expected to indicate either that the stellar 
population is metal rich, or old, or both.  Thus, in a passively evolving 
population, the relation should be weaker at high redshift.  
This is consistent with the trend we see.  The average value of Mg$_2$ 
decreases by about $(0.013\pm 0.004)$~mags between our lowest and highest 
redshift bins (a range of about 1.63$h^{-1}$Gyr).  
We will return to this shortly.  

The top panel of Figure~\ref{fig:lindices} shows that, at fixed redshift, 
Mg$b\propto \sigma^{0.28\pm 0.02}$, with a scatter of 0.028.  This is 
consistent with the scaling reported by Trager et al. (1998).  
[A plot of Mg$b$ versus Mg$_2$ is well fit by 
$\log_{10}{\rm Mg}b = (1.41\pm 0.18){\rm Mg}_2 + 0.26$; 
this slope is close to the value $0.28/0.18$ one estimates from the 
individual Mg$b-\sigma$ and Mg$_2-\sigma$ relations.  It is also 
consistent with Figure~58 in Worthey (1994).]  
As was the case for Mg$_2$, our data indicate that, at fixed velocity 
dispersion, Mg$b$ is weaker in the higher redshift population.  
The average difference between our lowest and highest redshift bins is 
$0.023\pm 0.009$.  This corresponds to a fractional change in Mg$b$ of 
0.05 over about 1.63$h^{-1}$Gyr.  In contrast 
Bender, Ziegler \& Bruzual (1996) find that Mg$b$ at $z=0.37$ is smaller 
by 0.3\AA\ compared to the value at $z=0$.  This is a fractional change 
of about 0.07 but over a redshift range which corresponds to a time 
interval of 4$h^{-1}$Gyr.  Bender et al. also reported weak evidence of 
differential evolution:  the low $\sigma$ population appeared to have 
evolved more rapidly.  Our Mg$_2-\sigma$ and Mg$b-\sigma$ relations also 
show some evidence of such a trend.  

Colless et al. (1999) define 
Mg$b' = -2.5\log_{10}(1 - {\rm Mg}b/32.5)$, and show that their 
data are well fit by 
Mg$b'\propto (0.131\pm0.017)\log_{10}\sigma - (0.131\pm 0.041)$ 
with a scatter around the mean relation of 0.022~mags.  
Kuntschner (2000) shows that the galaxies in the Fornax cluster follow 
this same scaling, although the scatter he finds is 0.011~mags.  
Our coadded spectra are also consistent with this:  we find 
Mg$b'\propto (0.11\pm 0.01)\log_{10}\sigma$, with a scatter of 0.011~mags.  
[A linear regression of the values of Mg$_2$ and Mg$b'$ in our coadded 
spectra is well fit by ${\rm Mg}_2 = (1.70\pm 0.30)\,{\rm Mg}b' - 0.01$; 
this is slightly shallower than the relation found by Colless et al.: 
${\rm Mg}_2 = 1.94{\rm Mg}b' - 0.05$.]  
The Mg$b'-\sigma$ relation in our data evolves:  in the highest 
redshift bins it is about $(0.022\pm 0.003)$~mags lower than in the lowest 
redshift bins.  Colless et al. find that in the single stellar population 
models of both Worthey (1994) and Vazdekis et al. (1996), changes in age 
or metallicity affect Mg$_2$ about twice as strongly as they do Mg$b'$.  
Figure~\ref{fig:Mg2sigma} suggests that Mg$_2$ has weakened by $-0.013$, 
so we expect Mg$b'$ to have decreased by about $-0.006$.  Therefore, 
this also suggests that the Mg$b$ (or Mg$b'$) evolution we see is large.  

The second panel in Figure~\ref{fig:lindices} shows that, at fixed 
redshift, ${\rm H}_\beta\propto \sigma^{-0.25\pm 0.02}$ with a scatter 
of 0.049.  This is consistent with J{\o}rgensen (1997), who found 
$\log_{10}{\rm H}_\beta = (-0.231\pm 0.082)\log_{10}\sigma +0.825$, 
with a scatter of 0.061.  At fixed $\sigma$, H$_\beta$ is stronger 
in the higher redshift spectra.  On average, the value of H$_\beta$ 
increases by about $0.055\pm 0.017$ between our lowest and highest 
redshift bins.  
An increase of star formation activity with redshift is consistent 
with a passively evolving population.  When a larger sample is available, 
it will be interesting to see if the scatter in H$_\beta$ at fixed 
$\sigma$ also increases with redshift.  

The third panel of Figure~\ref{fig:lindices} shows that, at fixed 
redshift, $\langle{\rm Fe}\rangle\propto \sigma^{0.10\pm 0.03}$ with 
a scatter of 0.034.  This lies between the $0.075\pm 0.025$ scaling 
and scatter of 0.041 found by J{\o}rgensen (1997), and that found 
by Kuntschner (2000):  
$\langle{\rm Fe}\rangle\propto \sigma^{0.209\pm 0.047}$.  
At fixed $\sigma$, $\langle{\rm Fe}\rangle$ may be slightly smaller 
at higher redshift, although the shift is not statistically significant:  
the change in $\log_{10}\langle{\rm Fe}\rangle$ is $0.011\pm 0.010$.  

The ratio Mg$b$/$\langle{\rm Fe}\rangle$ is sometimes used to 
constrain models of how early-type galaxies formed 
(e.g., Worthey, Faber \& Gonzalez 1992; Thomas, Greggio \& Bender 1999; 
but see Matteucci, Ponzone \& Gibson 1998).  In our coadded spectra, 
$\log_{10} {\rm Mg}b/\langle{\rm Fe}\rangle = 
(0.18\pm 0.07)\log_{10}\sigma - 0.23$ with a scatter of 0.035 
(bottom panel of Figure~\ref{fig:lindices}).  
The slope of this relation equals the difference between the slopes 
of the Mg$b-\sigma$ and $\langle {\rm Fe}\rangle - \sigma$ relations, 
and the evidence for evolution is not statistically significant.  
This correlation should be interpreted as evidence that the contribution 
of Fe to the total metallicity is depressed, rather than that alpha 
elements are enhanced, at high $\sigma$ (e.g., Worthey et al. 1992; 
Weiss, Peletier \& Matteucci 1995; Greggio 1997; Trager et al. 2000a).  

If the evolution in Mg and Fe is due to the same physical process, 
then one might have wondered if residuals from the Mg$b -\sigma$ 
relation are correlated with residuals from the 
$\langle{\rm Fe}\rangle - \sigma$ relation.  
This will be easier to address when the sample is larger.  
At the present time, we see no compelling evidence for such a 
correlation---we have not included a figure showing this explicitly.  
In addition, at any given redshift, galaxies which are richer in 
Mg$_2$ or $\langle{\rm Fe}\rangle$ than they should be (given 
their velocity dispersion), are neither more nor less likely to be 
richer in H$_\beta$ than expected---recent star formation is not 
correlated with metallicity.  

Having shown that the different line index--$\sigma$ relations are 
evolving, we use simple stellar population models to see what the 
evolution we see implies.  

\subsection{Comparison with stellar population models}\label{ssp}
The predictions of single age stellar population models 
(e.g., Bruzual \& Charlot 1993; Worthey 1994; Vazdekis et al. 1996; 
Tantalo et al. 1998) are often summarized as plots of H$_\beta$ versus 
Mg$b$ (or Mg$_2$) and $\langle{\rm Fe}\rangle$.  The usual caveats 
noted by these authors about the limitations of these models, and the 
assumption that all the stars formed in a single burst, apply.  
In addition, comparison with data is complicated because the models 
assume that the ratio of $\alpha$-elements to Fe~peak elements in 
early-type galaxies is the same as in the Sun, whereas, in fact, it 
differs from the solar value by an amount which depends on velocity 
dispersion (e.g., Worthey et al. 1992; see bottom panel of 
Figure~\ref{fig:lindices}).  We use a simplified version of the 
method described by Trager et al. (2000a) to account for this.  

\begin{figure}
\centering
\epsfxsize=\hsize\epsffile{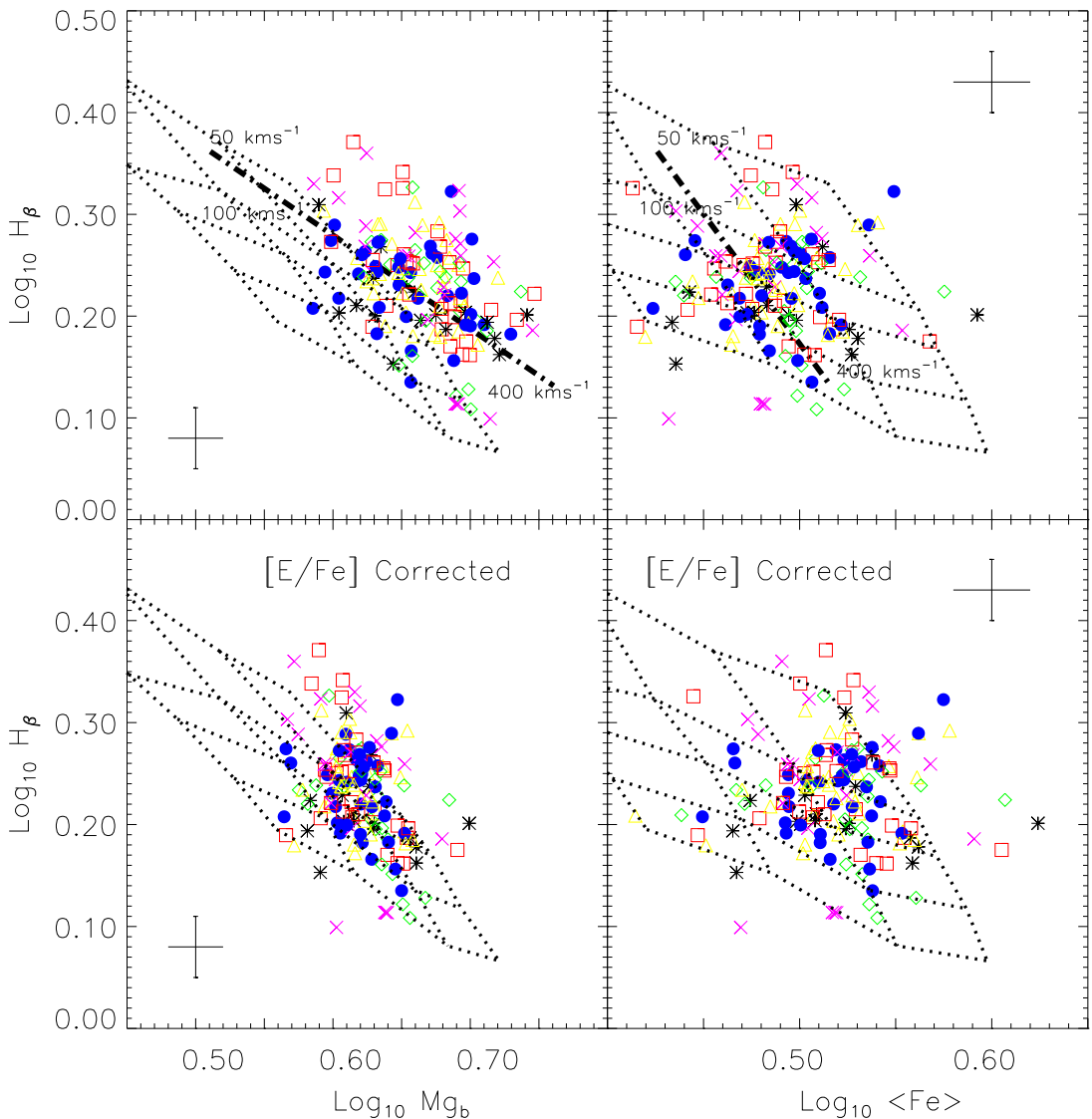}
\caption{H$_\beta$ versus Mg$b$ (left) and $\langle{\rm Fe}\rangle$ 
(right) for the coadded spectra in our sample.  Different symbols show 
different redshift bins;  the higher redshift population (squares 
and crosses) appears to show a larger range in H$_\beta$ compared 
to the low redshift population (stars and circles).  Cross in each 
panel shows the typical uncertainty on the measurements.  
Dotted grid shows a single age, solar abundance (i.e., [E/Fe] $=0$), 
stellar population model (from Worthey 1994); lines of constant age run 
approximately horizontally (top to bottom show ages of 2, 5, 8, 12 and 
17~Gyrs), lines of constant metallicity run approximately vertically 
(left to right show [Fe/H] $=-0.25$, 0, 0.25, 0.5).  
The two top panels provide different estimates of the age and metallicity, 
presumably because the [E/Fe] abundances in our data are different from 
solar.  In the bottom panels, this difference has been accounted for, 
and the age and metallicity estimates agree.  
Dot-dashed line in top panels shows what one 
expects if there is no scatter around the line index--$\sigma$ 
relations (solid lines in Figure~\ref{fig:lindices}).  }
\label{HbMgbFe}
\end{figure}

\begin{figure}
\centering
\epsfxsize=\hsize\epsffile{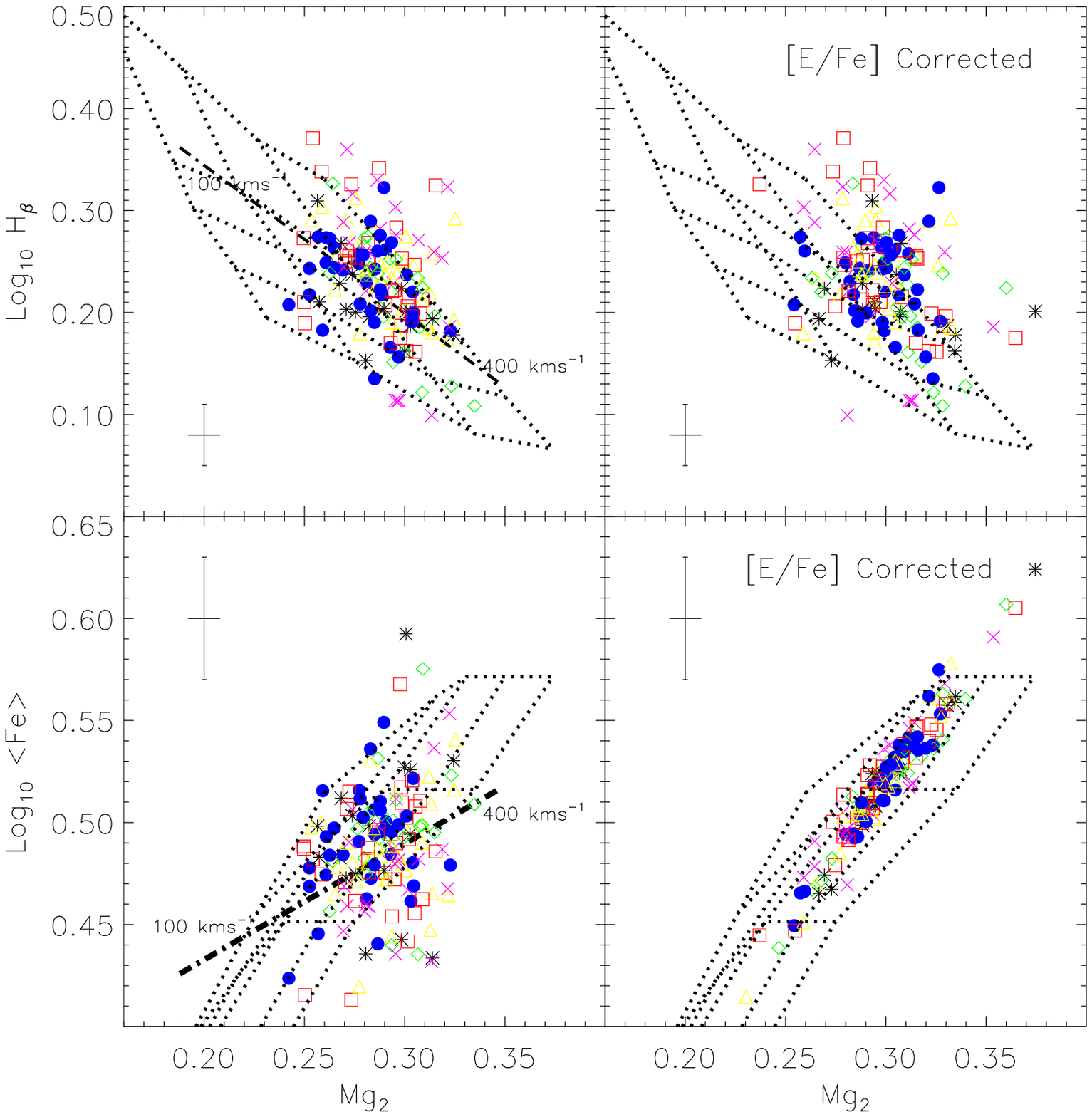}
\caption{Line indices H$_\beta$ and $\langle{\rm Fe}\rangle$ 
versus Mg$_2$ for the coadded spectra in our sample.  
Symbols (same as previous figure) show results for different redshift 
bins.  Dot-dashed line shows the relation one predicts if there were 
no scatter around the individual line index--$\sigma$ relations.  
Evolution is expected to move points upwards and to the left for 
H$_\beta$ versus Mg$_2$ (top panels), but downwards and left, and 
along the dot-dashed line in the case of $\langle{\rm Fe}\rangle$ and 
Mg$_2$ (bottom panels), although selection effects make these trends 
difficult to see.  Dotted grids show the same single stellar 
population model as in the previous figure (from Worthey 1994).
Age and metallicity estimates in the top and bottom panels 
are inconsistent if solar abundance is assumed (left panels), 
but the estimates agree once differences in abundances have been accounted 
for (right panels). }
\label{fig:HbFeMg2}
\end{figure}

Figure~\ref{HbMgbFe} shows such a plot.  The dotted grids (top and bottom 
left are the same, as are top and bottom right) show a single age, solar 
abundance (i.e., [E/Fe] $=0$), stellar population model (from Worthey 1994); 
lines of constant age run approximately horizontally (top to bottom show 
ages of 2, 5, 8, 12 and 17~Gyrs), lines of constant metallicity run 
approximately vertically (left to right show [Fe/H] $=-0.25$, 0, 0.25, 0.5).  
Points in the panels on the top show the values of H$_\beta$ and Mg$b$ 
(left) and $\langle{\rm Fe}\rangle$ (right) for the coadded spectra in 
our sample.  Different symbols show different redshift bins;  the higher 
redshift population (squares and crosses) appears to show a larger range 
in H$_\beta$ compared to the low redshift population (stars and circles).  
Cross in each panel shows the typical uncertainty on the measurements.  

The heavy dot-dashed lines in the top panels show the relation between 
H$_\beta$ and Mg$b$ or $\langle{\rm Fe}\rangle$ one predicts if there 
were no scatter around the individual line index--$\sigma$ relations 
(shown as solid lines in Figure~\ref{fig:lindices}):  
${\rm H}_\beta\propto {\rm Mg}b^{-0.89}$ and 
${\rm H}_\beta\propto \langle {\rm Fe}\rangle^{-2.5}$.  
We have included them to help disentagle the evolution we saw in the 
individual line index--$\sigma$ relations from the effect of the 
magnitude limit.  

The evolution in the Mg$b -\sigma$, $\langle {\rm Fe}\rangle - \sigma$, 
and H$_\beta - \sigma$ relations suggest that the higher redshift sample 
should be displaced upwards and to the left, with a net shift in 
zero-point of about 0.03 in both of the upper panels of 
Figure~\ref{HbMgbFe}.  We estimate these shifts as follows:  
Let $y_0 = s x_0 + c_0$ denote the mean relation between one line index 
$y$ and another $x$ at $z=0$.  Because we know that line indices 
correlate with $\sigma$, and we know how much the individual relations 
evolve, we can estimate the evolution in the index--index relation by 
setting 
$y(z) = y_0 + \Delta y = s x_0 + c_0 + \Delta y 
     = s x(z) + c_0 + \Delta y - s \Delta x$.
For $x={\rm Mg}b$ and $y={\rm H}_\beta$ we must set $\Delta y = 0.055$, 
$\Delta x = -0.023$, and $s = -0.25/0.28$ (from Figure~\ref{fig:lindices}).  
Thus the zero-point of a plot of H$_\beta$ versus Mg$b$ is expected to 
evolve by 0.034 between the lowest and highest redshift bins in our 
sample.  A similar analysis for $\langle{\rm Fe}\rangle$ shows that the 
zero-point of the H$_\beta - \langle{\rm Fe}\rangle$ 
relation is expected to evolve by 0.027.  
(These estimates assume that the slopes of the individual relations do 
not evolve.  Bender et al. (1996) present some evidence that Mg$b$ at 
high $\sigma$ evolves less than at low $\sigma$, suggesting that the 
slope of the Mg$b-\sigma$ relation was steeper in the past. A comparison 
of the $\log_{10}\sigma=2.3$ and 2.4 bins in Figure~\ref{fig:lindices} 
is in approximate agreement with this.  Because these estimates are of 
the order of the error in the measurements, we have not worried about 
the effects of a change in slope.)

Although the expected evolution is of the order of the error in the 
measurements, the top two panels in Figure~\ref{HbMgbFe} do appear 
to show that the high redshift population (squares and crosses) is 
displaced slightly upwards.  The shift to the left is not apparent, 
however, because of the selection effect:  evolution moves the high 
$\sigma$ objects of the high redshift sample onto the the lower $\sigma$ 
points of the low redshift sample, but the low $\sigma$ objects at 
higher redshift, which would lie clearly above and to the left, are not 
seen because of the selection effect.  Note that the selection effect 
works so that evolution effects are suppressed, rather than enhanced in 
plots like Figure~\ref{HbMgbFe}; therefore, a simple measurement of 
evolution in the upper panels should be interpretted as a lower limit 
to the true value.  

The top two panels show that our sample spans a range of about 0.3 
or more in metallicity, and a large range of ages.  However, notice 
that the two panels provide different estimates of the mean ages and 
metallicities in our sample.  This is because the [E/Fe] abundances 
in our data are different from solar.  Trager et al. (2000a) describe 
how to correct for this.  Our measurement  errors in Mg and Fe are 
larger than theirs, so we have adopted the following simplified version 
of their prescription.  

Let [Z(H$_\beta,\langle{\rm Fe}\rangle$)/H] denote the estimate of 
the metallicity given by the top right panel of Figure~\ref{HbMgbFe}:  
this estimate uses the observed values of H$_\beta$ and 
$\langle{\rm Fe}\rangle$, and the Worthey (1994) solar abundance 
ratio models.  
Trager et al. (2000a) argue that non-solar abundances change the 
relation between [Fe/H] and the true metallicity [Z/H]:  
${\rm [Fe/H]} = {\rm [Z/H]} - A{\rm [E/Fe]}$, where $A\approx 0.93$.  
Trager et al. (2000b) argue that 
 ${\rm [E/H]}\approx 0.33\log_{10}\sigma - 0.58$, 
and that the relation is sufficiently tight that one can substitute 
$\sigma$ for [E/H].  Therefore, we define a corrected metallicity 
[Z/H]$_{\rm corr}\approx$ 
[Z(H$_\beta,\langle{\rm Fe}\rangle$)/H] $+ 0.33\,A(\log_{10}\sigma - 0.58)$.
Trager et al. (2000a) also argue that correcting for nonsolar [E/H] makes 
a negligible change to H$_\beta$.  Therefore, we combine the measured 
value of H$_\beta$ with [Z/H]$_{\rm corr}$ to compute a corrected
age $\tau_{\rm corr}$.  We then use Worthey's model with these corrected 
ages and metallicities to obtain corrected values of Mg$b$ and 
$\langle{\rm Fe}\rangle$.  These are plotted in the bottom panels.  
By construction, the values of H$_\beta$ in all four panels are the 
same, and the age and metallicity estimates in the bottom two panels 
agree.  The differences between our top and bottom panels are similar 
to the differences between Figures~1 and~3 of Trager et al. (2000a),
suggesting that our simple approximate procedure is reasonably accurate.  

We apply the same correction procedure to plots of 
H$_\beta - {\rm Mg}_2$ and $\langle{\rm Fe}\rangle - {\rm Mg}_2$ 
in Figure~\ref{fig:HbFeMg2}.  
The dot-dashed lines in the panels on the left show 
$\log_{10}{\rm H}_\beta\propto -1.39\,{\rm Mg}_2$ and 
$\log_{10}\langle{\rm Fe}\rangle\propto 0.56\,{\rm Mg}_2$.  
Matteucci et al. (1998) report that a fit to a compilation of 
$\langle{\rm Fe}\rangle - {\rm Mg}_2$ data from various sources has 
slope 0.6.  The dot-dashed line does not appear to provide a good 
fit in either to top or the bottom panel.  In the bottom panel at 
least, this is because of a combination of evolution and selection 
effects:  fitting the relation separately for different redshift bins 
and averaging the results yields a line which is slightly steeper than 
the dot-dashed line.  

The expected evolution is upwards and to the left for 
H$_\beta - {\rm Mg}_2$ and down and left for 
$\langle{\rm Fe}\rangle - {\rm Mg}_2$, with net shifts in zero-points 
of $0.037$ and $-0.004$.  Thus, in the bottom panel, evolution moves 
points along the dot-dashed line.  As with the previous plot, the 
selection effect makes evolution difficult to see.  
The dotted grid shows the Worthey (1994) model for these relations.  
Comparison of the two bottom panels suggests that much of the scatter 
in the observed $\langle{\rm Fe}\rangle - {\rm Mg}_2$ relation is 
due to differences in enhancement ratios.  

We can now use the models to estimate the mean corrected ages and 
metallicities of the galaxies in our sample as a function of redshift.  
The mean metallicity is 0.33 and shows almost no evolution.  
The mean age in our lowest redshift bin (stars, median redshift 0.06) 
is 8~Gyrs, whereas it is 6~Gyrs in the highest redshift bin (crosses).  
The redshift difference corresponds to a time interval of 1.63~Gyr; 
if the population has evolved passively, this should equal the difference 
in ages from the stellar population models.  While the numbers are 
reasonably close, it is important to note that, because of the magnitude 
limit, our estimates of the typical age and metallicity at high redshift 
are biased towards high values, whereas our estimate of the evolution 
relative to the population at low redshift is probably biased low.  
Nevertheless, it is reassuring that this estimate of a formation time of 
8 or 9~Gyrs ago is close to that which we use to make our K-corrections.  

\begin{figure}
\centering
\epsfxsize=\hsize\epsffile{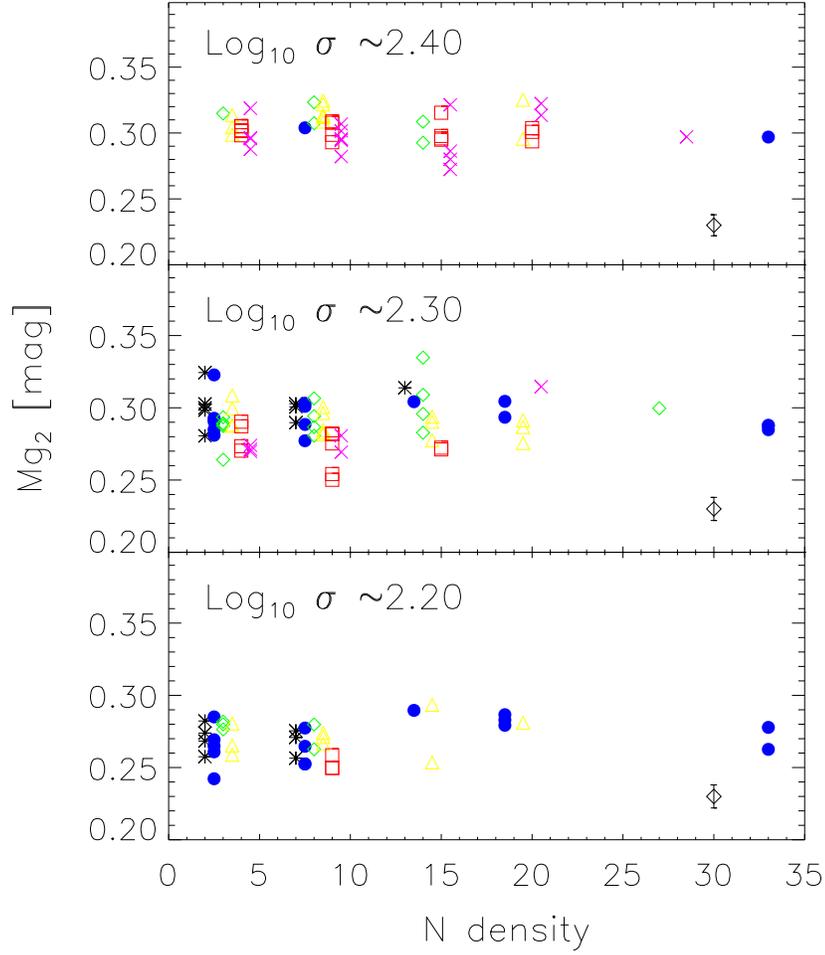}
\vspace{-2cm}
\caption{Mg$_2$--density relation for the galaxies in our sample.  
Symbols show the different redshift bins; higher redshift bins 
have been offset slightly to the right.  Symbol with bar in bottom 
right shows the typical uncertainty on the measurements. }
\label{densityMg2}
\end{figure}

\begin{figure}
\centering
\epsfxsize=\hsize\epsffile{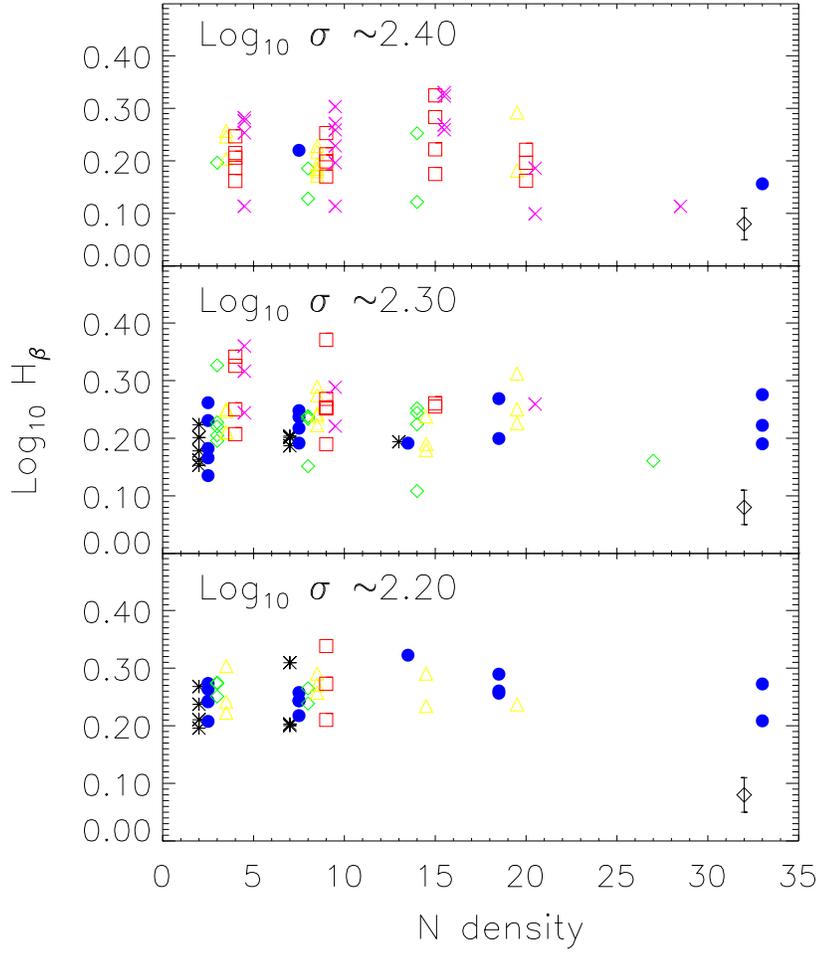}
\vspace{-2cm}
\caption{As for the previous figure, but for the H$_\beta$--density 
relation.  At fixed velocity dispersion, H$_\beta$ is slightly higher 
at high redshift, but there is no significant dependence on environment.}
\label{densityHb}
\end{figure}

\subsection{Dependence on environment}
We now turn to a study of how the coadded spectra depend on environment.  
Figures~\ref{densityMg2} and~\ref{densityHb} show the strength of Mg$_2$ 
and H$_\beta$ in a few small bins in $\sigma$, as a function of local 
density.  The different symbols in each density bin represent composite 
spectra from different redshifts---higher redshift bins have offset 
slightly to the right.  This allows us to separate the effects of 
evolution from those of environment.  
Figure~\ref{densityMg2} shows that, at fixed $\sigma$, Mg$_2$ 
decreases with redshift.  At any given redshift, the strength of Mg$_2$ 
is independent of local density.  (Our sample is not large enough 
to say with certainty if the evolution depends on environment.)  
A similar plot for H$_\beta$ also shows strong evolution with 
redshift, and no dependence on environment (Figure~\ref{densityHb}).  
Similar plots of $\langle{\rm Fe}\rangle$ and  [Mg$_2$/Fe] also 
show little if any dependence on redshift and no dependence on 
environment, so we have not shown them here.  

We caution that our definition of environment is limited, because 
it is defined by early-type galaxies only.  In addition, because 
we must divide our total sample up into bins in luminosity, size, 
radius, and redshift, and then by environment, the statistical 
significance of the results here would be greatly improved by 
increasing the sample size.  Analysis of environmental dependence 
using a larger sample is ongoing (Eisenstein et al. 2002).  

In conclusion, although we have evidence from the Fundamental Plane 
that early-type galaxies in dense regions are slightly different from 
their counterparts in less dense regions (Figure~\ref{FPenviron}), 
these differences are sufficiently small that the strengths of spectral 
features are hardly affected (Figures~\ref{densityMg2} and~\ref{densityHb}).  
However, the coadded spectra provide strong evidence that the 
chemical composition of the population at low and high redshifts 
is different (Figures~\ref{fig:Mg2sigma}--\ref{fig:HbFeMg2}).

\section{The color--magnitude and color$-\sigma$ relations}\label{cms}
The colors of early-type galaxies are observed to correlate with 
their luminosities, with small scatter around the mean relation 
(e.g., Baum 1959; de Vaucouleurs 1961; Bower, Lucey \& Ellis 1992a,b).  
In this section we examine these correlations using the model colors 
output by the SDSS photometric pipeline.  We show that, in fact, the 
primary correlation is color with velocity dispersion:  color--magnitude 
and color--size relations arise simply because magnitude and size are 
also correlated with velocity dispersion.  We also study the effects of 
evolution and environment on the color--velocity dispersion relation.  
The second subsection shows what happens if we use a different definition 
of galaxy color.  (Although the qualitative results of this section do 
not depend on the K-correction we apply, some of the quantitative results 
do.  See Appendix~\ref{kcorrs} for details.)

\subsection{Galaxy colors:  evolution and environment}

\begin{figure}
\centering
\epsfxsize=\hsize\epsffile{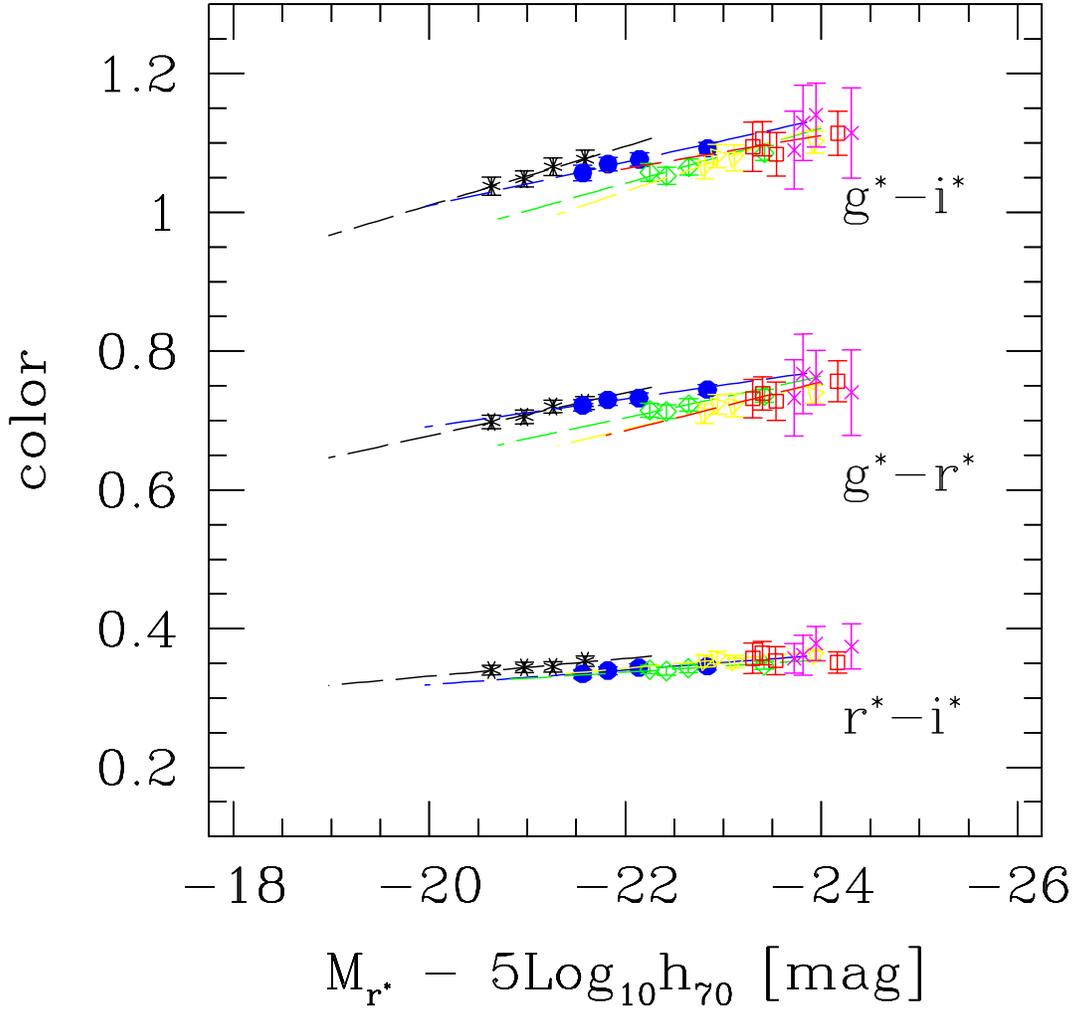}
\caption[]{Color versus $r^*$ magnitude in volume limited subsamples.  
Symbols show the median color at fixed luminosity as measured in the 
different volume limited subsamples, error bars show three times the 
uncertainty in this median.  Dashed lines show linear fits to the relation 
in each subsample.  The slope of the relation is approximately the same in 
all the subsamples, although the relations in the more distant subsamples 
are offset blueward.  This offset is greater for the $g^*-i^*$ colors than 
for $r^*-i^*$.  Because of this offset, the slope of a line which passes 
through the relation defined by the whole sample is very different from 
the slope in each of the subsamples.}
\label{cmag3}
\end{figure}

\begin{figure}
\centering
\epsfxsize=\hsize\epsffile{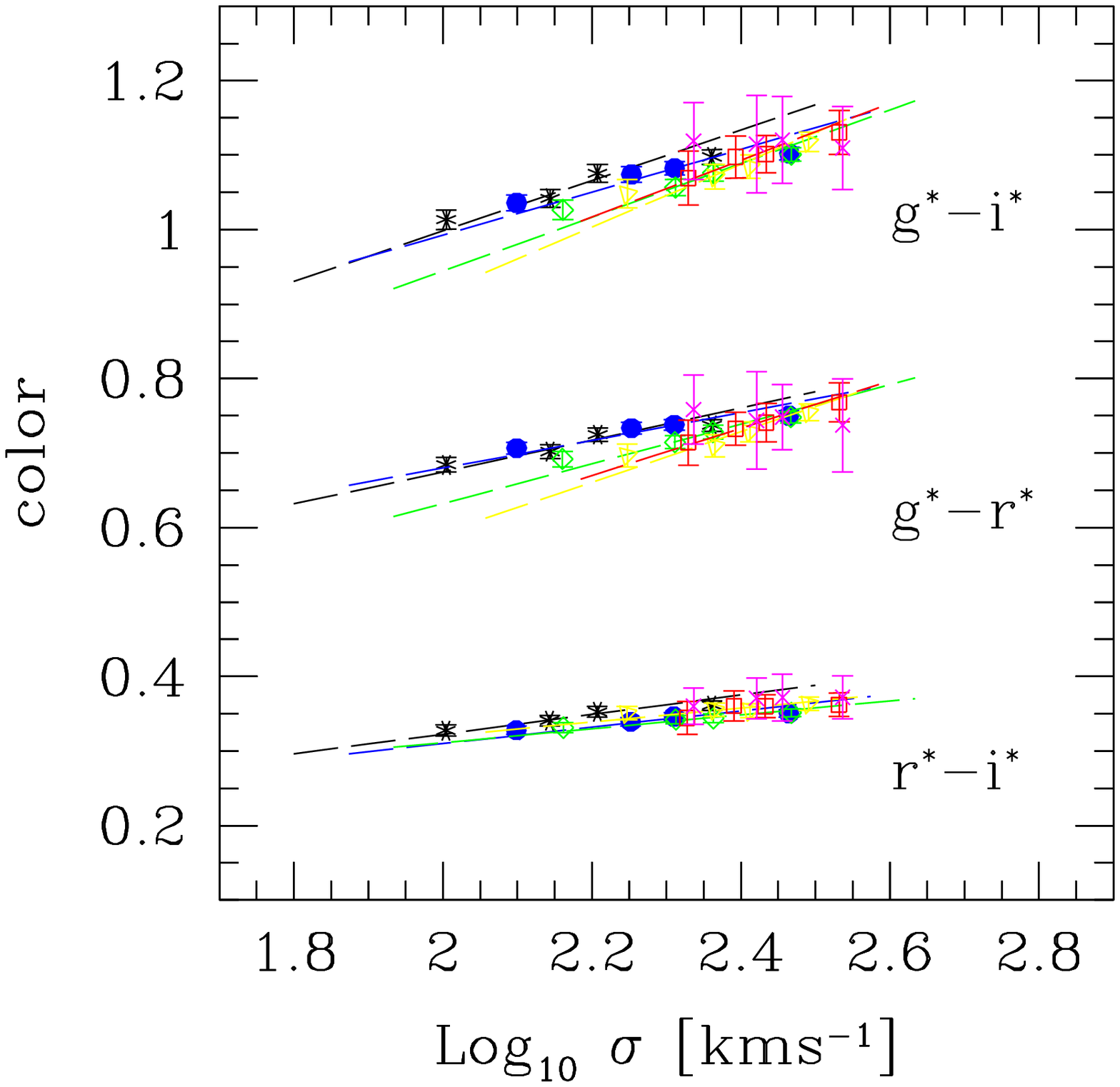}
\caption[]{Same as previous figure, but now showing color versus 
velocity dispersion.  Redder galaxies have larger velocity dispersions.  
Dashed lines show that the slope of the relation is approximately the 
same in all the subsamples, but that the relations in the more distant 
subsamples are offset blueward.  The offset is similar to that in the 
color--magnitude relations.}
\label{csig3}
\end{figure}

We begin with a study of the color--magnitude relation in our data set.  
Estimating the slope of this relation is complicated because our sample 
is magnitude limited and spans a relatively wide range of redshifts, and 
because the slope of the color--magnitude relation is extremely shallow.  
At any given redshift, we do not have a wide range of magnitudes over 
which to measure the relation.  If we are willing to assume that this 
relation does not evolve, then the different redshift bins probe different 
magnitudes, and we can build a composite relation by stacking together 
the relations measured in any individual redshift bin.  
However, the shallow slope of the relation means that small changes in 
color, whether due to measurement errors or evolution, result in large 
changes in $M$.  Thus, if the colors of early-type galaxies evolve even 
weakly, the slope of the composite color--magnitude relation is drastically 
affected.  We can turn this statement around, of course, and use the 
color--magnitude relation as a sensitive test of whether or not the colors 
of the galaxies in our sample have evolved.  

\begin{table}[t]
\centering
\caption[]{Maximum-likelihood estimates of the joint distribution of 
color, $r^*$ magnitude and velocity dispersion and its evolution.  
At redshift $z$, the mean values are $C_*-Pz$, $M_*-Qz$, and $V_*$, 
and the covariances are $\sigma^2_{CM}=\sigma_C\sigma_M\,\rho_{CM}$ etc.\\}
\begin{tabular}{cccccccccccc}
\tableline 
Color & $C_*$ & $\sigma_C$ & $V_*$ & $\sigma_V$ & $M_*$ 
& $\sigma_M$ & $\rho_{CM}$ & $\rho_{CV}$ & $\rho_{VM}$ & Q & P \\
\hline\\
$g^*-r^*$ & 0.736 & 0.0570 & 2.200 & 0.1112 & $-21.15$ & 0.841 & $-0.361$ &
0.516 & $-0.774$ & 0.85 & 0.30\\
$r^*-i^*$ & 0.346 & 0.0345 & 2.200 & 0.1112 & $-21.15$ & 0.841 & $-0.301$ &
0.401 & $-0.774$ & 0.85 & 0.10\\
$r^*-z^*$ & 0.697 & 0.0517 & 2.203 & 0.1114 & $-21.15$ & 0.861 & $-0.200$ &
0.346 & $-0.774$ & 0.85 & 0.15\\
\tableline
\end{tabular}
\label{MLcmag}
\end{table}

Figure~\ref{cmag3}, the relation between the absolute magnitude 
in $r^*$ and the $g^*-r^*$, $g^*-i^*$ and $r^*-i^*$ colors, illustrates 
our argument.  The figure was constructed by using the same volume limited 
subsamples we used when analyzing the $r^*$ luminosity function.  
Symbols with error bars show the median, and the error in this median, 
at fixed luminosity in each subsample.  Dashed lines show the mean color 
at fixed magnitude in each subsample; the slopes of these mean relations 
and the scatter around the mean are approximately the same (we will 
quantify the slopes of these relations shortly) but the zero-points are 
significantly different.  
All three color--magnitude relations show qualitatively similar trends, 
although the trend to shift blueward with increasing redshift is more 
dramatic for the relation involving $g^*-i^*$.  For example, $r^*-i^*$ 
is bluer by about 0.03~mags in the most distant subsample than in the 
nearest, whereas the shift in $g^*-i^*$ is closer to 0.09 mags.  
Because of the blueward shifts, the slope of a linear fit to the whole 
catalog, over the entire range in absolute magnitudes shown, is much 
shallower than the slopes of the individual subsamples.

How much of the evolution in Figure~\ref{cmag3} is due to changes in 
color, and how much to changes in luminosity?  
To address this, Figure~\ref{csig3} shows the same plot, but with $r^*$ 
magnitude replaced by velocity dispersion.  As before, the different 
dashed lines show fits to the color-$\sigma$ relations in the individual 
subsamples; the slopes of, and scatter around, the mean relations are 
similar but the zero-points are different.  The magnitude of the shift 
in color is similar to what we found for the color--magnitude relation, 
suggesting that the offsets are due primarily to changes in colors rather 
than luminosity.  

At first sight this might seem surprising, because single-burst models 
suggest that the evolution in the colors is about one-third that of 
the luminosities.  However, because the slope of the color--magnitude 
relation is so shallow, even a large change in magnitudes produces only 
a small shift in the zero-point of the colors.  To illustrate, let 
$(C - C_*) = -0.02(M - M_*)$ denote the color--magnitude relation at 
the present time.  Now let the typical color and magnitude change by 
setting $C_*\to C_* + \delta C$ and $M_*\to M_* + \delta M$, but assume 
that the slope of the color--magnitude relation does not.  This 
corresponds to a shift in the zero-point of $0.02\delta M + \delta C$,  
demonstrating that $\delta C$ dominates the change in the zero point 
even if it is a factor of ten smaller than $\delta M$.  

To account both for selection effects and evolution, we have computed 
maximum likelihood estimates of the joint color--magnitude--velocity 
dispersion distribution, allowing for evolution in the magnitudes and 
the colors but not in the velocity dispersions:  
i.e., the magnitudes and colors are assumed to follow Gaussian 
distributions around mean values which evolve, say $M_*(z) = M_* - Qz$ 
and $C_*(z) = C_* - Pz$, but the spread around the mean values, and the 
correlations between $C$ and $M$ do not evolve.  We have chosen to 
present results for $g^*-r^*$, $r^*-i^*$ and $r^*-z^*$ only, since 
the other colors are just linear combinations of these.  We chose to 
have $r^*$ in each of the colors we do explicitly analyze for two 
reasons: this is the band in which the SDSS spectroscopic sample is 
selected; it has a special status with regard to the SDSS `model' 
colors (see Section~\ref{cgrad}).  Also, we only present results for 
the color--$r^*$-magnitude relation, because, as we argue later, 
color correlates primarily with $\sigma$, which is independent of 
waveband.  The results are summarized in Table~\ref{MLcmag}.  
Notice that the colors at redshift zero are close to those of the 
Coleman, Wu \& Weedman (1980) templates; that the evolution in 
color is smaller than in magnitude, and consistent with the individual 
estimates of the evolution in magnitudes found earlier; 
and that the best fit distributions of $M_*$ and $V_*$ are the same for 
all three colors, and are similar to the values we found when studying 
the Fundamental Plane.

As was the case when we studied the Fundamental Plane, various combinations 
of the coefficients in Table~\ref{MLcmag} yield the maximum likelihood 
estimates of the slopes of linear regressions of pairs of variables; 
some of these are summarized in Table~\ref{CMVslopes}.  One interesting 
combination is the relation between color and magnitude at fixed velocity 
dispersion:  
\begin{displaymath}
{\Bigl\langle C- \langle C|V\rangle |M\Bigr\rangle\over\sigma_{C|V}} = 
{M-\langle M|V\rangle\over \sigma_{M|V}} \times 
{(\rho_{CM} - \rho_{CV}\rho_{VM})\over\sqrt{(1-\rho_{VM}^2)(1-\rho_{CV}^2)}}
\qquad {\rm at\ fixed\ } V=\log_{10}\sigma.
\end{displaymath}
Inserting the values from Table~\ref{MLcmag} shows that, at fixed 
velocity dispersion, there is little correlation between color and 
luminosity.  In other words, the color--magnitude relation is almost 
entirely due to the correlation between color and velocity dispersion.  

\begin{figure}
\centering
\epsfxsize=\hsize\epsffile{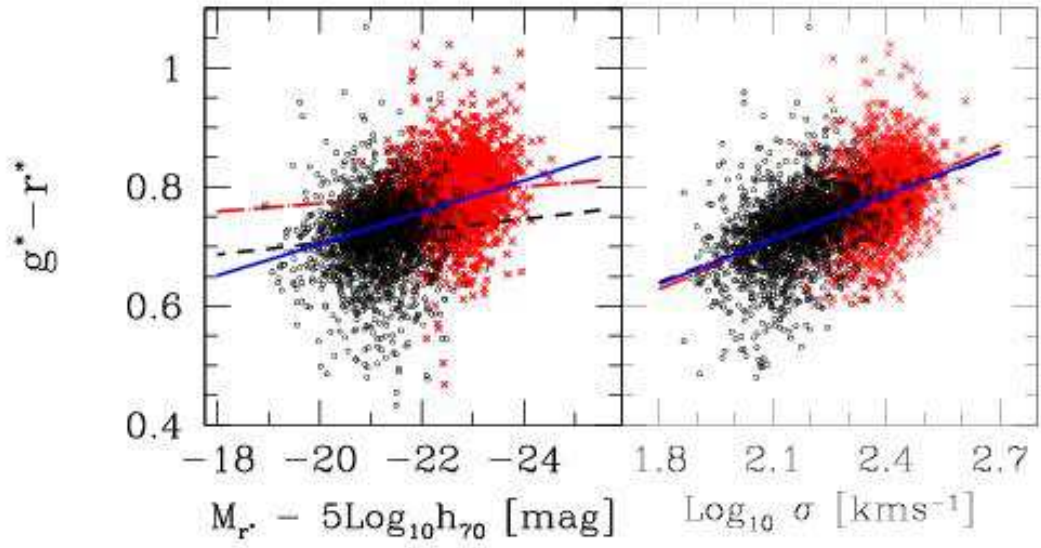}
\vspace{0cm}
\caption[]{Relation between color and magnitude at fixed velocity 
dispersion (left) and between color and velocity disperion at fixed 
magnitude (right). }
\label{clrmagv}
\end{figure}

Figure~\ref{clrmagv} shows this explicitly.  The dashed and dot--dashed 
lines show fits to the relation between color and magnitude at low 
(circles) and high (crosses) velocity dispersion (in the plots,
the maximum likelihood estimates of the evolution in color and magnitude 
have been removed).  The solid line shows the color--magnitude relation 
for the full sample which includes the entire range of $\sigma$; it is 
considerably steeper than the relation in either of the subsamples.  
The panel on the right shows the color$-\sigma$ relation at low (circles) 
and high (crosses) luminosity.  The individual fits to the two 
subsamples are indistinguishable from the fits to the whole sample.  

This is also true for the color--size relation, although we have not 
included a figure showing this.  One consequence of this is that 
residuals from the Faber--Jackson relation correlate with color, 
whereas residuals from the luminosity--size relation do not.  We will 
return to this later.  Because the primary correlation is color with 
velocity dispersion, in what follows, we will mainly consider the 
color--$\sigma$ relation, and residuals from it.  

\begin{table}[t]
\centering
\caption[]{Maximum-likelihood estimates of the slopes and zero-points 
of the color-at-fixed-magnitude and color-at-fixed-velocity dispersion 
relations, and the scatter around the mean relations.  \\}
\begin{tabular}{ccccccc}
\tableline 
Color & slope & zero-point & rms & slope & zero-point & rms \\
\hline
 &\multicolumn{3}{c}{color$-r^*$ magnitude} & 
  \multicolumn{3}{c}{color$-\log_{10}\sigma$} \\
$g^*-r^*$ & $-0.025\pm 0.003$ & 0.218 & 0.053 & $0.26\pm 0.02$ & 0.154 & 0.0488 \\
$r^*-i^*$ & $-0.012\pm 0.002$ & 0.085 & 0.033 & $0.12\pm 0.02$ & 0.072 & 0.0316 \\ 
$r^*-z^*$ & $-0.012\pm 0.003$ & 0.443 & 0.051 & $0.16\pm 0.02$ & 0.343 & 0.0485 \\
\tableline
\end{tabular}
\label{CMVslopes}
\end{table}

While the color--$\sigma$ provides clear evidence that the colors 
in the high redshift population in our sample are bluer than in the 
nearby population, quantifying how much the colors have evolved is 
more difficult, because the exact amount of evolution depends on the 
K-correction we assume.  Appendix~\ref{kcorrs} discusses this in more 
detail.  Because both the colors and the line indices are evolving, it 
is interesting to see if the evolution in color and in the indices is 
similar.  The line indices and color both correlate with $\sigma$, 
and we know how much the individual relations evolve, so we can estimate 
the evolution in the index--color relation just as we did when estimating 
the evolution in the index--index relations.  
Doing so shows that the slope of a plot of Mg$_2$ versus color should 
have a slope of $0.69$, and the zero-point should evolve by $0.02$ 
between the lowest and highest redshift bins in our sample.  
A similar analysis for $\log_{10}\langle{\rm Fe}\rangle$ versus color 
shows that the slope should be approximately $0.38$ and the offset 
should be $0.01$.  
And finally, the slope of the $\log_{10}{\rm H}_\beta$ versus color 
relation should be about $-0.96$ and the zero-point should evolve 
by $-0.012$.  

\begin{figure}
\centering
\epsfxsize=1.1\hsize\epsffile{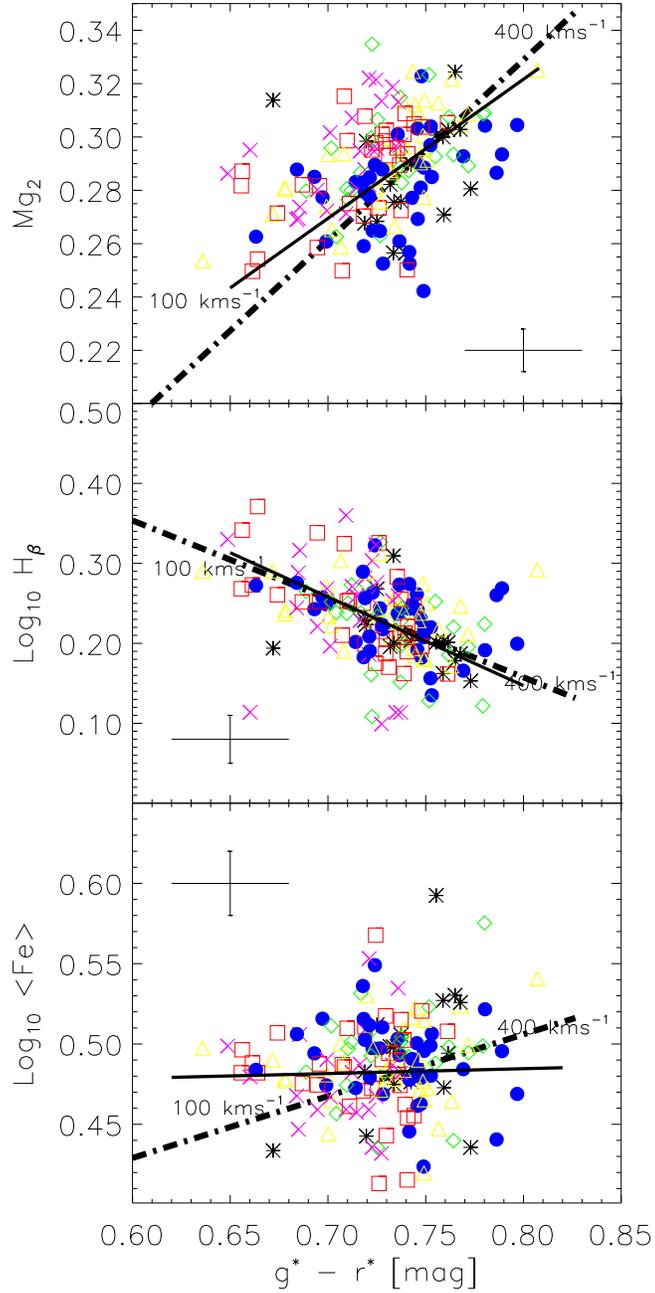}
\caption[]{Line indices versus color.  Dot-dashed lines show the slope 
one expects if there were no scatter around the mean color-$\sigma$ 
and lineindex--$\sigma$ relations, and solid lines show the linear 
relation which provides the best fit to the points.  Crosses in the bottom of 
each panel show the typical uncertainties.  The error in the color 
is supposed to represent the actual uncertainty in the color, rather 
than how well the mean color in each bin has been measured.  }
\label{lndxclr}
\end{figure}

To check the accuracy of these estimates Figure~\ref{lndxclr} shows 
plots of the line indices versus $g^*-r^*$ color.  (Recall that the 
line indices were computed from coadded spectra of galaxies which had 
the same $R_o$, $\sigma$ and $r^*$ luminosity.  The color here is the 
mean color of the galaxies in each of those bins.)  
The solid lines show best fits to the points contributed by the 
median redshift bins (triangles and diamonds).  
The dot-dashed lines in each panel show the slopes estimated above; 
they are not far off from the best-fit lines.  
The estimates of the evolution of the zero-point also appear to be 
reasonably accurate.  The higher redshift crosses in the Mg$_2$ panel 
appear to lie about 0.02~mags above the lower redshift stars; the 
difference between the low and high redshift populations is obvious.  
In contrast, the evolution in H$_\beta$ and color conspire so 
that there is little net offset between the low and high redshift 
populations (note that an offset of 0.02~mags in Mg$_2$ is much more 
obvious than an offset of 0.02 in H$_\beta$).  This suggests that 
the evolution in color and in $H_\beta$ are driven by the same process.  
And finally, if there is an offset between the low and high redshift 
bins in $\langle{\rm Fe}\rangle$ and $g^*-r^*$ (bottom panel), it is 
small and negative, as expected.

\begin{figure}
\centering
\epsfxsize=1.1\hsize\epsffile{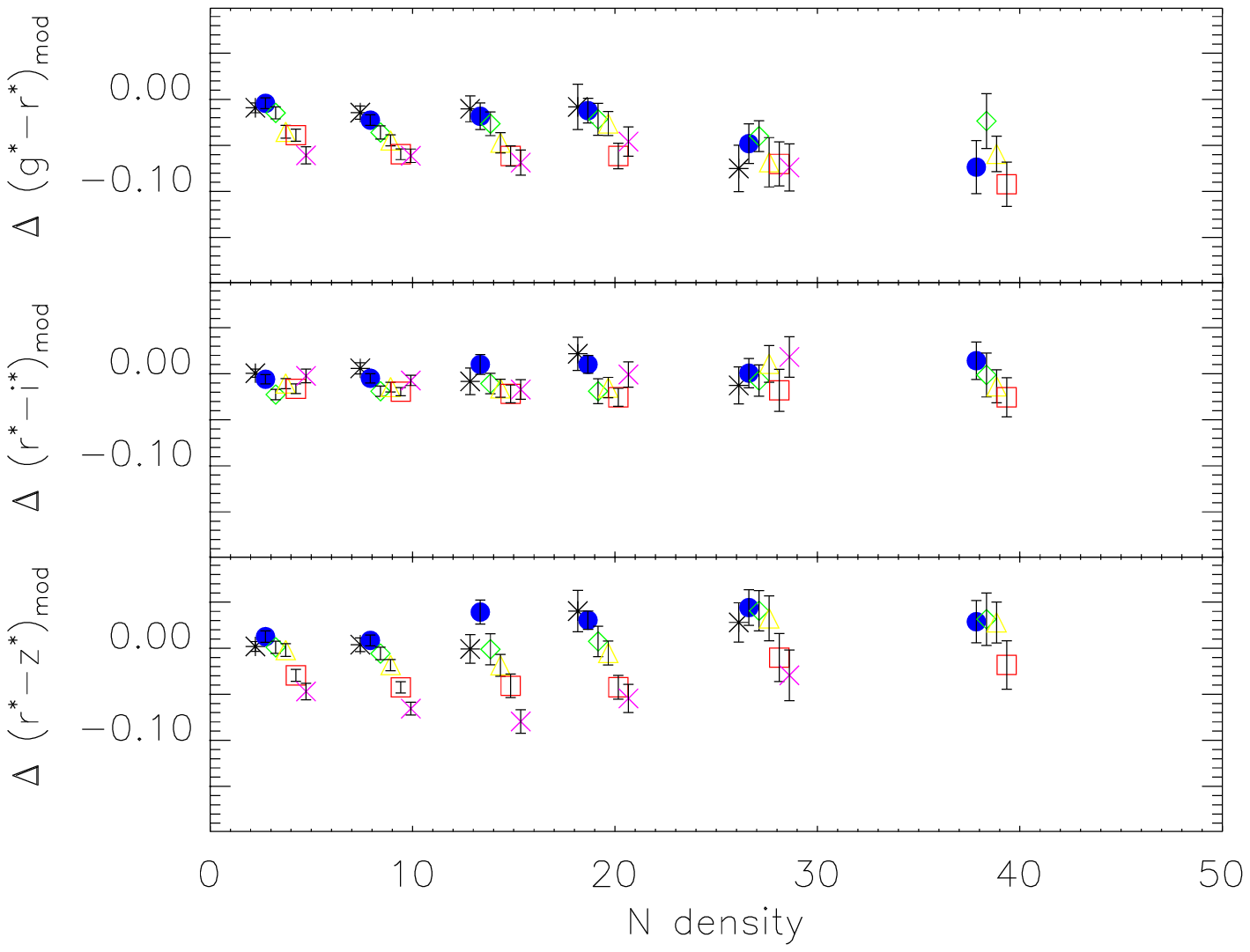}
\vspace{-3cm}
\caption[]{Residuals from the color-$\sigma$ relation as a function 
of local density.  At each bin in density, symbols showing results 
for higher redshifts have been offset slightly to the right.  
Galaxies at higher redshifts are bluer---hence the trend to slope 
down and to the right at fixed $N$.  The $(r^*-z^*)$ colors of galaxies 
in dense environments are redder than those of their counterparts in 
less dense regions, although the trend is weaker in the other colors.}  
\label{cmdensity}
\end{figure}
\begin{figure}
\centering
\epsfxsize=1.1\hsize\epsffile{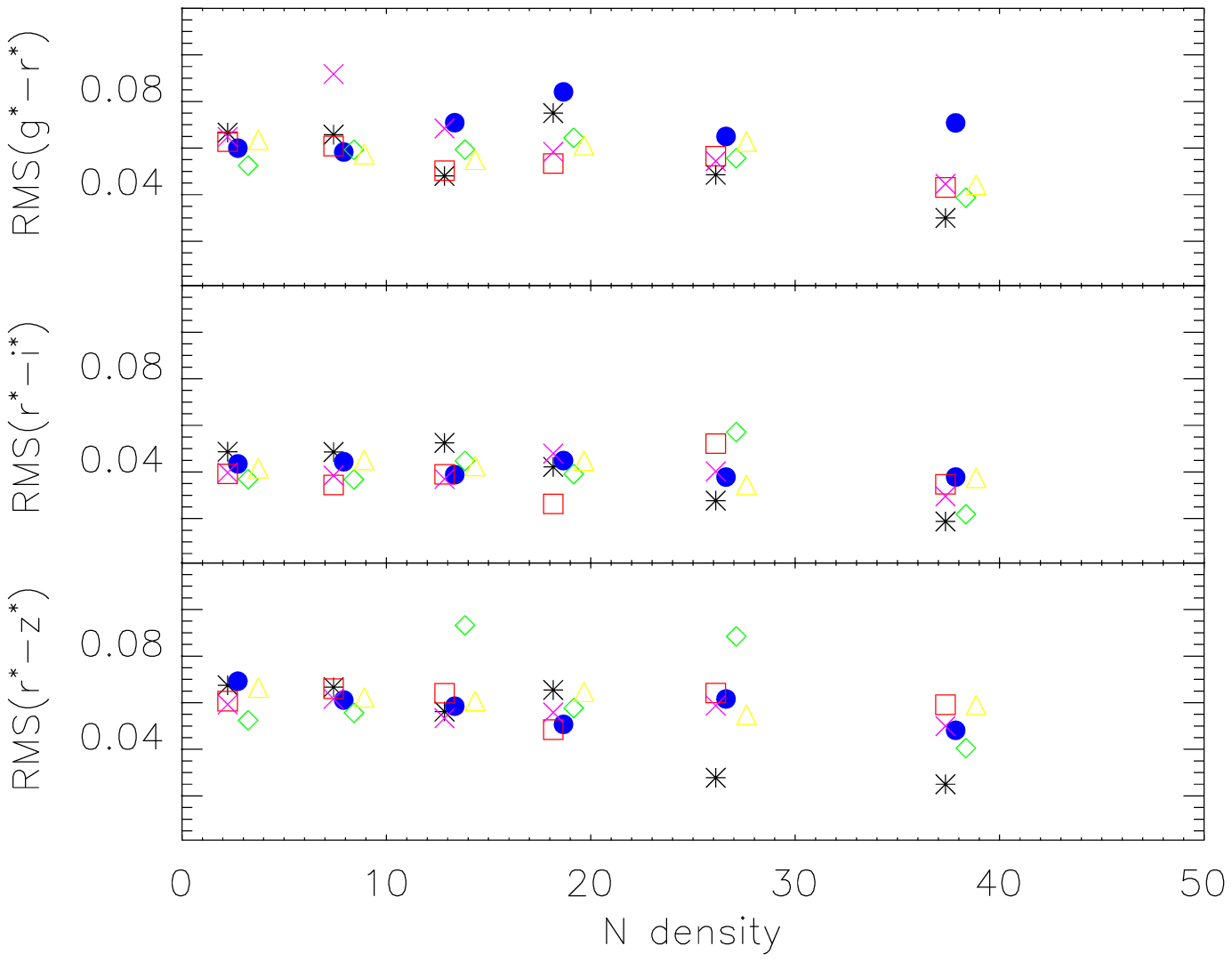}
\vspace{-3cm}
\caption[]{As for the previous figure, but now showing the thickness 
of the color-$\sigma$ relation as a function of local density.  }
\label{rmsdensity}
\end{figure}

Having shown that the colors are evolving, and that, to a reasonable 
approximation, this evolution affects the amplitude but not the slope 
of the color--magnitude and color--$\sigma$ relations, we now study 
how the colors depend on environment.  
To present our results, we assume that environmental effects affect 
the amplitude more strongly than the slope of the color--$\sigma$ 
relation.  Therefore, we assume that the slope is fixed, and fit for 
the shift in color which best describes the subsample.  
Figure~\ref{cmdensity} shows the results.  As in previous sections, 
galaxies were divided into different bins in local density, and then 
further subdivided by redshift.  Different symbols in each bin in local 
density show results for the different redshifts; higher redshifts are 
offset slightly to the right.  This makes trends with evolution easy 
to separate from those due to environment.  In addition to the 
evolutionary trends we have just discussed, the figure shows that 
the $r^*-z^*$ colors are redder in denser regions (bottom panel), 
but that this trend is almost completely absent for the other colors.

The tightness of the colour-magnitude relation of cluster early-types  
has been used to put constraints on the ages of cluster early-types 
(e.g., Kodama et al. 1999).  Figure~\ref{rmsdensity} shows how the 
thickness of this relation depends on environment.  The plot shows 
no evidence that the scatter around the mean relation decreases 
slightly with increasing density; a larger sample is needed to make 
conclusive quantitative statements about this, and about whether or 
not the scatter around the mean relation depends more strongly on 
environment at low than at high redshift.  

We have also checked if the residuals from the index--color relations 
shown in Figure~\ref{lndxclr} correlate with local density:  they do not.  
In short, we have shown that the color--magnitude and color--$\sigma$ 
relations in our sample provide strong evidence for evolution in the 
colors of early type galaxies, and no significant dependence on environment.  

\subsection{Color gradients and the color--magnitude relation}\label{cgrad}
\begin{figure}
\centering
\epsfxsize=\hsize\epsffile{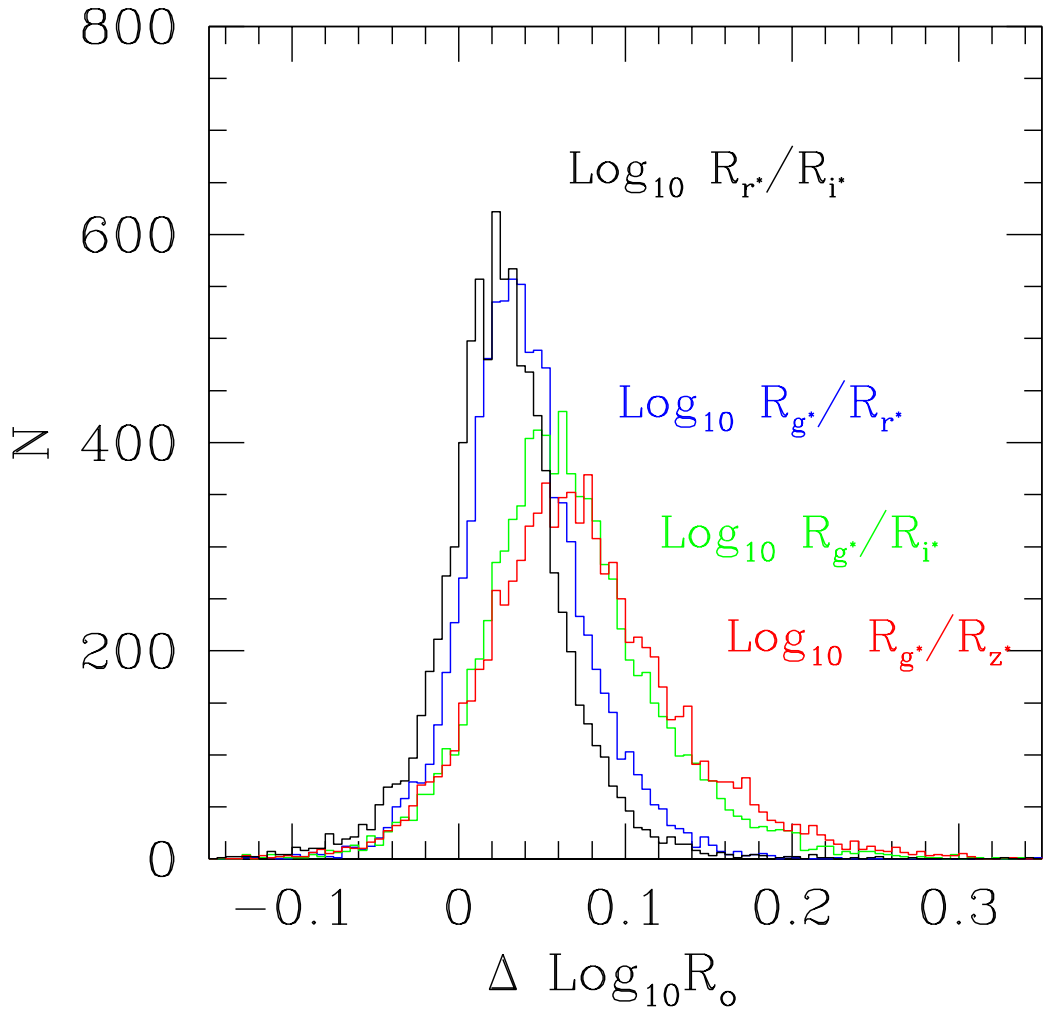}
\caption[]{Differences between the effective sizes of galaxies 
in different bands; the blue light is less centrally concentrated 
than the red light.  }
\label{rdiffs}
\end{figure}

It has been known for some time that giant early-type galaxies 
are reddest in their cores and become bluer toward their edges 
(e.g., de Vaucouleurs 1961; Sandage \& Visvanathan 1978).  
We showed previously that the angular sizes of galaxies were 
larger in the bluer bands (Figure~\ref{angre}).  
Figure~\ref{rdiffs} shows how the effective physical radii of the 
galaxies in our sample change in the four bands.  On average, 
early-type galaxies have larger effective radii in the bluer bands.  
This trend indicates that there are color gradients in early-type 
galaxies.  The distribution of size ratios does not correlate with 
luminosity.  However, the ratio of the effective size in the 
$g^*$ and $r^*$ bands is slightly larger for bluer galaxies 
than for redder ones, suggesting that color gradients are stronger 
in the galaxies which are bluer.  In addition the scatter around 
the mean ratio is slightly larger for the bluer galaxies.

As Scodeggio (2001) emphasizes, if the effective sizes of galaxies 
depend on waveband (cf. Figure~\ref{rdiffs}), then the strength of 
the color-magnitude relation depends on how the color is defined.  
Therefore, we have tried five different definitions for the color.  
The first uses the total luminosities one infers from fitting a 
de Vaucouleurs model to the light in a given band.  The `total' 
colors defined in this way are relatively noisy, because they depend 
on independent fits to the surface brightness distributions in each 
band.  

We have already shown that the half-light radius is larger in the 
bluer bands.  This means that a greater fraction of the light in the 
redder bands comes from regions which are closer to the center than 
for the bluer bands.  Therefore, the total color above can be quite 
different from that which one obtains with a fixed angular or physical 
aperture.  

To approximate fixed physical aperture colors, we have integrated the 
de~Vaucouleurs profiles in the different bands assuming a tophat 
filter (since this can be done analytically) of scale $f$ times the 
effective radius in $h^{-1}$kpc, $R_o(r^*)$, for a few choices of $f$.  
The resulting colors depend on $f$, and the associated color magnitude 
relation decreases as $f$ decreases.  We have arbitrarily chosen to 
present results for $f=2$.  
These are not quite fixed aperture colors, since the effective angular 
aperture size varies from one galaxy to another, but, for any given 
galaxy, the aperture size is the same in all the bands (i.e., it is 
related to the effective radius in $r^*$).  

A third color is obtained by using the light within a fixed angular 
aperture which is the same for all galaxies.  The `fiber' magnitudes 
output by the SDSS photometric pipeline give the integrated light 
within a three arsec aperture, and we use these to define the 
`fiber' color.  

A fourth color is that computed from the Petrosian magnitudes output 
by the SDSS photo pipeline.  

A fifth color uses the `model' magnitudes output by the SDSS photometric 
pipeline.  These are close to what one might call fixed aperture colors, 
because they are obtained by finding that filter which, given the 
signal-to-noise ratio, optimally detects the light in the $r^*$ band, 
and then using that same filter to measure the light in the other bands  
(which is one reason why they are less noisy than the total color defined 
above).  
(By definition, the model and total de~Vaucouleurs magnitudes are the 
same in $r^*$.  They are different in other bands because the effective 
radius is a function of wavelength.  We have verified that the difference 
between these two magnitudes in a given band correlates with the difference 
between the effective radius in $r^*$ and the band in question.)
In this respect, the model colors are similar to those one might get with 
a fixed physical aperture (they would be just like the fixed physical 
aperture colors if the optimal smoothing filter was a tophat).  These, 
also, are not fixed angular aperture colors, since the effective aperture 
size varies from one galaxy to another, but, for any given galaxy, the 
aperture is the same in all the bands.  

\begin{figure}
\centering
\epsfxsize=\hsize\epsffile{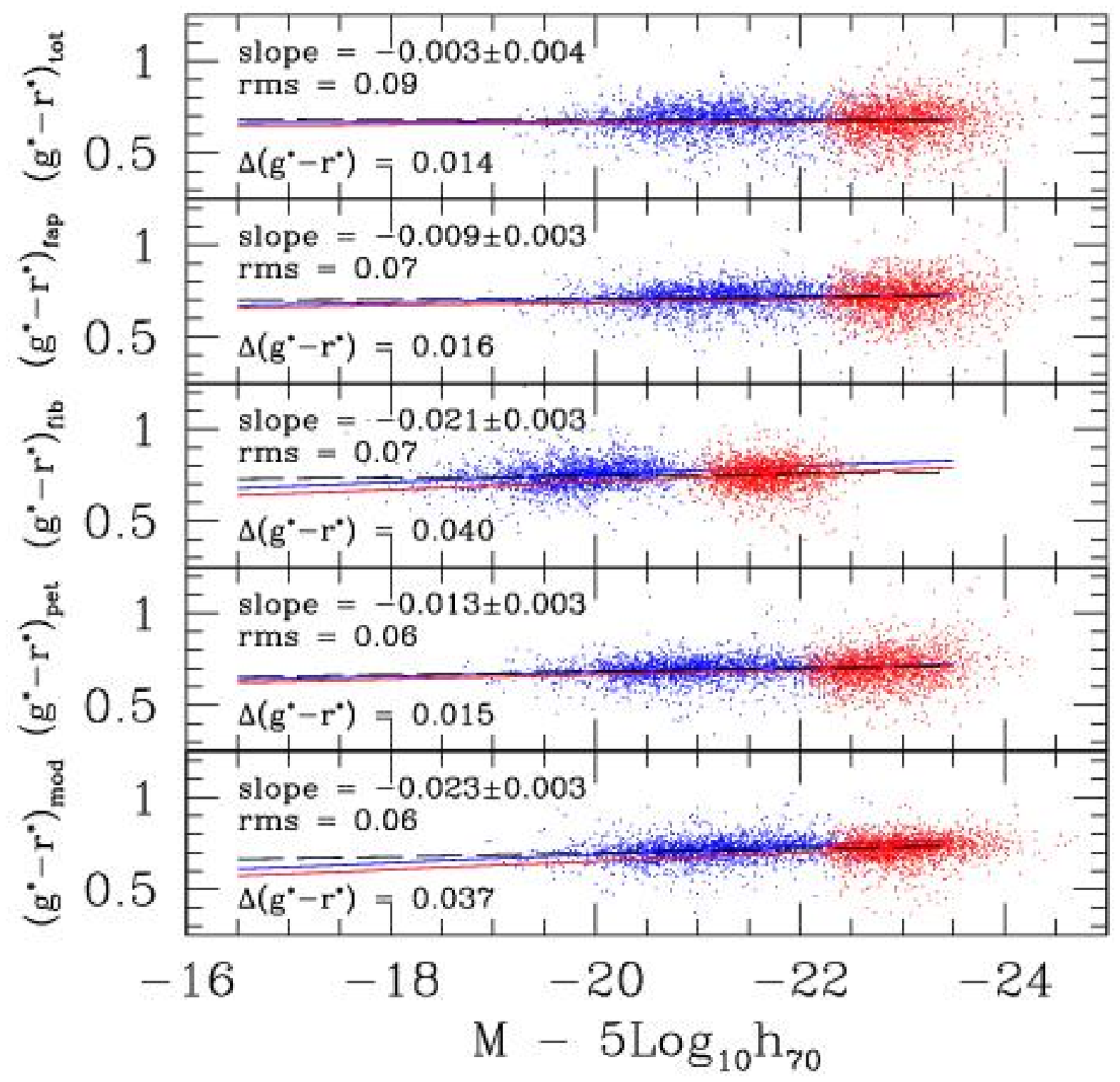}
\caption{Color-magnitude relations associated with various definitions 
of magnitude and $g^*-r^*$ color.  Top-left of each panel shows the 
slope determined from a low redshift subsample.  Fixed-aperture colors 
(bottom panel) give steeper color--magnitude relations; the correlation 
is almost completely absent if colors are defined using the total 
magnitudes (top panel).  Bottom left of each panel shows the zero-point 
shift required to fit the higher redshift sample.  This shift is an 
estimate of how the colors have evolved---it, too, depends on how the 
color was defined.  Dashed lines show fits to the whole sample; because 
they ignore the evolution of the colors, they are significantly 
shallower than fits which are restricted to a small range in redshifts, 
for which neglecting evolution is a better approximation. }  
\label{fig:cmag}
\end{figure}

\begin{figure}
\centering
\epsfxsize=\hsize\epsffile{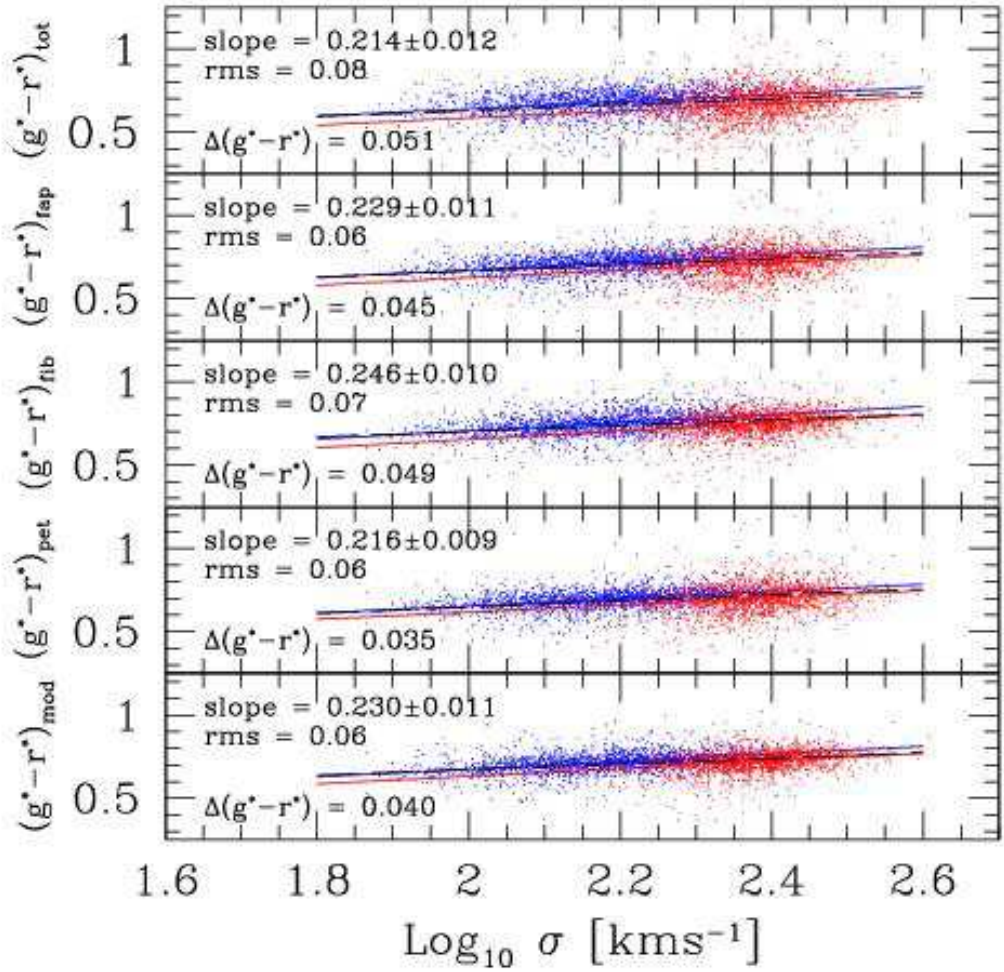}
\caption{As for the previous figure, but now showing how the 
color-$\sigma$ relation varies as the definition of $g^*-r^*$ color
changes.}
\label{fig:csig}
\end{figure}

A final possibility is to use `spectral magnitudes'; these can be made 
by integrating up the light in the spectrum of each galaxy, weighting by 
the different pass-band filters.  Whereas the other five colors require 
a good understanding of the systematics of the photometric data sets, 
this one requires a similar understanding of the spectroscopic data sets 
also.  We have yet to do this.  

The resulting $g^*-r^*$ color-magnitude relations are shown in 
Figure~\ref{fig:cmag}.  
The $x$-axis in the top two panels is the de Vaucouleurs magnitude in 
$r^*$, whereas it is the fiber magnitude in $r^*$ in the third panel, 
the Petrosian $r^*$ magnitude in the fourth panel, 
and the model $r^*$ magnitude in the bottom panel.  
For the reasons described in the previous subsection we must be 
careful that evolution effects do not combine with the magnitude limit 
of our sample to produce a shallow relation.  Our main goal here is to 
illustrate how the shape of the relation depends on the definition of 
color.  For this reason, we have chosen to divide our sample into two: 
a low-redshift sample, which includes all galaxies at $z\le 0.08$, 
and a high-redshift $z\ge 0.16$ sample.  For each definition of color, 
we computed the slope and amplitude of the color--magnitude relation 
in the low redshift sample.  This slope is shown in the top left corner 
of each panel.  We then required the slope of the high redshift sample 
to be the same (recall from Figure~\ref{cmag3} that this is a good 
approximation); the offset required to get a good fit is shown in 
the bottom left of each panel.  This is the quantity which provides 
an estimate of how much the colors have evolved.  The two thin solid 
lines in each panel show the low- and high-redshift color--magnitude 
relations computed in this way.  For comparison, the dashed line shows 
a fit to the full sample, ignoring evolution effects; in all the 
panels, it is obviously much flatter than the relation at low redshift.  

The figure shows clearly that the slope of the color-magnitude 
relation depends on how the color was defined:  
it is present when fixed-apertures are used (e.g., bottom panel), 
and it is almost completely absent when the total light within the 
de Vaucouleurs fit is used (top panel).  Our results are consistent 
with those reported by Okamura et al. (1998) and Scodeggio (2001).  
Note that one's inference of how much the colors have evolved,  
$\Delta (g^*-r^*)$, also depends on how the color was defined.  

A similar comparison for the correlation between color and velocity 
dispersion $\sigma$ is presented in Figure~\ref{fig:csig}.  
We have already argued that color$-\sigma$ is the primary correlation; 
this relation is also considerably less sensitive to the different 
definitions of color.  However, it is sensitive to evolution:  a fit 
to the full sample gives a slope of 0.14, compared to the value of 0.23 
for the low redshift sample.  Because the mean color$-\sigma$ relation 
is steeper than that between color and magnitude, the change to the 
slope of the relation is less dramatic.  The zero-point shifts, which 
estimate the evolution of the color, are comparable both for the 
color--magnitude and the color$-\sigma$ relations, provided the SDSS 
model colors are used (bottom panel).  

\section{Discussion and conclusions}\label{discuss}
We have investigated the properties of $\sim 9000$ early-type galaxies 
over the redshift range 
$0\le z\le 0.3$ using photometric (in the $g^*$, $r^*$, $i^*$ and $z^*$ 
bands) and spectroscopic observations.  
The intrinsic distributions of luminosity, velocity dispersion and 
size of the galaxies in our sample are each well described by Gaussians 
in absolute magnitude, $\log_{10}\sigma$, and $\log_{10}R_o$.  
At fixed luminosity, $\sigma\propto L^{1/4}$ and $R_o\propto L^{3/5}$ 
(see Table~\ref{lx4cov} for the exact coefficients), and galaxies which 
are slightly larger than expected (given their luminosity) have smaller 
velocity dispersions than expected.  This is expected if galaxies are 
in virial equilibrium.  The scatter around the mean $L-\sigma$ and $L-R_o$ 
relations is sufficiently large that it is a bad approximation to 
insert them into the distribution of luminosities to estimate the 
distribution of sizes or velocity dispersions.  

The $L-\sigma$ and $L-R_o$ relations are projections of a 
Fundamental Plane, in the space of $L$, $\sigma$ and $R_o$, which 
these galaxies populate.  If this Fundamental Plane is defined 
by minimizing the residuals orthogonal to it, then 
$R_o\propto \sigma^{1.5}I_o^{-0.77}$ (see Table~\ref{fpcoeffs} for 
the exact coefficients).  The Fundamental Plane is remarkably similar 
in the different bands (Figure~\ref{fig:FPgriz}), and appears to be 
slightly warped in the shorter wavebands (Figure~\ref{fig:warped}).  
Residuals with respect to the direct fit (i.e., the FP defined by 
minimizing the residuals in the direction of $\log_{10}R_o$) do not 
correlate with either velocity dispersion or color, whereas residuals 
from the orthogonal fit correlate with both (Figures~\ref{fig:FPresidR}).  
This correlation with $\sigma$ is simply a projection effect (see 
Figure~\ref{fig:FPsigR} and related discussion), whereas the correlation 
with color is mainly due to the fact that color and $\sigma$ are strongly 
correlated. The Fundamental Plane is intrinsically slightly thinner in 
the redder wavebands.  This thickness is sometimes expressed in 
terms of the accuracy to which the FP can provide redshift-independent 
distance estimates---this is about 20\%.  If the thickness is expressed 
as a scatter in the mass-to-light ratio at fixed size and velocity 
dispersion, then this scatter is about 30\%.  

The simplest virial theorem prediction for the shape of the 
Fundamental Plane is that $R_o\propto \sigma^2/I_o$.  
This assumes that the observed velocity dispersion $\sigma^2$ is 
proportional to the kinetic energy $\sigma_{vir}^2$ which enters the 
virial theorem.  Busarello et al. (1997) argue that in their data 
$\log_{10}\sigma = (1.28\pm 0.11)\log_{10}\sigma_{vir} - 0.58$, so 
that $\sigma^{1.5}\propto\sigma_{vir}^{1.92}$.  Since the coefficient 
of $\sigma$ in the Fundamental Plane we find in all four bands is 
$\sim 1.5$, it would be interesting to see if the kinetic energy for the 
galaxies in our sample scales as it did in Busarello et al.'s sample.  
To do this, measurements of the velocity dispersion profiles of 
(a subsample of) the galaxies in our sample are required.  

A plot of luminosity versus mass ($L$ versus $M_e\propto R_o\sigma^2/G$)  
has a slope which is slightly shallower than unity---on scales of a few 
kiloparsecs, $L\propto M_e^{0.86}$, approximately independent of waveband 
(Figure~\ref{lmass}).  This complements recent SDSS weak-lensing analyses 
(McKay et al. 2001) which suggest that mass is linearly proportional to 
luminosity in these same wavebands, but on scales which are two orders of 
magnitude larger ($\sim 260h^{-1}$kpc).  Together, these two measurements 
of the mass to light ratio can be used to provide a constraint on the 
density profiles of dark matter halos.  

A maximum likelihood analysis of the joint distribution of luminosities, 
sizes and velocity dispersions suggests that the population at higher 
redshifts is slightly brighter than the population nearby, and that 
the change with redshift is faster in the shorter wavebands:  
If $M_*(z) = M_*(0) - Qz$, then $Q=1.15$, $0.85$ and $0.75$ in 
$g^*$, $r^*$ and $i^*$.  
This evolution is sufficiently weak that, relative to their values at 
the median redshift ($z\sim 0.15$) of our sample, the sizes, surface 
brightnesses and velocity dispersions of the early-type galaxy population 
at lower and higher redshifts has evolved little.  Therefore, tests for 
passive evolution which use the Fundamental Plane only are severly affected 
by selection effects and the choice of the fiducial Fundamental Plane 
against which to measure the evolution 
(Figures~\ref{fig:FPzG}--\ref{fig:FPresidmuz}).  
Nevertheless, these tests also suggest that the surface brightnesses 
of galaxies at higher redshifts in our sample are brighter than those 
of similar galaxies nearby.  

The way in which galaxies scatter from the Fundamental Plane correlates 
weakly with their local environment (Figure~\ref{FPenviron}).  If this 
is caused entirely by differences in surface brightness, then galaxies 
in overdense regions are slightly fainter.  If so, then single-age 
stellar population models suggest that early-type galaxies in denser 
regions formed at higher redshift.  However, it may be that, the velocity 
dispersions are higher in denser regions (Figure~\ref{lrvden}).  A larger 
sample is necessary to make a more definitive statement.  

Additional tests of evolution and environment come from a study of the 
spectra of early-type galaxies.  The SDSS spectra of individual galaxies 
do not have extremely high values of signal-to-noise 
(typically $S/N\sim 15$; cf. Figure~\ref{fig:vmeth}).  
However, the dataset is so large that we were able to study stellar 
population indicators (Mg$_2$, Mg$b$, $\langle{\rm Fe}\rangle$ and 
H$_\beta$) by co-adding the spectra of early-type galaxies which have 
similar luminosities, sizes, velocity dispersions, environments and 
redshifts to create composite high $S/N$ spectra. 

All the line indices correlate with velocity dispersion 
(Section~\ref{lindices}): Mg$_2\propto\sigma^{0.18}$, 
Mg$b\propto\sigma^{0.28}$, $\langle {\rm Fe}\rangle\propto\sigma^{0.10}$, 
and H$_\beta\propto\sigma^{-0.25}$.  These correlations are consistent 
with those in the literature (note that the results from the literature 
were obtained from individual, as opposed to coadded, galaxy spectra).  
The coadded spectra show no signigicant dependence on environment.
However, the spectra show clearly that, at fixed velocity dispersion, 
the high redshift population is stronger in H$_\beta$ and weaker in 
Mg and Fe than the population at lower redshifts 
(Figures~\ref{fig:Mg2sigma}--\ref{fig:HbFeMg2}).  

The colors of the galaxies in our sample are strongly correlated with 
velocity dispersion---redder galaxies have larger velocity dispersions 
(Section~\ref{cms}).  
The color--magnitude and color--size relations are a consequence of the 
fact that $M$ and $R_o$ also correlate with $\sigma$ (Figure~\ref{clrmagv}).  
The strength of the color--magnitude relation depends strongly 
on whether or not fixed apertures were used to define the colors, 
whereas the color$-\sigma$ relation appears to be less sensitive to 
these differences (Figures~\ref{fig:cmag} and~\ref{fig:csig}).  
At fixed velocity dispersion, the population at high redshift is bluer 
than that nearby (Figures~\ref{cmag3} and~\ref{csig3}), and the 
evolution in colour is significantly less than that of the 
luminosities (Table~\ref{MLcmag}).  Color also correlates with the 
line-indices:  Figure~\ref{lndxclr} shows that the evolution in $g^*-r^*$ 
is more closely tied to evolution in H$_\beta$ than Mg$_2$.  
A larger sample, with well understood K-corrections, is required to 
quantify if galaxies in denser regions are slightly redder and more 
homogeneous (Figures~\ref{cmdensity} and~\ref{rmsdensity}) or not.  

Single burst stellar population models (e.g., Worthey 1994; 
Vazdekis et al. 1996) allow one to translate the evolution in the 
spectral features into estimates of the ages and metallicities of the 
galaxies in our sample (e.g., Trager et al. 2000).  In our sample, the 
$z\approx 0.05$ population appears to about 8~Gyrs old; 
the $z\approx 0.2$ population in our sample appears to be about 2~Gyrs 
younger; and the average metallicity appears to be similar in both 
populations.  The age difference is approximately consistent with the 
actual time difference in the $(\Omega_m,\Omega_\Lambda,h)=(0.3,0.7,0.7)$ 
world model we assumed throughout this paper, suggesting that the 
population is evolving passively.  Given a formation time, the single 
burst stellar population models also make predictions about how the 
luminosities and colors should evolve with redshift.  Our estimates of 
this evolution are also consistent with those of a population which 
formed the bulk of its stars 9~Gyrs ago.  

By the time the Sloan Digital Sky Survey is complete, the uncertainty 
in the K-corrections, which prevent us at the present time from making 
more precise quantitative statements about the evolution of the luminosities 
and colors, will be better understood.  In addition, the size of the sample 
will have increased by more than an order of magnitude.  This will allow 
us to provide a more quantitative study of the effects of environment than 
we are able to at the present time.  In addition, a larger sample will 
allow us to coadd spectra in finer bins; this will allow us to make 
maximum-likelihood estimates, rather than simple linear regression 
studies, of how features in the spectra correlate with other observables.

\vspace{1cm}

We would like to thank S. Charlot for making his stellar population 
synthesis predictions for the SDSS filters available to the 
collaboration and N. Benitez for making his package available.  
We thank D. Kelson, M. Pahre, M. Strauss and S. D. M. White for 
helpful discussions.  

The Sloan Digital Sky Survey (SDSS) is a joint project of 
The University of Chicago, Fermilab, the Institute for Advanced Study, 
the Japan Participation Group, The Johns Hopkins University, 
the Max-Planck-Institute for Astronomy (MPIA), 
the Max-Planck-Institute for Astrophysics (MPA), 
New Mexico State University, Princeton University, 
the United States Naval Observatory, and the University of Washington.
Apache Point Observatory, site of the SDSS telescopes, 
is operated by the Astrophysical Research Consortium (ARC). 

Funding for the project has been provided by 
the Alfred P. Sloan Foundation, 
the SDSS member institutions, 
the National Aeronautics and Space Administration, 
the National Science Foundation, the U.S. Department of Energy, 
the Japanese Monbukagakusho, and the Max Planck Society. 
The SDSS Web site is http://www.sdss.org/.

{}

\appendix 
\section{K-corrections}\label{kcorrs}
When converting the observed apparent magnitude to the rest-frame 
absolute magnitude of an object, we must account for the fact that 
the SDSS filters measure the light from a fixed spectral range in 
the observers rest-frame; therefore, they measure different parts of 
the rest-frame spectrum of galaxies at different redshifts.  
Correcting for this is known as the K-correction.  
One way to make this correction is to assume that all the galaxies at a 
given redshift are similar, and to use an empirically determined template 
spectrum, measured from a few accurately measured spectra, to estimate 
the K-correction.  Using a mean color to estimate the K-correction 
is not ideal.  When the survey is closer to completion it should become 
possible to make this correction on an object-by-object basis.  

Empirically determined template spectra for early-type galaxies at 
low redshifts exist (e.g. Coleman, Wu \& Weedeman 1980; 
Fukugita, Shimasaku \& Ichikawa 1995).  (We used N. Benitez's Bayesian 
Photometric Redshift package (Benitez 2000) to derive K-corrections 
for the Coleman, Wu \& Weedman early-type galaxy template in the SDSS 
passbands.)  If we were certain that evolution effects were not important, 
then these empirically determined K-corrections would allow us to work 
out the K-corrections we should apply to the high redshift population.  
However, if the stars in early-type galaxies formed at approximately 
the same time, and if the mass in the galaxies has remained constant, 
so the evolution is entirely due to the passive aging of the stellar 
population, then, the mass to light ratio of early-type galaxies is 
expected to vary approximately as $M/L\propto (t-t_{\rm form})^{-0.6}$ 
(e.g., Tinsley \& Gunn 1976), with the precise scaling being different in 
different bands.  Because our sample spans a relatively large range in 
redshift, we may be sensitive to the effects of this passive evolution.  
In addition, because the sample is large, we may be able to measure, and 
hence be sensitive to, even a relatively small amount of evolution.  
For this reason, it would be nice to have a prescription for making 
K-corrections which accounts for evolution.  Absent empirically determined 
templates for this evolution, we must use stellar population systhesis 
models to estimate this evolution, and so determine K-corrections for 
different bands.  

\begin{figure}[t]
\centering
\epsfxsize=\hsize\epsffile{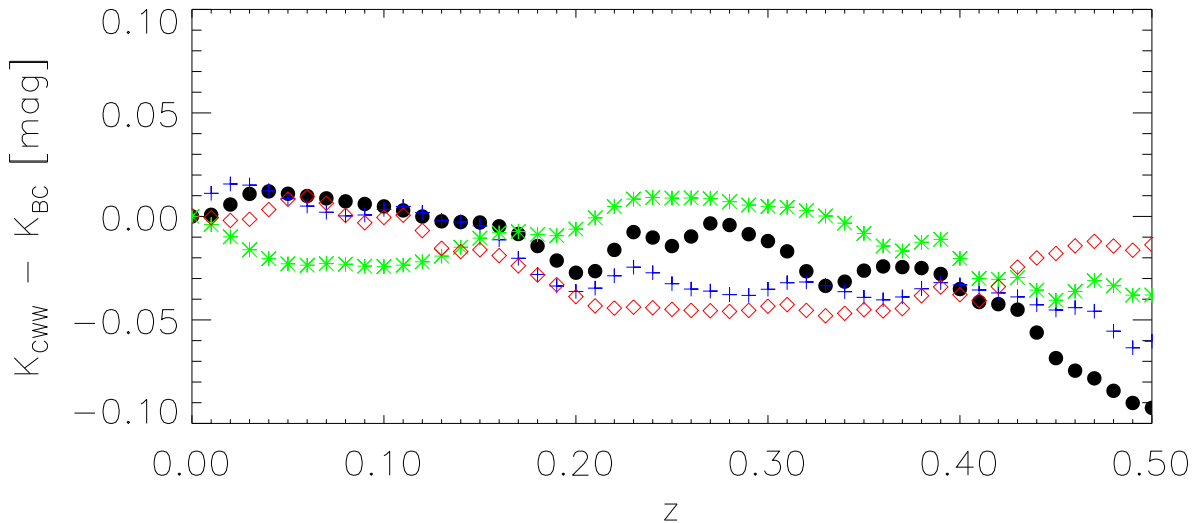}
\vspace{-8cm}
\caption[]{Difference between K-corrections based on two models 
(Coleman, Wu \& Weedman 1980 and Bruzual \& Charlot 2002) of the SDSS 
colors of early-type galaxies.  Filled circles, crosses, stars and 
diamonds are for the $g^*$, $r^*$, $i^*$ and $z^*$ bands.}
\label{kCWW-kBC}
\end{figure}

As a first example, we chose a Bruzual \& Charlot (2002) model for a 
$10^{11}M_\odot$ object which formed its stars with the universal 
IMF given by Kroupa (2000) in a single solar metallicity and abundance 
ratio burst 9~Gyr ago.  We then recorded how its colors, as observed 
through the SDSS filters, changed as it was moved through redshift without 
altering its age.  This provides what we will call the no-evolution 
K-correction.  Figure~\ref{kCWW-kBC} shows a comparison of this with 
the empirical Coleman, Wu \& Weedman (1980) nonevolving K-corrections.  
The two estimates are in good agreement in $g^*$ and $i^*$ out
to $z\sim 0.3$. They differ substantially at higher redshifts, 
but this is not a concern because none of the galaxies in our sample are 
so distant. In $r^*$ and $z^*$ the two estimates agree only at $z\le 0.15$.
Therefore, quantitative 
estimates of evolution in luminosity and/or color will depend on which 
K-correction we use.

\begin{figure}
\centering
\epsfxsize=\hsize\epsffile{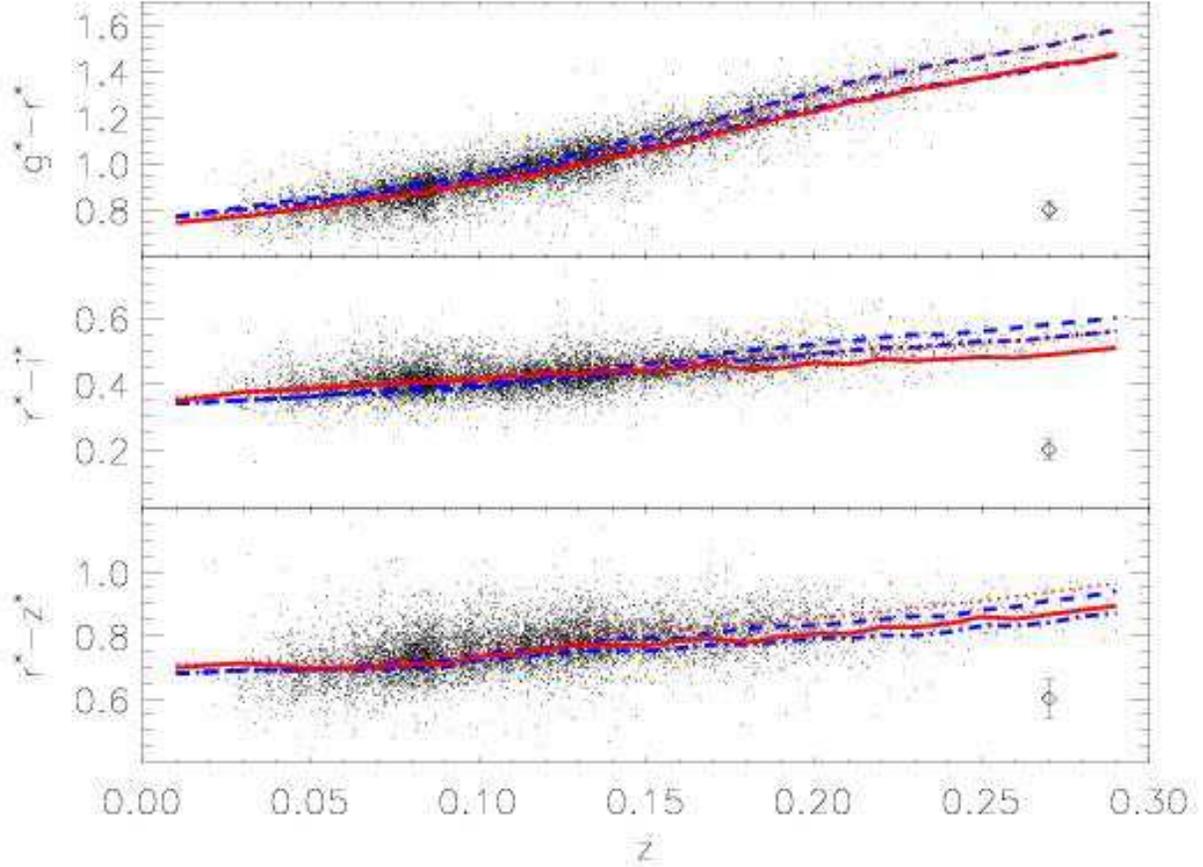}
\vspace{0cm}
\caption[]{Apparent colors of the galaxies in our sample.  In each panel, 
dotted and solid lines show the non-evolving and evolving Coleman et al. 
templates, whereas dashed and dot-dashed show Bruzual \& Charlot models.  
The upper set of curves in each panel show what one expects to see 
if the intrinsic colors of galaxies at higher redshifts are the same as 
they are nearby, whereas the lower sets of curves show the predictions 
if the higher redshift population is slightly younger, and so bluer.  
The magnitude limit of the sample makes it appear as though the no-evolution 
curves describe our data well.  In the main text, we use the lower 
solid curve to make K-corrections to the observed magnitudes. }
\label{fig:kcorr}
\end{figure}

Figure~\ref{fig:kcorr} compares both sets of nonevolving templates 
with the observed colors of the galaxies in our sample.  The upper 
set of curves in each panel show the colors associated with the 
nonevolving Coleman, Wu \& Weedman (1980) template (dotted) and the 
Bruzual \& Charlot (2002) no-evolution model (dashed).  
The figure shows that both predictions for $g^*-r^*$ are similar, but 
that they are different for $r^*-i^*$ and $r^*-z^*$, with the differences 
increasing with redshift.  
[In both cases, we have shifted the predicted $g^*-r^*$ blueward by 
0.08~mag at all $z$.  Such an offset appears to be required for the SDSS 
photometric calibrations in Stoughton et al. (2002) which we use here 
(also see Eisenstein et al. 2001 and Strauss et al. 2002), although the 
reason for it is not understood.]

\begin{figure}
\centering
\epsfxsize=\hsize\epsffile{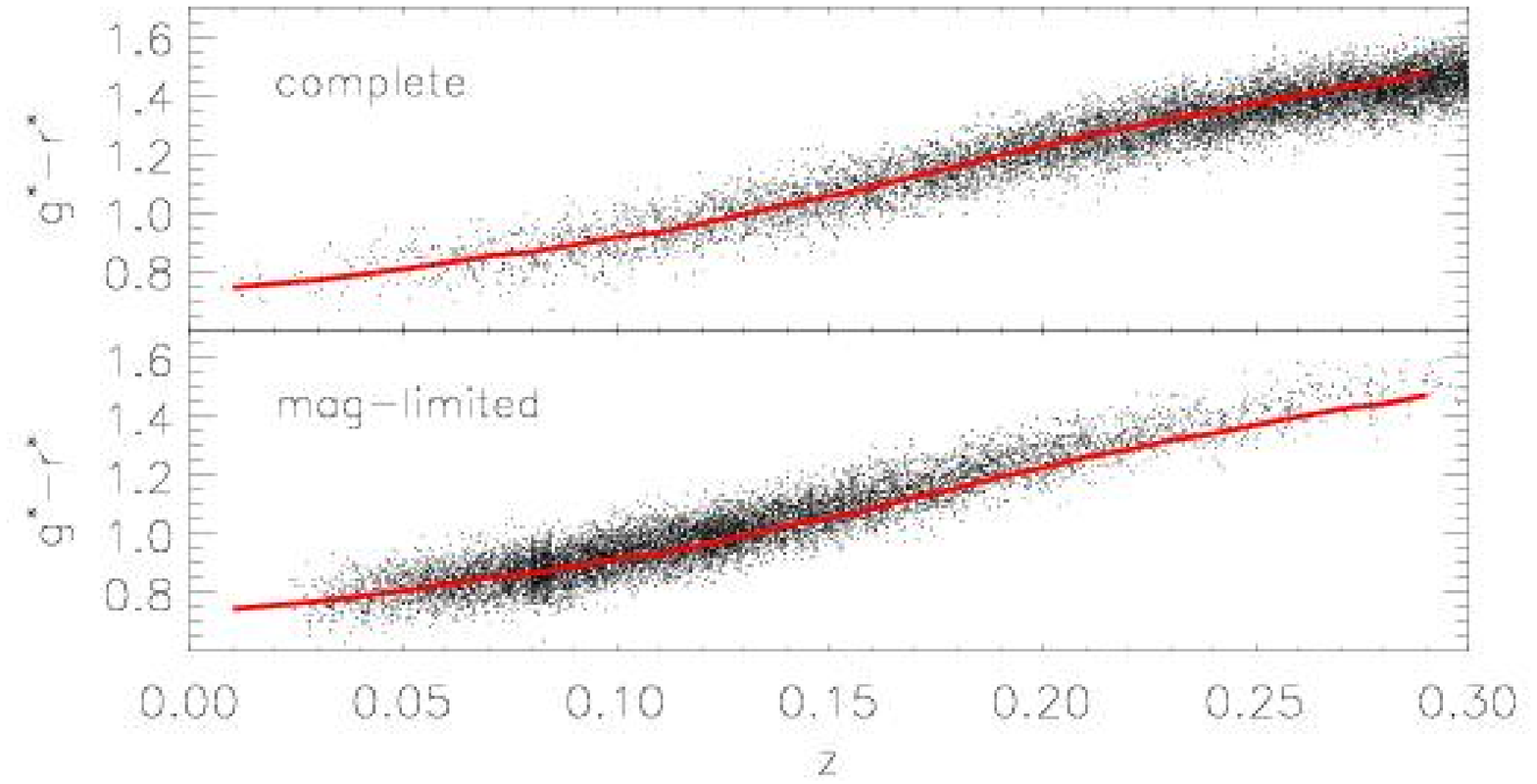}
\vspace{0cm}
\caption[]{Apparent $g^*-r^*$ colors of simulated galaxies in mock catalogs 
of a passively evolving population; the galaxies at higher redshift are 
younger and, in their rest-frame, bluer.  Top panel shows the expected 
distribution of observed colors if there were no magnitude limit; 
bottom panel shows the effect of imposing the same magnitude limit as in 
our SDSS sample.  Solid curves (same in both panels) show the trend of 
observed color with redshift of this evolving population.  
Although the smooth curve describes the complete catalog well, it is 
substantially bluer than the subset of objects which are included in the 
magnitude limited sample.  The difference between the curves and our 
magnitude limited mock catalogs is similar to that between the curves and 
the data (see previous Figure), suggesting that the colors of the 
galaxies in our data are evolving similarly to how we assumed in our 
simulations.}
\label{kcorrsim}
\end{figure}

We argued in the main text that the spectra of these objects show 
evidence for passive evolution: the higher redshift population appears 
to be slightly younger.  Therefore, our next step is to include the 
effects of evolution.  Because the predicted observed colors at redshift 
zero are in good agreement with our data, we took the same 
Bruzual \& Charlot (2002) model, 
but this time we recorded how its rest-frame colors evolve with redshift, 
and then computed what these evolved (i.e., $z$-dependent) colors look 
like when observed in the SDSS filters.  These evolving colors are shown 
as the dot-dashed lines in Figure~\ref{fig:kcorr}.  
In an attempt to include evolution in the CWW templates, we set 
$K_{\rm CWW}^{\rm evol}(z) = K_{\rm CWW}^{\rm noev}(z) + 
                             K_{BC}^{\rm evol}(z) - K_{BC}^{\rm noev}(z)$.  
The lower solid lines in each panel of Figure~\ref{fig:kcorr} show the 
observed $g^*-r^*$, $r^*-i^*$ and $r^*-z^*$ colors associated with these 
evolving models (and again, the predicted $g^*-r^*$ curves have been 
shifted blueward (downward) by 0.08~mag at all $z$).  
Comparing the evolving Bruzual--Charlot and the Coleman et al. colors 
with the upper set of no-evolution curves shows the evolution towards the 
blue at high redshift.  Although the data appear to be very well fit by 
the no-evolution curves, this agreement is slightly misleading.  
More luminous galaxies tend to be redder.  As a consequence, a magnitude 
limited catalog contains only the redder objects of the higher redshift 
population.  A curve which describes the colors of the population as a 
whole will therefore appear to be biased blue.  

Figure~\ref{kcorrsim} shows this explicitly.  The two panels were  
constructed by making mock catalogs of a passively evolving population 
(i.e., the higher redshift population is brighter and bluer) in which our 
estimates of the correlation between colors, luminosities, velocity 
dispersions and color and luminosity evolution were included
(see Table~\ref{MLcmag}).  The top panel shows the distribution of 
observed colors if there were no magnitude limit, and the bottom panel 
shows the observed colors of a magnitude limited sample.  The solid 
curve, the same in both panels, is the predicted trend of color with 
redshift which we use to make our K-corrections; i.e., 
$K_{\rm CWW}^{\rm evol}(z)$.  
Notice that although it describes the complete simulations well, 
it is bluer than the higher redshift galaxies in the magnitude limited 
sample.  Comparison with the previous figure shows that the difference 
here is similar to that seen in the real data, suggesting that our 
K-corrections and evolution estimates of the mean of the population 
are self-consistent.  

Of course, if we do not observe the mean of the high redshift population, 
but only the redder fraction, then we must decide whether it is realistic 
to use a K-correction which has been constructed to fit the truely typical 
galaxy at each redshift.  For example, if color is an indicator of age 
and/or metallicity, then the results above suggest that our sample contains 
the oldest and/or most metal rich part of the high redshift population.  
If the objects which satisfy our apparent magnitude limit are in fact, 
older than the typical high redshift galaxy, then it may be that those 
objects are similar in age to the average object at lower redshifts in our 
sample.  If so, then we are better-off using a nonevolving K-correction 
even though the higher redshift sample as a whole is younger.  None of 
the results presented in the main text change drastically if we use 
non-evolving rather than evolving K-corrections.  

\begin{figure}
\centering
\epsfxsize=1.1\hsize\epsffile{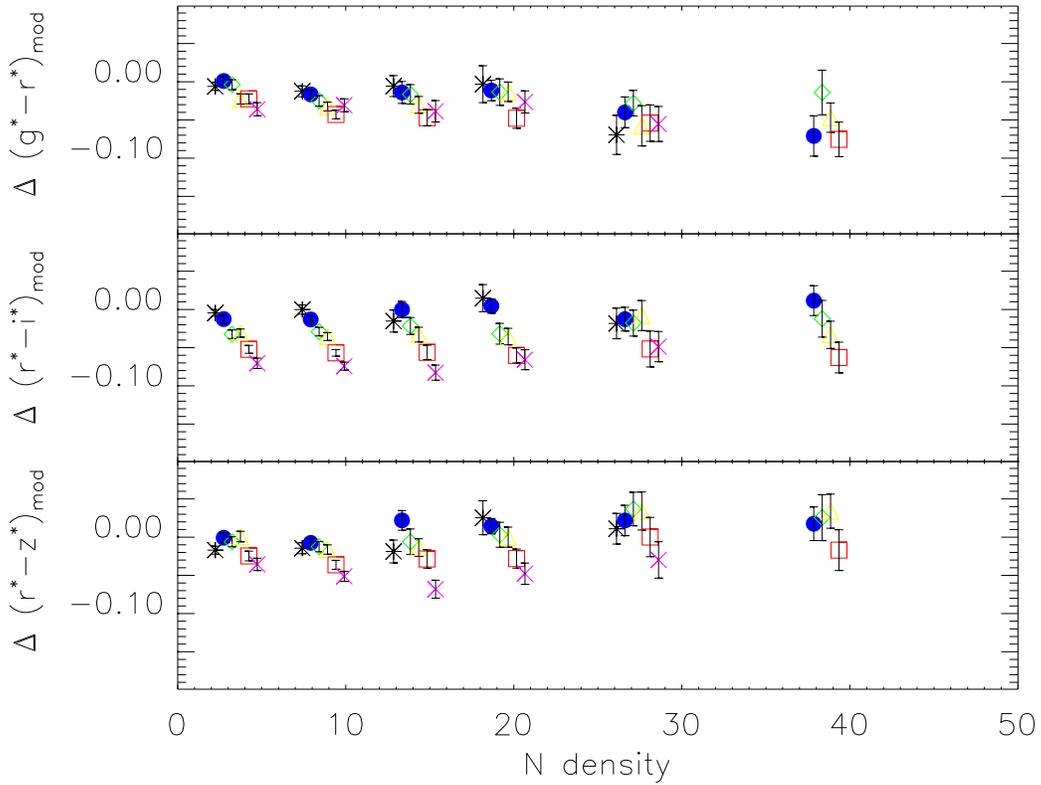}
\vspace{-3cm}
\caption[]{Residuals from the color-$\sigma$ relation as a 
function of local density.  At each bin in density, symbols showing 
results for higher redshifts have been offset slightly to the right.  
The trend for different symbols to slope down and to the right at 
fixed N indicates that galaxies at higher redshifts are bluer.  
This figure should be compared with Figure~\ref{cmdensity}.  
The two plots differ because here the K-corrections are based on 
the Bruzual \& Charlot, rather than Coleman et al., templates.  
The evidence for evolution in color is present in both plots, 
although for the Bruzual \& Charlot based corrections, $r^*-i^*$ 
appears to evolve more rapidly than $g^*-r^*$:  this is contrary 
to expectations.}
\label{BCcolorz}
\end{figure}

To decide whether to chose to use K-corrections based on the 
Coleman et al. (1980) template or the Bruzual \& Charlot (2002) models, 
we computed the color-magnitude and color-$\sigma$ relations in both 
cases.  The slopes of the mean relations, and the scatter around the mean, 
remained approximately the same for both K-corrections, so we have chosen 
to not show them here.  This suggests that our ignorance of the true 
K-correction does not strongly compromise our conclusions about how color 
correlates with magnitude and velocity dispersion.  Conclusions about 
evolution, however, do depend on the K-correction.  

Figure~\ref{BCcolorz} shows the result of remaking Figure~\ref{cmdensity}, 
but now using K$^{\rm evol}_{\rm BC}$, rather than K$^{\rm evol}_{\rm CWW}$.  
The three panels show the residuals from the color-$\sigma$ relation as a 
function of local density.  At each bin in density, symbols showing results 
for higher redshifts have been offset slightly to the right.  The trend for 
different symbols to slope down and to the right at fixed N indicates 
that galaxies at higher redshifts are bluer.  Comparison with 
Figure~\ref{cmdensity} shows that the evidence for evolution in color is 
present for both K-corrections.  However, K$^{\rm evol}_{\rm BC}$ 
yields evolution in $g^*-r^*$ of 0.04~mags, and in $r^*-i^*$ of 
0.07~mags.  In comparison, K$^{\rm evol}_{\rm CWW}$ has changes of 0.07 
and 0.03 respectively.  Thus, K$^{\rm evol}_{\rm BC}$ suggests 
that the evolution in $r^*-i^*$ is larger than in $g^*-r^*$.  
This is not the expected trend; the $g^*-r^*$ and $r^*-i^*$ wavelength 
baselines are about the same, so one expects more of the evolution to come 
in at the bluer color.  Using K$^{\rm evol}_{\rm CWW}$ instead suggests 
that most of the evolution is in $g^*-r^*$, which is more in line with 
expectations.  

We also tried K-corrections from Fukugita et al. (1995).  
At low redshifts, the predicted early-type colors are redder than those 
in our sample, the predicted S0 colors are bluer, and the differences 
depend on redshift.  A straight average of the two is an improvement, 
although the resulting low redshift $g^*-r^*$ is red by 0.05~mags.  
If we shift by this amount to improve the agreement at low redshifts, 
then the observed $g^*-r^*$ colors at $z=0.25$ are redder than the 
predicted no evolution curves by about 0.2 mags.  This is larger than 
the offset we expect for the selection effect introduced by the magnitude 
limit, so we decided against presenting further results from these 
K-corrections.  

\section{Velocity dispersion: methods and measurements}\label{vmethods}
This Appendix describes how we estimated the line-of-sight velocity 
dispersions $\sigma$ for the sample of galaxies selected for this paper.  

\begin{figure}[t]
\centering
\epsfxsize=\hsize\epsffile{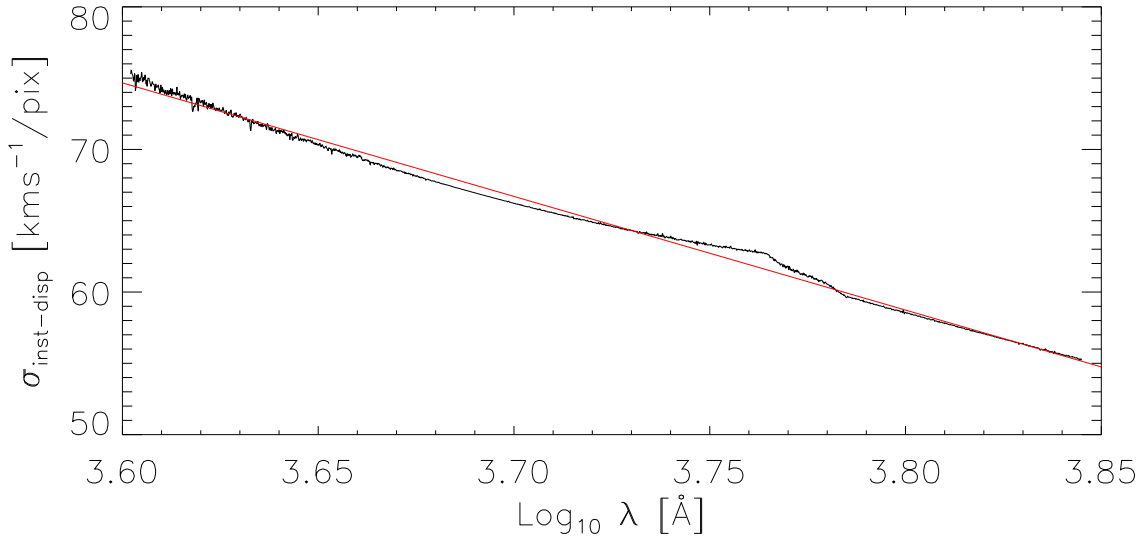}
\vspace{-8cm}
\caption[]{Variation of instrumental dispersion over the range in 
wavelengths used to measure velocity dispersions later in this paper.  
Solid line shows a linear fit. }
\label{fig:resol}
\end{figure}

Estimates of $\sigma$ are limited by the instrumental dispersion 
and resolution.  Recall that the instrumental dispersion of the SDSS 
spectrograph is 69 kms$^{-1}$ per pixel, and the resolution is 
$\sim 90$ kms$^{-1}$.  In addition, the instrumental dispersion may 
vary from pixel to pixel, and this can affect measurements of $\sigma$.
These variations are estimated for each fiber by using arc lamp 
spectra (upto 16 lines in the range 3800-6170~\AA\ and 39 lines 
between 5780-9230~\AA).  An example of the variation in instrumental 
dispersion for a single fiber is shown in Figure~\ref{fig:resol}.  The 
figure shows that a simple linear fit provides a good description of 
this variation.  This is true for almost all fibers, and allows us to 
remove the bias such variations may introduce when estimating galaxy 
velocity dispersions.  

A number of methods for making accurate and objective velocity dispersion 
measurements as have been developed (Sargent et al. 1977; 
Tonry \& Davis 1979; Franx, Illingworth \& Heckman 1989; Bender 1990; 
Rix \& White 1992).  These methods are all based on a comparison between 
the spectrum of the galaxy whose velocity dispersion is to be determined, 
and a fiducial spectral template. This can either be the spectrum of an 
appropriate star, with spectral lines unresolved at the spectra resolution 
being used, or a combination of different stellar types, or a high $S/N$ 
spectrum of a galaxy with known velocity dispersion. In this work, we use 
SDSS spectra of 32 K and G giant stars in M67 as stellar templates.  

Since different methods can give significantly different results, 
thereby introducing systematic biases especially for low $S/N$ spectra, 
we decided to use three different techniques for measuring the velocity 
dispersion.  These are 
1) the {\it cross-correlation} method (Tonry \& Davis 1979); 
2) the {\it Fourier-fitting} method (Tonry \& Davis 1979; 
Franx, Illingworth \& Heckman 1989; van der Marel \& Franx 1993); and 
3) a modified version of the {\it direct-fitting} method 
(Burbidge, Burbidge \& Fish 1961; Rix \& White 1992). Because a galaxy's 
spectrum is that of a mix of stars convolved with the
distribution of velocities within the galaxy, Fourier space is the natural
choice to estimate the velocity dispersions---the first two methods make 
use of this.  However, there are several advantages to treating the problem 
entirely in pixel space.  In particular, the effects of noise are much 
more easily incorporated in the pixel-space based {\it direct-fitting} 
method.  Because the $S/N$ of the SDSS spectra are relatively low, 
we assume that the observed absorption line profiles in early-type galaxies 
are Gaussian (see Rix \& White 1992 and Bender, Saglia \& Gerhard 1994 
for a discussion of how to analyze the line profiles of high $S/N$ spectra 
in the case of asymmetric profiles).  

It is well known that all three methods have their own particular biases, 
so that numerical simulations must be used to calibrate these biases.  
In our simulations, we chose a template stellar spectrum measured at high 
$S/N$, broadened it using a Gaussian with rms $\sigma_{input}$, added 
Gaussian noise, and compared the input velocity dispersion with the 
measured output value.  
The first broadening allows us to test how well the methods work as a 
function of velocity dispersion, and the addition of noise allows us to 
test how well the methods work as a function of $S/N$.

\begin{figure}
\centering
\epsfxsize=\hsize\epsffile{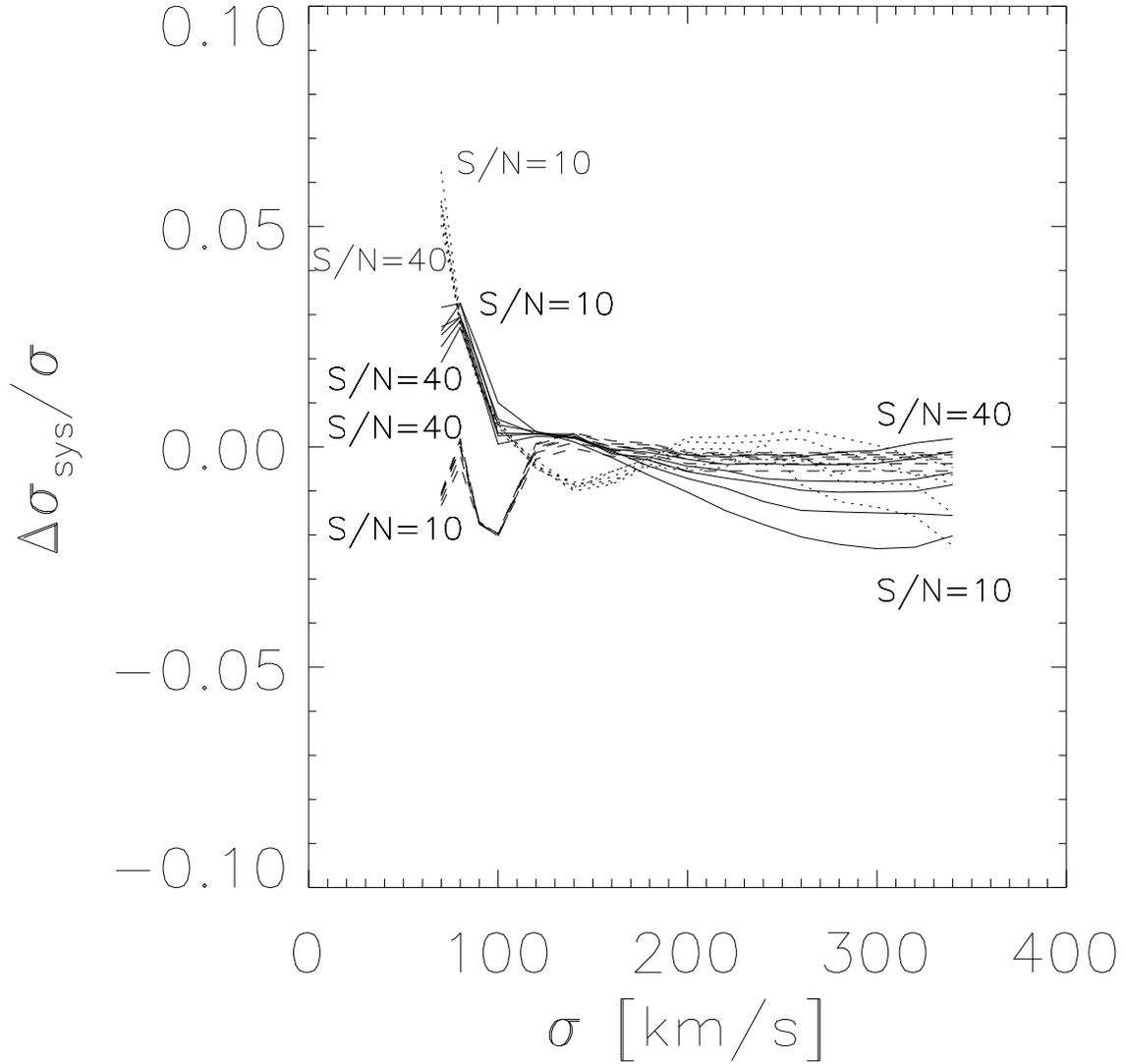}
\caption[]{Systematic biases in the three methods used to 
estimate the velocity dispersion.  Solid, dashed and dotted lines 
show the biases in the {\it Fourier-fitting}, {\it direct-fitting} and {\it
cross-correlation} 
methods, as a function of velocity dispersion and signal-to-noise.}
\label{fig:vsim}
\end{figure}

The best-case scenario is one in which there is no `template mismatch': 
the spectrum of the template star is exactly like that of the galaxy 
whose velocity dispersion one wishes to measure.  
Figure~\ref{fig:vsim} shows the fraction of systematic bias associated 
with each of the different methods in this best-case scenario.  
Slightly more realistic simulations, using a combination of stellar 
spectra as templates, were also done.  
The results are similar to those shown in Figure~\ref{fig:vsim}. 
With the exception of the {\it cross-correlation} method at low 
($\sigma < 100$ kms$^{-1}$) velocity dispersion, the systematic 
errors on the velocity dispersion measurements appear to be smaller 
than $\sim 3\%$. 

\begin{figure}
\centering
\epsfxsize=\hsize\epsffile{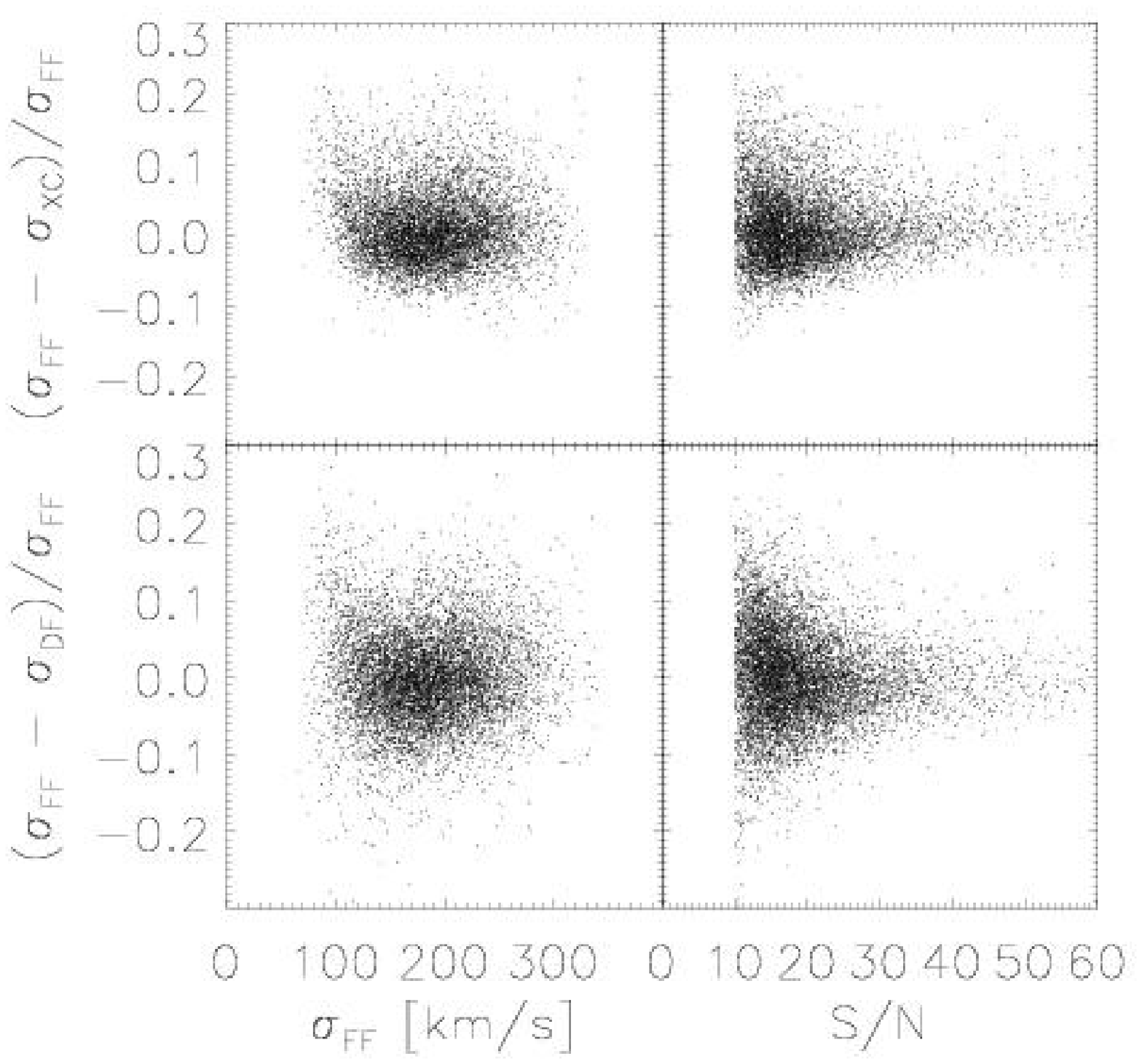}
\caption[]{Comparison of the various methods used to estimate 
the velocity dispersions; the agreement is quite good, with a 
scatter of about five percent.  Most of our spectra have 
$S/N \sim 15$, with approximately exponential tails on either 
side of this mean value.  }
\label{fig:vmeth}
\end{figure}

\begin{figure}
\centering
\epsfxsize=\hsize\epsffile{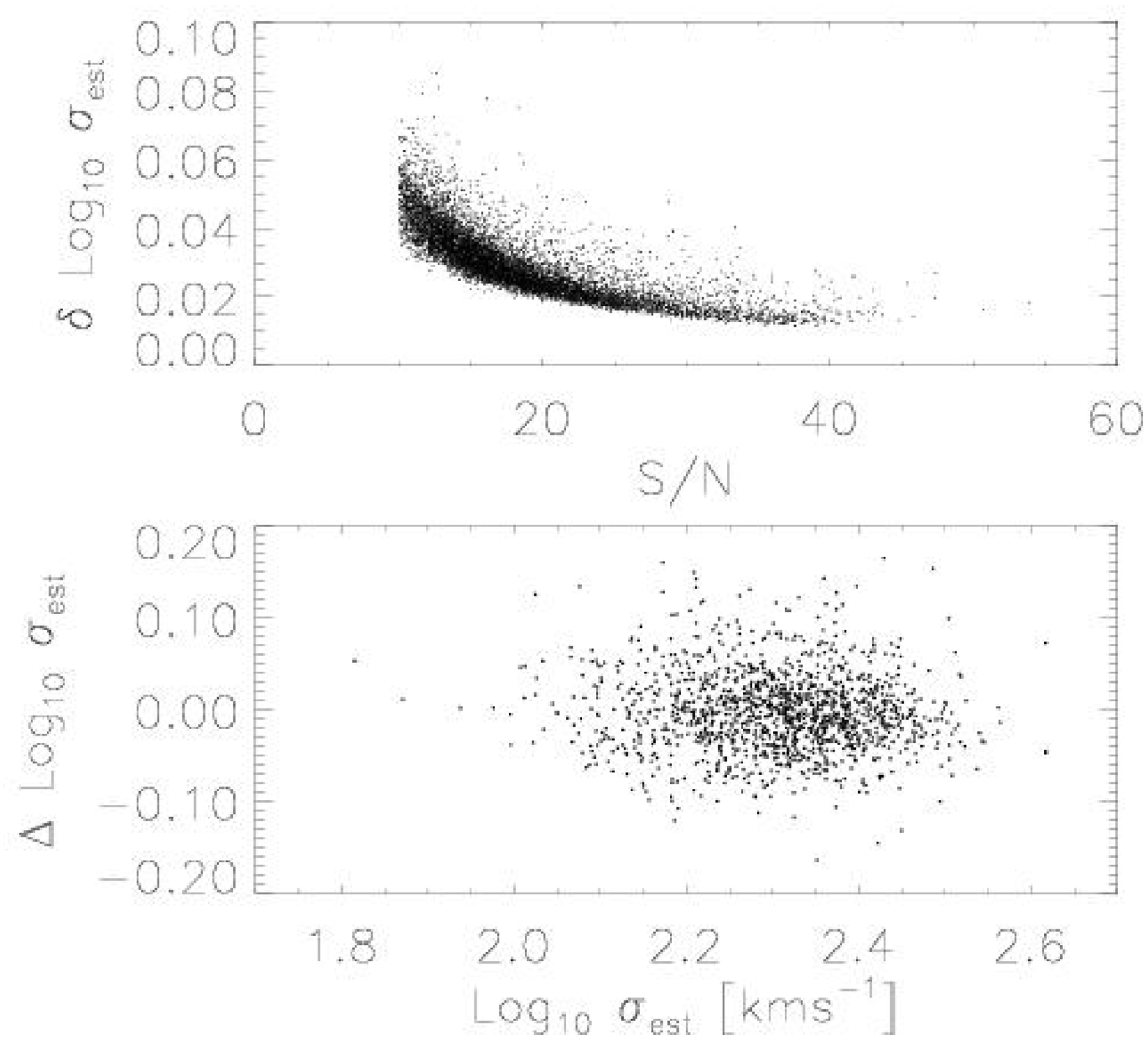}
\caption[]{Distribution of errors as a function of S/N (top) 
and comparison of estimates from repeated observations (bottom).  
Both panels suggest that, when the $S/N \ge 15$ then the typical 
error on an estimated velocity dispersion is 
$\delta \log_{10}\sigma < 0.04$.}
\label{fig:vrptd}
\end{figure}

Although the systematics are small, note that the measured velocity
dispersion is more biased at low velocity dispersions 
($\sigma < 100$ kms$^{-1}$).  For any given $S/N$ and resolution, 
there is a lower limit on the velocity dispersion measurable without 
introducing significant bias.  Since the $S/N$ of the SDSS spectra is 
not very high (see, e.g., Figure~\ref{fig:vmeth}),
and because the instrumental resolution is $\sim 90$ kms$^{-1}$, 
we chose 70 kms$^{-1}$ as a lower limit.  
Figure~\ref{fig:vmeth} shows a comparison of the 
velocity dispersion estimates obtained from the three different methods 
for the galaxies in our sample.  The median offsets are not 
statistically significant and the rms scatter is $\sim 0.05$.  
On the other hand, the top panels suggest that the cross-correlation 
method sometimes underestimates the velocity dispersion, particularly 
at low $S/N$.  

We evaluate the dependence of the velocity dispersion on the wavelength 
range by fitting the spectra in different intervals: $4000 - 5800$~\AA\ 
which is the usual wavelength range used in the literature; 
$3900 - 5800$~\AA\ to test the effect of including the Ca H and K 
absorption lines (e.g., Kormendy 1982); 
$4000 - 6000$~\AA\ to test the effect of including the NaD line 
(e.g., Dressler 1984); and 
$4000 - 7000$~\AA\  and $4000 - 9000$~\AA\ to test the effect of 
including longer wavelengths. 
The velocity dispersion obtained with the Ca H and K is $\sim 2 \%$ 
larger than that obtained using the standard wavelength region 
$4000 - 5800$~\AA, and the rms difference between the three different 
methods increases to $\sim 7\%$. 
Including the NaD line increases the velocity dispersion by $\sim 3\%$ 
but does not increase the scatter between the different methods. 
Using the wavelength range $4000 - 7000$~\AA\  only provides velocity
dispersions which are $\sim 3\%$ larger than the values obtained if 
only $4000 - 5800$~\AA\  range is used.  On the other hand, in this 
wavelength region, the different methods (and measurements from repeated
observations) are in better agreement; the scatter is $\sim 8\%$ 
smaller than in the $4000 - 5800$~\AA\  region.  
In the range $4000 - 9000$~\AA, the velocity dispersion estimates 
increase by $\sim 7\%$. This last effect is probably due to the 
presence of molecular bands in the spectra of early-type galaxies at 
long wavelengths  (i.e., to the presence of cool stars).  
Furthermore, the scatter in this wavelength region increases
dramatically ($\sim 15\%$).  Presumably  this is due to the presence 
of higher sky-line residuals and lower $S/N$.  

The estimated velocity dispersion we use in the main text are obtained 
by fitting the wavelength range $4000- 7000$~\AA\ and then using the 
average of the estimates provided by the {\it Fourier-fitting} and 
{\it direct-fitting} methods to define what we call $\sigma_{\rm est}$.  
We do not use the cross-correlation estimate because of its behavior 
at low $S/N$ as discussed earlier.  

The top panel of Figure~\ref{fig:vrptd} shows the distribution of 
the errors on the velocity dispersion as a function of the $S/N$ of the 
spectra.  The errors for each method were computed by adding in quadrature 
the statistical error due to the noise properties of the spectrum, and 
the systematic error associated with the template and galaxy mismatches.  
The final error on $\sigma_{\rm est}$ is got by adding in quadrature 
the errors on the two estimates (i.e., the Fourier-fitting and 
direct-fitting) which we average.  
The resulting errors range from $0.02\le\delta\log_{10}\sigma\le 0.06$~dex, 
depending on the $S/N$ of the spectra, with a median value of 
0.03~dex.  

A few galaxies in our sample have been observed more than once.  
The bottom panel shows a comparison of the velocity 
dispersion estimates from multiple observations. The scatter between 
different measurements is $\sim 0.04$~dex, consistent with the amplitude 
of the errors on the measurements.  

\section{Velocity dispersion:  profiles and aperture corrections}\label{vr}

\begin{figure}
\centering
\epsfxsize=\hsize\epsffile{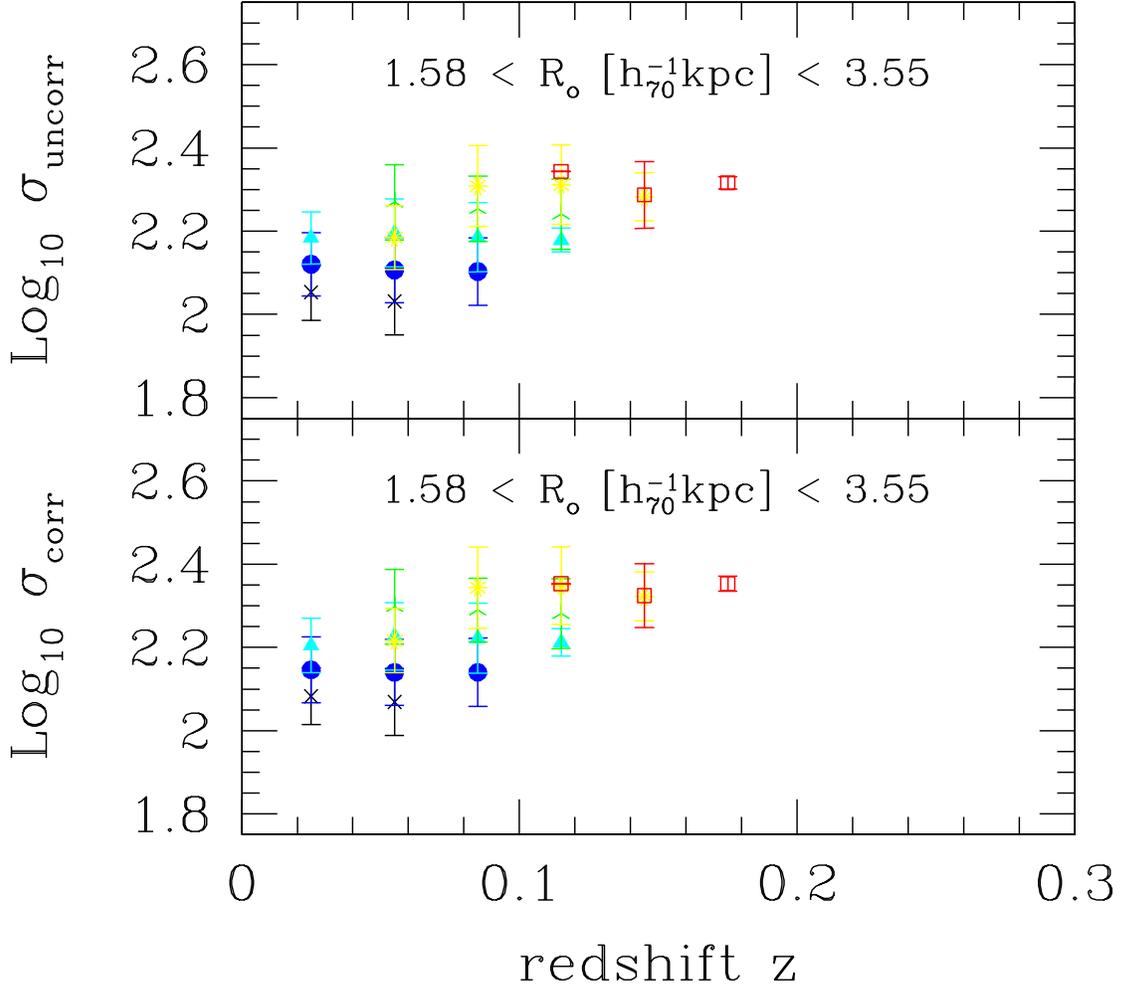}
\caption{Velocity dispersions of galaxies as a function of redshift.  
Top panel shows the estimated velocity dispersion, and bottom panel 
shows the values after correcting the estimate as described in the 
main text (equation~\ref{appcorr}).  
Different symbols show the result of averaging over 
volume-limited subsamples (same as in Figure~\ref{fig:XzR}) of 
galaxies having approximately the same luminosities and effective 
radii at each redshift.  (Error bars show the rms scatter around 
this mean value.)  The mean trends with redshift can be used to 
infer how, on average, the velocity dispersion changes with distance 
from the centre of the galaxy, and how this change depends on 
luminosity and effective radius.  }
\label{apcorr1}
\end{figure}

\begin{figure}
\centering
\epsfxsize=\hsize\epsffile{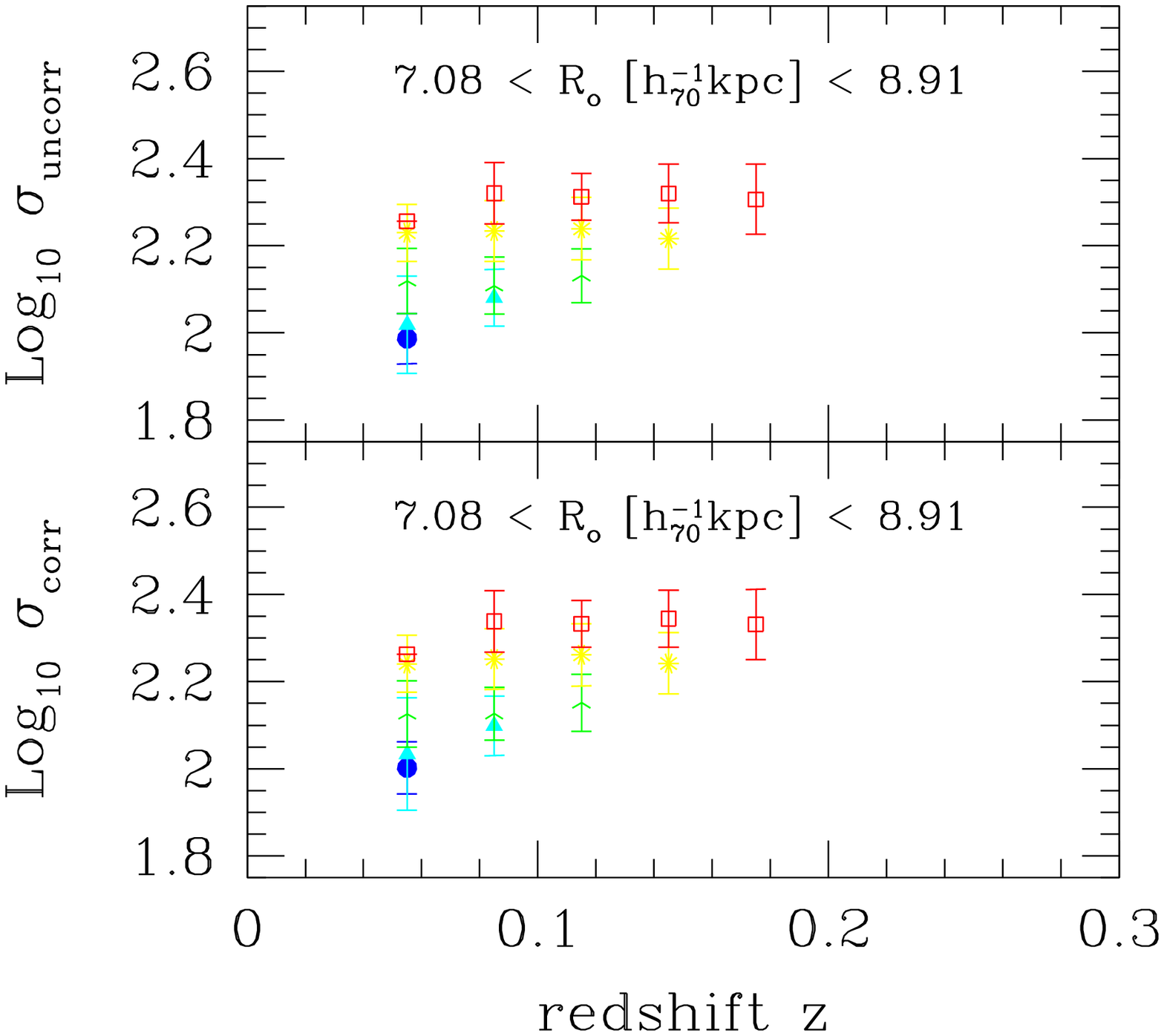}
\caption{As for previous figure, but for galaxies with larger radii.  }
\label{apcorr3}
\end{figure}

The SDSS spectra measure the light within a fixed aperture of radius 
1.5 arcsec.  Therefore, the estimated velocity dispersions of more 
distant galaxies are affected by the motions of stars at larger physical 
radii than for similar galaxies which are nearby.  If the velocity 
dispersions of early-type galaxies decrease with radius, then the 
estimated velocity dispersions (using a fixed aperture) of more distant 
galaxies will be systematically smaller than those of similar galaxies 
nearby.  

We have not measured the velocity dispersion profiles $\sigma(r)$ of 
any of the galaxies in our sample, so we cannot correctly account for 
this effect.  If we assume that the galaxies in our sample are similar 
to those for which velocity dispersion profiles have been measured, then 
we can use the published $\sigma(r)$ curves to correct for this effect.  
This is what equation~(\ref{appcorr}) in Section~\ref{spectro} does.  

An alternative procedure can be followed if evolution effects are not 
important for the velocity dispersions in our sample.  To illustrate the 
procedure, the galaxies in each of the volume-limited subsamples 
shown in Figure~\ref{fig:XzR} were further classified into small bins 
in effective physical radius (i.e., $R_o$ in kpc/$h$, not $r_o$ in arcsec).  
Figures~\ref{apcorr1}--\ref{apcorr3} show the result of plotting the 
velocity dispersions of these galaxies versus their redshifts.  
Since the galaxies at low and high redshift are supposed to be similar, 
any trend with redshift can be used to infer and average velocity 
dispersion profile, and how the shape of this profile depends on 
luminosity and effective radius.  In this way, the SDSS data themselves 
can, in principle, be used to correct for the effects of the fixed 
aperture of the SDSS spectrograph.  

In practice, because there is substantial scatter in the velocity 
dispersions at fixed luminosity and size (inserting the values from 
Table~\ref{MLcov} into the expression for $\sigma_{V|RM}$ in 
Appendix~\ref{simul} shows that this scatter is about 14\%), 
the trends in the present data set are relatively noisy.  
When the dataset is larger, it may be worth returning to this issue.  
For now, because the corrections are small anyway, we have chosen to 
use equation~(\ref{appcorr}) to correct the velocity dispersions.  
Nevertheless, curves like those presented above provide 
a novel way to study the velocity dispersion profiles of early-type 
galaxies.

\section{Maximum likelihood estimates of the correlations}\label{ML3d}
Section~\ref{lf} showed that, after accounting for the fact that the 
SDSS sample is magnitude-limited, the distribution of 
$M=-2.5\log_{10}L$ was well described by a Gaussian.  
We would also like to present the intrinsic distributions of $R_o$ 
and $\sigma$.  To do so, we must study how $R_o$ and $\sigma$ are 
correlated with luminosity, and with each other.  
In principle, we could do this by extending the 
Efstathiou, Ellis \& Peterson (1988) method (along the lines 
described by Sodr{\'e} \& Lahav 1993) to obtain a non-parametric 
maximum-likelihood estimate of the three-dimensional distribution 
of $L$, $R_o$ and $\sigma$.  The virtue of this approach is that it 
accounts for the fact that the observed sample is magnitude-limited, 
that there is also a cut at small velocity dispersions, and that there 
are correlated measurement errors associated with the luminosities, 
sizes and velocity dispersions.  

We chose not to make a non-parametric estimate of the joint distribution 
because just ten bins in each of $L$, $R_o$ and $\sigma$ yields $10^3$ 
free parameters to be determined from $10^4$ galaxies.  
In Appendix~\ref{lx} we show that, in each of the SDSS wavebands, the 
joint distribution of early-type galaxy luminosities, sizes, and velocity 
dispersions is well described by a tri-variate Gaussian distribution 
in the variables $M=-2.5\log_{10}L$, $R=\log_{10}R_o$ and 
$V=\log_{10}\sigma$.  Thus, we have a simple parametrization of the 
joint distribution for which, in each waveband, nine numbers suffice 
to describe the statistical properties of our sample:  
three mean values, $M_*$, $R_*$ and $V_*$, 
three dispersions, $\sigma^2_M$, $\sigma^2_R$ and $\sigma^2_V$, 
and three pairwise correlations, $\sigma_R\sigma_M\,\rho_{RM}$, 
$\sigma_V\sigma_M\,\rho_{VM}$, and $\sigma_R\sigma_V\,\rho_{RV}$.  

In addition, we will also allow for the possibility that the 
luminosities are evolving---a tenth parameter to be estimated from 
the sample.  The maximum likelihood technique allows us to estimate 
these ten numbers as follows.  We define the likelihood function  
\begin{eqnarray}
{\cal L} &=& \prod_i {\phi({\cal X}_i,{\cal C},{\cal E}_i)\over S(z_i)},
 \qquad{\rm where}\nonumber \\
{\cal X} &=& (M-M_*+Qz,R-R_*,V-V_*) , \nonumber\\
{\cal E} &=& \left( \begin{array}{ccc} 
                      \epsilon^2_{MM} & \epsilon^2_{RM} & \epsilon^2_{VM} \\ 
                      \epsilon^2_{RM} & \epsilon^2_{RR} & \epsilon^2_{RV} \\
                      \epsilon^2_{VM} & \epsilon^2_{RV} & \epsilon^2_{VV} \\
                    \end{array}\right),\nonumber\\
{\cal C} &=& \left( \begin{array}{ccc} 
     \sigma^2_M & \sigma_R\sigma_M\,\rho_{RM} & \sigma_V\sigma_M\,\rho_{VM}\\
     \sigma_R\sigma_M\,\rho_{RM} & \sigma^2_R & \sigma_R\sigma_V\,\rho_{RV}\\
     \sigma_V\sigma_M\,\rho_{VM} & \sigma_R\sigma_V\,\rho_{RV} & \sigma^2_V\\
           \end{array}\right)\qquad{\rm and} \nonumber\\
\phi({\cal X},{\cal C},{\cal E}) &=& 
{\phi_*\over (2\pi)^{3/2}\,|{\cal C}+{\cal E}|^{-1/2}}\, 
\exp\left(-{1\over 2}\,{\cal X}^{T}\,[{\cal C}+{\cal E}]^{-1} {\cal X}\right).
\end{eqnarray}
Similarly to when we discussed the luminosity function, $S(z_i)$ is defined 
by integrating over the range of absolute magnitudes, velocities and 
sizes at $z_i$ which make it into the catalog.  Here ${\cal X}$ is the 
vector of the observables, and ${\cal E}$ describes the errors in the 
measurements.

The elements of the error matrix ${\cal E}$ are obtained as follows.  
The photometric pipeline estimates the size $r_{\rm dev}$ and the 
apparent magnitude $m_{\rm dev}$ from the same fitting procedure.  
As a result, errors in these two quantities are correlated.  
Let $e_r$ denote the error in $\log_{10}r_{\rm dev}$, and $e_m$ 
the error in $m_{\rm dev}$.  The correlation means that we need three 
numbers to describe the errors associated with the fitting procedure, 
$\langle e_r e_r\rangle$, $\langle e_m e_m\rangle$, and 
$\langle e_m e_r\rangle$, but the pipeline only provides two.  
The error output by the pipeline in $r_{\rm dev}$, is correctly 
marginalized over the uncertainty in $m_{\rm dev}$, so it is 
essentially $\langle e_r e_r\rangle$.  
On the other hand, the quoted error in $m_{\rm dev}$, say 
$\langle e_{photo}^2\rangle$  is really 
$\langle e_m e_m\rangle -
\langle e_m e_r\rangle^2/\langle e_r e_r\rangle$.  
To estimate the values of $\langle e_m e_r\rangle$ and 
$\langle e_m e_m\rangle$ which we need, we must make an assumption 
about the correlation between the errors.  

Fortunately, this can be derived from the fact that, for a wide 
variety of galaxy profile shapes, the quantity 
$\xi\equiv e_r - \alpha e_\mu$, with $\alpha\approx 0.3$, 
has a very small scatter (e.g. Saglia et al. 1997).  Here 
$\mu \equiv m_{\rm dev} + 5\log_{10}r_{\rm dev} + 2.5\log_{10}(2\pi)$ 
is the surface brightness, and $e_\mu$ is the error in the surface 
brightness.  As a result, 
\begin{eqnarray}
\langle e_\mu e_\mu\rangle &=& {\langle e_r^2\rangle\over\alpha^2} + 
{\langle\xi^2\rangle\over\alpha^2}{(\alpha^2-1)\over (1+\alpha^2)} \nonumber\\
\langle e_\mu e_r\rangle &=& {\langle e_r^2\rangle\over\alpha} - 
{\langle\xi^2\rangle\over \alpha (1+\alpha^2)}
\end{eqnarray}
(Saglia et al. 1997).  This means that 
$\langle e_r\xi\rangle = \langle \xi^2\rangle/(1+\alpha^2)$, so that 
\begin{eqnarray}
\langle e_m e_r\rangle &=& 
      \langle e_r e_r\rangle \left(1-5\alpha\over\alpha\right) - 
      {\langle \xi^2\rangle\over \alpha(1+\alpha^2)} ,\nonumber\\
\langle e_m e_m\rangle &=& 
      \langle e_r e_r\rangle \left(1-5\alpha\over\alpha\right)^2 
      + {\langle \xi^2\rangle\over \alpha^2}\,
        \left[1 - 2\left(1-5\alpha\over 1+\alpha^2\right)\right],
\end{eqnarray}
and 
\begin{equation}
\langle e_m e_m\rangle - 
{\langle e_m e_r\rangle^2\over\langle e_r e_r\rangle} = 
 {\langle \xi^2\rangle\over \alpha^2}  
- {\langle \xi^2\rangle\over\alpha^2} 
  {\langle \xi^2\rangle/\langle e_r e_r\rangle \over(1+\alpha^2)^2} = 
   \langle e_{photo}^2\rangle.  
\end{equation}
The final equality shows that the error output from the pipeline 
provides an estimate of $\langle \xi^2\rangle$ which we can insert into 
our expressions for $\langle e_m e_m\rangle$, $\langle e_m e_r\rangle$, 
and $\langle e_\mu e_\mu\rangle$.  
(Notice that if $\langle \xi^2\rangle\ll \langle e_r e_r\rangle$, then 
it would be a good approximation to set 
 $\langle e_{photo}^2\rangle\approx \langle \xi^2\rangle/\alpha^2$.  
Since this is not always the case for our dataset, we must solve 
the quadratic.)  Once this has been done, we set 
\begin{eqnarray}
\epsilon_{MM}^2 &=& \langle e_m e_m\rangle  ,\nonumber\\
\epsilon_{RR}^2 &=& \langle e_r e_r\rangle + 
                    {\langle e_{ab}\,e_{ab}\rangle\over 4} ,\nonumber\\
\epsilon_{RM}^2 &=& \langle e_m e_r\rangle ,\nonumber\\
\epsilon_{VM}^2 &=& 0 ,\nonumber\\
\epsilon_{VV}^2 &=& \langle e_v e_v\rangle + (0.04\,\epsilon_{RR})^2,
 \nonumber\\
\epsilon_{RV}^2 &=& -0.04\,\epsilon_{RR}^2 ,
\end{eqnarray}
That is, we compute the error in the absolute magnitude by assuming 
that there are no errors in the determination of the redshift which 
would otherwise propagate through.  

The main text works with a circularly averaged radius $R_o$, so 
the errors in it are given by adding the errors in the size 
$r_{\rm dev}$ to those which come from the error on the shape $b/a$.  
We assume that the errors in $b/a$ are neither correlated with those 
in $\log_{10}r_{\rm dev}$ nor with those in the absolute magnitude.  
Finally, we assume that errors in magnitudes are not correlated 
with those in velocity dispersion, so $\langle\epsilon^2_{VM}\rangle$ 
is set to zero, and that errors in size and velocity dispersion are 
only weakly correlated because of the aperture correction we apply.  
Here $\langle e_v e_v\rangle$ is the error in what was called 
$\log_{10} \sigma_{\rm est}$ in the main text.  

The covariance matrix ${\cal C}$ contains six of the ten free parameters 
we are seeking.  It is these parameters, along with the three mean values, 
$M_*$, $R_*$ and $V_*$, and the evolution parameter $Q$ which are varied 
until the likelihood is maximized.  The maximum-likelihood estimates of 
these parameters in each band are given in Table~\ref{MLcov} of 
Section~\ref{fp}.  
Notice that although the luminosity and size distributions differ from 
band to band, the velocity distributions do not.  This is reassuring, 
because the intrinsic distribution of velocity dispersions, estimated 
from the spectra, should not depend on the band in which the photometric 
measurements were made.  
As an additional test, we also computed maximum-likelihood estimates 
of the $2\times 2$ covariance matrices of the bivariate Gaussians for 
the pairs $(M,R)$ and $(M,V)$.  These estimates of, e.g., $\rho_{RM}$ 
and $\rho_{VM}$ were similar to those in Table~\ref{MLcov}.  

In Section~\ref{fp}, we use the fact that the surface brightnesses of 
the galaxies in our sample are defined using their luminosities and 
sizes.  This allows us to transform the covariance matrix ${\cal C}$ 
into the one which describes the Fundamental Plane.  
In Appendix~\ref{prjct} we show how knowledge of ${\cal C}$ allows 
us to estimate various pairwise correlations.  
And in Appendix~\ref{simul} we use our knowledge of ${\cal C}$ 
and the evolution parameter $Q$ to generate mock galaxy catalogs which 
are similar to our data set.  

\section{Distributions at fixed luminosity}\label{lx}
This Appendix presents scatter plots between different observables 
$X$ and luminosity.  This is done because, except for a cut at small 
velocity dispersions, our sample was selected by luminosity alone.  
This means that the distributions of $X$ at fixed luminosity are not 
biased by the selection cut (e.g., Schechter 1980).  
The distribution of $X$ at fixed $L$ is shown to be reasonably well 
described by a Gaussian for all the choices of $X$ we consider.  
This simplifies the maximum likelihood analysis in 
Appendix~\ref{ML3d} which we use to estimate the parameters of the 
Fundamental Plane (Section~\ref{fp}), and the various projections of 
it (Appendix~\ref{prjct}).  

The best way to think of any absolute magnitude $M$ versus $X$ 
scatter plot is to imagine that, at fixed absolute magnitude $M$, 
there is a distribution of $X$ values.  The scatter plot then shows 
the joint distribution 
\begin{equation}
\phi(M,X|z)\,{\rm d}M\,{\rm d}X = {\rm d}M\,\phi(M|z)\ p(X|M,z){\rm d}X ,
\label{plx}
\end{equation}
where $\phi(M,X|z)$ denotes the density of galaxies with $X$ and $M$ at 
$z$, and $\phi(M|z)$ is the luminosity function at $z$ which we computed 
in Section~\ref{lf}.  One of the results of this section is to show 
that the shape of $p(X|M,z)$ is simple for most of the relations of 
interest.  

The mean value of $X$ at fixed $M$ is independent of the fact that our 
catalogs are magnitude limited.  Therefore, we estimate the parameters 
of linear relations of the form:  
\begin{equation}
(X-X_*) = {-0.4\,(M-M_*)\over S},
\label{xlregress}
\end{equation}
where $M=-2.5\log_{10}L$ is the absolute magnitude 
and $X$ is the observable (for example, we will study 
$X=\log_{10}\sigma$, $\log_{10}R_o$ or $\mu_o=-2.5\log_{10}I_o$).  
For each volume limited catalog, we fit for the slope $S$ and 
zero-point of the linear relation.  If there really were a 
linear relation between $M$ and $X$, and neither $X$ nor $M$ 
evolved, then the slopes and zero-points of the different 
volume limited catalogs would be the same.  

\begin{figure}
\centering
\epsfxsize=\hsize\epsffile{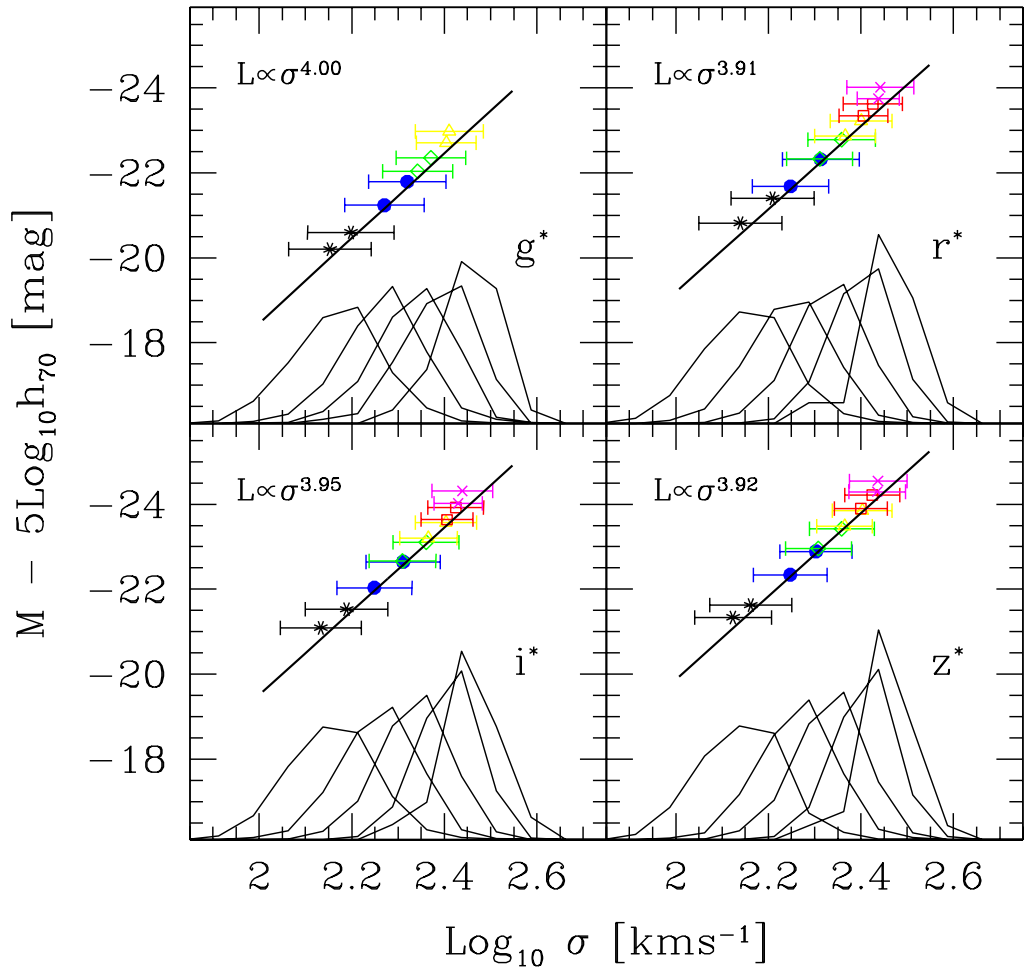}
\vspace{-1.cm}
\caption{Relation between luminosity $L$ and velocity dispersion $\sigma$.  
Stars, circles, diamonds, triangles, squares and crosses show the 
error-weighted mean value of $\log_{10}\sigma$ for a small range in 
luminosity in each volume limited catalog (see text for details).  
(Only catalogs containing more than one hundred galaxies are shown.)  
Error bars show the rms scatter around this mean value.  
Solid line shows the maximum-likelihood estimate of this relation 
computed in Appendix~\ref{prjct}, and the label in the top left 
shows the scaling it implies.  Histograms show the distribution of 
$\log_{10}\sigma$ in small bins in luminosity.  They were obtained 
by stacking together non-overlapping volume limited catalogs to construct 
a composite catalog, and then dividing the composite catalog into five 
equal size bins in luminosity.}
\label{ls}
\end{figure}

To illustrate, the different symbols in Figure~\ref{ls} show 
$\langle\log_{10}\sigma|M\rangle$ computed in each of the different 
volume-limited subsamples.  Stars, circles, diamonds, triangles, squares 
and crosses shows successively higher redshift catalogs (redshift limits 
are the same as in Figure~\ref{lfev}).  The galaxies in each subsample 
were divided into two equal-sized parts based on luminosity.  The symbols 
with error bars show the mean $\log_{10}\sigma$ for each of these small 
bins in $M$, and the rms spread around it (note that the error on the mean 
is smaller than the size of the symbols in all but the highest redshift 
catalogs).  The solid line shows the maximum-likelihood estimate of the 
slope of this relation at $z=0$, which we describe in Appendix~\ref{prjct}.  
The slope of this line is shown in the top of each panel:  
$\sigma\propto L^{1/4}$, approximately, in all the bands.  
The figure shows that, at fixed velocity dispersion, the higher redshift 
population is brighter.  

\begin{figure}
\centering
\epsfxsize=\hsize\epsffile{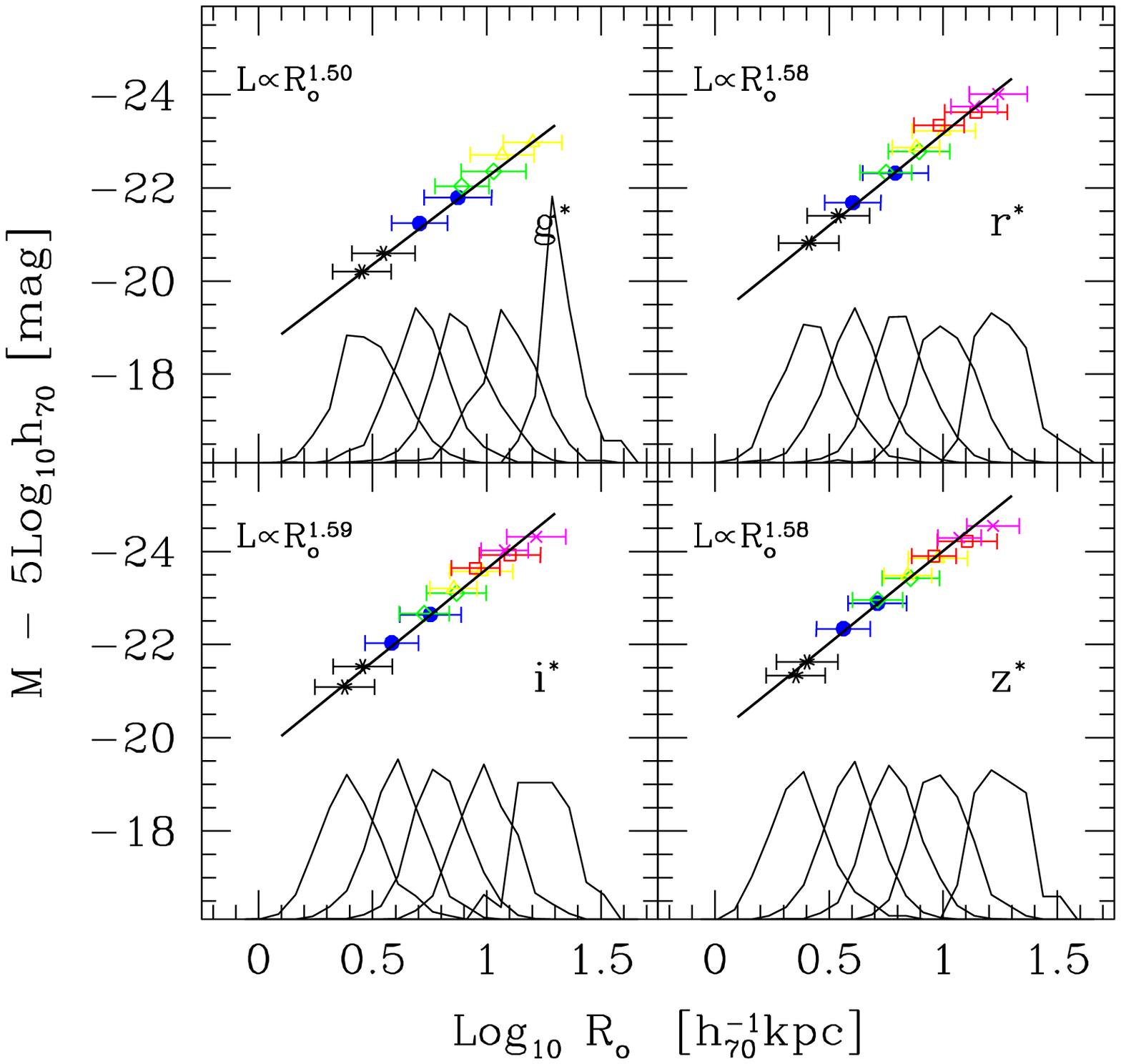}
\vspace{-1.cm}
\caption{Same as previous figure, but for the relation between 
luminosity $L$ and effective radius $R_o$.}
\label{lre}
\end{figure}

We have enough data that we can actually do more than simply measure the 
mean $X$ at fixed $M$; we can also compute the distribution around the 
mean.  If we do this for each catalog, then we obtain distributions 
which are approximately Gaussian in shape, with dispersions which 
depend on the range of luminosities which are in the subsample.  Rather 
than showing these, we created a composite catalog by stacking together 
the galaxies from the nonoverlapping volume limited catalogs, and we 
then divided the composite catalog into five equal sized bins in 
luminosity.  The histograms in the bottom of the plot show the shapes 
of the distribution of velocities in the different luminosity bins.  
Except for the lowest and highest redshift catalogs for which the 
statistics are poorest, the different distributions have almost the 
same shape; only the mean changes.  

One might have worried that the similarity of the distributions is a 
signature that they are dominated by measurement error.  This is not 
the case:  the typical measurement error is about a factor of two smaller 
than the rms of any of these distributions.  If we assume that the 
measurement errors are Gaussian-distributed, then the distributions we 
see should be the true distribution broadened by the Gaussian from the 
measurement errors.  The fact that the observed distributions are well 
approximated by Gaussians suggests that the true intrinsic distributions 
are also Gaussian.  The fact that the width of the intrinsic distribution 
is approximately independent of $M$ considerably simplifies the maximum 
likelihood analysis presented in the main text.  

We argued in the main text that color was strongly correlated with velocity 
dispersion.  One consequence of this is that residuals from the $\sigma-L$ 
relation shown in Figure~\ref{ls} correlate strongly with color:  at fixed 
magnitude, the redder galaxies have the highest velocity dispersions.  
In addition, as a whole, the reddest galaxies populate the high $\sigma$ 
part of the relation.  
Forbes \& Ponman (1999) reported that residuals from the Faber--Jackson 
relation correlate with age.  If color is an indicator of age and/or 
metallicity, then our finding is qualitatively consistent with theirs:  
the typical age/metallicity varies along the Faber--Jackson relation.  

A similar study of the relation between the luminosities and sizes 
of galaxies is shown in Figure~\ref{lre}.  The distribution 
$p(\log_{10}R_o|M)$ is also reasonably well fit by a Gaussian, 
with a mean which increases with luminosity, and a dispersion which 
is approximately independent of $M$.  The rms around the mean is about 
one and a half times larger than the rms around the mean $\sigma-L$ 
relation.  We argued in the main text that the color--magnitude and 
color--size relations were a consequence of the color$-\sigma$ correlation.  
If this is correct, then residuals from the $R_o-L$ relation, should not 
correlate with size or magnitude.  We have checked that this is correct, 
although we have not included a plot showing this explicitly.  

\begin{figure}
\centering
\epsfxsize=\hsize\epsffile{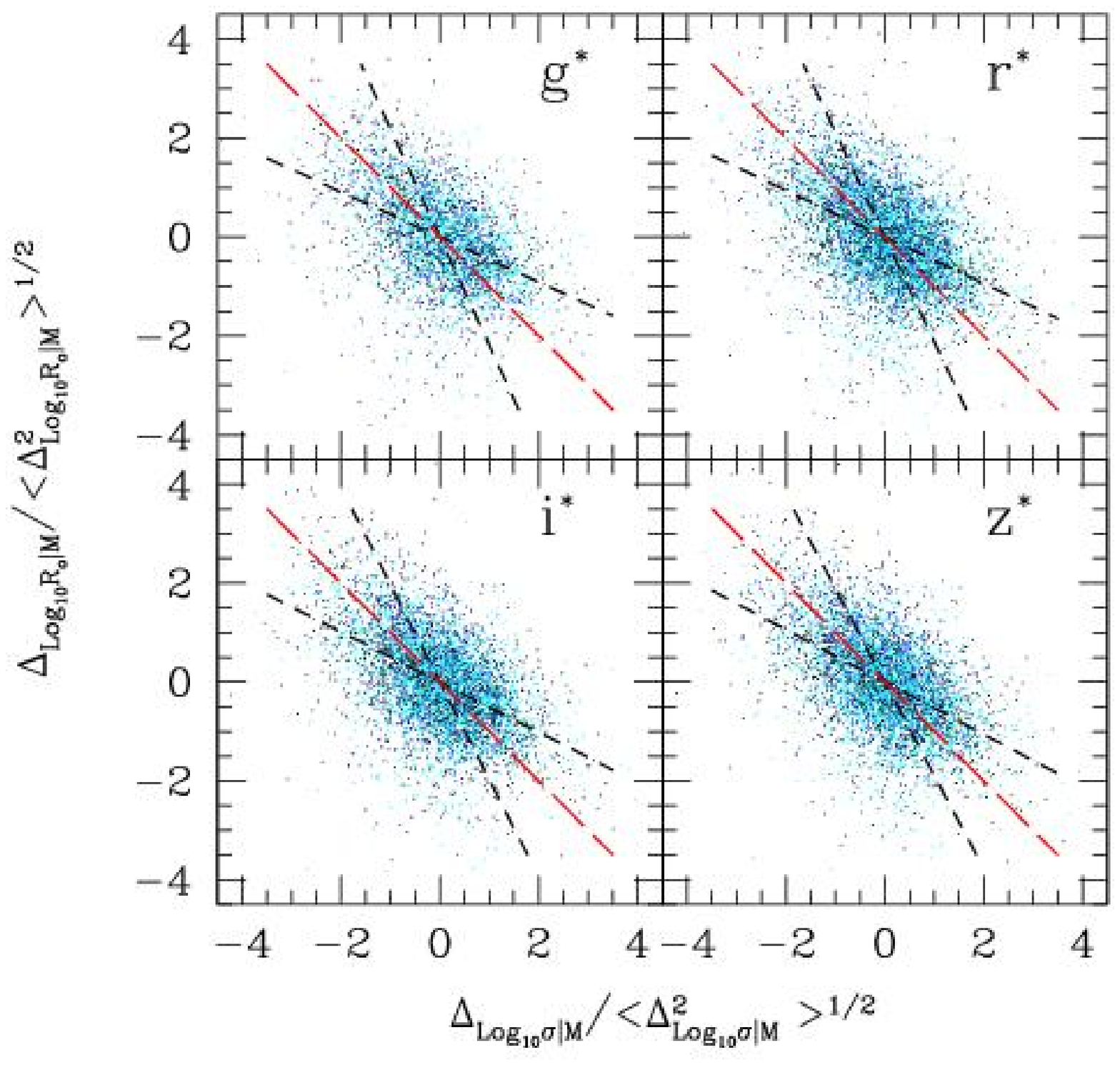}
\caption{Residuals of the $R_o-L$ relation are anti-correlated with 
residuals of the $\sigma-L$ relation; galaxies of the same luminosity 
which are smaller than expected have larger velocity dispersions than 
expected.  Plot shows the residuals normalized by their rms value.  
Short-dashed lines show forward and inverse fits to the scatter plots, 
and long-dashed line in between the other two shows  
$\Delta_{R_o|M}/\sigma_{R|M} = -\Delta_{\sigma-M}/\sigma_{V|M}$}.  
\label{resid}
\end{figure}

\begin{figure}
\centering
\epsfxsize=\hsize\epsffile{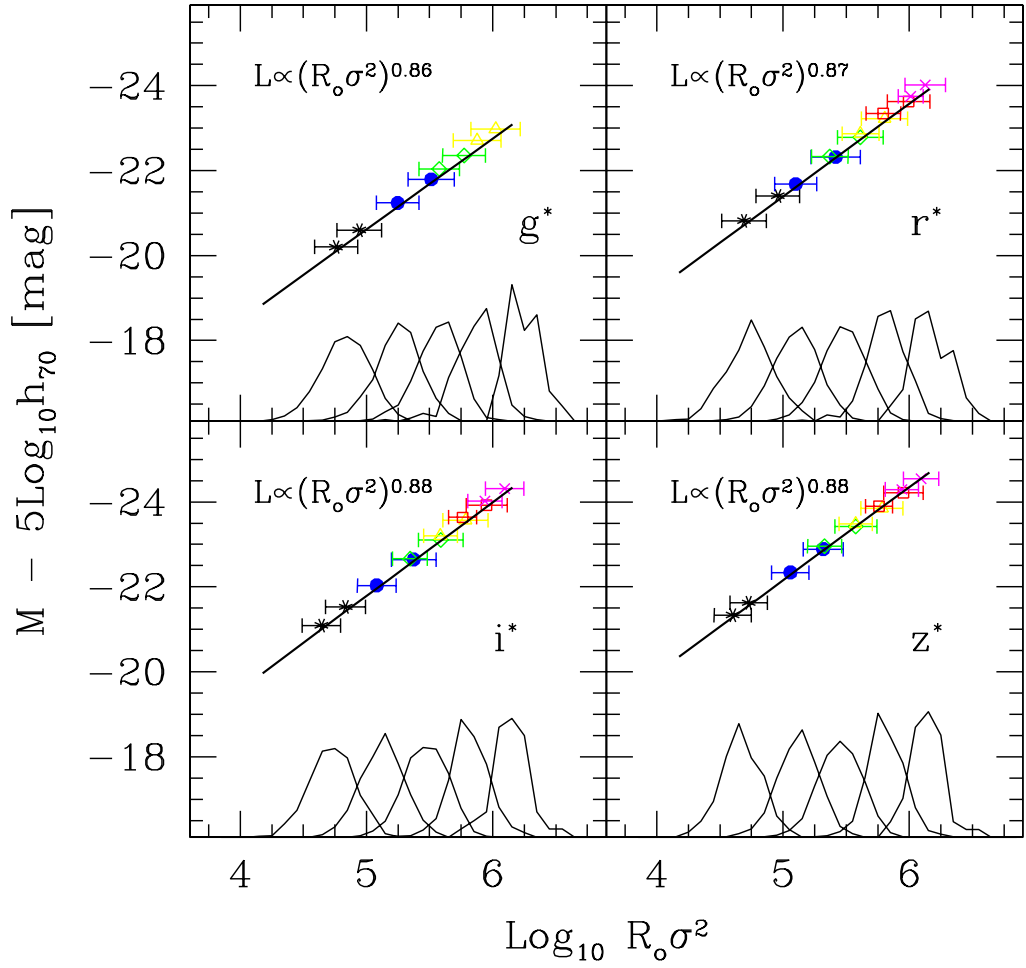}
\vspace{-1.cm}
\caption{Same as Figure~\ref{ls}, but for the relation between 
luminosity $L$ and the combination $R_o\sigma^2$, which is supposed 
to be a measure of mass.}
\label{lmass}
\end{figure}

\begin{figure}
\centering
\epsfxsize=\hsize\epsffile{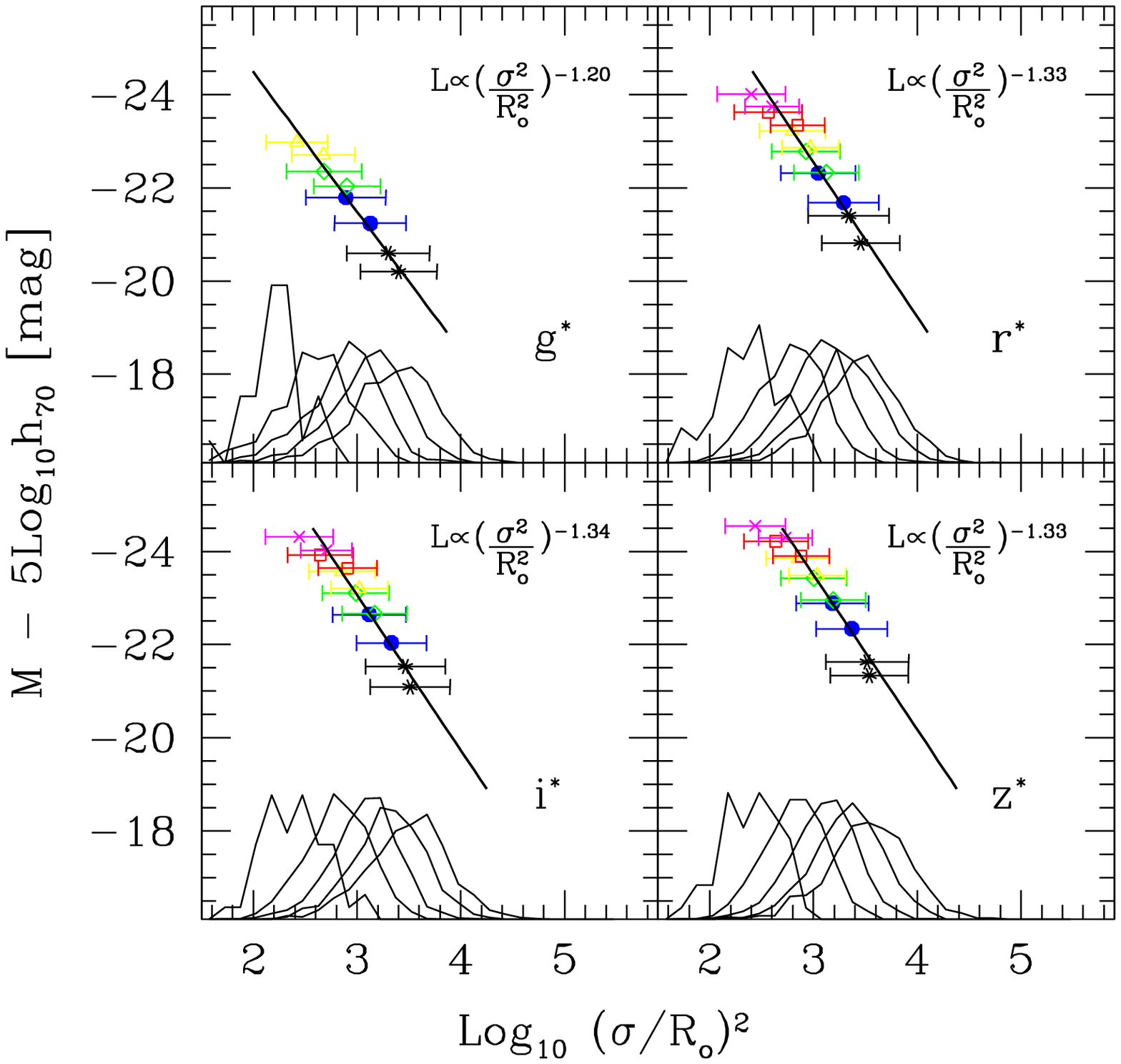}
\vspace{-1.cm}
\caption{Same as previous figure, but for the relation between 
luminosity $L$ and the combination $(\sigma/R_o)^2$, which 
is supposed to be a measure of density.}
\label{lden}
\end{figure}

There is, however, an interesting correlation between the residuals of 
the Faber--Jackson and $R_o-L$ relations.  At fixed luminosity, galaxies 
which are larger than the mean $\langle R_o|L\rangle$ tend to have smaller 
velocity dispersions.  
This is shown in Figure~\ref{resid}, which plots the residuals from 
the $\sigma-L$ relation versus the residuals from the $R_o-L$ 
relation.  The short dashed lines show the forward and inverse fits to 
this scatter plot.  The long-dashed line in between the other two shows 
$\Delta_{R|M}/\sigma_{R|M} = -\Delta_{V|M}/\sigma_{V|M}$, 
where $\Delta_{X|M}$ denotes the residual from the mean relation at 
fixed $M$, and $\sigma_{X|M}$ denotes the rms of this residual.  
The anti-correlation is approximately the same for all $L$.  

This suggests that a plot of $L$ versus some combination of $R_o$ and 
$\sigma$ should have considerably less scatter than either of the two 
individual relations.  To illustrate, Figure~\ref{lmass} shows the 
distribution of the combination $R_o\sigma^2$ at fixed $L$.  The scatter 
in $L$ is significantly reduced, making the mean trend of increasing 
$R_o\sigma^2$ with increasing $L$ quite clean.  (The combination of 
observables for which the scatter is minimized is discussed in 
Appendix~\ref{prjct}.)  This particular combination defines an effective 
mass:  $M_o \equiv 2R_o\sigma^2/G$.  In slightly more convenient units, 
this mass is  
\begin{equation}
\left({M_o\over 10^{10}h^{-1}M_\odot}\right) = 
 0.465\,\left({R_o\over h^{-1}{\rm kpc}}\right)
 \left({\sigma\over 100\,{\rm kms}^{-1}}\right)^2 .
\end{equation}
(Because many of our galaxies are not spherical, some of their support 
must come from rotation, and so ignoring rotation as we are doing is 
likely to mis-estimate the true mass.  See Bender, Burstein \& Faber 1992 
for one way to account for this.)  
Inserting the mean values of Table~\ref{MLcov} into this relation 
yields $M_o\approx 10^{10.56}h_{70}^{-1}M_\odot$ (we used the parameters 
for the $r^*$ band, for which $R_*+2V_* \approx 4.89$).  
The corresponding total absolute magnitude is 
$M_*-5\log_{10}h_{70} \approx -21.15$.  
The luminosity of the sun in $r^*$ is $4.62$~mags, so 
$L_*\approx 10^{10.31}h_{70}^{-2}L_\odot$.  
The luminosity within the effective radius is half this value, so that 
the effective mass-to-light ratio within the effective radius of an 
$L_*$ object is $2h_{70}\times 10^{10.56-10.31}\approx 3.57h_{70}$ times 
that of the sun.  Figure~\ref{lmass} shows that the effective mass-to-light 
ratio is not constant:  at fixed luminosity $M_o/L\propto L^{0.15}$.  
At larger radii, the luminosity can double at most, whereas, if the 
galaxy is embedded in a dark matter halo, the mass at large radii may 
continue to increase.  For this reason one might expect the mass-to-light 
ratios to be significantly larger at larger radii.  

Since $R_o\sigma^2/G$ defines an effective mass, the combination 
$(\sigma/R_o)^2/G$ defines an effective density.  If we set 
 $3M_o/4\pi R_o^3 = \Delta_o\,\rho_{\rm crit}$, 
with $\rho_{\rm crit} \equiv 3H^2/8\pi G$, then 
\begin{equation}
\Delta_o = 4\times 10^6\times 0.7^2
 \left({\sigma\over 100\,{\rm kms}^{-1}}
       {h_{70}^{-1}{\rm kpc}\over R_o}\right)^2.
\end{equation}
Again, inserting mean values for $\sigma$ and $R_o$ yields 
$\Delta_o\approx 5.16\times 10^5$ in $r^*$.  
Figure~\ref{lden} shows that this density decreases with increasing 
luminosity, although the scatter in densities at fixed luminosity is 
quite large ($\sim 0.32$).  It is interesting that such a trend is 
qualitatively similar to that seen in numerical simulations of 
dissipationless gravitational clustering:  the central densities of 
virialized halos in such simulations are smaller in the more massive 
halos (Navarro, Frenk \& White 1997; Bullock et al. 2001).  

\begin{figure}
\centering
\epsfxsize=\hsize\epsffile{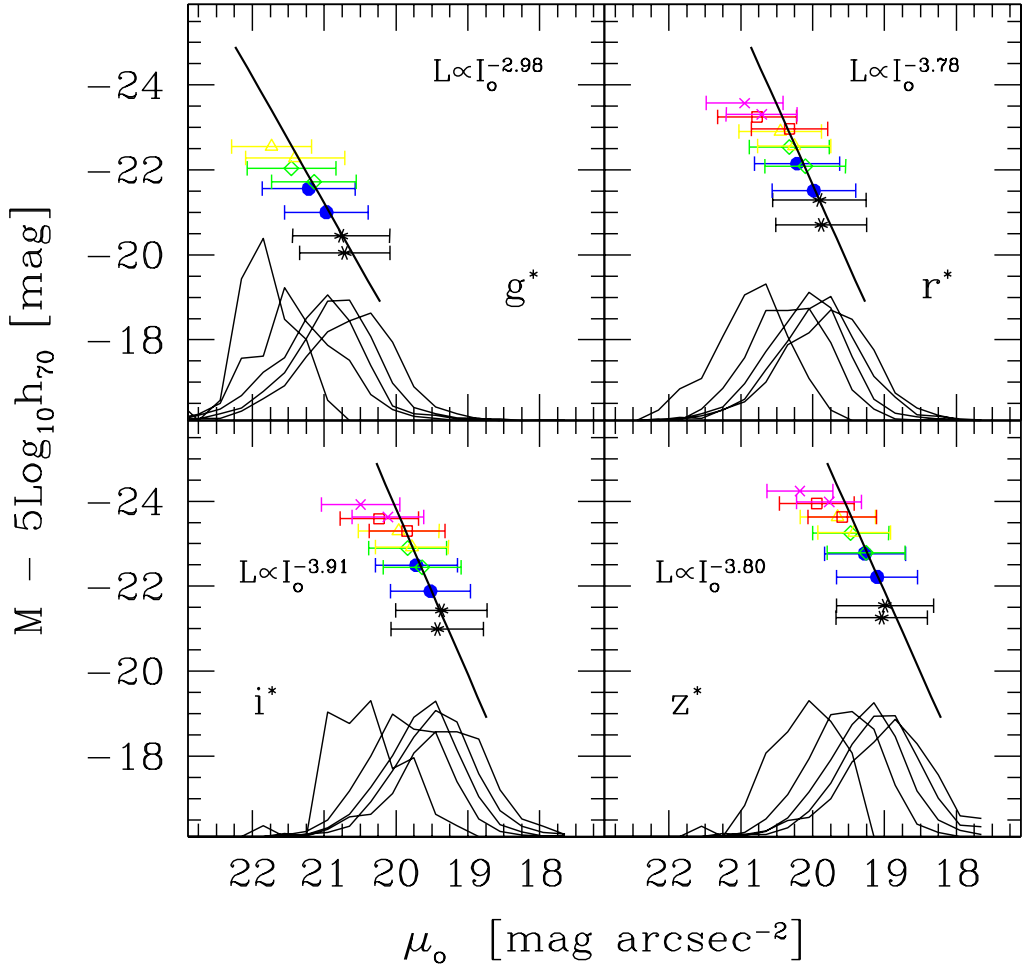}
\vspace{-1.cm}
\caption{Relation between luminosity $L$ and surface brightness $\mu_o$ 
in different volume limited catalogs (higher redshift catalogs contribute 
points to the upper-left corners of each plot).  
Passive evolution of luminosities would shift points upwards and to 
the right of the zero-redshift relation, but, the slope of the relation 
should remain unchanged.  This shift has been subtracted.}
\label{lmu}
\end{figure}

Figure~\ref{lmu} shows a final relation at fixed luminosity:  the 
surface-brightness$-L$ relation.  In such a plot, luminosity evolution 
moves objects upwards and to the right, so that the higher redshift 
population should be obviously displaced from the zero-redshift 
relation.  The plot shows the distribution of $\mu_o$ and $M$ after 
subtracting the maximum likelihood estimate of the evolution from both 
quantities.  The solid line shows the maximum likelihood value of 
the slope of this relation:  it is steeper than the local slope reported 
by Sandage \& Perelmuter (1990).  A careful inspection of the figure 
suggests that the relation is becoming shallower at high redshift; this 
is the subject of work in progress.  

\section{Projections of the Fundamental Plane}\label{prjct}
\subsection{Maximum-likelihood estimates}
Section~\ref{fp} describes the Fundamental Plane and its dependence 
on redshift and environment.  As we now describe, the shapes of 
all the correlations in Appendix~\ref{lx} can be estimated by 
marginalizing the covariance matrices ${\cal C}$ or ${\cal F}$.  
In this respect, these correlations may be thought of as views of 
the Fundamental Plane from different projections.  

Before we present these projections, it is worth remarking that 
because the trivariate Gaussian is a good description of the data, 
our results indicate that, in addition to the intrinsic distribution 
of absolute magnitudes, the intrinsic distributions of (the logarithms 
of) early-type galaxy sizes and velocity dispersions are also well 
fit by Gaussian forms.  
The means and dispersions of these Gaussians are given by 
$(R_*,\sigma^2_R)$ and $(V_*,\sigma^2_V)$ in Table~\ref{MLcov} of 
Section~\ref{fp}.  Note that the width of the distribution of 
$\log_{10}\sigma$ happens to be about twice that of the width of the 
distribution of $\log_{10}R_o$.  

As we describe below, appropriate combinations of the numbers in 
Table~\ref{MLcov} provide maximum likelihood estimates of various 
linear regressions between pairs of observables which are often 
studied; these are summarized in Table~\ref{lx4cov}.  
Plots comparing the linear regressions themselves with the 
maximum likelihood estimates are shown in Appendix~\ref{lx}.  

The Faber--Jackson relation (Faber \& Jackson 1976) describes the 
correlation between luminosity and velocity dispersion.  
This relation depends weakly on wavelength, although most datasets 
are consistent with the scaling $L\propto\sigma^4$.  For example, 
Forbes \& Ponman (1999), using a compilation of data from 
Prugniel \& Simien (1996) report that $L\propto\sigma^{3.92}$ in 
the B-band.  At longer wavelengths Pahre et al. (1998) report 
$L_K\propto\sigma^{4.14\pm 0.22}$ in the K-band, with a scatter of 
0.93 mag.  
In our data set, the mean velocity dispersion at fixed $M$ is 
\begin{equation}
\Bigl\langle V-V_*|M-M_*\Bigr\rangle = 
{(M-M_*)\over\sigma_M}\ \sigma_V\,\rho_{VM},
\label{meanvm}
\end{equation}
and the dispersion around this mean is 
$\sigma^2_{V|M}\equiv \sigma^2_{VV}(1 - \rho^2_{VM})$.  
Writing this as $(V/V_*)\propto (L/L_*)^{1/S_{VL}}$, 
shows that $S_{VL} = -0.4\sigma_M/(\sigma_V\rho_{VM})$.  
Inserting the values in Table~\ref{MLcov} into these expressions for 
$S_{VL}$ and $\sigma^2_{V|M}$ provides the maximum likelihood 
estimate of the slope and thickness of this relation.  These are 
shown in the second column of Table~\ref{lx4cov}.
(The errors we quote on the slopes of this, and the other relations 
in the Table, were obtained using subsamples just as we did for the 
Fundamental Plane.  Note that the errors we find in this way are 
comparable to those sometimes quoted in the literature, even though 
each of the subsamples we selected is an order of magnitude larger than 
any sample available in the literature.)

\begin{table}[t]
\centering
\caption[]{Maximum-likelihood slopes of the relations between the mean 
velocity dispersion, effective radius, effective mass, effective density, 
and effective surface brightness, and luminosity at fixed luminosity, 
$S_{VL}$, $S_{RL}$, $S_{ML}$, $S_{DL}$, and $S_{IL}$, and the slope 
of the relation between the mean size at fixed surface brightness, $S_k$.\\}
\begin{tabular}{ccccccc}
\tableline
Band  & $S_{VL}$ & $S_{RL}$ & $S_{ML}$ & $S_{DL}$ & $S_{IL}$ & $S_k$ \\
\tableline\\
$g^*$ & 4.00$\pm 0.25$ & 1.50$\pm 0.06$ & 0.86$\pm 0.02$ & $-1.20\pm 0.08$ & $-2.98\pm 0.16$ & $-0.73\pm 0.02$ \\
$r^*$ & 3.91$\pm 0.20$ & 1.58$\pm 0.06$ & 0.87$\pm 0.02$ & $-1.33\pm 0.07$ & $-3.78\pm 0.17$ & $-0.75\pm 0.02$ \\
$i^*$ & 3.95$\pm 0.15$ & 1.59$\pm 0.06$ & 0.88$\pm 0.02$ & $-1.34\pm 0.08$ & $-3.91\pm 0.18$ & $-0.76\pm 0.02$ \\
$z^*$ & 3.92$\pm 0.15$ & 1.58$\pm 0.06$ & 0.88$\pm 0.02$ & $-1.33\pm 0.07$ & $-3.80\pm 0.17$ & $-0.76\pm 0.01$ \\
\tableline \\
\end{tabular}
\label{lx4cov}
\end{table}

Notice that $\sigma_{V|M}$ is not negligible compared to $\sigma_V$.  
This has an important consequence.  The distribution of velocity 
dispersions is sometimes (e.g., when spectroscopic data are not available) 
approximated by substituting the mean Faber--Jackson relation in the 
expression for the luminosity function.  This simple change of variables 
is only accurate if the scatter in the Faber--Jackson relation is 
negligible---for the galaxies in our dataset, this is not the case
(see Figure~\ref{ls}).  
Because the simple change of variables underestimates the width of the 
velocity dispersion distribution, it greatly underestimates the number 
density of galaxies which have large velocity dispersions.  

The mean size at fixed absolute luminosity $M$, and the dispersion around 
this mean, are obtained by replacing all $V$'s with $R$'s in 
equation~(\ref{meanvm}).  
The third column in Table~\ref{lx4cov} gives the maximum likelihood 
value of the slope $S_{RL}$, of the size-at-fixed-luminosity relation 
in the four bands.  This fit is shown in Figure~\ref{lre}.  
For comparison, Schade et al. (1997) find $L_B\propto R_o^{1.33}$ in 
the B band, whereas, at longer wavelengths, Pahre et al. (1998) find 
$L_K\propto R_o^{1.76\pm 0.10}$ with an rms of 0.88~mag.  
As was the case with the velocities, approximating the distribution of 
sizes by using the size-luminosity relation to change variables in the 
luminosity function is not particularly accurate, although, because 
$\rho_{RM}$ is larger than $\rho_{VM}$, the approximation is slightly 
better for the sizes than for the velocities.  

Similarly, one can show that the slopes of the mean $L$-mass and 
$L$-density relations shown in Figures~\ref{lmass} and~\ref{lden} 
are $S_{ML} = (2/S_{VL} + 1/S_{RL})^{-1}$ and 
$S_{DL} = 1/(2/S_{VL} - 2/S_{RL})^{-1}$.  These are the fourth and 
fifth columns of Table~\ref{lx4cov}.  The dispersions around these 
mean $L$-mass and $L$-density relations can be written in terms of 
the elements of ${\cal C}$ we define in Appendix~\ref{ML3d}, though 
we have not included the expressions here.  
Even though these relations are made from linear combinations of $R$ 
and $V$, they may be tighter than either the $L$--$\sigma$ or $L$--$R_o$ 
relations because the correlation coefficients $\rho_{RM}$, $\rho_{VM}$ 
and $\rho_{RV}$ are different from zero.  

In contrast to the Faber--Jackson and size--luminosity relations, 
the luminosity--mass and luminosity--density relations involve three 
variables.  Is there some combination of these variables which provides 
the least scatter?  Just as the eigenvectors of the $3\times 3$ 
covariance matrix ${\cal F}$ provided information about the shape and 
scatter around the Fundamental Plane, the eigenvectors and eigenvalues 
of the matrix ${\cal C}$ (from Appendix~\ref{ML3d}) give the directions 
of the principle axes of the ellipsoid in $(M,R,V)$ space which the 
early-type galaxies populate.  As was the case for ${\cal F}$, one of 
the eigenvalues of ${\cal C}$ is considerably smaller than the others, 
suggesting that the galaxies populate a two-dimensional plane in $(M,R,V)$ 
space.  The eigenvectors of ${\cal C}$ show that the plane is viewed 
edge-on in the projection 
\begin{equation}
-0.4(M-M_*) = \alpha\,(R-R_*) + \beta\,(V-V_*),
\end{equation}
where $M_*$, $R_*$ and $V_*$ were given in Table~\ref{MLcov} of 
Section~\ref{fp}, and the coefficients $\alpha$ and $\beta$, and the 
thickness of the plane in this projection, $\sigma_{MRV}$, are given 
in Table~\ref{FPlvr}.  This plane is only about 10\% thicker than the 
Fundamental Plane.  
Appendix~\ref{lx} shows that a scatter plot of luminosity versus 
mass is considerably tighter than plots of $M$ versus $\log_{10}R_o$ 
or $\log_{10}\sigma$.  The eigenvectors of ${\cal C}$ show that the 
$M$ versus $R_o + 2V$ projection is actually quite close to the edge-on 
projection.  

The surface brightnesses of the galaxies in our sample are defined 
by $(\mu_o-\mu_*) \equiv (M-M_*) + 5 (R-R_*)$, so the dispersion is 
$\sigma_{\mu}^2 = 
 \sigma_M^2 + 10\sigma_M\sigma_R\rho_{RM} + 25\sigma_R^2$.  
The mean surface brightness at fixed luminosity is obtained by 
replacing all $V$s with $\mu$s in the expressions above.  
This means that we need $\rho_{\mu M}$, which we can write in 
terms of $\sigma_M$, $\sigma_R$ and $\rho_{RM}$.  
The sixth column in Table~\ref{lx4cov} gives the slope of the surface 
brightness $I$ at fixed luminosity relation, $I\propto L^{1/S_{IL}}$, 
in the four bands.  It is shallower than the relation $I\propto L^{-0.45}$ 
for giant galaxies with $M_B<-20$ reported by Sandage \& Perelmuter (1990), 
although the scatter around the mean relation of $\sim 0.58$ mags is 
similar.  

\begin{table}[t]
\centering
\caption[]{Coefficients $\alpha$ and $\beta$ which define the projection 
of minimum scatter, $\sigma_{MRV}$, in the space defined by absolute 
magnitude, and the logarithms of the size and velocity dispersion.  
Notice that the scatter orthogonal to the plane is about 10\% larger 
than it is for the Fundamental Plane.\\}
\begin{tabular}{cccc}
\tableline
Band & $\alpha$ & $\beta$ & $\sigma_{MRV}$ \\
\tableline\\
$g^*$ & 0.76 & 1.94 & 0.063 \\
$r^*$ & 0.79 & 1.93 & 0.058 \\
$i^*$ & 0.82 & 1.89 & 0.054 \\
$z^*$ & 0.81 & 1.90 & 0.054 \\
\tableline \\
\end{tabular}
\label{FPlvr}
\end{table}

Kormendy (1977) noted that the surface brightnesses of early-type 
galaxies decrease with increasing effective radius.  
The mean size at fixed surface brightness in our sample is 
\begin{equation}
\Bigl\langle R-R_*\Big|\mu-\mu_*\Bigr\rangle = 
{(\mu-\mu_*)\over \sigma_{\mu}}\ \sigma_R\,\rho_{\mu R} \equiv 
    -0.4\,S_k\,(\mu-\mu_*).
\end{equation}
where $\rho_{\mu R}$ can be written in terms of $\sigma_M$, $\sigma_R$ 
and $\rho_{RM}$, and the final equality defines $S_k$.  
The seventh column in Table~\ref{lx4cov} gives the slope of this 
relation in the four bands.  For comparison, Kormendy (1977) found 
that $\log_{10}I_o \propto 1.29 \log_{10} R_o$ in the B-band, and 
Pahre et al. (1998) find $R_o\propto I_o^{-0.61}$ in the K-band.  

For the reasons described in Appendix~\ref{lx}, when presented with a 
magnitude limited catalog, correlations at fixed luminosity are useful 
because they are unbiased by the selection.  When luminosity is not one 
of the variables then forward and inverse correlations may be equally 
interesting, and equally biased.  For example, in the Kormendy (1977) 
relation, $\langle R-R_*\Big|\mu-\mu_*\rangle$ may be just as interesting 
as $\langle\mu-\mu_*|R-R_*\rangle$.  The slopes of the two relations 
are, of course, simply related to each other.  In fact, it may be 
preferable to study the relations which are defined by the principle 
axes of the ellipse in $(R,\mu)$ space which the galaxies populate.  
The directions of these axes are obtained by computing the eigenvalues 
and vectors of the covariance matrix associated with the sizes and 
surface brightnesses.  To illustrate, the eigenvalues of the $2\times 2$ 
covariance matrix associated with the Kormendy relation are 
\begin{displaymath}
\sigma^2_{\pm} = \Bigl( \sigma_R^2 + \sigma_\mu^2 \pm\sqrt{D_{\mu R}} \Bigr)
\Big/2,
\end{displaymath}
where we have set 
$D_{\mu R} = 
 (\sigma_R^2 - \sigma_\mu^2)^2 + (2\sigma_R\sigma_\mu\rho_{\mu R})^2$.
The mean relation is $(R-R_*) = S_K\,(\mu_o - \mu_*)$, where 
\begin{displaymath}
S_K = {\sigma_R^2 - \sigma_\mu^2 + \sqrt{D_{\mu R}}\over 
               2\sigma_R\sigma_\mu\,\rho_{\mu R}}.
\end{displaymath}
The $+/-$ eigenvalues give the dispersions along and perpendicular to 
the major axis of the ellipse.  

\begin{figure}
\centering
\epsfxsize=\hsize\epsffile{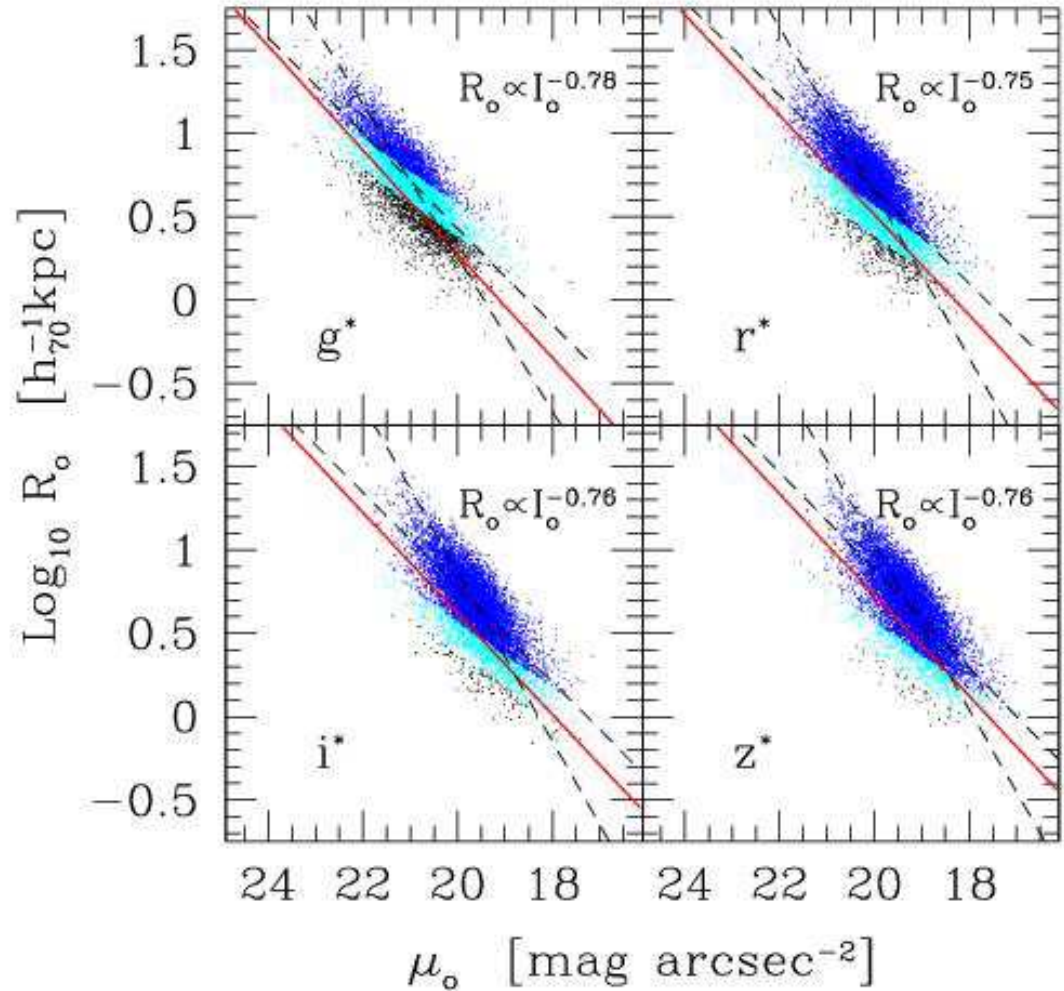}
\vspace{0.cm}
\caption{Relation between effective radius and surface brightness.  
Short dashed lines show forward and inverse fits to this relation.  
The zero-points of these fits are strongly affected by the magnitude 
limit of our sample.  To illustrate, solid line shows the 
maximum-likelihood estimate of the relation in the simulated 
complete catalog from which the magnitude limited catalog, shown 
by the dotted line, was drawn.  }
\label{krmndy}
\end{figure}

The Kormendy relation in our sample is shown in Figure~\ref{krmndy}.  
The dashed lines show forward and inverse fits to the data:
i.e., the mean size at fixed surface brightness, and the mean 
surface brightness at fixed size.  The parameters of the fits are 
affected by the magnitude limit of the catalog.  
To estimate the effect of the magnitude limit cut on this relation,
we compute the direct and inverse fits to the Kormendy relation in 
the simulated complete and magnitude-limited samples we describe in 
Appendix~\ref{simul}.  The dotted line in Figure~\ref{krmndy} shows 
the direct fit to the magnitude limited simulations (it can hardly 
be distinguished from the fit to the data).  

In comparison, the maximum-likelihood estimate of the true direct 
relation provides a very good description of the relation in the 
complete simulations in which there is no magnitude limit: it is 
shown as the solid line.  Notice that the dashed and dotted lines 
have approximately the same slope as the solid line:  the magnitude 
limit hardly affects the slope, although it changes the zero-point 
dramatically.  Lines of constant luminosity run downwards and to the 
right with slope -1/5, so that changes in luminosity act approximately 
perpendicular to the relation.  This is why the slope of the relation 
is hardly affected by the magnitude limit, but the zero-point is 
affected drastically; at fixed surface brightness, the typical $R_o$ 
is significantly larger in the magnitude limited sample than in the 
complete sample.  

This demonstrates that whereas linear regression fits to the data 
provide good estimates of the true slope of the Kormendy relation, 
they provide bad estimates of the true zero-point.  In comparison, 
the maximum-likelihood technique, which accounts for the selection 
on apparent magnitudes, is able to estimate the slope and the 
zero-point correctly.  

Although we have only chosen to present the argument for the Kormendy 
relation, the matrices ${\cal C}$ and ${\cal F}$ allow one to work out 
the maximum likelihood estimates of the principle axes and thicknesses 
of the ellipses associated with other combinations of observables.

\subsection{The $\kappa$-space projection}\label{kpspace}
Bender et al. (1992) suggested three simple combinations of the three 
observables:
\begin{eqnarray}
\kappa_1 &=& {\log_{10} (R_o\,\sigma^2)\over\sqrt{2}} ,\nonumber\\
\kappa_2 &=& {\log_{10} (I_o^2\sigma^2/R_o)\over\sqrt{6}},
\qquad {\rm and}\nonumber\\
\kappa_3 &=& {\log_{10} (I_o^{-1}\sigma^2/R_o)\over\sqrt{3}},
\end{eqnarray}
which, they argued, correspond approximately to the FP viewed 
face-on ($\kappa_2$--$\kappa_1$), and the two edge-on projections 
($\kappa_3$--$\kappa_1$ and $\kappa_3$--$\kappa_2$).  
They also argued that their parametrization was simply related to the 
underlying physical variables.  For example, $\kappa_1\propto$ mass 
and $\kappa_3\propto$ the mass-to-light ratio.  
The $\kappa_1$--$\kappa_2$ projection would view the FP face on 
if $R_o\propto \sigma^2/I_o$; recall that we found 
$R_o\propto (\sigma^2/I_o)^{0.75}$.  

\begin{figure}
\centering
\epsfxsize=\hsize\epsffile{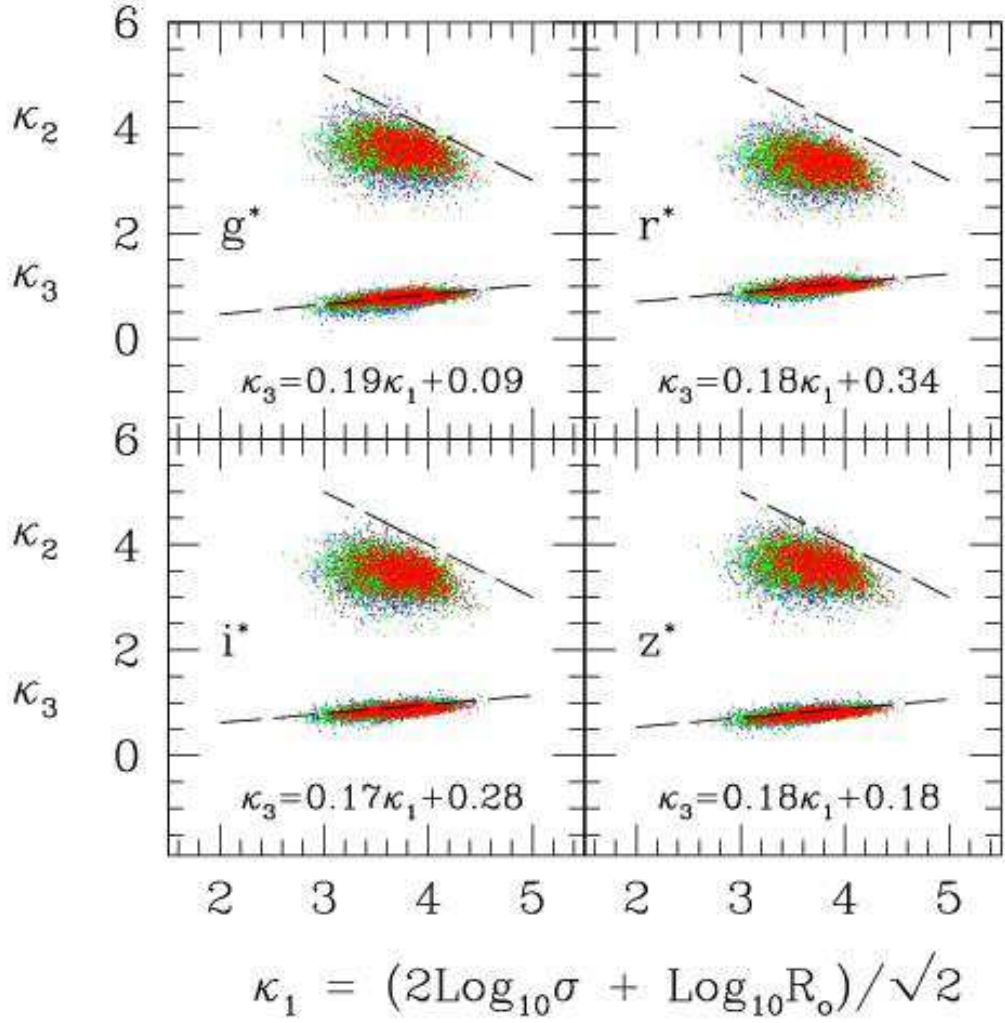}
\vspace{0.cm}
\caption{The early-type sample in the four SDSS bands viewed 
in the $\kappa$-space projection of Bender et al. (1992).  
The dashed line in the upper right corner of each panel shows 
$\kappa_1+\kappa_2=8$, what Burstein et al. (1997) termed the 
`zone of exclusion'.  }
\label{fpk4}
\end{figure}

Bender et al.'s choice of parameters was criticized by Pahre et al. (1998) 
on the grounds that if the effective radius $R_o$ is a function 
of wavelength, then the `mass' becomes a function of wavelength, which 
is unphysical.  On average, the effective radii of the galaxies in our 
sample do increase with decreasing wavelength (see Figure~\ref{rdiffs} in 
Section~\ref{cms}), so one might conclude 
that Pahre et al.'s objections are valid.  However, recall that we 
do not use the measured velocity dispersion directly; rather, $\sigma$ 
represents the value the dispersion would have had at some fixed 
fraction of $R_o$.  If $R_o$ depends on wavelength, and we wish to 
measure the velocity dispersion at a fixed fraction of $R_o$, then 
one might argue that we should also correct the measured velocity 
dispersion differently in the different bands.  The velocity dispersion 
decreases with increasing radius.  So, if $R_o$ is larger in the blue 
band than the red, then the associated velocity dispersion we should 
use in the blue band should be smaller than in the red.  One might 
imagine that the combination $R_o\sigma^2$ remains approximately 
constant after all.  For this reason, we have chosen to present the 
SDSS early-type sample in the $\kappa$-space projection introduced 
by Bender et al.  

Figure~\ref{fpk4} shows the results for the four SDSS bands.  
Because the mean surface brightness depends on waveband, we set 
$\log_{10}I = 0.4(27 - \mu_o + \langle\mu_o\rangle - \langle\mu_g\rangle)$
when making the plots, so as to facilitate comparison with 
Bender et al. (1992) and Burstein et al. (1997).  
The dashed line in the upper right corner of each panel shows 
$\kappa_1+\kappa_2=8$, what Burstein et al. termed the `zone of 
exclusion'.  Had we not accounted for the fact that the mean surface 
brightness is different in the different bands, then the galaxies 
would populate this zone.  

The magnitude limit is clearly visible in the lower right corner of 
the $\kappa_3$--$\kappa_1$ projection; we have not made any correction 
for it.  We have color-coded the points to represent the actual $(g^*-i^*)$ 
colors of the galaxies:  blue, red and green points are for colors 
bluer than 1.1, redder than 1.2, and in between.  
The redder galaxies appear to follow a tighter relation than the 
blue.  They also tend to lie slightly closer to the zone of 
avoidance.  We leave interpretation of these trends to future work.  

\section{Simulating a complete sample}\label{simul}
This Appendix describes how to simulate mock galaxy samples which have 
the same correlated observables as the data.  We will use these mock 
samples to estimate the effect of the magnitude limit cut on the  
relations we wanted to measure in the main text.

The observed parameters $L$, $R_o$ and $\sigma$ of each galaxy in our 
sample are drawn from a distribution, say, $\phi(M,R,V|z)$, where 
$M$ is the absolute magnitude, $R=\log_{10}R_o$ and $V=\log_{10}\sigma$.  
We show in Appendix~\ref{lx} that $\phi(M,R,V|z) = p(R,V|M,z)\,\phi(M|z)$, 
where $\phi(M|z)$ is the luminosity function at redshift $z$, and the 
distribution of $R$ and $V$ at fixed luminosity is, to a good approximation, 
a bivariate Gaussian.  The maximum likelihood estimates of the parameters 
of the luminosity function and of the bivariate distribution at fixed 
luminosity can be obtained from Table~\ref{MLcov}.

To make the simulations we must assume that, when extrapolated down to 
luminosities which we do not observe, these relations remain accurate.  
Assuming this is the case, we draw $M$ from the Gaussian distribution 
that we found was a good fit to $\phi(M|z)$ (Section~\ref{lf}).  
We then draw $R$ from the Gaussian distribution with mean 
$\langle R|M\rangle$ and dispersion $\sigma_{R|M}^2$.  
Finally, we draw $V$ from a Gaussian distribution with mean and variance 
which accounts for the correlations with both $M$ and $R$.  In practice 
we draw three zero mean unit variance Gaussian random numbers:  
$g_0$, $g_1$, and $g_2$, and then set 
\begin{eqnarray*}
M &=& M_* + \sigma_M\,g_0,\nonumber\\
R &=& R_* + {(M-M_*)\over\sigma_M}\ \sigma_R\,\rho_{RM} +
g_1\,\sigma_R\sqrt{1-\rho_{RM}^2} \qquad {\rm and} \nonumber\\
V &=& V_* + {(M-M_*)\over\sigma_M}\,\xi_{MV} 
          + {(R-R_*)\over\sigma_R}\,\xi_{RV} + g_2\,\sigma_{V|RM},
          \qquad {\rm where} \nonumber\\
\xi_{MV} &=& \sigma_V
{(\rho_{VM} - \rho_{RM}\rho_{RV})\over (1 - \rho_{RM}^2)},\nonumber\\
\xi_{RV} &=& \sigma_V {(\rho_{RV}-\rho_{RM}\rho_{VM})\over (1-\rho_{RM}^2)}, 
\qquad {\rm and} \nonumber\\
\sigma_{V|RM} &=& \sigma_V \sqrt{{1 - \rho_{RM}^2 - \rho_{RV}^2 - \rho_{VM}^2 +
2\rho_{RV}\rho_{VM}\rho_{RM}\over 1 - \rho_{RM}^2}}.
\end{eqnarray*}
Because each simulated galaxy is assigned a luminosity and size, 
its surface brightness is also fixed:  $\mu = M + 5R + $ constant.  

If we generate a catalog in $r^*$, then we can also generate colors 
using the parameters given in Table~\ref{MLcmag}.  Specifically, generate 
a Gaussian variate $g_3$, and then set 
 $C = C_* + \xi_{CM}(M-M_*)/\sigma_M + \xi_{CV}(V-V_*)/\sigma_V + 
      g_3\,\sigma_{C|MV}$, 
where $\xi_{CM}$, $\xi_{CV}$ and $\sigma_{C|MV}$ are defined analogously 
to $\xi_{MV}$, $\xi_{RV}$ and $\sigma_{V|RM}$ above.  Inserting the 
values from Table~\ref{MLcmag} shows that $\xi_{CM}\approx 0$, 
and $\sigma_{C|MV}\approx \sigma_{C|V}=\sigma_C\sqrt{1-\rho_{VM}^2}$:  
in practice, the mean color is determined by the velocity dispersion 
and not by the absolute magnitude.  

Passive evolution of the luminosities and colors is incorporated by adding 
the required $z$ dependent shift to $M$ and $C$ after the sizes and velocity 
dispersions have been generated.  

This complete catalog can be used to simulate a magnitude limited 
catalog if we assign each mock galaxy a redshift, assuming a world 
model and homogeneity.  
Let $m_{\rm min}$ and $m_{\rm max}$ denote the apparent magnitude 
limits of the observed sample.  Let $M_{\rm Bright}$ denote the absolute 
magnitude of the most luminous galaxy we expect to see in our catalog.  
Because the luminosity function cuts off exponentially at the bright end, 
we can estimate this by setting $M_{\rm Bright}\approx M_* + 5\sigma_M$.  
This means that the most distant object which can conceivably make 
it into the magnitude limited catalog lies at a luminosity distance of 
about  $d_{\rm Lmax} = 10^{(m_{\rm max}-M_{\rm Bright}-25)/5}$, 
from which the maximum redshift $z_{\rm max}$ can be determined.  
If the comoving number density of mock galaxies is to be independent 
of redshift, we must assign redshifts as follows.  
Draw a random variate $u_1$ distributed uniformly between zero and one, 
and set $d_{\rm Com} = u_1^{1/3}\,d_{\rm Lmax}/(1+z_{\rm max})$.  
The redshift $z$ can be obtained by inverting the 
$d_{\rm Com}(z;\Omega,\Lambda)$ relation.  The apparent magnitude of 
this mock galaxy is 
 $m = M + 5 {\rm Log_{10}}d_{\rm L} + 25 + K(z)$, 
where $K(z)$ is the K-correction.  
If $m_{\rm min}\le m\le m_{\rm max}$, then this galaxy would have been 
observed; add it to the subset of galaxies from the complete catalog 
which would have been observed in the magnitude limited catalog.  

If our simulated catalogs are accurate, then plots of magnitude, 
size, surface-brightness and velocity dispersion versus redshift 
made using our magnitude-limited subset should look very similar 
to the SDSS dataset shown in Figure~\ref{fig:XzR}.  In addition, 
${\rm d}N/{\rm d}z$ in the simulated magnitude limited subset should 
be similar to that in Figure~\ref{nz4}.  Furthermore, any correlations 
between observables in the magnitude limited subset should be just 
like those in the actual SDSS dataset.  If they are, then one has 
good reason to assume that similar correlations measured in the 
complete, rather than the magnitude-limited simulation, represent the 
true correlations between the parameters of SDSS galaxies, corrected 
for selection effects.  In this way, the simulations allow one to 
estimate the impact that the magnitude-limited selection has when 
estimating correlations between early-type galaxy observables.
  
We have verified that our simulated magnitude limited catalogs have 
similar $dN/dz$ distributions to those observed, and the simulated 
$\sigma$ and $R_o$ versus $z$ plots show the same selection cuts at low 
velocities and sizes as do the observed data.  The distribution of 
apparent magnitudes, angular sizes, and velocity dispersions in the 
magnitude limited simulations are very similar to those in the real 
data.  The simulated parameters also show the same correlations at 
fixed luminosity as the data.  Maximum likelihood analysis on the 
simulations produces an estimate of the covariance matrix 
which is similar to that of the data.  Therefore, we are confident that 
our simulated complete catalogs have correlations between luminosity, 
size, and velocity dispersion which are similar to the data.  
(Because they do not allow for differential evolution of the luminosities, 
they do not show the redshift dependent $I_o-L$ or $R_o-L$ slopes 
discussed in Section~\ref{difflz}.)

\section{Composite volume-limited catalogs}\label{compcat}
Our parent sample is magnitude limited; this might introduce a bias 
into the relations we present in this paper.  
For this reason, we thought it useful to present results for 
a few volume limited subsamples.  
Because of the cuts at both the faint and the bright ends of the 
catalog, each volume-limited subsample used in the main text spans 
only a small range in luminosity.  However, because the galaxies in 
our sample luminosity show little or no evolution relative to the 
values at the median redshift of the sample, we can extend this range 
in either of three ways.  

One method is to construct a composite volume-limited catalog 
by stacking together smaller volume-limited subsamples as follows.  
First, select a set of volume-limited subsamples which are adjacent 
in redshift and in luminosity, but which do not overlap at 
all.  This can be done by drawing rectangles in the top left panel 
of Figure~\ref{fig:XzR} which touch only at the bottom-left and 
top-right corners, and using only (a subset of) the galaxies which 
lie in these rectangles.  In general, the volumes of the individual 
subsamples will be different.  Let $V_i$ 
denote the volume of the $i$th subsample, and let $N_i$ denote the 
number of galaxies in it.  A conservative approach is to randomly 
choose the galaxies in $V_i$ with probability proportional to 
min$(V_i)/V_i$, where min$(V_i)$ denotes the volume of the smallest 
of the subsamples.  This has the disadvantage of removing much of the 
data, but, because our data set is so large, it may be that we can 
afford this luxury.  
A more cavalier approach is to choose all the galaxies in the largest 
$V_i$, all the galaxies in the other $V_j$, and to generate a set of 
additional galaxies by randomly choosing one of the $N_j$ galaxies 
in $V_j$, adding to each of its observed parameters a Gaussian random 
variate with dispersion given by the quoted observational error, and 
repeating this $N_j\times [{\rm max}(V_j)/V_j -1]$ times.  
A final possibility is to weight all the galaxies in $V_i$ (even 
those which were not in the volume limited subsample) by the inverse 
of the volume in which they could have been observed 
$(V_{\rm max}-V_{\rm min})$.  
We chose the first, most conservative option.  

By piecing together three volume limited subsamples, we were able to 
construct composite catalogs of about $10^3$ objects each.  Because 
the completeness limits are different in the different bands, the 
composite catalogs are different for each band.  In addition, because 
any one composite catalog is got by subsampling the set of eligible 
galaxies, by subsampling many times, we can generate many realizations 
of a composite catalog.  This allows us to estimate the effects of 
sample variance on the various correlations we measure.


\begin{thebibliography}{}

\bibitem[]{} Baum, W. A. 1959, PASP, 71, 106

\bibitem[]{} Bender, R. 1990, A\&A, 229, 441

\bibitem[]{} Bender, R., Burstein, D., \& Faber, S. M. 1992, ApJ, 399, 462

\bibitem[]{} Bender, R., Saglia, R. P., \& Gerhard, O. E. 1994, MNRAS, 269,785

\bibitem[]{} Bender, R., Ziegler, B., Bruzual, G. 1996, ApJL, 463, 51

\bibitem[]{} Benitez, N. 2000, ApJ, 536, 571

\bibitem[]{} Bernardi, M., Renzini, A., da Costa, L. N., Wegner, G.,
Alonso, M. V., Pellegrini, P. S., Rit\'e, C. \& Willmer, C. N. A. 
1998, ApJL, 508, 143

\bibitem[]{} Bernardi, M., Alonso, M. V., da Costa, L. N., Willmer, C. N. A.,
Wegner, G., Pellegrini, P. S., Rit\'e, C. \& M. A. G. Maia 2001a, AJ, {\it submitted} 

\bibitem[]{} Bernardi, M., Alonso, M. V., da Costa, L. N., Willmer, C. N. A.,
Wegner, G., Pellegrini, P. S., Rit\'e, C. \& M. A. G. Maia 2001b, AJ, {\it submitted}

\bibitem[]{} Bingelli, B., Sandage, A., Tarenghi, M. 1984, AJ, 89, 64

\bibitem[]{} Binney, J., \& Tremaine, S.  1987, 
Galactic Dynamics (Princeton:  Princeton Univ. Press)

\bibitem[]{} Blakeslee, J. P., Lucey, J. R., Barris, B. J., Hudson, M. J., \& 
Tonry, J. L.  2001, MNRAS, in press (astro-ph/0108194)

\bibitem[]{} Blanton, M. R., et al., 2001, AJ, 121, 2358

\bibitem[]{} Bower, R. G., Lucey, J. R., \& Ellis, R. S. 1992a, MNRAS, 254, 589

\bibitem[]{} Bower, R. G., Lucey, J. R., \& Ellis, R. S. 1992b, MNRAS, 254, 601

\bibitem[]{} Bressan, A., Chiosi, C., \& Fagotto, F.  1994, ApJS, 94, 63

\bibitem[]{} Brodie, J. P., \& Hanes, D. A. 1986, ApJ, 300, 258

\bibitem[]{} Bruzual, G., \& Charlot, S.  1993, ApJ, 405, 538

\bibitem[]{} Bruzual, G., \& Charlot, S.  2002, in preparation

\bibitem[]{} Burbidge, E. M., Burbidge, G. R., \& Fish, R. A. 1961, 
ApJ, 134,251

\bibitem[]{} Burstein, D., Bender, R., Faber, S. M., \& Nolthenius, R. 
1997, AJ, 114, 1365

\bibitem[]{} Busarello, G., Capaccioli, M., Capozziello, S., Longo, G., 
\& Puddu, E.  1997, A\&A, 320, 415 

\bibitem[]{} Colberg, J. M., White, S. D. M., Yoshida, N., MacFarland, T., 
Jenkins, A., Frenk, C. S., Pearce, F. R., Evrard, A. E., Couchman, H. M. P., 
Efstathiou, G., Peacock, J., Thomas, P.  (The Virgo Consortium)
2000, MNRAS, 319, 209

\bibitem[]{} Colless, M., Burstein, D., Davies, R. L., McMahan, R. K., 
Saglia, R. P., \& Wegner, G. 1999, MNRAS, 303, 813

\bibitem[]{} Colless, M., Saglia, R. P., Burstein, D., Davies, R. L., 
McMahan, R. K., \& Wegner, G. 2001, MNRAS, 321, 277

\bibitem[]{} Coleman, G. D., Wu, C.-C., \& Weedman, D. W. 1980, ApJS, 43, 393

\bibitem[]{} Connolly, A. J., \& Szalay, A. S. 1999, AJ, 117, 2052

\bibitem[]{} da Costa, L. N., Bernardi, M., Alonso, M. V., Wegner, G.,
Willmer, C. N. A., Pellegrini, P. S., Rit\'e, C., \& Maia, M. A. G. 
2000, AJ, 120, 95

\bibitem[]{} de Vaucouleurs, G. 1948, Annales d'Astrophysique, 11, 247

\bibitem[]{} de Vaucouleurs, G. 1961, ApJS, 5, 233

\bibitem[]{} Diaz, A. I., Terlevich E., \& Terlevich, R. 1989, MNRAS, 239, 325

\bibitem[]{} Djorgovski, S. \& Davis, M. 1987, \apj, 313, 59

\bibitem[]{} Dressler, A.  1984, ApJ, 286, 97

\bibitem[]{} Dressler, A., Lynden-Bell, D., Burstein, D., Davies, R. L., Faber,
S. M., Terlevich, R. J., \& Wegner, G. 1987, ApJ, 312, 42

\bibitem[]{} Efstathiou, G., Ellis, R. S., \& Peterson, B. S. 1988, MNRAS, 232,
431

\bibitem[]{} Eisenstein, D., et al. 2001, AJ, accepted

\bibitem[]{} Eisenstein, D., et al. 2002, {\it in preparation}

\bibitem[]{} Ellis, R. S., Smail, I., Dressler, A., Couch, W. J., Oemler, A.,
Butcher, H., \& Sharples, R. M.  1997, ApJ, 483, 582

\bibitem[]{} Faber, S. M., Dressler, A., Davies, R. L., Burstein, D.,
Lynden-Bell, D., Terlevich, R. J., \& Wegner, G. 
1987, in Nearly Normal Galaxies, From the Planck Time to the Present, 
ed. S. M. Faber (NY:Springer), 175

\bibitem[]{} Faber, S. M., \& Jackson, R. 1976, ApJ, 204, 668

\bibitem[]{} Faber, S. M., Wegner, G., Burstein, D., Davies, R. L.,
Dressler, A., Lynden-Bell, D., \& Terlevich, R. J. 1989, ApJS, 69, 763

\bibitem[]{} Fish, R. A. 1964, ApJ, 139, 284

\bibitem[]{} Forbes, D. A., \& Ponman, T. J. 1999, MNRAS, 309, 623

\bibitem[]{} Franx, M., Illingworth, G. D., \& Heckman, T. 1989, ApJ, 344, 613


\bibitem[]{} Fukugita, M., Shimasaku, K., \& Ichikawa, T.  1995, PASP, 107, 945

\bibitem[]{} Fukugita, M., Ichikawa, T., Gunn, J. E., Doi, M., 
Shimasaku, K., \& Schneider, D. P. 1996, AJ, 111, 1748

\bibitem{} Gibbons, R. A., Fruchter, A. S. \& Bothun, G. D. 2001, AJ, 121, 649

\bibitem[]{} Giovanelli, R., Haynes, M. P., Herter, T., Vogt, N.P., 
da Costa, L.N., Freudling, W., Salzer, J.J. and Wegner, G. 1997, AJ, 113, 53

\bibitem[]{} Greggio, L.  1997, MNRAS, 285, 151

\bibitem[]{} Gunn, J. E. et al. 1998, AJ, 116, 3040

\bibitem[]{} Guzm{\'a}n, R., Lucey, J. R., \& Bower, R. G. 1993, MNRAS, 265, 731

\bibitem[]{} Hernquist, L. 1990, ApJ, 356, 359

\bibitem[]{} Hogg, D. 1999, (astro-ph/9905116)

\bibitem{} Hudson, M.J., Lucey, J. R., Smith, R. J. \& Steel, J. 1997, MNRAS, 291, 488

\bibitem[]{} Im, M., Griffiths, R. E., Ratnatunga, K. U., Sarajedini, V. L. 
1996, ApJL, 461, 791

\bibitem[]{} J{\o}rgensen, I. 1997, MNRAS, 288, 161

\bibitem[]{} J{\o}rgensen, I., Franx, M. \& Kj{\ae}rgaard, P. 
1995, MNRAS, 276, 1341 

\bibitem[]{} J{\o}rgensen, I., Franx, M., \& Kj{\ae}rgaard, P. 
1996, MNRAS, 280, 167

\bibitem[]{} J{\o}rgensen, I., Franx, M., Hjorth, J., \& van Dokkum, P. G. 
1999, MNRAS, 308, 833

\bibitem[]{} Kauffmann, G. 1996, MNRAS, 281, 487

\bibitem[]{} Kauffmann, G. \& Charlot, S. 1998, MNRAS, 294, 705

\bibitem[]{} Kelson, D. D., Illingworth, G. D., van Dokkum, P. G., 
\& Franx, M. 2000, ApJ, 531, 184

\bibitem[]{} Kochanek, C. S., Pahre, M. A., \& Falco, E. E. 2000, (astro-ph/0011458)

\bibitem[]{} Kodama, T., Bower, R. G., \& Bell, E. F.  1999, MNRAS, 306, 561

\bibitem[]{} Kormendy, J. 1977, ApJ, 218, 333

\bibitem[]{} Kormendy, J. 1982, Saas-Fee Lectures 12, p. 115

\bibitem[]{} Kuntschner, H.  2000, MNRAS, 2000, 315, 184

\bibitem[]{} Kuntschner, H., Lucey, J. R., Smith, R. J., Hudson, M. J., \& 
Davies, R. L.  2001, MNRAS, 323, 615

\bibitem[]{} Larson, R. B.  1975, MNRAS, 173, 671

\bibitem[]{} Lucey, J. R., Bower, R. G., \& Ellis, R. S. 1991, MNRAS, 249, 755

\bibitem[]{} Lupton, R., Gunn, J. E., Ivezi\'c, Z., Knapp, G. R., 
\& Kent, S. 2001, in  
ASP Conf. Ser. 238, Astronomical Data Analysis Software and
Systems X, ed. F. R. Harnden, Jr., F.~A.~Primini, and H. E. Payne 
(San Francisco: Astr. Spc. Pac.), in press (astro-ph/0101420)

\bibitem[]{} Madgwick, D. S., et al. 2001, MNRAS, submitted (astro-ph/0107197)

\bibitem[]{} McKay, T. A., Sheldon, E. S., Racusin, J., et al. 2001, 
ApJ, submitted, astro-ph/0108013

\bibitem[]{} Navarro, J. F., Frenk, C. S., \& White, S. D. M. 1997, 
ApJ, 490, 493

\bibitem[]{} Okamura, S., Doi, M., Kashikawa, N., Kawasaki, W., 
Komiyama, Y., Sekiguchi, M., Shimasaku, K., Yagi, M., \& Yasuda, N. 
1998, IAUS, 179, 2870

\bibitem[]{} Pahre, M. A., Djorgovski, S. G., \& de Carvalho, R. R. 1996,
ApJL, 456, 79

\bibitem[]{} Pahre, M. A., 1998, PASP, 110, 1249

\bibitem[]{} Pahre, M. A., Djorgovski, S. G., \& de Carvalho, R. R. 1998a, 
AJ, 116, 1591

\bibitem[]{} Pahre, M. A., de Carvalho, R. R., \& Djorgovski, S. G. 1998b, 
AJ, 116, 1606

\bibitem[]{} Petrosian, V. 1976, ApJL, 209, 1

\bibitem[]{} Prugniel, P., \& Simien, F. 1996, A\&A, 309, 749

\bibitem[]{} Rix, H.-W., \& White, S. D. M. 1992, MNRAS, 254, 389

\bibitem[]{} Roberts M. S. \& Haynes M. P., 1994, ARA\&A, 32, 115

\bibitem[]{} Saglia, R. P., Bertschinger, E., Baggley, G., Burstein, D., 
Colless, M., Davies, R. L., McMahan, Jr., R. K. \& Wegner, G. 1997,
ApJS, 109, 79

\bibitem[]{} Saglia, R. P., Colless, M., Burstein, D., Davies, R. L., 
McMahan, R. K., \& Wegner, G. 2001, MNRAS, 324, 389

\bibitem[]{} Sandage, A., \& Visvanathan, N., 1978, ApJ, 223, 707

\bibitem[]{} Sandage, A., \& Perelmuter, 1990, ApJ, 361, 1

\bibitem[]{} Sargent, W. L. W., Schechter, P. L., Boksenberg, A., Shortridge,
K. 1977, ApJ, 212, 326

\bibitem[]{} Schade, D., Barientos, L. F., \& L{\'o}pez-Cruz, O., 
1997, ApJL, L17

\bibitem[]{} Schade, D., Lilly, S. J., Crampton, D., Ellis, R. S., 
Le F{`}evre, O., Hammer, F., Brinchmann, J., Abraham, R., Colless, M., 
Glazebrook, K., Tresse, L., \& Broadhurst, T. 1999, ApJ, 525, 31

\bibitem[]{} Schlegel, D. J., Finkbeiner, D. P., \& Davis, M. 
1998, ApJ, 500, 525

\bibitem[]{} Scodeggio, M. 1997, Ph.D. Thesis, Cornell University

\bibitem[]{} Scodeggio, M. 2001, AJ, 121, 2413

\bibitem[]{} Scodeggio, M., Giovanelli, R. \&  Haynes, M. P. 1997,
AJ, 113, 101

\bibitem[]{} Scodeggio, M., Gavazzi, G., Belsole, E., Pierini, D.,
 \&  Boselli, A.  1998, MNRAS, 301, 1001

\bibitem[]{} Schechter, P. L. 1980, AJ, 85, 801

\bibitem[]{} Simard, L., Koo, D. C., Faber, S. M., Sarajedini, V. L., 
Vogt, N. P., Phillips, A. C., Gebhardt, K., Illingworth, G. D., \& Wu, K. L.
1999, ApJ, 519, 563

\bibitem[]{} Smith, J. A., et al. 2002, {\it in preparation}

\bibitem[]{} Sodr{\'e}, L. Jr. \& Lahav, O. 1993, MNRAS, 260, 285

\bibitem[]{} Stoughton, C. et al. 2002, AJ, {\it submitted}

\bibitem[]{} Strauss, M. A., et al. 2002, {\it in preparation}

\bibitem[]{} Tantalo, R., Chiosi, C., \& Bressan, A.  1998, A\& A, 333, 490

\bibitem[]{} Thomas, D., Greggio, L., \& Bender, R., 1999, MNRAS, 302, 537
\bibitem[]{} Tinsley, B. M., \& Gunn, J. E. 1976, ApJ, 206, 525

\bibitem[]{} Tonry, J., \& Davis, M.  1979, AJ, 84, 1511

\bibitem[]{} Trager, S. C., Worthey, G., Faber, S. M., Burstein, D., 
\& Gonz\'alez, J. J. 1998, ApJS, 116, 1

\bibitem[]{} Trager, S. C., Faber, S. M., Worthey, G., \& Gonz\'alez,
J. J. 2000a, AJ, 119, 1645

\bibitem[]{} Trager, S. C., Faber, S. M., Worthey, G., \& Gonz\'alez,
J. J. 2000b, AJ, 120, 165

\bibitem[]{} Treu, T., Stiavelli, M., Casertano, S., M{\o}ller, P., \&
Bertin, G. 1999, MNRAS, 308, 1037

\bibitem[]{} Treu, T., Stiavelli, M., M{\o}ller, P., Casertano, S., \&
Bertin, G. 2001a, MNRAS, 326, 221

\bibitem[]{} Treu, T., Stiavelli, M., Bertin, G., Casertano, S., \&
 M{\o}ller, P. 2001b, MNRAS, 326, 237

\bibitem[]{} Tully, R. B. \& Fisher, J. R., 1977, A\&A, 54, 661

\bibitem[]{} Uomoto, A. et al. 2002, {\it in preparation}

\bibitem[]{} van Albada, T. S. 1982, MNRAS, 201, 939

\bibitem[]{} van der Marel, R. P., \& Franx, M.  1993, ApJ, 407, 525

\bibitem[]{} van Dokkum, P. G. \& Franx M., 1996, \mnras, 281, 985

\bibitem[]{} van Dokkum, P. G. \& Franx M., 2001, ApJ, 553, 90

\bibitem[]{} van Dokkum, P. G., Franx M., Kelson, D. D., Illingworth,
G. D. 1998, ApJL, 504, 17

\bibitem[]{} van Dokkum, P. G., Franx, M., Kelson, D. D., \&
Illingworth, G. D. 2001, ApJL, {\it submitted}, (astro-ph/0005558)

\bibitem[]{} Vazdekis, A., Casuso, E., Peletier, R. F., \& Beckman, J. E. 
1996, ApJS, 106, 307


\bibitem[]{} Wegner, G., Colless, M., Saglia, R. P., McMahan, R. K., 
Davies, R. L., Burstein, D., Baggley, G., 1999, MNRAS, 305, 259

\bibitem[]{} Weinberg, S., 1972, Gravitation, and Cosmology: Principles and
Applications of the General Theory of Relativity, John Wiley \& Sons, New York

\bibitem[]{} Weiss, A., Peletier, R. F., \& Matteucci, F. 1995, A\&A, 296, 73

\bibitem[]{} White, S. D. M., \& Rees, M. J.  1978, MNRAS, 183, 341

\bibitem[]{} Worthey, G., Faber, S. M., \& Gonzales, J. 1992, ApJ, 398, 69

\bibitem[]{} Worthey, G. 1994, ApJS, 95, 107

\bibitem[]{} York, D. G.\ et al.\ 2000, \aj, 120, 1579

\bibitem[]{} Ziegler, B. L., Saglia, R. P., Bender, R., Belloni, P., Greggio,
L., \& Seitz, S.  1999, A\& A, 346, 13

\bibitem[]{} Ziegler, B. L., Bower, R. G., Smail, I., Davies, R. L., 
\& Lee, D.  2001, MNRAS, in press

\end{thebibliography}
\end{document}